\documentclass[11pt,a4paper]{article}
\usepackage{authblk}
\usepackage{lmodern}
\usepackage{jheppub}
\usepackage{amsmath}
\usepackage{mathtools}
\usepackage{tikz}
\usepackage{empheq}
\usepackage{enumitem}
\usepackage{graphics}
\usepackage{wrapfig}
\usepackage{caption}
\usepackage{natbib}
\usepackage{xcolor}
\usepackage{framed}
\definecolor{shadecolor}{gray}{0.925}

\usepackage[cbgreek]{textgreek}

\def\sideremark#1{\ifvmode\leavevmode\fi\vadjust{\vbox to0pt{\vss
 \hbox to 0pt{\hskip\hsize\hskip1em
 \vbox{\hsize3cm\tiny\raggedright\pretolerance10000
 \noindent #1\hfill}\hss}\vbox to8pt{\vfil}\vss}}}%

\newcommand{\bi}{\begin{itemize}}
\newcommand{\ei}{\end{itemize}}
\newcommand{\bea}{\begin{align}}
\newcommand{\eea}{\end{align}}
\newcommand{\be}{\begin{equation}}
\newcommand{\ee}{\end{equation}}

\newcommand{\pl}{{\partial}}

\newcommand{\tcr}{\textcolor{red}}

\newcommand{\tcb}{\textcolor{blue}}

\makeatletter
\renewcommand*\env@matrix[1][\arraystretch]{%
  \edef\arraystretch{#1}%
  \hskip -\arraycolsep
  \let\@ifnextchar\new@ifnextchar
  \array{*\c@MaxMatrixCols c}}
\makeatother

\author[\ensuremath{\eta}]{Charlotte SLEIGHT\footnote{Also at the Universit\'e Libre de Bruxelles and International Solvay Institutes, Belgium.}}
\author[\ensuremath{\mathsf{s}},\ensuremath{\mathsf{t}},\ensuremath{\mathsf{u}}]{\quad Massimo TARONNA}

\affiliation[\ensuremath{\eta}]{School of Natural Sciences, Institute for Advanced Study,\\
1 Einstein Drive, Princeton, NJ 08540}

\affiliation[\ensuremath{\mathsf{s}}]{Department of Physics, Princeton University,\\
Jadwin Hall, Princeton, NJ 08544}

\affiliation[\ensuremath{\mathsf{t}}]{Dipartimento di Fisica ``Ettore Pancini'', Universit\`a degli Studi di Napoli Federico II, \\Monte S. Angelo, Via Cintia, 80126 Napoli, Italy}

\affiliation[\ensuremath{\mathsf{u}}]{INFN, Sezione di Napoli, Monte S. Angelo, Via Cintia, 80126 Napoli, Italy}

\emailAdd{csleight@ias.edu, mtaronna@princeton.edu}

\title{\centering \huge Bootstrapping Inflationary Correlators in Mellin Space}

\abstract{We develop a Mellin space approach to boundary correlation functions in anti-de Sitter (AdS) and de Sitter (dS) spaces. Using the Mellin-Barnes representation of correlators in Fourier space, we show that the analytic continuation between AdS$_{d+1}$ and dS$_{d+1}$ is encoded in a collection of simple relative phases. This allows us to determine the late-time tree-level three-point correlators of spinning fields in dS$_{d+1}$ from known results for Witten diagrams in AdS$_{d+1}$ by multiplication with a simple trigonometric factor. At four point level, we show that Conformal symmetry fixes exchange four-point functions both in AdS$_{d+1}$ and dS$_{d+1}$ in terms of the dual Conformal Partial Wave (which in Fourier space is a product of boundary three-point correlators) up to a factor which is determined by the boundary conditions. In this work we focus on late-time four-point correlators with external scalars and an exchanged field of integer spin-$\ell$. The Mellin-Barnes representation makes manifest the analytic structure of boundary correlation functions, providing an analytic expression for the exchange four-point function which is valid for general $d$ and generic scaling dimensions, in particular massive, light and (partially-)massless fields. It moreover naturally identifies boundary correlation functions for generic fields with multi-variable Meijer-G functions. When $d=3$ we reproduce existing explicit results available in the literature for external conformally coupled and massless scalars. From these results, assuming the weak breaking of the de Sitter isometries, we extract the corresponding correction to the inflationary three-point function of general external scalars induced by a general spin-$\ell$ field at leading order in slow roll. These results provide a step towards a more systematic understanding of de Sitter observables at tree level and beyond using Mellin space methods.}

\begin{document}

\begin{flushright}    
\texttt{PUPT-2590}
\end{flushright}

\maketitle

\section{Introduction}\label{sec::Intro}

Holography has by now inspired a great number of tools to understand boundary observables for theories on asymptotic anti-de Sitter (AdS) space in terms of simple consistency requirements on correlators in Conformal Field Theory (CFT) \cite{Maldacena:1997re,Gubser:1998bc,Heemskerk:2009pn,ElShowk:2011ag}. CFT correlators have, step-by-step, acquired the flavour of actual ``S-matrix''-like observables for scattering processes in AdS. The Mellin space representation of conformal correlators \cite{Mack:2009mi,Mack:2009gy,Penedones:2010ue,Paulos:2011ie,Fitzpatrick:2011ia,Fitzpatrick:2011dm,Fitzpatrick:2011hu,Costa:2012cb,Goncalves:2014rfa} and Harmonic Analysis for the Euclidean Conformal Group \cite{Mack:1974sa,Dobrev:1975ru,Dobrev:1977qv,Costa:2012cb,Caron-Huot:2017vep} have proven instrumental in making this connection manifest, which encode bulk physics in a way which shares key similarities with the flat-space scattering amplitudes.

In contrast, our understanding of boundary correlators in de Sitter space is in a primordial stage. As opposed to scattering amplitudes, these are spatial correlations at late times which encode the imprints of past scattering processes. Because of this, correlators in the dual Euclidean CFT are not bound to satisfy the Osterwalder-Schrader axioms such as reflection positivity. Currently we do not have a complete grasp on the rules that the corresponding late-time correlators have to obey, in particular how they encode consistent bulk time evolution. 

In recent years there has been a drive to refine our understanding of late-time correlators in de Sitter space, which has been largely motivated by inflationary cosmology \cite{Guth:1980zm,Linde:1981mu,Albrecht:1982wi,Starobinsky:1982ee}. Cosmological observations can be traced back to spatial correlations of primordial fluctuations at the end of inflation, which lie on the boundary of an (approximate) de Sitter space-time. Non-Gaussianities in primordial correlation functions encode the physics of inflation, including interactions and field content, and the ultimate goal of the ``Cosmological Collider Physics" programme \cite{Chen:2009zp,Maldacena:2011nz,Baumann:2011nk,Assassi:2012zq,Chen:2012ge,Noumi:2012vr,Assassi:2013gxa,Arkani-Hamed:2015bza,Lee:2016vti,An:2017hlx,Kehagias:2017cym,Kumar:2017ecc,Baumann:2017jvh,Franciolini:2017ktv,Arkani-Hamed:2018kmz,Goon:2018fyu,Anninos:2019nib,noumi} is to classify the possible shapes of non-Gaussianities for comparison with observations. To this end it is important to develop new tools to systematically carve out the shapes of non-Gaussianities generated by a given spectrum and couplings.

In this work we propose a new framework for the computation of late-time correlators in de Sitter space, which is tailored to bridge the gap with our (comparably better) understanding of boundary correlators in anti-de Sitter space. This is based on the Mellin-Barnes representation of boundary correlators in Fourier space, in which the analytic continuation from anti-de Sitter to de Sitter turns out to be encoded in a collection of simple relative phases.\footnote{The relation between correlators in AdS and dS through analytic continuation has been considered in previous works \cite{Maldacena:2002vr,Ghosh:2014kba,Anninos:2014lwa}. In this work we propose a slightly different approach to the analytic continuation, which we discuss in detail in \S\ref{dS2pt}.} This observation allows the systematic derivation of late-time correlators in dS from the Fourier transform of boundary correlators in AdS, which are simpler to obtain and, in many cases, already known. This in particular includes correlators involving scalars and totally symmetric fields of arbitrary integer spin, for which there are few results available to date in de Sitter, where in AdS the complete kinematic map between bulk cubic couplings for any triplet of spinning fields and boundary three-point conformal structures has been worked out explicitly \cite{Sleight:2017fpc}. 

\begin{figure}
    \centering
    \captionsetup{width=0.95\textwidth}
    \includegraphics[width=0.9\textwidth]{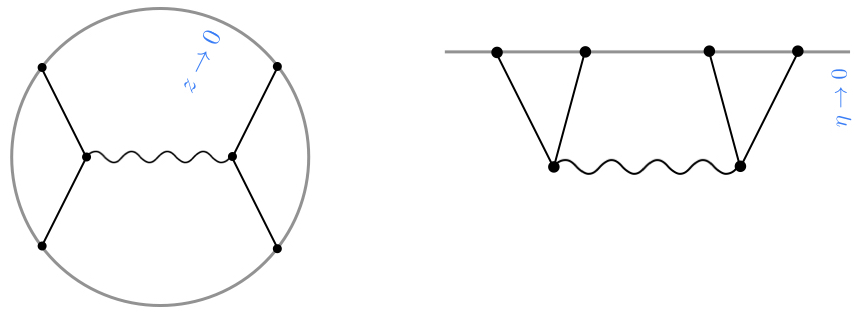}
    \caption{Comparison between a scattering process in AdS and dS. In AdS the boundary condition is imposed at the conformal boundary at $z\to 0$, while in dS it is imposed at the infinite past $\eta \to -\infty$.}
    \label{fig:dS_vs_AdS}
\end{figure}

Within this framework, bulk tree-level exchange four-point functions both in AdS and dS are naturally expressed directly in terms of the boundary Conformal Partial Wave (CPW) that is dual to the exchanged single-particle in the bulk.\footnote{See also \cite{Charlotte} where the same form for de Sitter exchange four-point functions was obtained taking a direct bulk approach which also employs the Mellin-Barnes representation.} The CPW is the basic kinematic object which is fixed by conformal symmetry,\footnote{In particular, Conformal Partial Waves are single-valued Eigenfunctions of the Casimir invariants of the Conformal Group $SO\left(1,d+1\right)$.} and in Fourier space it is factorised into the two three-point boundary correlators generated by the cubic couplings that participate in the exchange. In the Mellin-Barnes representation of the exchange four-point function, the CPW is dressed by a factor whose poles that generate the Effective Field Theory (EFT) expansion and a second factor consisting only of zeros that implement the boundary conditions. The EFT expansion is also constrained by conformal symmetry up to a finite number of bulk contact terms, which is related to the freedom of adding higher-derivative improvement terms to the cubic vertices (i.e. higher-derivative terms which vanish on-shell).\footnote{This is the usual contact term ambiguity of exchange Witten diagrams.} We show that there is a natural basis of contact interactions in which the factor in the Mellin-Barnes representation that encodes the EFT expansion is given by a simple $\csc$-function. In particular, for the exchange of a spin-$\ell$ field of scaling dimension $\Delta=\frac{d}{2}+i\nu$ between fields $\phi^{\left(\nu_j\right)}$ of scaling dimension $\Delta_j=\frac{d}{2}+i\nu_j$, we have the Mellin-Barnes representation:
\begin{shaded}
\noindent \emph{Mellin-Barnes representation of exchange four-point functions in (A)dS$_{d+1}$ }
\begin{multline}
 \hspace*{-0.7cm}   \langle \phi^{\left(\nu_1\right)}_{\vec{k}_1}\phi^{\left(\nu_2\right)}_{\vec{k}_2}\phi^{\left(\nu_3\right)}_{\vec{k}_3}\phi^{\left(\nu_4\right)}_{\vec{k}_4} \rangle^\prime = {\cal N}_4 \int^{i\infty}_{-i\infty}\prod^4_{j=1}\frac{ds_j}{2\pi i} \int_{-i\infty}^{+i\infty}\frac{du\,d\bar{u}}{(2\pi i)^2}\,4 \underbrace{\csc(\pi(u+\bar{u}))}_{\text{EFT}}\underbrace{\delta(u,\bar{u})}_{\text{b.c.}} \underbrace{{\cal F}^\prime_{\nu,\ell}(s_i;u,{\bar u}|\vec{k}_i;\vec{k})}_{\text{CPW}}\,, \nonumber
\end{multline}
\end{shaded}
\noindent where the Mellin variables $s_j$ and the momenta $\vec{k_j}$, $j=1,...,4$, are associated to the four external legs, while the Mellin variables $u$, ${\bar u}$ and the momentum $\vec{k}$ are associated to the exchanged field. The Mellin variables in the argument of the CPW indicate that we are using its Mellin-Barnes representation. The prime indicates that the momentum-conserving delta function has been stripped off and the constant ${\cal N}_4$ is fixed in section \ref{dS2pt}. 

The above form of the exchange four-point function, which holds both in AdS$_{d+1}$ and dS$_{d+1}$, makes manifest how the exchange four-point function can be completely specified by the CPW, which is recovered upon evaluating the discontinuity of the correlator in ${\sf s}=k^2$,
\begin{equation}
    2i\,\text{Disc}[f(\mathsf{s})]=f(e^{+i\pi}\mathsf{s})-f(e^{-i\pi}\mathsf{s})\,,
\end{equation}
where
\begin{multline}\nonumber
     \text{Disc}_{\mathsf{s}}\left[\langle \phi^{\left(\nu_1\right)}_{\vec{k}_1}\phi^{\left(\nu_2\right)}_{\vec{k}_2}\phi^{\left(\nu_3\right)}_{\vec{k}_3}\phi^{\left(\nu_4\right)}_{\vec{k}_4} \rangle^\prime\right]= {\cal N}_4 \int^{i\infty}_{-i\infty}\prod^4_{j=1}\frac{ds_j}{2\pi i} \int_{-i\infty}^{+i\infty}\frac{du\,d\bar{u}}{(2\pi i)^2}\,4 \delta(u,\bar{u}) {\cal F}^\prime_{\nu,\ell}(s_i;u,{\bar u}|\vec{k}_i;\vec{k}).
\end{multline}
The zeros of the Mellin integrand, which are encoded in the function $\delta\left(u,{\bar u}\right)$, are not fixed by conformal symmetry and are determined by the boundary conditions. Accordingly, this factor takes different forms in AdS and dS, where in AdS it implements the Dirichlet/Neumann boundary conditions at the conformal boundary, while in dS it implements the Bunch-Davies vacuum condition at early times. 

Aside from providing a convenient framework which places anti-de Sitter and de Sitter scattering processes on an equal footing, the Mellin-Barnes representation is advantageous on various other levels. It makes manifest the analytic structure, which allows to efficiently study analytic continuations of the correlator with respect to all of its parameters. This not only includes the momenta, but also both the boundary dimension $d$, the scaling dimensions $\Delta$ of the fields and their spin $\ell$. Interestingly, this also allows to establish simple recursion relations among boundary correlators with fields of different scaling dimensions and spin, as we shall demonstrate. On top of this, well-established Mellin-Barnes methods allow to straightforwardly derive asymptotic expansions of boundary correlators in the momenta for regimes of interest.

In this work we focus on late-time exchange four-point functions in dS$_{d+1}$ with general external scalars and a general exchanged spin-$\ell$ field, though the above expression for the exchange holds more generally. As an intermediate step, we also derive the Mellin-Barnes representation for the late-time tree-level three-point function of two general scalars and a general spin-$\ell$ field. Our results are valid both for massive, light and massless external scalars, and for massive and (partially-)massless exchanged field of arbitrary integer spin-$\ell$. For (partially-)massless exchanged fields, extra care needs to be taken due to divergences that emerge in the general expression for the exchange four-point function at those values of the scaling dimension, which can be treated systematically in the Mellin framework -- as we shall demonstrate. 

From the result for general external scalars, we can obtain closed form analytic expressions for the leading slow-roll correction to inflationary three-point functions induced by the exchange of a spin-$\ell$ field. This is extracted by taking one of the external scalars in the de Sitter exchange four-point function to soft momentum and a small mass \cite{Kundu:2014gxa,Kundu:2015xta}, which is straightforward to implement at the level if the Mellin-Barnes representation by taking the appropriate residues. In the squeezed limit, where $k_{3}/k_{1}\ll 1$ (if we took the momentum $k_4$ to be soft), we find:
\begin{shaded}
\noindent \emph{Squeezed limit of the correction to the inflationary 3pt function from a spin-$\ell$ exchange}
\begin{multline}
    \langle \phi^{\left(\nu_1\right)}_{\vec{k}_1}\phi^{\left(\nu_2\right)}_{\vec{k}_2}\phi^{\left(\nu_3\right)}_{\vec{k}_3} \rangle^\prime_{(\text{infl.})}\sim -\frac{\epsilon\, \mathcal{N}_3}{32 \sqrt{\pi }}\left(\frac{k_1}{2}\right)^{-\frac{d}{2}+i (\nu_1+\nu_2)}\left(\frac{k_3}{2}\right)^{+i \nu_3}  \\\times\,\Bigg[\left(\frac{k_3}{k_1}\right)^{i \nu }\frac{\Gamma (-i \nu )\left(1-\tfrac{\ell}{2}+\tfrac{i \nu }{2}\pm\tfrac{i \nu_3}{2}\right)_{\ell-1}}{ \left(\tfrac{d}{2}+i \nu -1\right)_\ell \Gamma \left(\frac{d}{2}+\ell+i \nu \right)}\sin \left(\tfrac{\pi}{4}  (d+2 \ell+2 i (\nu +\nu_1+\nu_2))\right) \\
    \times\, \text{csc}\left(\tfrac{\pi}{2}  (\ell+i(\nu_3- \nu))\right)\prod_{\pm\,{\hat \pm}} \Gamma \left(\tfrac{d+2 \ell+2 i (\nu \pm\nu_1{\hat \pm}\nu_2)}{4}\right)\\
    + \nu \to -\nu \Bigg]\frac{(-2)^\ell\ell! }{(\tfrac{d}{2}-1)_\ell}\,C_{\ell}^{\left(\frac{d-2}{2}\right)}(\cos\theta),
\end{multline}
\end{shaded}
\noindent with slow-roll parameter $\epsilon$. This exhibits the characteristic power-law behaviour in $k_{3}/k_{1}$ for a particle exchange \cite{Chen:2009we,Byrnes:2010xd,Baumann:2011nk,Assassi:2012zq,Noumi:2012vr,Arkani-Hamed:2015bza}, which is oscillatory for massive exchanged particles on the Principal Series, $\nu \in \mathbb{R}$. The phase of the oscillatory behaviour arises from the quantum interference between two processes \cite{Noumi:2012vr,Arkani-Hamed:2015bza} (the expansion of the universe and particle creation), which turns out to be determined by the factor $\delta\left(u,{\bar u}\right)$ in the de Sitter exchange four-point function. The fact that we are exchanging a spin-$\ell$ particle is encoded in the angular dependence of the Gegenbauer polynomial, which reduces to the usual Legendre polynomial for the $d=3$ case considered in \cite{Arkani-Hamed:2015bza}. The ease at which this result could be obtained for general external scalars, which is new even for $d=3$ (as far as we are aware), is testament to the strength of the Mellin formalism.

\paragraph{Outline.}
This paper is organised as follows. We begin in section \ref{dS2pt} with a discussion on propagators of scalar and spinning fields in (A)dS$_{d+1}$. We show that Wightman functions in dS and Harmonic functions in EAdS are related by analytic continuation. In Fourier space, this analytic continuation is encoded in a simple phase at the level of the Mellin-Barnes representation. This observation allows us to establish a dictionary to obtain late-time correlators in dS$_{d+1}$ from Witten diagrams in AdS$_{d+1}$. In section \ref{sec::3pt} we apply this dictionary to obtain late-time tree-level three-point functions for two general scalars and a general field of integer spin $\ell$ in dS$_{d+1}$ from the Fourier transform of the known corresponding result in AdS$_{d+1}$. In section \ref{sec:FourPoint} we consider late-time exchange four-point functions in dS$_{d+1}$. We show how a Mellin-Barnes representation for the exchange can be obtained by dressing the dual Conformal Partial Wave with appropriate factors which encode the EFT expansion and the boundary conditions. We show how to extract both the OPE and EFT expansion from the Mellin-Barnes representation, and moreover how it reproduces existing results in the literature when $d=3$ and the external scalars are conformally coupled or massless. In section \ref{sec:InflationaryCorr} we show how our results for exchange four-point functions in de Sitter can be used to extract the correction at leading order in slow roll to inflationary three-point functions induced by the exchange of a spin-$\ell$ field.

Various technical details and brief reviews of relevant material are relegated to the appendices.

\subsection{Notations and Conventions}

We primarily work in $\left(d+1\right)$-dimensional de Sitter space with ``mostly plus" metric signature $\left(-\,+\,+\,...\,+\right)$. Greek letters denote space-time indices, $\mu=0,1,...,d$, lower-case Latin letters denote spatial indices, $i=1,...,d$, while Ambient space indices are denoted by $M,N=0,1,...,d+1$. Bulk scalar fields of scaling dimension $\Delta=\frac{d}{2}+i\nu$ are denoted by $\phi^{\left(\nu\right)}$ and spin-$\ell$ fields by $\varphi^{\left(\nu\right)}_\ell$. Momentum vectors are represented either by $\vec{k}$ or $k^i$, with magnitude $k=|\vec{k}|$. The spatial auxiliary vectors $\vec{\xi}$ (or equivalently $\xi^i$) encode spatial tensor indices. The momentum of the $n$-th external leg in a correlation function is denoted by $\vec{k}_n$, while we use $\vec{k}$ for the exchanged momentum. It is sometimes convenient to express three-point correlators in terms of the combinations $p=\frac{k_1+k_2}{k_3}$ and $q=\frac{k_1-k_2}{k_3}$, while exchange four-point functions in terms of $p_{mn}=\frac{k_m+k_n}{k}$ and $q_{mn}=\frac{k_m-k_n}{k}$. The symbols $s$, $u$ and ${\bar u}$ are reserved for Mellin-variables. 

\section{Propagators}\label{dS2pt}

This section is dedicated to the propagators of scalar fields $\phi$ and fields $\varphi_\ell$ of integer spin-$\ell$ in dS$_{d+1}$. After reviewing some basics of classical geometry in (anti-)de Sitter space in \S \ref{kinematics}, we begin with the scalar propagators in \S \ref{subsec::scalar2pt}, demonstrating how the Wightman two-point function in the Bunch-Davies vacuum can be obtained via analytic continuation of Harmonic functions in Euclidean AdS$_{d+1}$. This gives a so-called ``split representation'' of the dS Wightman function, in which it is expressed as a product of bulk-to-boundary propagators. In \S \ref{MSMBrepprops} we present a Mellin-Barnes representation in Fourier space, where the analytic continuation from EAdS$_{d+1}$ is encoded in a simple phase. We give the extension to spinning fields $\varphi_\ell$ in \S \ref{subsec::spin2pt}. In \S \ref{subsec::SKdic} we derive the corresponding Keldysh propagators, giving the dictionary of phases required to go from the Mellin-Barnes representation of Harmonic functions in EAdS$_{d+1}$ to a given branch of the in-in contour. In \S \ref{2pt} we use this framework to derive the late-time limit of scalar and spinning two-point functions in dS$_{d+1}$. For clear pedagogical reviews for some of the topics touched upon in this section, e.g. \cite{Bros:1995js,Spradlin:2001pw,Joung:2006gj,Baumann:2009ds,Anninos:2012qw,Akhmedov:2013vka,Chen:2017ryl}.

\subsection{Classical geometry of (anti)-de Sitter space}
\label{kinematics}

It is often convenient to realise $\left(d+1\right)$-dimensional de Sitter space dS$_{d+1}$ as the following embedding:
\begin{equation}
    -(X^0)^2+(X^1)^2+\cdots +(X^{d+1})^2=L^2\,,
\end{equation}
into a $\left(d+2\right)$-dimensional ambient Minkowski space with metric
\begin{equation}
    ds^2=\eta_{MN}dX^M dX^N, \qquad \eta_{MN}=\text{diag}\left(-\,+\,...\,+\,+\right),
\end{equation}
and $M, N=0,...,d+1$. The constant $L$ is the de Sitter radius. It is manifest that the de Sitter embedding can be obtained from that of the Euclidean sphere $S^{d+1}$
\begin{equation}
    (X^1)^2+\cdots+(X^{d+1})^2+(X^{d+2})^2=L^2\,,
\end{equation}
through the analytic continuation
\begin{equation}
    X^{d+2}\to \pm i X^0,\label{analysphere}
\end{equation}
or from Euclidean AdS space EAdS$_{d+1}$ via
\begin{equation}
    X^M\to \pm i X^M. \label{analyads}
\end{equation}
If one considers complexified de Sitter space
\begin{equation}
   \{Z=X+i Y\in \mathbb{C}^{d+2}\quad |\quad Z^2=L^2\},
\end{equation}
both dS$_{d+1}$ and EAdS$_{d+1}$ can be obtained as appropriate real submanifolds. To focus on de Sitter space we set $Z=X$. Throughout we shall work in the expanding Poincar\'e patch,
\begin{equation}\label{dSm}
ds^2=\frac{L^2}{\eta^2}\left(-d\eta^2+dx^idx^i\right)\,,
\end{equation}
which solves the embedding constraints as: 
\begin{align}\label{exppp}
    X^M=\frac{L}{\eta}\left(\frac{L^2-\eta^2+x^ix^i}{2L},x^i,\frac{L^2+\eta^2-x^ix^i}{2L}\right),
\end{align}
where $\eta\in(-\infty,0]$ is the conformal time and the $x^i$ parameterise the spatial slices of dS space, including the conformal boundary at late-times $\eta=0$. This patch only covers $X^->0$ and is therefore not geodesically complete. When considering instead Euclidean anti-de Sitter space we set $Z=iY$, and in the Poincar\'e patch we have
\begin{align}\label{adspp}
 Y^M=\frac{L}{z}\left(\frac{L^2+z^2+x^ix^i}{2L},x^i,\frac{L^2-z^2-x^ix^i}{2L}\right)\,,
\end{align}
where $z \in \left[0,\infty\right)$ is the AdS radial co-ordinate and here the $x^i$ parameterise the AdS conformal boundary at $z=0$. Note that the parameterisations \eqref{exppp} and \eqref{adspp} are related under $z=\pm i\eta$ and changing the sign of the metric. The condition $Y^0>0$ selects one of the two disconnected branches of the hyperboloid. From this point onward we set the de Sitter radius to one, $L=1$.

The conformal boundary is identified with light rays:
\begin{equation}
    P^2=0, \qquad P \sim \lambda P, \qquad \lambda \ne 0,
\end{equation}
where the boundary points are parameterised by:
\begin{align}\label{boundamb}
    Z^M&\to P^M=\frac{1}{2}\left(1+x^2,2x^i,1-x^2\right)\,.
\end{align}
(Anti)-de Sitter invariant two point functions are functions of the geodesic distance $D$ \cite{Allen:1985ux}, 
\begin{equation}
   \cos\left(D/L\right)=2\sigma-1
\end{equation}
which is convenient to express through the chordal distance $\sigma$, where:
\begin{subequations}
\begin{align}
   \sigma_{\text{AdS}}&= \frac{1+Y_1\cdot Y_2}2=-\frac{(z_1-z_2)^2+|\vec{x}_1-\vec{x}_2|^2}{4z_1z_2}\,,\\
   \sigma_{\text{dS}}&= \frac{1+X_1\cdot X_2}2=\frac{(\eta_1+\eta_2)^2-|\vec{x}_1-\vec{x}_2|^2}{4\eta_1\eta_2}\,.
\end{align}
\end{subequations}
The dS chordal distance $\sigma_{\text{dS}}$ can be obtained from the AdS chordal distance $\sigma_{\text{AdS}}$ by taking opposite analytic continuations for $\eta_1$ and $\eta_2$:
\begin{align}\label{epsilon_p2}
    z_1=-\,\eta_1\,e^{{\color{red}\pm} i\tfrac{\pi}2}\,,\qquad z_2=-\,\eta_2\,e^{\tcb{\mp} i\tfrac{\pi}2}\,,
\end{align}
which is equivalent to\footnote{In position space we shall refer to \eqref{ambanalconwight} as equivalent to \eqref{epsilon_p2}.} 
\begin{align}\label{ambanalconwight}
   Y_1 = \tcr{\mp}iX_1, \qquad Y_2 = \tcb{\pm}iX_2.
\end{align}
These correspond to the possible Euclidean orderings of two operators in EAdS$_{d+1}$ associated to the following $\epsilon$-prescriptions:
\begin{equation}\label{epsilonpresc}
   (\eta_1-\eta_2)^2\pm i\,\text{sgn}(\eta_1-\eta_2)\,\epsilon\, \approx (\eta_1-\eta_2\pm \tfrac{i\epsilon}{2})^2.
\end{equation}
Further details about out-of-time ordered correlators are given in appendix \ref{Wick Rotation}. In the following sections we shall employ \eqref{epsilon_p2} to consider Wightman two-point functions in dS$_{d+1}$ as an analytic continuation from EAdS$_{d+1}$.

\begin{figure}[t]
    \centering
    \captionsetup{width=0.95\textwidth}
    \includegraphics[width=0.5\textwidth]{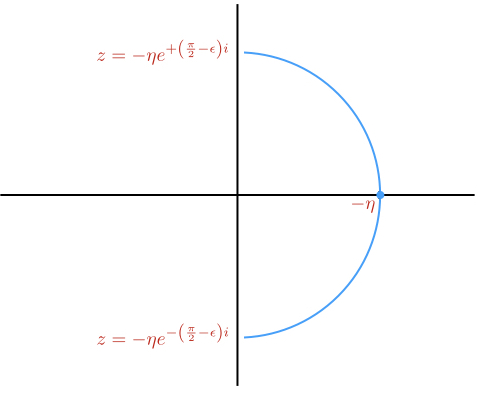}
    \caption{Analytic continuation from Euclidean anti-de Sitter space to de Sitter space. Here $z$ is the radial coordinate in EAdS while $\eta$ is conformal time in dS and we display the two possible analytic continuations from complexified dS.}
\end{figure}

\subsection{Review: Scalar Fields}
\label{subsec::scalar2pt}

Let us consider a scalar field $\phi$ with mass $m$
\begin{equation}\label{kgsc}
    \left(\nabla^2_{\text{dS}}-m^2\right) \phi=0,
\end{equation}
which at late times $\eta \to 0$ behaves as 
\begin{equation}
    \phi\left(\eta,\vec{x}\right)\,\sim\,O_{\Delta_+}\left(\vec{x}\right)\eta^{\Delta_+}+O_{\Delta_-}\left(\vec{x}\right)\eta^{\Delta_-},
\end{equation}
where the mass is related to the scaling dimensions via
\begin{equation}
    \Delta_{\pm}=\frac{d}{2}\pm i\nu, \qquad m^2=\left(\tfrac{d}{2}\right)^2+\nu^2.
\end{equation}
The Wightman function
\begin{equation}\label{scalarwightman}
    G\left(X_1,X_2\right) = \langle 0 | \phi\left(X_1\right)\phi\left(X_2\right) | 0 \rangle, 
\end{equation}
obeys the free field equation 
\begin{equation}
    \left(\nabla^2_{\text{dS}}-m^2\right) G\left(X_1,X_2\right)=0,
\end{equation}
whose solution in the standard Bunch-Davies vacuum \cite{Gibbons:1977mu,Bunch:1978yq} reads \cite{Burges:1984qm,Mottola:1984ar,Allen:1985ux}:
\begin{equation}\label{BD2pt}
    G\left(X_1,X_2\right) = \frac{\Gamma \left(\tfrac{d}{2}+i \nu \right)\Gamma \left(\tfrac{d}{2}-i \nu \right)}{(4\pi)^{(d+1)/2}\Gamma \left(\frac{d+1}{2}\right)} {}_2F_1\left(\begin{matrix}\frac{d}{2}+i\nu ,\frac{d}{2}-i\nu \\\frac{d+1}{2}\end{matrix};\sigma_{\text{dS}}\right)\,.
\end{equation}
 The Hypergeometric function has a singularity at $\sigma_{\text{dS}} = 1$ (i.e. at short distances) and a branch cut for $\sigma_{\text{dS}} \in \left[1,\infty\right)$. The two possible $i\epsilon$ prescriptions for going around the singularity in the complex plane are given in \eqref{epsilonpresc} for the flat slicing of de Sitter, in particular:
\begin{equation}\label{eps2pt}
    \sigma^\pm_{\text{dS}} = 1+\frac{(\eta_1-\eta_2\pm\tfrac{i\epsilon}{2})^2-|\vec{x}_1-\vec{x}_2|^2}{4\eta_1\eta_2},
\end{equation}
where
\begin{subequations}\label{wightmantwopresc}
\begin{align}
   G_{+-}\left(X_1,X_2\right)&=\langle 0 | {\hat \phi}\left(X_2\right){\hat \phi}\left(X_1\right) | 0 \rangle =G(\sigma^+_{\text{dS}}),\\
    G_{-+}\left(X_1,X_2\right)&=\langle 0 | {\hat \phi}\left(X_1\right){\hat \phi}\left(X_2\right) | 0 \rangle=G(\sigma^-_{\text{dS}}),
\end{align}
\end{subequations}
and the $\mp \pm$ subscripts refer to the analytic continuations of the two points as in \eqref{ambanalconwight}.
The Wightman two-point function serves as the basic object from which other de Sitter two-point functions (retarded, advanced, Feynman,...) can be obtained, as we shall discuss in \S \ref{subsec::SKdic} when we introduce the Schwinger-Keldysh formalism. 

The corresponding object in $(d+1)$-dimensional Euclidean anti-de Sitter space is the Harmonic function 
\begin{equation}
    \left(\nabla^2_{\text{AdS}}-m^2\right) \Omega_{\nu}\left(Y_1,Y_2\right)=0,
\end{equation}
where (see e.g. appendix 4.C of \cite{Penedones:2007ns})
\begin{equation}\label{harmfuncomeg}
    \Omega_\nu(Y_1,Y_2)=\frac1{\Gamma(i\nu)\Gamma(-i\nu)}\,\frac{\Gamma \left(\tfrac{d}{2} + i \nu \right)\Gamma \left(\tfrac{d}{2} - i \nu \right)}{(4\pi)^{\tfrac{d+1}{2}}\Gamma \left(\tfrac{d+1}{2}\right)}\,{}_2F_1\left(\begin{matrix}\frac{d}{2}+i\nu ,\frac{d}{2}-i\nu \\ \frac{d+1}{2}\end{matrix};\sigma_{\text{AdS}}\right)\,.
\end{equation}
The short distance limit in this case corresponds to $\sigma_{\text{AdS}} \rightarrow 0$, which is non-singular. It is straightforward to see that through the analytic continuations \eqref{epsilon_p2} we can obtain the de Sitter Wightman function \eqref{BD2pt} from the Harmonic function \eqref{harmfuncomeg}. In particular,
\begin{subequations}\label{wightmanharm}
\begin{align}
G_{+-}\left(X_1,X_2\right)&=\Gamma(i\nu)\Gamma(-i\nu)\,\Omega_\nu(-i X_1,+i X_2)\,,\\
G_{-+}\left(X_1,X_2\right)&=\Gamma(i\nu)\Gamma(-i\nu)\,\Omega_\nu(+i X_1,-i X_2)\,.
\end{align}
\end{subequations}
The Harmonic function \eqref{harmfuncomeg} admits the following useful representation (see e.g. \cite{Moschella:2007zza,Costa:2014kfa}):
\begin{equation}\label{scalarharm}
    \Omega_{\nu}(Y_1,Y_2)=\frac{\nu^2}{\pi}\int dP\,K_{\tfrac{d}2+i\nu}(Y_1,P)\,K_{\tfrac{d}2-i\nu}(Y_2,P)\,,
\end{equation}
which is an integrated product of EAdS$_{d+1}$ bulk-to-boundary propagators,
\begin{equation}\label{scbuboamb}
    K_{\Delta}(Y,P)=\frac{C_{\Delta,0}}{(-2Y\cdot P)^\Delta}\,, \qquad C_{\Delta,0} = \frac{\Gamma\left(\Delta\right)}{2\pi^{d/2}\Gamma\left(\Delta+1-\tfrac{d}{2}\right)}.
\end{equation}
In the AdS/CFT literature, the representation \eqref{scalarharm} for the Harmonic function is often referred to as the ``split-representation". From the analytic continuations \eqref{wightmanharm}, a split representation for the de Sitter Wightman function \eqref{BD2pt} naturally follows: 
\begin{shaded}
\noindent \emph{Split representation for the scalar Wightman two-point function in dS$_{d+1}$}
\begin{equation}\label{wightmanscalar}
    G_{\pm \mp}\left(X_1,X_2\right)=\int dP\, {\cal K}_{\tfrac{d}2+i\nu}(\mp iX_1,P)\,{\cal K}_{\tfrac{d}2-i\nu}(\pm iX_2,P)\,. 
\end{equation}
\end{shaded}
\noindent where, in going to dS, it is convenient to adopt the normalisation:
\begin{equation}\label{dsbubonorm}
    {\cal K}_{\Delta}=\frac{\Gamma\left(\Delta-\tfrac{d}{2}+1\right)}{\sqrt{\pi}}K_{\Delta},
\end{equation}
which we shall use henceforth. The discussion of this section naturally extends to spinning fields, which we consider in \S \ref{subsec::spin2pt}.

The split representation has proven to be an instrumental tool in the evaluation of Witten diagrams in EAdS \cite{Hartman:2006dy,Penedones:2010ue,Giombi:2011ya,Paulos:2011ie,Fitzpatrick:2011ia,Costa:2014kfa,Bekaert:2014cea,Sleight:2016hyl,Sleight:2017fpc,Chen:2017yia,Tamaoka:2017jce,Giombi:2017hpr,Sleight:2017cax,Yuan:2017vgp,Giombi:2018vtc,Yuan:2018qva,Nishida:2018opl,Costa:2018mcg,Carmi:2018qzm,Zhou:2018sfz,Jepsen:2019svc} and is particularly suitable to obtain the Conformal Partial Wave decomposition of tree-level exchange Witten diagrams, which factorise on Harmonic functions into an integrated product of three-point Witten diagrams. In this work we show that the split representation is also useful in de Sitter space, where late-time tree-level exchange diagrams in dS$_{d+1}$ can be obtained from existing results for EAdS$_{d+1}$ three-point Witten diagrams through the analytic continuations \eqref{wightmanscalar}.

\subsubsection*{Mellin-Barnes representation in Fourier space}
\label{MSMBrepprops}

Cosmological correlators are generally studied in Fourier space. In Fourier space the split representation \eqref{scalarharm} for the EAdS Harmonic function conveniently factorises as a consequence of the Convolution theorem:
\begin{equation}\label{scalarharmfourier}
   \Gamma\left(i\nu\right)\Gamma\left(-i\nu\right) \Omega_{\nu,\vec{k}}(z_1,z_2)={\cal K}_{\tfrac{d}2+i\nu}(z_1,\vec{k})\,{\cal K}_{\tfrac{d}2-i\nu}(z_2,-\vec{k})\,.
\end{equation}
The Fourier transform of the EAdS bulk-to-boundary propagator is given by a modified Bessel function of the second kind \cite{Gubser:1998bc}, which admits a convenient representation as a Mellin-Barnes integral:
\begin{subequations}\label{MBbubosc}
\begin{align}
    {\cal K}_{\tfrac{d}2+i\nu}(z,\vec{k}) &= \int^{i\infty}_{-i\infty}\,\frac{du}{2\pi i}\,{\cal K}_{\tfrac{d}2+i\nu}(z,\vec{k}|u) \\
    &=\frac{z^{\tfrac{d}{2}-i\nu}}{2\sqrt{\pi}}\int^{i\infty}_{-i\infty}\,\frac{du}{2\pi i}\,\Gamma\left(u+\tfrac{i\nu}2\right)\Gamma\left(u-\tfrac{i\nu}2\right)\left(\frac{z k}2\right)^{-2u+i\nu}.
\end{align}
\end{subequations}
This implies the following Mellin-Barnes representation for the Harmonic function:
\begin{subequations}
\begin{align}
  \hspace*{-0.3cm}  \Omega_{\nu,\vec{k}}(z_1,z_2)&=\int \left[du\right]_2 \,\Omega_{\nu,\vec{k}}(z_1,z_2|u_1,u_2),\\
   \hspace*{-0.3cm}  \Gamma\left(i\nu\right)\Gamma\left(-i\nu\right)\Omega_{\nu,\vec{k}}(z_1,z_2|u_1,u_2)&=\frac{\left(z_1z_2\right)^{\tfrac{d}{2}}}{4\pi}\,\prod^2_{j=1}\Gamma\left(u_j+\tfrac{i\nu}2\right)\Gamma\left(u_j-\tfrac{i\nu}2\right)\left(\frac{z_j k}2\right)^{-2u_j}.\label{MBomega}
\end{align}
\end{subequations}
\noindent At the level of the Mellin-Barnes representation \eqref{MBomega}, the analytic continuations \eqref{epsilon_p2} to the flat slicing of de Sitter space are encoded into simple phases owing to the power-law dependence on the AdS radial co-ordinate. In particular, for the Wightman two-point function \eqref{wightmanscalar} in Fourier space we have:
\begin{subequations}\label{MBwightman}
\begin{align}
    G_{+-,\vec{k}}\left(\eta_1,\eta_2\right)&=\Gamma\left(-i\nu\right)\Gamma\left(i\nu\right)\int \left[du\right]_2 e^{\delta_{\prec}\left(u_1,u_2\right)}\,\Omega_{\nu,\vec{k}}(-\eta_1,-\eta_2|u_1,u_2),\\
    G_{-+,\vec{k}}\left(\eta_1,\eta_2\right)&=\Gamma\left(-i\nu\right)\Gamma\left(i\nu\right)\int \left[du\right]_2 e^{\delta_{\succ}\left(u_1,u_2\right)}\,\Omega_{\nu,\vec{k}}(-\eta_1,-\eta_2|u_1,u_2),
\end{align}
\end{subequations}
with phases: 
\begin{subequations}\label{intlegphase}
\begin{align}
    \delta_\prec\left(u_1,u_2\right)&=-i\pi \left(u_1-u_2\right),\\ \delta_\succ\left(u_1,u_2\right)&=+i\pi \left(u_1-u_2\right).
\end{align}
\end{subequations}

\paragraph{Late-time limit and bulk-to-boundary propagators.} Within the Mellin-Barnes representation \eqref{MBwightman} the late-time limits of the de Sitter Wightman function are encoded in the residues of the leading $\Gamma$-function poles. For example, the limit $\eta_2 \to 0$ with $\eta_1$ fixed is given by the leading poles in the corresponding Mellin variable $u_2$, which are at $u_2=\pm\frac{i\nu}{2}$. This gives
\begin{equation}\label{ltlimoneleg}
    \lim_{\eta_2 \to 0}G_{\pm \mp,\vec{k}}\left(\eta_1,\eta_2\right) = F^{\left(\nu\right)}_{\pm,\vec{k}}\left(\eta_1,\eta_2\right)+F^{\left(-\nu\right)}_{\pm,\vec{k}}\left(\eta_1,\eta_2\right),
\end{equation}
where we introduced the de Sitter bulk-to-boundary propagator
\begin{equation}\label{dsbuboMB}
     F^{\left(\nu\right)}_{\pm,\vec{k}}\left(\eta_1,\eta_2\right)={\cal N}_{\nu}\left(\eta_2\right)\underbrace{ \int^{+i\infty}_{-i\infty} \frac{ds}{2\pi i}\, e^{\delta^{\pm}_{\nu}\left(s\right)} {\cal K}_{\tfrac{d}2+i\nu}(-\eta_1,\vec{k}|s)}_{e^{\mp \tfrac{i\pi}{2}\left(\frac{d}2+i\nu\right)}\mathcal{K}_{\frac{d}2+i\nu}\Big(-e^{\pm \frac{i\pi}2}\eta_1,\vec{k}\Big)},
\end{equation}
and the overall constant
\begin{equation}\label{Ncalconst}
  {\cal N}_{\nu}\left(\eta_2\right)= \left(-\eta_2\right)^{\tfrac{d}{2}+i\nu}\frac{\Gamma\left(-i\nu\right)}{2\sqrt{\pi}}.
\end{equation}
Like for the Wightman function, the analytic continuation of the bulk-to-boundary propagator from EAdS to dS is encoded in a simple phase:
\begin{equation}
    \delta^{\pm}_{\nu}\left(s\right)=\mp i\pi \left(s+\tfrac{i\nu}{2}\right).
\end{equation}
Above we relabelled $u_1 \to s$ and we henceforth use the variable $s$ to denote external legs connected to the boundary (or $s_j$ when there is more than one).

\begin{figure}[t]
    \centering
    \captionsetup{width=0.95\textwidth}
    \includegraphics[width=\textwidth]{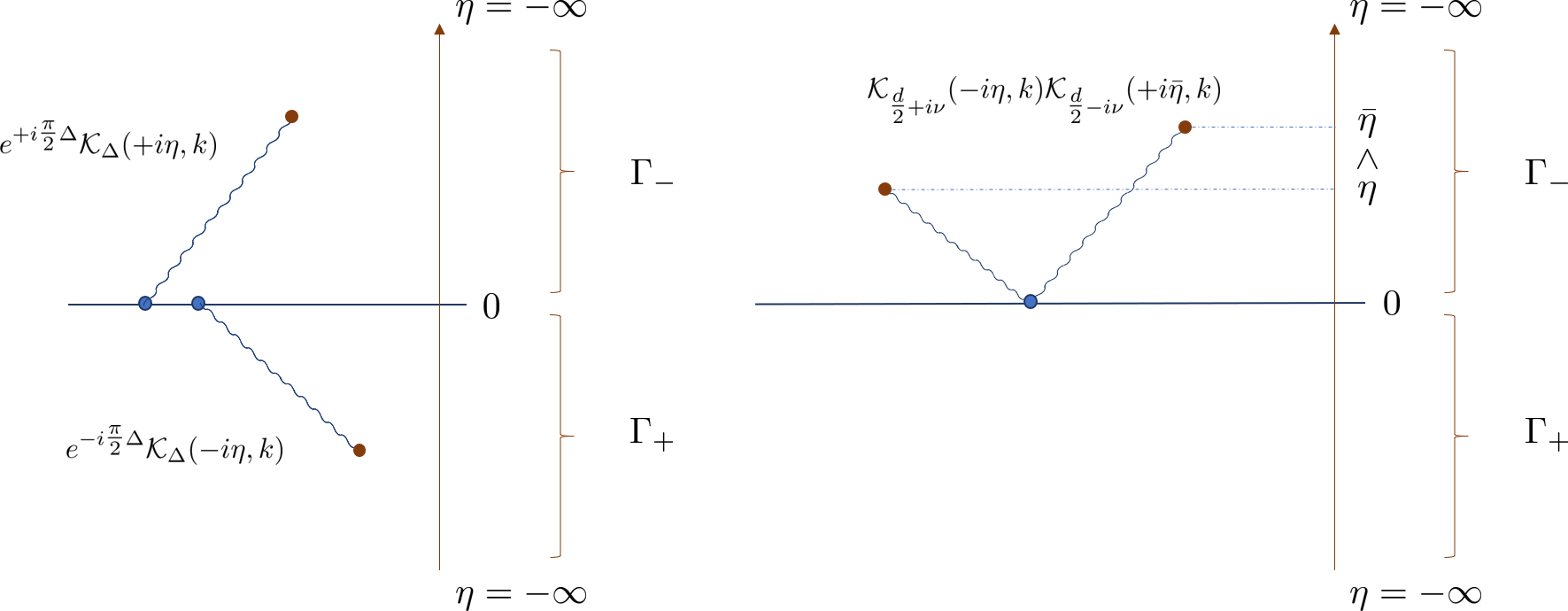}
    \caption{A pictorial representation of the dS in-in contours, of  bulk-to-boundary propagators on the left and of the split representation for bulk-to-bulk propagators on the right. The horizontal direction parameterise the momentum space coordinates while $\Gamma_{\pm}$ represent the (anti-)time ordered contours which we unfolded above the late time de Sitter horizon to distinguish the path ordering along the contour from the actual time-ordering relations.}
    \label{fig:Propagators}
\end{figure}

\paragraph{Relation to the mode functions.} To gain some further intuition it is instructive to review the standard derivation of two-point functions in Fourier space. One expands each Fourier mode of the field in creation and annihilation operators:
\begin{align}\label{scmodeexp}
    \phi_{\vec{k}}(\eta)=f_k(\eta)a_{\vec{k}}^\dagger+{\bar f}_k(\eta)a_{-\vec{k}}\,,
\end{align}
where the Klein-Gordon equation
\begin{equation}\label{waveeq}
    (\nabla^2_{\text{dS}}-m^2)\phi_{\vec{k}}(\eta)e^{i\vec{k}\cdot\vec{x}}=0\,,
\end{equation}
implies that the mode functions satisfy the equation:
\begin{equation}
    \eta  \left((d-1) f'(\eta )-\eta  f''(\eta )\right)+ \left(\Delta(\Delta-d)-\eta ^2 k^2\right)f(\eta )=0\,.
\end{equation}
For the Bunch-Davies vacuum the solution is given by the following combination of Hankel functions 
\begin{align}
    f_k(\eta)&=(-\eta)^{\tfrac{d}{2}}\,\underbrace{\frac{\sqrt{\pi}}{2}\,e^{\pi \nu/2}H_{i\nu}^{(2)}(-k\eta)}_{h_{i\nu}(k\eta)}\,,\\
    {\bar f}_k(\eta)&=(-\eta)^{\tfrac{d}{2}}\,\underbrace{\frac{\sqrt{\pi}}{2}\,e^{\pi \nu/2}H_{i\nu}^{(1)}(-k\eta)}_{\bar{h}_{i\nu}(k\eta)},
\end{align}
where the overall normalisation is fixed by requiring that $a_{-\vec{k}}$ and $a^\dagger_{\vec{k}}$ satisfy canonical commutation relations. From the Mellin-Barnes representation for the Hankel functions various similarities between the mode functions and the bulk-to-boundary propagators become manifest:
\begin{subequations}\label{mellinhmode}
\begin{align}
    h_{i\nu}(k\eta)&=+\frac{i}{2\sqrt{\pi }}\int_{-i\infty}^{+i\infty}\frac{ds}{2\pi i}\,\Gamma \left(s+\tfrac{i\nu}{2}\right) \Gamma \left(s-\tfrac{i\nu}{2}\right) \left(-\tfrac{\eta  k}2\,e^{+i\tfrac{\pi}2}\right)^{-2s}\,,\\
    \bar{h}_{i\nu}(k\eta)&=-\frac{i}{2\sqrt{\pi }}\int_{-i\infty}^{+i\infty}\frac{ds}{2\pi i}\,\Gamma \left(s+\tfrac{i\nu}{2}\right) \Gamma \left(s-\tfrac{i\nu}{2}\right) \left(-\tfrac{\eta  k}2\,e^{-i\tfrac{\pi}2}\right)^{-2 s}\,,
\end{align}
\end{subequations}
from which one can also read off the following relations:
\begin{subequations}
\begin{align}
h_{i\nu}(k\eta)&=h_{-i\nu}(k\eta),\\{\bar h}_{i\nu}(k\eta)&={\bar h}_{-i\nu}(k\eta),\\
    [h_{i\nu}(k\eta)]^\star&=-h_{i\nu}(e^{-i\pi}k\eta)=\bar{h}_{i\nu}(k\eta)\,.
\end{align}
\end{subequations}
In terms of the mode functions the Wightman two-point functions are
\begin{subequations}
\begin{align}
G_{+-,\vec{k}}\left(\eta_1,\eta_2\right)&= \langle 0| \phi_{-\vec{k}}\left(\eta_2\right) \phi_{\vec{k}}\left(\eta_1\right)|0\rangle = f_{k}\left(\eta_1\right){\bar f}_{k}\left(\eta_2\right),\\
G_{-+,\vec{k}}\left(\eta_1,\eta_2\right)&= \langle 0| \phi_{\vec{k}}\left(\eta_1\right) \phi_{-\vec{k}}\left(\eta_2\right)|0\rangle = {\bar f}_{k}\left(\eta_1\right)f_{k}\left(\eta_2\right),
\end{align}
\end{subequations}
from which one can identify:
\begin{align}
   f_{k}\left(\eta_1\right)f^*_{k}\left(\eta_2\right)&= \Gamma\left(i\nu\right)\Gamma\left(-i\nu\right) \Omega_{\nu,\vec{k}}\left(-e^{+ \tfrac{i\pi}{2}}\eta_1,-e^{-\tfrac{i\pi}{2}}\eta_2\right),\\
    f^*_{k}\left(\eta_1\right)f_{k}\left(\eta_2\right)&=  \Gamma\left(i\nu\right)\Gamma\left(-i\nu\right)\Omega_{\nu,\vec{k}}\left(-e^{- \tfrac{i\pi}{2}}\eta_1,-e^{+\tfrac{i\pi}{2}}\eta_2\right),
\end{align}
by comparing with equation \eqref{MBwightman}. These relations generalise to mode functions for fields of non-zero spin, which we consider in the following section.

\subsection{Fields of arbitrary integer spin}
\label{subsec::spin2pt}

The discussion of the previous section carries over to fields of non-trivial spin. In the following we consider a totally symmetric spin-$\ell$ field $\varphi_{\mu_1...\mu_\ell}$ of generic mass $m$, which at zeroth order in interactions satisfies the Fierz-Pauli conditions: 
\begin{subequations}\label{FPconds}
\begin{align}
    \left(\nabla^2-m^2\right)\varphi_{\mu_1...\mu_\ell}&=0,\\
    \nabla^{\mu_1}\varphi_{\mu_1...\mu_\ell}&=0,\\ \label{tless}
   g^{\mu_1\mu_2} \varphi_{\mu_1...\mu_{\ell}}&=0.
\end{align}
\end{subequations}
The boundary behaviour of the spin-$\ell$ field is
\begin{equation}
    \varphi_{i_1...i_\ell}\left(\eta,\vec{x}\right)\, \sim \, O_{\Delta_+,i_1...i_\ell}\left(\vec{x}\right)\eta^{\Delta_+-\ell}+O_{\Delta_-,i_1...i_\ell}\left(\vec{x}\right)\eta^{\Delta_--\ell},
\end{equation}
where the mass is related to the scaling dimensions by:\footnote{In this work ``mass" is defined by the Fierz-Pauli system \eqref{FPconds}. Another convention for mass is such that $m^2=0$ for gauge fields, which is used e.g. in \cite{Arkani-Hamed:2015bza}, in which case:
\begin{equation}
    m^2_{\text{there}}=\nu^2+\left(\ell+\tfrac{d}{2}-2\right)^2,
\end{equation}
where
\begin{equation}
    m^2_{\text{there}}=m^2_{\text{here}}-\left(\ell+d-2\right)\left(2-\ell\right)+\ell.
\end{equation}} 
\begin{equation}
    \Delta_{\pm}=\frac{d}{2}\pm i\nu, \qquad m^2=\left(\tfrac{d}{2}\right)^2+\nu^2-\ell.
\end{equation}
When considering fields of arbitrary spin it is convenient to use an operator notation in which fields are represented by generating functions:
\begin{align}
    \varphi_{\mu_1...\mu_\ell} \to \varphi\left(x;w\right)=\frac{1}{\ell!}\,\varphi_{\mu_1...\mu_\ell}\left(x\right)w^{\mu_1}...w^{\mu_\ell},
\end{align}
where $w^\mu$ is a constant $(d+1)$-dimensional auxiliary vector. In this formalism the Fierz-Pauli conditions read 
\begin{subequations}\label{FPcondsgenfunc}
\begin{align}
    \left(\nabla^2-m^2\right)\varphi\left(x;w\right)&=0,\\
    (\nabla \cdot {\hat \partial}_{w})\varphi\left(x;w\right)&=0,
\end{align}
\end{subequations}
where requiring that the auxiliary vector is null, $w^2=0$, implements the trace constraint \eqref{tless}. The Thomas-D operator
\cite{Thomas352} (see also \cite{Dobrev:1975ru}):
\begin{equation}
{\hat \partial}^\mu_w=\left(\tfrac{d-1}{2}+w\cdot \pl_w\right)\,\pl^\mu_w-\frac12\, w^\mu\,(\pl_w\cdot\pl_w)\,,
\end{equation}
implements the trace-less contraction of indices. For boundary operators we instead use $\,\xi^i\,$ to denote the corresponding null auxiliary vectors, i.e. 
\begin{equation}\label{boundgenfunc}
    O_{\Delta,i_1...i_\ell}\left(\vec{y}\right) \to O_{\Delta}(\vec{y};\vec{\xi}\,)=\frac{1}{\ell!}O_{\Delta,i_1...i_\ell}\left(\vec{y}\right)\xi^{i_1}...\xi^{i_\ell}, \qquad \vec{\xi}^2=0,
\end{equation}
and the corresponding boundary Thomas-D operator reads
\begin{equation}\label{thomasdboundary}
{\hat \partial}^i_\xi=\left(\tfrac{d}{2}-1+\vec{\xi}\cdot \vec{\pl}_{\xi}\right)\,\pl^i_\xi-\frac12\, \xi^i\,(\vec{\pl}_\xi\cdot\vec{\pl}_\xi)\,.
\end{equation}
Following \S \ref{subsec::scalar2pt}, to obtain the corresponding Wightman two-point function
\begin{equation}
    G_{\ell}\left(X_1,X_2;W_1,W_2\right)=\langle 0| \varphi_\ell\left(X_1;W_1\right)\varphi_\ell\left(X_2;W_2\right)|0\rangle,
\end{equation}
we first consider the corresponding spin-$\ell$ Harmonic function in EAdS$_{d+1}$, \begin{subequations}
\begin{align}
    \left(\nabla^2_{\text{AdS}}-m^2\right)\Omega_{\nu,\ell}(Y_1,Y_2;W_1,W_2)&=0,\\
    \left(\nabla_{\text{AdS}} \cdot {\hat \partial}_{W_1}\right)\Omega_{\nu,\ell}(Y_1,Y_2;W_1,W_2)&=0.
\end{align}
\end{subequations}
This admits the split representation (see e.g. \cite{Leonhardt:2003qu,Costa:2014kfa}):
\begin{multline}\label{spinlharm}
    \Omega_{\nu,\ell}(Y_1,Y_2;W_1,W_2)\\=\frac{1}{\ell!\left(\tfrac{d}{2}-1\right)_\ell}\frac{\nu^2}{\pi}\int dP\,{K}_{\tfrac{d}2+i\nu,\ell}(Y_1,P;W_1,{\hat \partial}_\Xi){K}_{\tfrac{d}2-i\nu,\ell}(Y_2,P;W_2,\Xi)\,,
\end{multline}
where the spin-$\ell$ EAdS$_{d+1}$ bulk-to-boundary propagators read \cite{Costa:2014kfa}:
\begin{equation}\label{bulk-to-boundary-spin}
    {K}_{\Delta,\ell}(Y,P;W,\Xi)=\frac{C_{\Delta,\ell}}{(-2Y\cdot P)^{\Delta}}\,\left[W \cdot I\left(Y,P\right) \cdot \Xi  \right]^\ell,
\end{equation}
where
\begin{subequations}
\begin{align}
  I_{MN}\left(Y,P\right)&= \eta_{MN}-\frac{P_M Y_N}{P \cdot Y},
    \\
    C_{\Delta,\ell}&=\frac{\Delta +\ell-1}{\Delta-1}\frac{\Gamma (\Delta )}{2 \pi ^{d/2} \Gamma \left(\Delta-\frac{d}{2} +1\right)}\,.\label{Cnorm}
\end{align}
\end{subequations}
The vectors $W^M$ and $\Xi^M$ are the ambient space representatives of the bulk and boundary auxiliary vectors $w^\mu$ and $\xi^i$ (see e.g. \cite{Costa:2011dw,Joung:2011ww,Joung:2012fv,Taronna:2012gb,Sleight:2017krf}):
\begin{align}\label{bulkambrepaux}
    W^{M}= w^{\mu}\frac{\partial Y^M}{\partial x^{\mu}},  \qquad \Xi^{M}= \xi^{i}\frac{\partial P^M}{\partial x^{i}}.
\end{align}
While it shall not be used explicitly in this work, it may be useful to note that in Poincar\'e co-ordinates \eqref{adspp} the bulk-to-boundary propagator \eqref{bulk-to-boundary-spin} reads \cite{Mikhailov:2002bp}:\footnote{This can be derived from the ambient space expression \eqref{bulk-to-boundary-spin} using that 
\begin{equation}
    -2Y\cdot P = \frac{z^2+\left(\vec{x}-\vec{x}^\prime\right)^2}{z},
\end{equation}
combined with
\begin{align}
   W \cdot I\left(Y,P\right) \cdot \Xi &= \left(P \cdot Y\right) \left(\Xi \cdot \partial_P\right)\left(W \cdot \partial_Y\right)\log\left(Y \cdot P\right)=\left(P \cdot Y\right) \left(\vec{\xi} \cdot \partial_{\vec{x}^\prime}\right)\left(w \cdot \partial_{x}\right)\log\left(Y \cdot P\right),
\end{align}
where in the second equality we used the relation \eqref{bulkambrepaux}. Evaluating the derivatives recovers \eqref{invtensintr}.}
\begin{subequations}\label{poinspinlbubo}
\begin{align}
     {K}_{\Delta,\ell}\left(x;\vec{x}^\prime\right) 
     &= C_{\Delta,\ell}\left(\frac{z}{z^2+\left(\vec{x}-\vec{x}^\prime\right)^2}\right)^\Delta \left(w \cdot I\left(z,\vec{x}-\vec{x}^\prime\right) \cdot \vec{\xi}\,\right)^\ell, \\ \label{invtensintr}
     w \cdot I\left(z,\vec{x}-\vec{x}^\prime\right) \cdot \vec{\xi}&=\frac1{z}\left[\vec{w}\cdot\vec{\xi}-\frac{2\vec{w}\cdot(\vec{x}-\vec{x}^\prime)\vec{\xi}\cdot(\vec{x}-\vec{x}^\prime)}{z^2+(\vec{x}-\vec{x}^\prime)^2}+\frac{2w^ z\,z\,\vec{\xi}\cdot(\vec{x}-\vec{x}^\prime)}{z^2+(\vec{x}-\vec{x}^\prime)^2}\right], 
\end{align}
\end{subequations}
where $w=\left(w^z,\vec{w}\right)$. 

The ambient space auxiliary vectors \eqref{bulkambrepaux} are unaffected by the analytic continuations from EAdS to dS. The corresponding de Sitter Wightman functions are therefore 
\begin{align}\label{wightmanharmspinl}
 G_{\pm\mp,\ell}\left(X_1,X_2;W_1,W_2\right)&=\Gamma\left(i\nu\right)\Gamma\left(-i\nu\right) \Omega_{\nu,\ell}(\mp i X_1,\pm i X_2;W_1,W_2),
\end{align}
where, as before, the $\mp \pm$ subscripts refer to the analytic continuations of the two points as in \eqref{ambanalconwight} and the coefficient of the Harmonic function is the same as for the scalar Wightmann function \eqref{wightmanharm}.\footnote{This is fixed by the normalisation of the short-distance behaviour, where the short distance limit of the spin-$\ell$ Harmonic function \eqref{spinlharm} is given in \cite{Costa:2014kfa}.} Equation \eqref{wightmanharmspinl} combined with \eqref{spinlharm} provides a split representation for spin-$\ell$ Wightman functions in de Sitter space.

\paragraph{Fourier space.} Notice that the tensorial structure of the bulk-to-boundary propagator \eqref{bulk-to-boundary-spin} is invariant under the above analytic continuations. A useful consequence of this observation is that the phase factor in the Mellin-Barnes representation for the Fourier-space Wightman function is independent from the spin $\ell$. In other words,
\begin{align}
 G_{+-,\ell,\vec{k}}\left(\eta_1,\eta_2;w_1,w_2\right)&=\Gamma\left(-i\nu\right)\Gamma\left(i\nu\right)\int \left[du\right]_2 e^{\delta_{\prec}\left(u_1,u_2\right)}\Omega_{\nu,\ell,\vec{k}}(-\eta_1, -\eta_2;w_1,w_2|u_1,u_2), \nonumber \\
  G_{-+,\ell,\vec{k}}\left(\eta_1,\eta_2;w_1,w_2\right)&=\Gamma\left(-i\nu\right)\Gamma\left(i\nu\right)\int \left[du\right]_2 e^{\delta_{\succ}\left(u_1,u_2\right)}\Omega_{\nu,\ell,\vec{k}}(-\eta_1,-\eta_2;w_1,w_2|u_1,u_2), \label{spinlharmwightanal}
\end{align}
where the phases are the same as those for the spin $\ell=0$ Wightman function which were given in \eqref{intlegphase}. 

\subsection{Late-time Two-Point Functions}\label{2pt}

Before discussing interactions it is convenient to consider the late-time limit of the bulk two point function with respect to both points, i.e. $\eta_{1}$, $\eta_{2} \to 0$.

Using the relation \eqref{wightmanharmspinl}, one way to obtain the late-time limit of the bulk two point function is to analytically continue the boundary limit $z_1,z_2 \to 0$ of the Harmonic function in EAdS$_{d+1}$.\footnote{Another way was outlined in \S \ref{MSMBrepprops} using directly the Mellin-Barnes representation for the Wightman function.} The latter can be straightforwardly obtained in position space using the identity \cite{Costa:2014kfa}:
\begin{align}
    \Omega_{\nu,\ell}\left(x_1,x_2\right)=\frac{i\nu}{2\pi}\,\Pi_{\nu,\ell}\left(x_1,x_2\right)+(\nu\to-\nu)\,,
\end{align}
which expresses the Harmonic function as a sum of spin-$\ell$ bulk-to-bulk propagators in EAdS$_{d+1}$. The boundary limit of the Harmonic function is then fixed by the boundary limit of the bulk-to-bulk propagators, which is 
\begin{align}\label{spin2ptpos}
    \lim_{z_1,z_2 \to 0} \Pi_{\nu,\ell}(z_1,\vec{x}_1,z_2,\vec{x}_2)\,=&\,C_{\tfrac{d}{2}+i\nu,\ell}\frac{(z_1z_2)^{\tfrac{d}{2}+i\nu-\ell}}{(\vec{x}_{12}^2)^{\tfrac{d}{2}+i\nu}}\,\left(\vec{\xi}_1\cdot\vec{\xi}_2+\frac{2\vec{\xi}_1\cdot\vec{x}_{12}\vec{\xi}_1\cdot\vec{x}_{12}}{\vec{x}_{12}^2}\right)^{\ell},
\end{align}
where the $\vec{\xi}_{1,2}$ are the boundary auxiliary vectors \eqref{boundgenfunc} and the coefficient $C_{\tfrac{d}{2}+i\nu,\ell}$ is the coefficient of the bulk-to-boundary propagator \eqref{Cnorm}. This has the structure required by conformal symmetry \cite{Polyakov:1974gs}.

In Appendix \ref{Appendix: Fourier Transform0} we derive the Fourier transform of the above conformal structure, which gives the following expression for the boundary limit of the Harmonic function in Fourier space:
\begin{multline}\label{GenD_2pt}
   \Omega_{\nu,\ell}(z_1,\vec{k},z_2,-\vec{k})\sim\frac14\,(z\bar{z})^{\tfrac{d}{2}+i\nu-\ell}\,\frac{\text{csch}(\pi  \nu ) \Gamma (-i \nu )}{\left(\frac{d}{2}+i \nu -1\right)_\ell \Gamma\left(1-\ell+i \nu\right)}\,\\\times\,\left(\frac{k}{2}\right)^{2i\nu}\left(-\frac{2\,\vec{\xi}_1\cdot \vec{k}\,\vec{\xi}_2\cdot \vec{k}}{k^2}\right)^\ell\, {}_2F_1\left(\begin{matrix}-\ell,\tfrac{d}{2}+i \nu -1\\1-\ell+i \nu \end{matrix};\frac{k^2 \vec{\xi}_1\cdot\vec{\xi}_2}{2\vec{\xi}_1\cdot k\, \vec{\xi}_2\cdot k}\right)+(\nu\to -\nu)\,,
\end{multline}
\noindent where we have dropped analytic terms in $k$. It is then straightforward to obtain the corresponding late-time two-point function in dS$_{d+1}$ through the analytic continuation \eqref{epsilon_p2}:
\begin{shaded}
\noindent \emph{Spin-$\ell$ late-time two-point function in dS$_{d+1}$}
\begin{multline}\label{spinlLT2pt}
   \lim_{\eta_1,\eta_2 \to 0}  \left\langle\varphi_{\ell,\vec{k}}(\eta_1;\vec{\xi}_1)\varphi_{\ell,-\vec{k}}(\eta_2;\vec{\xi}_2)\right\rangle^\prime=\frac14\,(\eta_1\eta_2)^{\tfrac{d}{2}+i\nu-\ell}\,\frac{\text{csch}(\pi  \nu ) \Gamma (-i \nu )}{\left(\frac{d}{2}+i \nu -1\right)_\ell\Gamma\left(1-\ell+i \nu\right)}\,\\\times\,\left(\frac{k}{2}\right)^{2i\nu}\left(-\frac{2\,\vec{\xi}_1\cdot \vec{k}\,\vec{\xi}_2\cdot \vec{k}}{k^2}\right)^\ell\, {}_2F_1\left(\begin{matrix}-\ell,\tfrac{d}{2}+i \nu -1\\1-\ell+i \nu \end{matrix};\frac{k^2 \vec{\xi}_1\cdot\vec{\xi}_2}{2\vec{\xi}_1\cdot \vec{k}\, \vec{\xi}_2\cdot \vec{k}}\right)+(\nu\to -\nu)\,.
\end{multline}
\end{shaded}

\paragraph{Helicity decomposition.} In the following we derive the helicity decomposition of the two-point function \eqref{spinlLT2pt}, which is the projection onto spherical harmonics in the plane orthogonal to the exchanged momentum $\vec{k}$. In general $d$ these are the Gegenbauer polynomials
\begin{equation}
    \Xi_{m}\left(z\right)=\frac{m!}{2^m\left(\frac{d-3}{2}\right)_m}C^{\left(\frac{d-3}{2}\right)}_m\left(z\right),
\end{equation}
where $z={\hat \xi}^{\perp}_1 \cdot {\hat \xi}^{\perp}_2$ is the contraction of the transverse polarisations\footnote{The transverse mode can be obtained through the application of a simple projector
\begin{align}
    \Pi_{ij}(k)=\delta_{ij}-\hat{k}_{i}\hat{k}_j\,,
\end{align}
where $\hat{k}=\vec{k}/k$, yielding
\begin{equation}
    \xi^{i}_{\perp}=\Pi^{ij}(k)\xi^j\,,
\end{equation}
where $\xi_\perp$ is the transverse component of the polarisation with respect to the momentum $\vec{k}$ and therefore satisfies $\xi_\perp\cdot \vec{k}=0$. The longitudinal component is instead proportional to $\hat{k}$ and simply reads
\begin{equation}
    \xi_{\parallel}=i\,\hat{k}\,,
\end{equation}
where we have normalised for convenience $\vec{\xi}\cdot \hat{k}=i$ so that $\xi_\perp^2=1$.}
\begin{equation}
    \xi^{\perp}_1 \cdot \vec{k}=0, \qquad \xi^{\perp}_2 \cdot \vec{k}=0.
\end{equation}
In particular, we can choose
\begin{subequations}
\begin{align}
    \vec{\xi}_1&=({\hat \xi}_{\perp,1},i)\,,\\
    \vec{\xi}_2&=({\hat \xi}_{\perp,2},-i)\,,\\
    \vec{k}&=(0,k)\,,
\end{align}
\end{subequations}
so that
\begin{equation}
  \vec{\xi}_1 \cdot \vec{\xi}_2=z+1, \qquad \vec{\xi}_1 \cdot \vec{k}=ik, \qquad   \vec{\xi}_2 \cdot \vec{k}=-ik.
\end{equation}
The helicity decomposition of the two-point function
\begin{equation}
    \lim_{\eta_1,\eta_2 \to 0}  \left\langle\varphi_{\ell,\vec{k}}(\eta_1;\xi_1)\varphi_{\ell,-\vec{k}}(\eta_2;\xi_1)\right\rangle^\prime=(\eta_1\eta_2)^{\tfrac{d}{2}+i\nu-\ell}\sum^\ell_{m=0}c^{\left(\ell\right)}_{m}\,\Xi_m\left(z\right)+(\nu\to -\nu),
\end{equation}
can then be obtained using the standard inversion formula to extract the coefficients $c^{\left(\ell\right)}_{m}$:
\begin{subequations}\label{gegenbauerinv}
\begin{align}
    f(z)&=\sum_{m=0}^\ell c_m\,\Xi_m(z)
    ,\\
    c_n&=\frac1{\mathcal{N}_n}\,\int_{-1}^{1}dz\,(1-z^2)^{\tfrac{d-4}{2}}f(z)\,C_n^{\left(\frac{d-3}{2}\right)}\left(z\right),\\
    \mathcal{N}_n&=\frac{2^{d+n-4} \Gamma \left(\frac{d}{2}-\frac{3}{2}\right) \Gamma \left(\frac{d}{2}+n-\frac{1}{2}\right)}{\pi  \Gamma (d+n-3)}\,.
\end{align}
\end{subequations}
This gives
\begin{multline}\label{HelicityNorm}
    c_m^{(\ell)}=\binom{\ell}{m}\frac{ \Gamma \left(\tfrac{d-2 \Delta}{2}\right)^2  \Gamma \left(\frac{d}{2}+\ell-1\right)\Gamma (d+\ell-\Delta -1) }{2^{5- d -\ell-m}\pi ^{3/2}  \Gamma (\ell+\Delta -1)}\\ \times \frac{\Gamma \left(\frac{d-1}{2}+m\right) \Gamma (m+\Delta -1)}{\Gamma (d+\ell+m-2) \Gamma (d+m-\Delta -1)}\,,
\end{multline}
where $\Delta=\tfrac{d}{2}+i\nu$. The divergences of the coefficients $c_m^{(\ell)}$ are associated with the emergence of gauge symmetries in the bulk, where some of the helicity components decouple (see also \cite{Deser:2001us,Dolan:2001ih,Joung:2012hz,Joung:2014aba,Arkani-Hamed:2015bza,Joung:2019wwf}). 

\subsection{Dictionary from EAdS$_{d+1}$ to dS$_{d+1}$}
\label{subsec::SKdic}

In this section we summarise the Mellin-space dictionary which allows us to go from Witten diagrams involving totally symmetric bosonic fields in EAdS$_{d+1}$ to the corresponding late-time correlators in dS$_{d+1}$. 

In time-dependent backgrounds like de Sitter, the standard approach to compute vacuum expectation values is the Schwinger-Keldysh (or in-in) formalism \cite{doi:10.1063/1.1703727,kadanoff1962quantum,Keldysh:1964ud}. The first applications of this formalism to the calculation of cosmological correlation functions include \cite{Maldacena:2002vr,Weinberg:2005vy}; for clear pedagogical reviews see \cite{Baumann:2009ds,Akhmedov:2013vka,Chen:2017ryl}. In this formalism one carries out a time-ordered integral from the initial time ($\eta=-\infty$) to the time of interest $\eta_0$, followed by an anti-time ordered integral back to the initial time. To this end one introduces propagators with points along different parts of the contour, which in the usual way are given in terms of the Wightman functions \eqref{wightmantwopresc}:
\begin{subequations}\label{SKprop}
\begin{align}
  G_{++}\left(X_1,X_2\right)&=\theta\left(\eta_1-\eta_2\right)G_{-+}\left(X_1,X_2\right) + \theta\left(\eta_2-\eta_1\right)G_{+-}\left(X_1,X_2\right), \\
    G_{--}\left(X_1,X_2\right)&=\theta\left(\eta_2-\eta_1\right)G_{-+}\left(X_1,X_2\right) + \theta\left(\eta_1-\eta_2\right)G_{+-}\left(X_1,X_2\right),  \\
    G_{+-}\left(X_1,X_2\right)&=\langle 0| {\hat \phi}\left(X_2\right) {\hat \phi}\left(X_1\right)| 0\rangle,
     \\
    G_{-+}\left(X_1,X_2\right)&=\langle 0| {\hat \phi}\left(X_1\right) {\hat \phi}\left(X_2\right)| 0\rangle,   
\end{align}
\end{subequations}
where the $+(-)$ subscripts correspond to the (anti-)time ordered part of the integration (``in-in") contour respectively, with the $T$ and ${\bar T}$ denoting time and anti-time ordered products. The expression \eqref{wightmanharmspinl} for the Wightman functions provides a split representation for the above Keldysh propagators via \eqref{spinlharm}.

As we saw in the preceding sections, at the level of the Mellin-Barnes representation the analytic continuation \eqref{spinlharmwightanal} of EAdS Harmonic functions to de Sitter two-point functions is encoded in a simple phase \eqref{intlegphase}, which for the Keldysh propagators above depends on the path ordering of $\eta_1$ and $\eta_2$ along the in-in contour. This can be summarised by
\begin{equation}\label{bubumbgen}
    G_{\bullet \bullet,\ell, \vec{k}}\left(\eta_1;\eta_2\right)=\Gamma\left(i\nu\right)\Gamma\left(-i\nu\right) \int \left[du\right]_2e^{\delta\left(u_1,u_2\right)}\Omega_{\nu,\ell,\vec{k}}\left(-\eta_1;-\eta_2|u_1,u_2\right),
\end{equation}
where 
\begin{subequations}\label{IntLegPh}
\begin{align}
  \text{for} \quad \eta_1 \prec \eta_2 &: \qquad \delta\left(u_1,u_2\right)=\delta_\prec\left(u_1,u_2\right)=-i\pi(u_1-u_2),\\
   \text{for} \quad \eta_1 \succ \eta_2 &: \qquad \delta\left(u_1,u_2\right)=\delta_\succ\left(u_1,u_2\right) = +i\pi(u_1-u_2).
\end{align}
\end{subequations}
The $\bullet$ serve as place holders for the labels which denote the branch of the in-in contour. Note that the $+-$ and $-+$ propagators are described by a definite phase (as in \eqref{spinlharmwightanal}) since $\eta_1$ and $\eta_2$ lie on different branches of the in-in contour and so have a definite path ordering.\footnote{E.g. if $\eta_1$ lies on the $+$ branch and $\eta_2$ on the $-$ branch then $\eta_1 \prec \eta_2$, since one first traverses $+$ branch of the in-in contour before the $-$ branch.} For the $++$ and $--$ propagators, where $\eta_1$ and $\eta_2$ lie on the same branch of the in-in contour, there are two phases (corresponding to the two theta functions in \eqref{SKprop}) which depend on whether $\eta_1$ is ahead or behind $\eta_2$.

For the spin-$\ell$ bulk-to-boundary propagators we instead have:
\begin{equation}\label{buboFsc}
     F^{\left(\nu\right)}_{\pm,\ell,\vec{k}}\left(\eta,\eta_0;w,\vec{\xi}\right)={\cal N}_{\nu,\ell}\left(\eta_0\right) \int^{i\infty}_{-i\infty} \frac{ds}{2\pi i}\, e^{\delta^{\pm}_{\nu}\left(s\right)} {\cal K}_{\tfrac{d}2+i\nu,\ell}(-\eta,\vec{k};w,\vec{\xi}|s),
\end{equation}
at some late time $\eta_0 \sim 0$, where the subscripts refer to the branch of the in-in contour, and
\begin{subequations}\label{ExtLegPh}
\begin{align}
    +&:\quad  \delta^+_\nu\left(s\right) = -i\pi s+\tfrac{\pi\nu}2,\\
    -&:\quad  \delta^-_\nu\left(s\right) = +i\pi s-\tfrac{\pi\nu}2.
\end{align}
\end{subequations}
For spin-$\ell$ fields the overall constant \eqref{Ncalconst} is 
\begin{equation}\label{overallconstspinl}
    {\cal N}_{\nu,\ell}\left(\eta_0\right)=\left(-\eta_0\right)^{\tfrac{d}{2}+i\nu-\ell}\frac{\Gamma\left(-i\nu\right)}{2\sqrt{\pi}},
\end{equation}
while, as for the spin-$0$ case \eqref{dsbubonorm}:
\begin{equation}\label{dsbubonormspinl}
    {\cal K}_{\Delta,\ell}=\frac{\Gamma\left(\Delta-\tfrac{d}{2}+1\right)}{\sqrt{\pi}}K_{\Delta,\ell}.
\end{equation}
which is the same as the spin-$0$ case \eqref{dsbubonorm} as a consequence of equation \eqref{wightmanharmspinl}.

With the above dictionary we can straightforwardly translate the results of \cite{Sleight:2017fpc} for the tree-level three-point Witten diagrams of a generic triplet of totally symmetric spinning fields in EAdS$_{d+1}$ into late-time three-point functions for the same triplet of spinning fields in dS$_{d+1}$. We need only work out the Fourier transform of the spinning three-point conformal structures appearing in each Witten diagram, which is straightforward using the Mellin-Barnes representation.\footnote{This is demonstrated in appendix \ref{Appendix: Fourier Transform} for the spinning three-point conformal structures considered in this work.} The contributions to the corresponding late-time three-point function from the $+$ and $-$ branches of the in-in contour are then obtained simply by multiplying with the appropriate phase \eqref{ExtLegPh} and re-normalising the three-point function coefficient (e.g. equation (3.29) in \cite{Sleight:2017fpc}) with \eqref{overallconstspinl} and \eqref{dsbubonormspinl}:
\begin{subequations}
\begin{align}\label{BAdS}
    {\sf B}\left(\ell_j,n_j,\Delta_j\right) &\to {\cal N}_3\,{\cal B}\left(\ell_j,n_j,\Delta_j\right),\\ \label{Btilde}
    {\cal B}\left(\ell_j,n_j,\Delta_j\right)&={\sf B}\left(\ell_j,n_j,\Delta_j\right)\prod^3_{j=1}\frac{\Gamma\left(\Delta_j-\tfrac{d}{2}+1\right)}{\sqrt{\pi}},\\
   {\cal N}_N&=\prod^N_{j=1}{\cal N}_{\nu_j,\ell_j}\left(\eta_0\right).
\end{align}
\end{subequations}
This is considered in detail in section \ref{sec::3pt}. The $\left\{\Delta_j,\ell_j\right\}$ with $j=1,2,3$ denote the scaling dimensions and spins of the triplet fields participating in the three-point interaction. The variables $n_j$ label the three-point conformal structure concerned (see \cite{Sleight:2017fpc}) and will not play a role in this work since we focus on three-point Witten diagrams involving only a single spin-$\ell$ field, for which $n_1=n_2=n_3=0$. 

As we shall see in section \ref{sec:FourPoint}, the split representation \eqref{spinlharm} of the spin-$\ell$ Harmonic function allows us, via the analytic continuation \eqref{bubumbgen}, to obtain expressions for the late-time exchange four-point functions of spinning fields in dS$_{d+1}$ simple from the above results for tree-level three-point Witten diagrams.

\section{Three-point correlators}
\label{sec::3pt}

In this section we consider late-time three-point functions in dS$_{d+1}$ at tree-level. We show that they can be obtained solely from the knowledge of the corresponding Witten diagrams in EAdS$_{d+1}$ using the dictionary detailed in \S \ref{subsec::SKdic}. In particular, from the Mellin-Barnes representation of the Fourier-transformed Witten diagram, the result for the de Sitter late-time correlator in Fourier space can be obtained by multiplying with the appropriate interference factor. We carry out this analysis both for correlators involving only scalar fields (in \S \ref{subsec::genextsca}) and for correlators involving a single spin-$\ell$ field and two scalar fields (in \S \ref{subsec::spinningads3pt}). We shall present the more general case of correlators with more than one spinning fields in \cite{ToAppear2}, which simply requires to Fourier transform the three-point Witten diagrams given in \cite{Sleight:2017fpc}. In \S \ref{subsec::3ptexamples} we consider some examples in which the three-point functions simplify, which includes conformally coupled and massless scalars. In \S \ref{subsec::softlimit2pt} we demonstrate the utility of the Mellin-Barnes representation in taking the soft limit of both scalar and spinning external legs.

\subsection{General External Scalars}
\label{subsec::genextsca}

In this section we consider the cubic interaction $\phi_1\phi_2\phi_3$ of general scalars $\phi_k$ with scaling dimension $\Delta_k=\tfrac{d}{2}+i\nu_k$, which is unique on-shell up to total derivatives. We shall demonstrate how to obtain the late-time three-point correlator from the corresponding three-point Witten diagram from Euclidean anti-de Sitter space, using the dictionary given in \S \ref{subsec::SKdic}. Strictly speaking, in the following we assume that $\Delta_k$ lie on the Principal Series, i.e. $\nu_k \in \mathbb{R}$, though results for other representations\footnote{Complementary series results are connected to the principal series and can be obtained with no major problem. Some additional subtleties arise for discrete and exceptional series, as we shall discuss.} can be obtained with due care about the analytic continuation of $\nu_k$, as we shall discuss in detail in \S\ref{subsec::someparticcases} and touch upon briefly in \S \ref{subsec::3ptexamples}.

In position space, the three-point Witten diagram in EAdS$_{d+1}$ reads \cite{Muck:1998rr,Freedman:1998tz}: 
\begin{subequations}\label{scwd3pt}
\begin{align}
    \langle \mathcal{O}_{\Delta_1}(\vec{x}_1)\mathcal{O}_{\Delta_2}(\vec{x}_2)\mathcal{O}_{\Delta_3}(\vec{x}_3)\rangle&={\cal B}\left({\bf 0};{\bf 0};\Delta_1,\Delta_2,\Delta_3\right)\,I_{\Delta_1,\Delta_2,\Delta_3}\left(\vec{x}_1,\vec{x}_2,\vec{x}_3\right),\label{scwd3pta}\\
   I_{\Delta_1,\Delta_2,\Delta_3}\left(\vec{x}_1,\vec{x}_2,\vec{x}_3\right)&=\frac{1}{(x_{12}^2)^{\tfrac{\Delta_1+\Delta_2-\Delta_3}2}(x_{23}^2)^{\tfrac{\Delta_2+\Delta_3-\Delta_1}2}(x_{31}^2)^{\tfrac{\Delta_3+\Delta_1-\Delta_2}2}}\,. \label{scwd3ptb}
\end{align}
\end{subequations}
where the function \eqref{scwd3ptb} is fixed by conformal symmetry while its coefficient in \eqref{scwd3pta} arises from the integration over the volume of EAdS$_{d+1}$ and, in the view of the analytic continuation to dS$_{d+1}$, we used the normalisation \eqref{Btilde}. In appendix \ref{Appendix: Fourier Transform} it is explained how to derive the Fourier transform of the above, which for general $\Delta_i$ is given by the following Mellin-Barnes integral:
\begin{subequations}
\begin{align}
 \hspace*{-0.4cm}   \langle \mathcal{O}_{\Delta_1}(\vec{k}_1)\mathcal{O}_{\Delta_2}(\vec{k}_2)\mathcal{O}_{\Delta_3}(\vec{k}_3)\rangle &= \left(2\pi\right)^d \delta^{(d)}\left(\vec{k}_1+\vec{k}_2+\vec{k}_3\right) \langle \mathcal{O}_{\Delta_1}(\vec{k}_1)\mathcal{O}_{\Delta_2}(\vec{k}_2)\mathcal{O}_{\Delta_3}(\vec{k}_3)\rangle^\prime,\\
  \hspace*{-0.4cm}  \langle \mathcal{O}_{\Delta_1}(\vec{k}_1)\mathcal{O}_{\Delta_2}(\vec{k}_2)\mathcal{O}_{\Delta_3}(\vec{k}_3)\rangle^\prime&= {\cal B}\left({\bf 0};{\bf 0};\Delta_1,\Delta_2,\Delta_3\right)I_{\Delta_1,\Delta_2,\Delta_3}(\vec{k}_1,\vec{k}_2,\vec{k}_3) \label{mb3ptwitten} \\
     &\hspace{-50pt}=\int [ds]_3\,i\pi \delta\left(\tfrac{d}{4}-s_1-s_2-s_3\right)\,\rho_{\nu_1,\nu_2,\nu_3}(s_1,s_2,s_3)\prod^3_{j=1}\left(\frac{k_j}{2}\right)^{-2s_j+i\nu_j},\nonumber
\end{align}
\end{subequations}
where we defined 
\begin{align}\label{rho3}
    \rho_{\nu_1,\nu_2,\nu_3}(s_1,s_2,s_3)= \prod^3_{j=1}\frac{1}{2\sqrt{\pi}}\,\Gamma(s_j+ \tfrac{i\nu_j}2)\Gamma(s_j-\tfrac{i\nu_j}2)\,,
\end{align}
and employed the shorthand notation
\begin{align}\label{rho3}
    \int [ds]_n=\int^{i\infty}_{-i\infty}  \frac{ds_1}{2\pi i}\,... \frac{ds_n}{2\pi i}.
\end{align}
The above two variables Mellin-Barnes integral is Appell's function $F_4$ \cite{appell1880series,AppelletKampe} which, up to a constant coefficient, can be \emph{defined} by the Mellin-Barnes integral above. Conformal symmetry in fact requires momentum-space scalar correlators to be given by Appell's $F_4$ function up to a coefficient \cite{Coriano:2013jba,Bzowski:2013sza}, which was automatically implemented in the above by starting from the conformal structure \eqref{scwd3ptb} in position space. The above Mellin form is advantageous over the Appel representation since at the level of the Mellin-Barnes representation \eqref{mb3ptwitten}, conformal symmetry fixes uniquely the locations of the poles in the Mellin variables, including the ones associated to the Dirac delta distribution. Furthermore, this representation for the correlator also follows from the bulk calculation for the Witten diagram in Fourier space (see \cite{Charlotte}):
\begin{equation}
     \langle \mathcal{O}_{\Delta_1}(\vec{k}_1)\mathcal{O}_{\Delta_2}(\vec{k}_2)\mathcal{O}_{\Delta_3}(\vec{k}_3)\rangle = \int^{\infty}_{0} \frac{dz}{z^{d+1}} {\cal K}_{\Delta_1}(z;\vec{k}_1){\cal K}_{\Delta_2}(z;\vec{k}_2){\cal K}_{\Delta_3}(z;\vec{k}_3),
\end{equation}
 by employing the Mellin-Barnes representation \eqref{MBbubosc} of the bulk-to-boundary propagators, which combine into the function \eqref{rho3} with the Mellin variable $s_j$ associated to the propagator of the scalar field $\phi_j$. The Dirac delta function in \eqref{mb3ptwitten} is generated from the integral over the radial co-ordinate $z$ of EAdS$_{d+1}$:\footnote{To be precise, because the Mellin integration contours run over the imaginary axis, the identity to use would be:
 \begin{equation}
     \int_0^\infty x^{s-1}=2\pi \delta(i s)\,,
 \end{equation}
 where along the integration contour from $-i\infty$ to $+i\infty$ one indeed recovers $is\in\mathbb{R}$. However, since up to change of variables we have:
 \begin{equation}
     \int_{-i\infty}^{+i\infty}\frac{ds}{2\pi i}\,[2\pi\delta(i s)]\,f(s)=f(0)\,,
 \end{equation}
 for simplicity we shall often write $\delta(i s)=i\,\delta(s)$.}
{\allowdisplaybreaks\begin{subequations}\label{Mellin Delta}
\begin{align}
    i\pi \delta\left(\tfrac{d}4-s_1-s_2-s_3\right)&=\lim_{z_0 \to 0}\left[\int^{\infty}_{z_0}\frac{dz}{z^{d+1}}\,z^{\frac{3d}{2}-2(s_1+s_2+s_3)}\right]\\&=\lim_{z_0\to0}\left[-\frac{1}{\tfrac{d}{2}-2 (s_1+s_2+s_3)}\,z_0^{\tfrac{d}{2}-2 (s_1+s_2+s_3)}\right]\,,
\end{align}
\end{subequations}}
\!where convergence of the $z$-integral restricts where the integration contours intersect the real axis. In particular: 
\begin{align}
        \mathfrak{Re}\left[s_1+s_2+s_3\right]&>\tfrac{d}{4}\,,
\end{align}
which requires the integration contour in $s_i$ passes on the right of the pole at $\tfrac{d}{4}-(s_1+s_2+s_3)\sim0$, which encodes the leading contribution in the limit $z_0 \to 0$ while $\Delta_k$ lie on the Principal Series.

The presence of the Dirac delta function \eqref{Mellin Delta} implies a freedom to add terms proportional to positive powers of $\tfrac{d}{4}-(s_1+s_2+s_3)$ in the Mellin-Barnes representation \eqref{mb3ptwitten}. In the bulk, this corresponds to the freedom of adding terms to a cubic vertex which vanish on-shell -- i.e. improvements -- which thus do not contribute to the three-point Witten diagram at tree level.

\paragraph{de Sitter late-time correlator.}

Given the Mellin-Barnes representation \eqref{mb3ptwitten} of the Witten diagram, using the dictionary detailed in \S \ref{subsec::SKdic} we can immediately write down the corresponding late-time correlator in dS$_{d+1}$. The Schwinger-Keldysh formalism prescribes that we sum over the time-ordered $\left(+\right)$ and anti-time-ordered $\left(-\right)$ branches of the in-in contour: 
\begin{equation}\label{0003ptdS}
    \langle \phi^{\left(\nu_1\right)}_{\vec{k}_1}\phi^{\left(\nu_2\right)}_{\vec{k}_2}\phi^{\left(\nu_3\right)}_{\vec{k}_3} \rangle^\prime =  {\cal N}_3\underbrace{\left[\mathcal{A}_{+|\nu_1,\nu_2,\nu_3}(\vec{k}_1,\vec{k}_2,\vec{k}_3)+\mathcal{A}_{-|\nu_1,\nu_2,\nu_3}(\vec{k}_1,\vec{k}_2,\vec{k}_3)\right]}_{\mathcal{A}_{\nu_1,\nu_2,\nu_3}(\vec{k}_1,\vec{k}_2,\vec{k}_3)},
\end{equation}
where 
\begin{multline}
\label{Apm_scalar}
    \mathcal{A}_{\pm|\nu_1,\nu_2,\nu_3}(\vec{k}_1,\vec{k}_2,\vec{k}_3) =  \pm i \int[ds]_3 \,i\pi \delta\left(\tfrac{d}{4}-s_1-s_2-s_3\right)\rho_{\nu_1,\nu_2,\nu_3}(s_1,s_2,s_3)\\ \times \prod^3_{j=1}\left(\tfrac{k_j}{2}\right)^{-2s_j+i\nu_j}e^{\delta^\pm_{\nu_j}\left(s_j\right)}\,.
\end{multline}
This was obtained from the Mellin-Barnes representation \eqref{mb3ptwitten} of the corresponding EAdS$_{d+1}$ Witten diagram by dressing each propagator with the appropriate phase factor as prescribed in equation \eqref{ExtLegPh} at the level of the Mellin-integrand. The $\pm$ factor multiplying the integral comes from inverting the range of integration $\left[0,-\infty\right] \rightarrow \left[-\infty,0\right]$ of $\eta$ for the anti-time-ordered ($-$) branch of the in-in contour. The factors of $i$ naturally arise from the analytic continuation of the volume form. 

Combining the contributions from the $+$ and $-$ contours, which differ only by a phase, gives
\begin{subequations}\label{000latetimedS}
\begin{align}
    \langle \phi^{\left(\nu_1\right)}_{\vec{k}_1}\phi^{\left(\nu_2\right)}_{\vec{k}_2}\phi^{\left(\nu_3\right)}_{\vec{k}_3} \rangle^\prime &= {\cal N}_3 \int [ds]_3 \, i\pi \delta\left(\tfrac{d}{4}-s_1-s_2-s_3\right)\rho_{\nu_1,\nu_2,\nu_3}(s_1,s_2,s_3) \\ & \hspace*{1cm} \times 2\sin \left(\pi \left(s_1+s_2+s_3+\tfrac{\left(\nu_1+\nu_2+\nu_3 \right)i}{2}\right)\right)\prod^3_{j=1}\left(\tfrac{k_j}{2}\right)^{-2s_j+i\nu_j}\nonumber \\ 
    &= {\cal N}_3\,2\sin \left(\pi \left(\tfrac{d}{4}+\tfrac{i\left(\nu_1+\nu_2+\nu_3 \right)}{2}\right)\right)\\ & \hspace*{1cm} \times \int[ds]_3 \, i\pi \delta\left(\tfrac{d}{4}-s_1-s_2-s_3\right)\rho_{\nu_1,\nu_2,\nu_3}(s_1,s_2,s_3)\prod^3_{j=1}\left(\tfrac{k_j}{2}\right)^{-2s_j+i\nu_j} \nonumber
\end{align}
\end{subequations}
where in the second equality we used the Dirac delta distribution to translate the analytic continuations \eqref{ExtLegPh} from EAdS$_{d+1}$ into an overall phase for the $\pm$ contributions \eqref{Apm_scalar}. At the end, the sinusoidal function nicely encodes the interference pattern. 
We note that, while conformal symmetry fixes the location of the poles in the Mellin integrand, the zeros, encoded in the sine function in \eqref{000latetimedS}, are fixed by the early time boundary conditions (Bunch-Davis in our case).

The expression \eqref{000latetimedS} is also obtained by simply evaluating the late-time correlator directly in de Sitter space using the Mellin-Barnes representation for the propagators \cite{Charlotte}. Here, we directly evaluated the Fourier transform of the known result \cite{Muck:1998rr,Freedman:1998tz} for tree-level three-point Witten diagrams of scalar fields in Euclidean anti-de Sitter space (which is most naturally given by a Mellin-Barnes integral) and applying the dictionary spelled out in \S \ref{subsec::SKdic} to each propagator at the level of the Mellin integrand to obtain the corresponding late-time correlator in de Sitter space. This approach also straightforwardly extends to correlators of spinning fields, where it is readily applicable for totally symmetric fields in general $d$ using the results derived in \cite{Sleight:2017fpc} for their tree-level three-point Witten diagrams. We shall demonstrate this in the following section for correlators involving a single totally symmetric spin-$\ell$ fields in general $d$, and discuss the more general spinning case in a forthcoming work \cite{ToAppear2}. 

\subsection{Two General Scalars and a Spin-$\ell$ field}
\label{subsec::spinningads3pt}

In this section we consider late-time correlators involving two general scalar fields $\phi_{1,2}$ and a field $\varphi_{\mu_1...\mu_\ell}$ of integer spin-$\ell$ and scaling dimension $\Delta_3$, whose tensor structure is fixed uniquely by conformal symmetry \cite{Ferrara:1973yt,Polyakov:1974gs,Osborn:1993cr,Erdmenger:1996yc,Costa:2011mg} up to a coefficient.\footnote{For three-point correlators involving a generic triplet of spinning fields, there are various tensor structures consistent with conformal symmetry \cite{Osborn:1993cr,Erdmenger:1996yc,Maldacena:2011nz,Costa:2011mg}.} In  position space it reads
\begin{equation}\label{00lposdb}
    \langle \langle \mathcal{O}_{\Delta_1}(\vec{x}_1)\mathcal{O}_{\Delta_2}(\vec{x}_2)\mathcal{O}_{\Delta_3}(\vec{x}_3;\vec{\xi}\,)\rangle \rangle =\,I_{\Delta_1,\Delta_2,\Delta_3-\ell}\left(\vec{x}_1,\vec{x}_2,\vec{x}_3\right)\underbrace{\left(\frac{\vec{\xi}\cdot \vec{x}_{31}}{x_{31}^2}-\frac{\vec{\xi}\cdot \vec{x}_{32}}{x_{23}^2}\right)^\ell}_{{\sf Y}_3^\ell}\,,
\end{equation}
where, following \cite{Sleight:2017fpc}, the notation $\langle \langle\bullet\rangle \rangle$ indicates that we have stripped off the Operator Product Expansion (OPE) coefficient. This is the scalar conformal structure \eqref{scwd3ptb} dressed with the conformally covariant tensor structure ${\sf Y}_3$. The Witten diagram generated by the vertex (which, up to total derivatives, is unique on-shell):
\begin{equation}\label{00lvertex}
    V_{0,0,\ell} = \phi_1 \nabla^{\mu_1} ... \nabla^{\mu_\ell} \phi_2\, \varphi_{\mu_1 ... \mu_\ell},
\end{equation}
is given by multiplying the three-point structure \eqref{00lposdb} by the coefficient \eqref{Btilde}: 
\begin{multline}\label{00leadswitten}
   \langle \mathcal{O}_{\Delta_1}(\vec{x}_1)\mathcal{O}_{\Delta_2}(\vec{x}_2)\mathcal{O}_{\Delta_3}(\vec{x}_3;\vec{\xi}\,)\rangle\\={\cal B}\left(0,0,\ell;{\bf 0};\Delta_1,\Delta_2,\Delta_3-\ell\right)  \langle \langle \mathcal{O}_{\Delta_1}(\vec{x}_1)\mathcal{O}_{\Delta_2}(\vec{x}_2)\mathcal{O}_{\Delta_3}(\vec{x}_3;\vec{\xi}\,)\rangle \rangle.
\end{multline}
As we detail in appendix \ref{Appendix: Fourier Transform}, through the replacement 
\begin{equation}\label{diffop}
   \vec{\xi}\cdot \vec{x}_{ij}\rightarrow i \vec{\xi} \cdot\vec{\pl}_{k_{ij}}\equiv i \vec{\xi}\cdot(\vec{\pl}_{k_i}-\vec{\pl}_{k_j})\,,
\end{equation}
the Fourier transform of the three-point confromal structure \eqref{00lposdb} can be expressed in the form of a differential operator acting on the Fourier transform of the scalar conformal structure \eqref{scwd3ptb}. The differential operator generates the tensorial structure, which is given by a polynomial in $\vec{\xi} \cdot k_i$, $i=1,2,3$. The naive application of the differential operator gives a cumbersome expression involving numerous terms, but the Mellin-Barnes representation \eqref{mb3ptwitten} affords some useful simplifications which we detail in appendix \ref{Appendix: Fourier Transform}. The final result for the Mellin-Barnes representation of the Witten diagram \eqref{00lposdb} in Fourier space is:\footnote{The tensorial structure, which is fixed by conformal symmetry, re-produces existing expressions for spinning conformal structures in Fourier space e.g. \cite{Mata:2012bx,Bzowski:2013sza,Arkani-Hamed:2015bza,Anninos:2017eib,Isono:2018rrb,Isono:2019ihz}.}
\begin{shaded}
\noindent \emph{Mellin-Barnes repesentation of the $0$-$0$-$\ell$ Witten diagram in Fourier space} \begin{multline}\label{00lmbfseads}
    \langle \mathcal{O}_{\Delta_1}(\vec{k}_1)\mathcal{O}_{\Delta_2}(\vec{k}_2)\mathcal{O}_{\Delta_3}(\vec{k}_3;\vec{\xi}\,)\rangle\\=\int[ds]_3\underbrace{\int^{\infty}_{0}\frac{dz}{z^{d+1}}\,z^{\tfrac{3d}{2}-2(s_1+s_2+s_3)+\ell}}_{i\pi \delta\left(\frac{d+2\ell}4-s_1-s_2-s_3\right)}\,\rho_{\nu_1,\nu_2,\nu_3}(s_1,s_2,s_3)\prod^3_{j=1}\left(\tfrac{k_j}{2}\right)^{-2s_j+i\nu_j}\\\times\underbrace{\sum_{\alpha=0}^\ell\binom{\ell}{\alpha}(-\vec{\xi}\cdot \vec{k}_3)^\alpha\sum_{\beta=0}^\alpha\binom{\alpha}{\beta}\, H_{\nu_1,\nu_2,\nu_3|\alpha,\beta}(s_1,s_2,s_3)\, \mathcal{Y}^{(\ell)}_{\nu_1,\nu_2,\nu_3|\alpha,\beta}(\vec{\xi}\cdot \vec{k}_1,\vec{\xi}\cdot \vec{k}_2)}_{{\sf p}^{\left(\ell\right)}_{\nu_1,\nu_2,\nu_3}(\vec{\xi} \cdot \vec{k}_1,\vec{\xi} \cdot \vec{k}_2,\vec{\xi} \cdot \vec{k}_3|s_1,s_2,s_3)}\,.
\end{multline}
\end{shaded}
\noindent This expression has some useful similarities to the analogous expression for the scalar three-point correlator \eqref{mb3ptwitten}. In particular, the first line is of the same form as the scalar correlator but with $d \rightarrow d+2\ell$. This shift originates from the factor of $z^{-\ell}$ in the spin-$\ell$ propagator \eqref{poinspinlbubo}, which when combined in the cubic vertex \eqref{00lvertex} is accompanied by a factor of $\left(z^{2}\right)^{\ell}$ coming from the index contraction. The second line is the tensor structure generated by the action of the differential operators \eqref{diffop}, which is encoded in the polynomial ${\sf p}^{\left(\ell\right)}_{\nu_1,\nu_2,\nu_3}(\vec{\xi}\cdot \vec{k}_i|s_i)$ in $\vec{\xi}\cdot \vec{k}_i$. The dependence on $\vec{\xi} \cdot \vec{k}_{1}$ and $\vec{\xi} \cdot \vec{k}_{2}$ given by the two-variable polynomial:
\begin{multline}\label{yab}
    \mathcal{Y}^{(\ell)}_{\nu_1,\nu_2,\nu_3|\alpha,\beta}(\vec{\xi}\cdot \vec{k}_1,\vec{\xi} \cdot \vec{k}_2)=(-i)^\ell\frac{\left(\tfrac{4-d-2 \ell-2i (\nu_1-\nu_2+\nu_3)}{4} \right)_\beta \left(\tfrac{d-4 \alpha+4 \beta+2 \ell-2 i (\nu_1-\nu_2-\nu_3)}{4}\right)_{\alpha-\beta}}{\left(\tfrac{d}{2}+i \nu_3-1\right)_{\ell-\alpha}\left(\frac{d}{2}+i \nu_3+\ell-\alpha-1\right)_\alpha}\\\times\sum_{n=0}^{\ell-\alpha}\left(\tfrac{d-4 \beta+2 \ell-4 n+2 i \nu_1-2 i \nu_2+2 i \nu_3}{4}\right)_n \left(\tfrac{d+4 \beta-2 \ell+4 n-2 i \nu_1+2 i \nu_2+2 i \nu_3}{4}\right)_{\ell-\alpha-n} \\\times \binom{\ell-\alpha}{n}\,(\vec{\xi}\cdot \vec{k}_1)^{\ell-\alpha-n}\,(-\vec{\xi}\cdot \vec{k}_2)^n.
\end{multline}
The dependence of the polynomial ${\sf p}^{\left(\ell\right)}_{\nu_1,\nu_2,\nu_3}(\vec{\xi}\cdot \vec{k}_i|s_i)$ on the Mellin-variables is given by the function
\begin{align}\label{HHpochh}
    H_{\nu_1,\nu_2,\nu_3|\alpha,\beta}(s_1,s_2,s_3)&=\frac{\left(s_1+\frac{i \nu_1}{2}\right)_{\alpha-\beta} \left(s_2+\frac{i \nu_2}{2}\right)_\beta}{\left(s_3+\frac{i \nu_3}{2}-\alpha\right)_\alpha}.
\end{align}

\paragraph{Recursion relations.} Interestingly, the Pochhammer factors in $H_{\nu_1,\nu_2,\nu_3|\alpha,\beta}(s_1,s_2,s_3)$ are precisely of the right form to telescopically combine with the function $\rho_{\nu_1,\nu_2,\nu_3}(s_1,s_2,s_3)$: 
\begin{align}\label{scdecomp}
  \mathfrak{A}_{\nu_1,\nu_2,\nu_3|\alpha,\beta}^{(x)}\left(\vec{k}_1,\vec{k}_2,\vec{k}_3\right)&=\int[ds]_3\,i\pi \delta\left(\tfrac{x}4-s_1-s_2-s_3\right)  \rho_{\nu_1,\nu_2,\nu_3}(s_1,s_2,s_3)\\\nonumber
  &\hspace{120pt}\times H_{\nu_1,\nu_2,\nu_3|\alpha,\beta}(s_1,s_2,s_3)\prod^3_{j=1}\left(\frac{k_j}{2}\right)^{-2s_j+i\nu_j} \\\nonumber
  &= \int[ds^\prime]_3\,i\pi \delta\left(\tfrac{x}4-s^\prime_1-s^\prime_2-s^\prime_3\right) \rho_{\nu^\prime_1,\nu^\prime_2,\nu^\prime_3}(s^\prime_1,s^\prime_2,s^\prime_3)\prod^3_{j=1}\left(\frac{k_j}{2}\right)^{-2s^\prime_j+i\nu^\prime_j},
\end{align}
\noindent where 
\begin{subequations}
\begin{align}\label{ext3pt1}
\nu^\prime_1&=\nu_1-i\left(\alpha-\beta\right), 
\\ \nu^\prime_2&=\nu_2-i\beta,\label{ext3pt2}\\ \nu^\prime_3&=\nu_3+i\alpha, \label{ext3pt3}
\end{align}
\end{subequations}
and we have performed the change of variables $s^\prime_1= s_1+\tfrac{\alpha-\beta}{2}$, $s^\prime_2=s_2+\tfrac{\beta}{2}$ and $s^\prime_3=s_3-\tfrac{\alpha}{2}$, noting that $s^\prime_1+s^\prime_2+s^\prime_3=s_1+s_2+s_3$. In this way the Fourier transform of the Witten diagram \eqref{00leadswitten} can be expressed as a sum of scalar Witten diagrams \eqref{scwd3pt} with integer-shifted scaling dimensions, each of which is dressed with a given tensor structure:
\begin{multline}\label{00lscalardecomp}
    \langle \mathcal{O}_{\nu_1}(\vec{k}_1)\mathcal{O}_{\nu_2}(\vec{k}_2)\mathcal{O}_{\nu_3,\ell}(\vec{k}_3;\vec{\xi}\,)\rangle = \sum_{\alpha=0}^\ell\binom{\ell}{\alpha}(-\vec{\xi}\cdot \vec{k}_3)^\alpha\sum_{\beta=0}^\alpha\binom{\alpha}{\beta}\mathcal{Y}^{(\ell)}_{\nu_1,\nu_2,\nu_3|\alpha,\beta}(\vec{\xi}\cdot \vec{k}_1,\vec{\xi}\cdot \vec{k}_2)\\
    \times \underbrace{\langle \mathcal{O}_{\nu_1-i\left(\alpha-\beta\right)}(\vec{k}_1)\mathcal{O}_{\nu_2-i\beta}(\vec{k}_2)\mathcal{O}_{\nu_3+i\alpha}(\vec{k}_3)\rangle\big|_{d\to d+2\ell}}_{\mathfrak{A}_{\nu_1,\nu_2,\nu_3|\alpha,\beta}^{(x)}\left(\vec{k}_1,\vec{k}_2,\vec{k}_3\right)}.
\end{multline}

The telescopic property of the Mellin-Barnes integrals \eqref{scdecomp} makes manifest various recursion relations that exist among them. For example, the shifts in the scaling dimensions \eqref{ext3pt1} and \eqref{ext3pt2} associated to the scalar fields can be simply lifted from the Mellin-Barnes integral via
\begin{multline}\label{alphabeta_fromgamma0}
   {\mathfrak{A}}_{\nu_1,\nu_2,\nu_3|\alpha,\beta}^{(x)}(\vec{k}_1,\vec{k}_2,\vec{k}_3)\\=(-1)^{\alpha} k_1^{2 (\alpha-\beta+i \nu_1)}k_2^{2 (\beta+i \nu_2)}\pl_{k_1^2}^{\alpha-\beta}\pl_{k_2^2}^{\beta}\left[k_1^{-2i\nu_1}k_2^{-2i\nu_2}{\mathfrak{A}}_{\nu_1,\nu_2,\nu_3}^{(x,\alpha)}(\vec{k}_1,\vec{k}_2,\vec{k}_3)\right]\,,
\end{multline}
where 
\begin{multline}\label{scdecomp2}
  \mathfrak{A}_{\nu_1,\nu_2,\nu_3}^{(x,\alpha)}(\vec{k}_1,\vec{k}_2,\vec{k}_3) \equiv \int[ds]_3\,i\pi \delta\left(\tfrac{x}4-s_1-s_2-s_3\right)\\\times  \frac{\rho_{\nu_1,\nu_2,\nu_3}(s_1,s_2,s_3)}{\left(s_3+\frac{i\nu_3}2-\alpha\right)_\alpha}\prod^3_{j=1}\left(\frac{k_j}{2}\right)^{-2s_j+i\nu_j},
\end{multline}
which shifts only the scaling dimension associated to the spinning field. This in turn can be generated from the expression with fixed $\alpha=\ell$ via
\begin{align}\label{seed_recursion_full0}
    {\mathfrak{A}}_{\nu_1,\nu_2,\nu_3}^{(x,\alpha)}(\vec{k}_1,\vec{k}_2,\vec{k}_3)=\pl_{\lambda_1}^{\ell-\alpha}\Big[\lambda ^{-\alpha+\frac{i \nu_3 }{2}-\frac{i \nu_1}{2}-\frac{i \nu_2}{2}+\frac{x}{4}-1}{\mathfrak{ A}}_{\nu_1,\nu_2,\nu_3}^{(x,\ell)}(\lambda^{1/2} \vec{k}_1,\lambda^{1/2} \vec{k}_2,\vec{k}_3)\Big]_{\lambda=1}\,.
\end{align}
In other words, the full correlator \eqref{00lscalardecomp} can be generated from the Mellin-Barnes integral \eqref{scdecomp2} with $\alpha=\ell$ through the recursion relations \eqref{alphabeta_fromgamma0} and \eqref{seed_recursion_full0}. This is useful for scaling dimensions where the integral \eqref{scdecomp2} simplifies with respect to \eqref{scdecomp}.

Similarly we can write down recursion relations which raise and lower the scaling dimensions of the scalar fields by integer units. In particular, the operations
\begin{subequations}\label{raisingop0}
\begin{align} 
   & {\mathfrak{A}}_{\nu_1-i,\nu_2,\nu_3|\alpha,\beta}^{(x)}(\vec{k}_j)=-\frac{1}{2} k_1^{1+2 i \nu_1+2(\alpha-\beta)}\partial_{k_1}\left[k_1^{-2i\nu_1-2(\alpha-\beta)}{\mathfrak{A}}_{\nu_1,\nu_2,\nu_3|\alpha,\beta}^{(x-2)}(\vec{k}_j)\right]\,,\\ 
   & {\mathfrak{A}}_{\nu_1,\nu_2-i,\nu_3|\alpha,\beta}^{(x)}(\vec{k}_j)=-\frac{1}{2} k_2^{1+2 i \nu_2+2\beta}\partial_{k_2}\left[k_2^{-2i\nu_2-2\beta}{\mathfrak{A}}_{\nu_1,\nu_2,\nu_3|\alpha,\beta}^{(x-2)}(\vec{k}_j)\right]\,,
\end{align}
\end{subequations}
increase the external scaling dimensions by an integer, while the lowering operators are given by:
\begin{subequations}\label{recursion10}
\begin{align}
    {\mathfrak{A}}_{\nu_1+i,\nu_2,\nu_3|\alpha,\beta}^{(x)}(\vec{k}_j)&=-\frac{2}{k_1}\partial_{k_1}{\mathfrak{A}}_{\nu_1,\nu_2,\nu_3|\alpha,\beta}^{(x-2)}(\vec{k}_j)\,,\\
    {\mathfrak{A}}_{\nu_1,\nu_2+i,\nu_3|\alpha,\beta}^{(x)}(\vec{k}_j)&=-\frac{2}{k_2}\partial_{k_2}{\mathfrak{A}}_{\nu_1,\nu_2,\nu_3|\alpha,\beta}^{(x-2)}(\vec{k}_j)\,.
\end{align}
\end{subequations}

The recursion relations discussed in this section are useful for scaling dimensions where the initial or ``seed" Mellin-Barnes integral simplifies. We shall consider some examples of this type in section \ref{subsec::3ptexamples}. These recursion relations also carry over at the four-point level, which we shall discuss in further detail in section \ref{subsec::recursion}.

\paragraph{de Sitter late-time correlator.} The expression \eqref{00lscalardecomp} for the $0$-$0$-$\ell$ Witten diagram as a sum of scalar Witten diagrams allows us to immediately write down the corresponding late-time correlator in dS$_{d+1}$ from the result \eqref{000latetimedS} for general scalars. In particular, the tensor structure on the first line of \eqref{00lscalardecomp} is unchanged in going from EAdS$_{d+1}$ to dS$_{d+1}$, which can also be seen at the level of the propagators due to the independence of the phases \eqref{ExtLegPh} from the spin of the field. For the contributions from the $+$ and $-$ branches of the branches of the in-in contour, this gives
\begin{multline}
    \mathcal{A}_{\pm|\nu_1,\nu_2,\nu_3}\left(\vec{k}_1,\vec{k}_2,\vec{k}_3;\vec{\xi}\,\right) =\pm i e^{\mp i \pi \left(\tfrac{d}{4}+\tfrac{(\nu_1+\nu_2+\nu_3+\ell)i}{2}\right)} \int \left[ds\right]_3
   i\pi \delta\left(\tfrac{d+2\ell}4-s_1-s_2-s_3\right)\\\times {\sf p}^{\left(\ell\right)}_{\nu_1,\nu_2,\nu_3}(\vec{\xi} \cdot \vec{k}_1,\vec{\xi} \cdot \vec{k}_2,\vec{\xi} \cdot \vec{k}_3|s_1,s_2,s_3) \rho_{\nu_1,\nu_2,\nu_3}(s_1,s_2,s_3)\prod^3_{j=1}\left(\tfrac{k_j}{2}\right)^{-2s_j+i\nu_j},
\end{multline}
where the phase factor now also depends on the spin $\ell$ due to the Dirac delta-function.

The full late-time correlator is therefore:
\begin{shaded}
\noindent \emph{Mellin-Barnes representation of the $0$-$0$-$\ell$ late-time correlation function in dS$_{d+1}$}
\begin{multline}\label{00lltmb}
    \langle \phi^{\left(\nu_1\right)}_{\vec{k}_1}\phi^{\left(\nu_2\right)}_{\vec{k}_2}\varphi^{\left(\nu_3\right)}_{\vec{k}_3} \rangle^\prime=2{\cal N}_3 \sin\left(\pi \left(\tfrac{d}{4}+\tfrac{i(\nu_1+\nu_2+\nu_3+\ell)}{2}\right)\right)
    \int \left[ds\right]_3
   i\pi \delta\left(\tfrac{d+2\ell}4-s_1-s_2-s_3\right)\\\times {\sf p}^{\left(\ell\right)}_{\nu_1,\nu_2,\nu_3}(\vec{\xi} \cdot \vec{k}_1,\vec{\xi} \cdot \vec{k}_2,\vec{\xi} \cdot \vec{k}_3|s_1,s_2,s_3) \rho_{\nu_1,\nu_2,\nu_3}(s_1,s_2,s_3)\prod^3_{j=1}\left(\tfrac{k_j}{2}\right)^{-2s_j+i\nu_j},
\end{multline}
\end{shaded}
\noindent Equivalently the result can be manifestly expressed as a sum of scalar correlators \eqref{mb3ptwitten} as in \eqref{00lscalardecomp}:
\begin{multline}\label{00lscdecomp}
    \langle \phi^{\left(\nu_1\right)}_{\vec{k}_1}\phi^{\left(\nu_2\right)}_{\vec{k}_2}\varphi^{\left(\nu_3\right)}_{\vec{k}_3} \rangle^\prime=2{\cal N}_3 \sin\left(\pi \left(\tfrac{d}{4}+\tfrac{i(\nu_1+\nu_2+\nu_3+\ell)}{2}\right)\right)
    \sum_{\alpha=0}^\ell\sum_{\beta=0}^\alpha\binom{\ell}{\alpha}\binom{\alpha}{\beta}(-\vec{\xi}\cdot \vec{k}_3)^\alpha\\
    \times\mathcal{Y}^{(\ell)}_{\nu_1,\nu_2,\nu_3|\alpha,\beta}(\vec{\xi}\cdot \vec{k}_1,\vec{\xi}\cdot \vec{k}_2) \underbrace{\langle \mathcal{O}_{\nu_1-i\left(\alpha-\beta\right)}(\vec{k}_1)\mathcal{O}_{\nu_2-i\beta}(\vec{k}_2)\mathcal{O}_{\nu_3+i\alpha}(\vec{k}_3)\rangle\big|_{d\to d+2\ell}}_{\mathfrak{A}_{\nu_1,\nu_2,\nu_3|\alpha,\beta}^{(d+2\ell)}(\vec{k}_1,\vec{k}_2,\vec{k}_3)},
\end{multline}
and the recursion relations discussed in the previous section continue to apply. In the following we discuss the helicity decomposition.

\paragraph{Expansion into Helicity components.} Before concluding this section let us discuss the helicity decomposition of the correlator \eqref{00lltmb}, which can be obtained along the same lines as for the two-point functions in section \S \ref{2pt}. For the $0$-$0$-$\ell$ conformal structures \eqref{00lmbfseads} we have a single polarization vector $\vec{\xi}$, which we can parameterise as:
\begin{align}
    \vec{\xi}&=(\vec{\xi}_\perp,i)\,, \qquad \vec{k}_3=(0,k_3)\,, \qquad (\xi_\perp)^2=1,
\end{align}
so that
\begin{align}
    \vec{\xi}&=\vec{\xi}_\perp+i\,\hat{k}_3\,,& \vec{\xi}_\perp\cdot\vec{k}_3&=0\,,& \vec{\xi}\cdot\vec{k}_3&=i k_3\,.
\end{align}
Employing momentum conservation one can then expand:
\begin{subequations}\label{kinematic_helicity}
\begin{align}
    \vec{\xi}\cdot \vec{k}_1&=+\frac12\,\vec{\xi}_\perp\cdot \vec{q}_{12}+i\, \hat{k}_3\cdot \vec{k}_1=+\frac{1}{2}\left[z-i k_3(1+pq)\right]\,,\\
    \vec{\xi}\cdot \vec{k}_2&=-\frac12\,\vec{\xi}_\perp\cdot \vec{q}_{12}+i \hat{k}_3\cdot \vec{k}_2=-\frac{1}{2}\left[z+i k_3(1-pq))\right]\,,
\end{align}
\end{subequations}
where
\begin{equation}
    \vec{q}_{12}=\vec{k}_1-\vec{k}_2, \qquad z={\xi}_\perp\cdot \vec{q}_{12},
\end{equation}
and for convenience (following \cite{Arkani-Hamed:2015bza}) we introduced
\begin{align}\label{pqdef}
    p_{12}=\frac{k_1+k_2}{k_3}\,,\qquad q_{12}=\frac{k_1-k_2}{k_3}\,.
\end{align}
The helicity decomposition of the correlator \eqref{00lltmb} can be obtained by expanding the polynomials \eqref{yab} in powers of $z$ (which is straightforward using the above replacements): 
\begin{align}
     \mathcal{Y}^{(\ell)}_{\nu_1,\nu_2,\nu_3|\alpha,\beta}(\vec{\xi}\cdot \vec{k}_1,\vec{\xi}\cdot \vec{k}_2)&=\sum^\ell_{r=0}\mathfrak{y}^{\left(\ell\right)}_{\nu_1,\nu_2,\nu_3|\alpha,\beta}\left(p,q\right) z^r,
\end{align}
and then decomposing each power in terms of Gegenbauer polynomials using the inversion formula \eqref{gegenbauerinv}:
\begin{subequations}
\begin{align}
    z^r&=\sum^r_{m=0} \mathfrak{Z}_{r,m} \Xi_{m}\left(z\right),\\
   \mathfrak{Z}_{r,m} &=\frac{2^{d+m-4} \Gamma \left(\frac{d}{2}-\frac{3}{2}\right) \Gamma \left(\frac{d}{2}+m-\frac{1}{2}\right)}{\pi  \Gamma (d+m-3)}\,\int_{-1}^{1}dz\,(1-z^2)^{\tfrac{d-4}2}\,z^r\,C_{m}^{(\frac{d-3}2)}(z)\\&=\frac{2^{m-r-1}r! \left(1+(-1)^{r+m}\right) \Gamma \left(\frac{d-1}{2}+m\right)}{m!\, \Gamma \left(\tfrac{2+r-m}{2}\right) \Gamma \left(\tfrac{d+r+m-1}{2}\right)}.
\end{align}
\end{subequations}
This gives the helicity decomposition
\begin{align}\label{HD00l}
    {\sf p}^{\left(\ell\right)}_{\nu_1,\nu_2,\nu_3}(\vec{\xi} \cdot \vec{k}_1,\vec{\xi} \cdot \vec{k}_2,\vec{\xi} \cdot \vec{k}_3|s_1,s_2,s_3)&=\sum^\ell_{m=0}\mathfrak{p}^{\left(\ell\right)}_m\left(p,q,k_3|s_1,s_2,s_3\right) \Xi_m\left(z\right),
\end{align}
where 
\begin{multline}
    \mathfrak{p}_m^{(\ell)}(p,q|s_1,s_2,s_3)=\sum_{\alpha=0}^\ell\sum_{\beta=0}^{\alpha} \binom{\ell}{\alpha}\binom{\alpha}{\beta}\,(-i k_3)^\alpha\,H_{\nu_1,\nu_2,\nu_3|\alpha,\beta}(s_1,s_2,s_3)\\ \times \sum_{r=0}^\ell \mathfrak{y}_{\nu_1,\nu_2,\nu_3|\alpha,\beta}^{(\ell)}(p,q)\,\mathfrak{Z}_{r,m}\,,
\end{multline}
where $z=\sigma\cos(\theta)={\xi_\perp}\cdot \vec{q}_{12}$ with $\sigma^2=-k_3^2(1-p_{12}^2)(1-q_{12}^2)$. The highest helicity component receives contributions only from the $\alpha=\beta=0$ term, and takes the simple form:
\begin{equation}
   \mathfrak{p}_\ell^{(\ell)}(p,q|s_1,s_2,s_3)=\left(-\frac{i}{2}\right)^\ell\,.
\end{equation}
In the following we give a couple of lower spin examples. We moreover set $\nu_1=\nu_2=\mu$ and $\nu_3=\nu$ just to simplify the expressions.

\paragraph{$\ell=1$:}
\begin{align}
\mathfrak{p}_0^{(1)}(p_{12},q_{12}|s_1,s_2,s_3)=-\frac{k_3\, p_{12}\, q_{12}}{2}+\frac{k_3 (s_2-s_1)}{i \nu +2 s_3-2}\,.
\end{align}

\paragraph{$\ell=2$:}
\begin{subequations}
\begin{align}
    \mathfrak{p}_1^{(2)}(p_{12},q_{12}|s_1,s_2,s_3)&=\frac{ik_3 p_{12} q_{12}}{2}+\frac{i\,k_3 (s_1-s_2)}{i\nu +2  s_3-2 }\,,\\
    \mathfrak{p}_0^{(2)}(p_{12},q_{12}|s_1,s_2,s_3)&=\frac{k_3^2}{8}\left(-\frac{4}{2 i\nu +1}+3p_{12}^2 q_{12}^2-p_{12}^2-q_{12}^2+1\right)\\\nonumber
    &+\frac{k_3^2 (-2i \mu +2 i\nu  p_{12} q_{12} (s_1-s_2)+ s_1 (p_{12} q_{12}-2)- s_2 (p_{12} q_{12}+2))}{(2 i\nu +1) (i \nu +2 s_3-2)}\\\nonumber
    &+\frac{k_3^2 \left(2 \mu ^2+2 i\nu  \left(i \mu-2 s_1 s_2+s^2_1+s_1+s^2_2+s_2\right)\right)}{(2 i\nu +1) (i \nu +2 s_3-4) (i \nu +2 s_3-2)}\,\nonumber\\
    &-\frac{\left(6 s_1 s_2+s^2_1+s_1+s^2_2+s_2\right)k^2_3}{(2 i\nu +1) (i \nu +2 s_3-4) (i \nu +2 s_3-2)}\nonumber\\
    &-\frac{i\mu  (4 s_1+4 s_2+1)k^2_3}{(2 i\nu +1) (i \nu +2 s_3-4) (i \nu +2 s_3-2)}.\nonumber
\end{align}
\end{subequations}

\subsection{Examples}
\label{subsec::3ptexamples}

In the preceding sections we considered three-point correlators of two scalars and a spin-$\ell$ field with generic scaling dimensions, which are given by the Mellin-Barnes integral \eqref{00lltmb}. For certain special scaling dimensions that are away from the Principal Series, the Mellin representation simplifies. In the following we illustrate some examples of this type.

\paragraph{Two conformally coupled scalars.} The Mellin representation simplifies when one or more of the fields is conformally coupled, which corresponds to $\nu=\frac{i}{2}$. This is because when $\nu=\frac{i}{2}$ the two $\Gamma$-functions in the Mellin-Barnes representation of the propagator \eqref{MBbubosc} are replaced with a single $\Gamma$-function by virtue of the Legendre duplication formula, which allows to lift the corresponding Mellin integral. 

Let us suppose that the two scalars in the three-point function \eqref{00lltmb} are conformally coupled,  $\nu_{1,2}=\tfrac{i}2$. The seed Mellin-Barnes integral \eqref{scdecomp2} in this case reads
\begin{align} \nonumber 
    \hspace*{-0.5cm} \mathfrak{A}_{\frac{i}{2},\frac{i}{2},\nu_3}^{(x,\alpha)}(\vec{k}_1,\vec{k}_2,\vec{k}_3) &= \frac{2}{\sqrt{\pi}} \int \left[ds\right]_3\,i\pi \delta \left(\tfrac{x}{4}-s_1-s_2-s_3\right)\Gamma \left(2 s_1-\tfrac{1}{2}\right) \Gamma \left(2 s_2-\tfrac{1}{2}\right)\,\Gamma \left(s_3-\tfrac{i \nu_3}{2}\right)\\ & \hspace*{2.5cm}\times\, \Gamma \left(s_3-\alpha+\tfrac{i \nu_3}{2}\right)\,k_1^{-2s_1-\frac12}k_2^{-2s_2-\frac12}\left(\frac{k_3}2\right)^{-2s_3+i\nu_3}\\ \nonumber
    & = \frac{2}{\sqrt{\pi}} \frac{p_{12}}{k_1k_2} \left(\frac{k_3}{2}\right)^{i\nu_3-\tfrac{x}{2}+1} \int^{i\infty}_{-i\infty} \frac{ds}{2\pi i}\, \Gamma (2 s-1)  \Gamma \left(\tfrac{x}{4}-s-\tfrac{i \nu_3}{2}\right) \Gamma \left(\tfrac{x}{4}-s-\alpha+\tfrac{i \nu_3}{2}\right) \\ 
    & \hspace*{9cm} \times \left(2p_{12}\right)^{-2 s},
\end{align}
where in the second equality we eliminated one of the Mellin variables using the Dirac delta function and then evaluated one of the two leftover Mellin integrals using Cauchy's residue theorem. The remaining Mellin-Barnes integral in fact represents a Gauss Hypergeometric function with argument $z=\frac{1-p_{12}}{2}$, so that
\begin{multline}\label{ccspinl3ptseed}
    \mathfrak{A}_{\tfrac{i}2,\tfrac{i}2,\nu_3}^{(x,\alpha)}(\vec{k}_1,\vec{k}_2,\vec{k}_3)=\frac{2^{2 \alpha-x+3} }{k_1 k_2}\,\left(\frac{k_3}{2}\right)^{i \nu_3-\frac{x}{2}+1}\,\frac{\Gamma \left(\tfrac{x}{2}-i \nu_3-1\right) \Gamma \left(\tfrac{x}{2}-2\alpha+i \nu_3-1\right)}{\Gamma\left(\tfrac{x-1}{2}-\alpha\right)}\\ \, _2F_1\left(\begin{matrix}\tfrac{x}{2}-1 -i\nu_3,\tfrac{x}{2}-2\alpha+i \nu_3-1\\\tfrac{x-1}{2}-\alpha\end{matrix};\frac{1-p_{12}}{2}\right).
\end{multline}
From the above expression, all terms \eqref{00lscdecomp} in the correlator \eqref{00lltmb} are generated by acting with the differential operator \eqref{alphabeta_fromgamma0}. 

When the spin-$\ell$ field is massless, $\nu_3=\frac{i}{2}\left(d-4+2\ell\right)$, the term with $\alpha=\beta=0$ gives the physical helicity-$\ell$ component. When $d=3$ this reads:
\begin{align}
    \mathfrak{A}_{\tfrac{i}2,\tfrac{i}2,\tfrac{i}2(2\ell-1)|0,0}^{(3+2\ell)}(\vec{k}_1,\vec{k}_2,\vec{k}_3)&=\,\frac{\Gamma (2 \ell)}{\Gamma\left(\ell+1\right)}\frac{\mathcal{N}_3}{k_1 k_2 k_3^{2 \ell}}  \, _2F_1\left(1,2 \ell;\ell+1;\frac{1-p_{12}}{2}\right)\,.
\end{align}
Note that when $\ell=0$ we have a divergence. This divergence is however cancelled upon including the sinusoidal factor in \eqref{00lltmb} which arises from combining the contributions from the $+$ and $-$ branches of the in-in contour, giving \cite{Arkani-Hamed:2015bza}:
\begin{align}
    \langle \phi^{\left(i/2\right)}_{\vec{k}_1}\phi^{\left(i/2\right)}_{\vec{k}_2}\varphi^{\left(i/2\right)}_{\vec{k}_3} \rangle^\prime&=\,\frac{\pi\,\mathcal{N}_3}{k_1 k_2 k_3}\,.
\end{align}

\paragraph{Two massless scalars in $d=3$.} From the simplified result \eqref{ccspinl3ptseed} for two conformally coupled scalars, using the raising operators \eqref{raisingop0} we can obtain expressions for when the two scalars have scaling dimension $\nu=\frac{i}{2}+n$ for any $n \in \mathbb{N}$. The simplest application is when $n=1$, which for $d=3$ corresponds to a massless scalar. In this case we have
\begin{equation}\label{mfccrecur}
    \mathfrak{A}^{(x,\alpha)}_{\tfrac{3i}2,\tfrac{3i}2,\nu_3}(\vec{k}_1,\vec{k}_2,\vec{k}_3)=\frac{4}{k_1 k_2}\pl_{k_1}\pl_{k_2}\mathfrak{A}^{(x-4,\alpha)}_{\tfrac{i}2,\tfrac{i}2,\nu_3}(\vec{k}_1,\vec{k}_2,\vec{k}_3)\,.
\end{equation}
A nice application of this formula is for the graviton three-point function with two massless scalars in $d=3$, which corresponds to $\nu_3=\frac{3i}{3}$ and $x=d+2\ell=7$. The physical helicity-2 component is given by \eqref{mfccrecur} with $\alpha=0$. Inserting \eqref{ccspinl3ptseed} for these values and evaluating the derivatives straightforwardly gives
\begin{align}\label{gravitonRes}
    \mathfrak{A}^{(7,0)}_{\tfrac{3i}2,\tfrac{3i}2,\nu_3}(\vec{k}_1,\vec{k}_2,\vec{k}_3)=-\frac{16}{(k_1k_2 k_3)^3}\left[-\frac{k_1 k_2 k_3}{k_t^2}-\frac{k_1 k_2+k_1 k_3+k_2 k_3}{k_t}+k_t\right]\,,
\end{align}
where $k_t=k_1+k_2+k_3$, which matches the result in \cite{Bzowski:2013sza}.

Although the relation \eqref{mfccrecur} gave the result with little effort from the conformally coupled scalar case, it is instructive consider the simple graviton example in more detail directly at the level of the Mellin-Barnes representation \eqref{scdecomp2}. 

Since results for scaling dimensions away from the Principal Series are defined by analytic continuation it is wise to set $d=3+\epsilon$, for which we have
\begin{multline}
    \mathfrak{A}_{\frac{3i}{2},\frac{3i}{2},\frac{3i}{2}}^{\left(7+\epsilon,0\right)}(\vec{k}_1,\vec{k}_2,\vec{k}_3)=\frac{32k^{-\frac{\epsilon }{2}-4}_3}{k_1^3 k_2^3}\int_{-i\infty}^{i\infty}\frac{ds}{2\pi i}\,  (1-s) (4 s-\epsilon -6) \Gamma \left(2-2s+\tfrac{\epsilon}{2}\right)\\\times \left(k^2_3 p_{12}^2 \Gamma (2 s-3)+k_1 k_2 \Gamma (2 s-2)\right)p_{12}^{1-2 s}\,.
\end{multline}
To evaluate the Mellin integral it is simplest to close the integration contour to the right, which encloses the sequence of poles
\begin{equation}
  2 s=2+\frac{\epsilon}{2}+n, \qquad n \in \mathbb{N}_0,
\end{equation}
which gives the helicity-2 component of the three-point function as the following series
\begin{multline}\label{gravitonsummand}
     \mathfrak{A}_{\frac{3i}{2},\frac{3i}{2},\frac{3i}{2}}^{\left(7+\epsilon,0\right)}(\vec{k}_1,\vec{k}_2,\vec{k}_3)=\frac{1}{k_{1}^3 k_{2}^3}\sum_{n=0}^\infty\frac{8 (-1)^n}{n!} (1-n) k_3^{-\frac{\epsilon }{2}-4} (2 n+\epsilon ) p_{12}^{-n-\frac{\epsilon }{2}-1}\\\times \left[k_3^2 p_{12}^2 \Gamma \left(n+\frac{\epsilon }{2}-1\right)+k_{1} k_{2} \Gamma \left(n+\frac{\epsilon }{2}\right)\right]\,.
\end{multline}
From the above it is interesting to notice how the $n=1$ term does not contribute $\forall \,\epsilon\neq0$, while exactly at $\epsilon=0$ it gives a non-vanishing contribution $\frac{16\,k_3}{k_1^3 k_2^3 k_3^3}$. This example therefore exhibits how the limit $\epsilon\to0$ does not in general commute with the integration over the Mellin variables. The result \eqref{gravitonRes} is only obtained by taking the limit $\epsilon\to0$ after the Mellin integration has been performed (i.e. after re-summing the series in $n$). This illustrates the importance of keeping $d$ arbitrary in the calculation in order to keep these subtleties under control. 

\subsection{Soft Limit and Inflationary two-point function}
\label{subsec::softlimit2pt}

The Mellin-Barnes representation of correlators in Fourier space is a convenient tool to extract kinematic limits in the phase space of momenta. When considering cosmological correlators we are often interested in soft momentum limits $k \rightarrow 0$, which for a scalar of small mass $\nu = i \left(\tfrac{d}{2}-\epsilon \right)$ gives the leading slow-roll correction where $\epsilon$ is related to the slow-roll parameter \cite{Arkani-Hamed:2015bza,Kundu:2015xta}. In the following, for the $0$-$0$-$\ell$ correlator \eqref{00lltmb} we detail how to extract the soft limits of both scalar and spinning legs within the Mellin formalism. From the expression for the soft limit of one scalar leg we also give the corresponding inflationary two-point function of a scalar field and a spin-$\ell$ field at leading order in slow roll.

\paragraph{Soft limit of scalar legs.} Let us consider the soft limit $k_2 \rightarrow 0$ of the scalar field $\phi^{\left(\nu_2\right)}$. Assuming $\nu_2\in i\mathbb{R}_+$, the dominant term as $k_2 \rightarrow 0$ is encoded in the residue of the pole at $s_2=-\tfrac{i\nu_2}{2}$, which is the leading Gamma-function pole that generates non-analytic terms in the momentum $k_2$ in the Mellin-Barnes representation \eqref{buboFsc} of the corresponding propagator. Momentum conservation implies that $\vec{\xi} \cdot \vec{k}_1 \sim - \vec{\xi} \cdot \vec{k}_3$ as $k_2 \rightarrow 0$, for which the polynomial ${\sf p}_\ell$ encoding the tensor structure in \eqref{00lltmb} becomes proportional to a single monomial $(\vec{\xi}\cdot \vec{k}_3)^\ell$:\footnote{In deriving this expression note that the pole at $s_2=-\tfrac{i\nu_2}{2}$ is only present in the $\beta=0$ contribution to the correlator \eqref{00lmbfseads}.} 
\begin{align}\label{softlimitpl}
   & {\sf p}^{\left(\ell\right)}_{\nu_1,\nu_2,\nu_3}(-\vec{\xi}\cdot \vec{k}_3,0,\vec{\xi}\cdot \vec{k}_3|s_1,s_2,s_3)\Big|_{s_2=-\tfrac{i\nu_2}{2}}\\& \hspace*{0.5cm} = (-i \vec{\xi} \cdot \vec{k}_3)^\ell\,\sum_{\alpha=0}^\ell \binom{\ell}{\alpha} \frac{\left(s_1+\frac{i \nu_1}{2}\right)_\alpha}{\left(\frac{i \nu_3 }{2}-\alpha+s_3\right)_\alpha}\,\frac{\left(\tfrac{d-2 \ell+2 i \nu_3 -2 i \nu_1+2 i \nu_2}{4} \right)_{\ell-\alpha} \left(\tfrac{d+2 \ell-4 \alpha+2 i \nu_3 -2 i \nu_1+2 i \nu_2}{4} \right)_\alpha}{\left(\tfrac{d}{2}+i \nu_3 -1\right)_{\ell-\alpha} \left(\tfrac{d}{2}+\ell-\alpha+i \nu_3 -1\right)_\alpha} \nonumber
    \\&\hspace*{0.5cm}=(i\,\vec{\xi}\cdot \vec{k}_3)^\ell\,\frac{\left(\tfrac{d-2 \ell-2 i (\nu_1- \nu_2- \nu_3)}{4}\right)_\ell \left(\tfrac{2s_1+2s_3-2\ell+i (\nu_1+\nu_3)}{2}\right)_\ell}{\left(\tfrac{d}{2}+i \nu_3-1\right)_\ell \left(s_3+\tfrac{i \nu_3}{2}-\ell\right)_\ell}. \nonumber
\end{align}
This implies that only the zero helicity component contributes in the soft limit $k_2 \rightarrow 0$, since components with non-zero helicity are orthogonal to $\vec{k}_3$. Using this expression, upon eliminating the Dirac delta function in \eqref{00lltmb} the soft limit is given by a single Mellin-Barnes integral:
\begin{multline}
  \langle \phi^{\left(\nu_1\right)}_{\vec{k}_1}\phi^{\left(\nu_2\right)}_{\vec{k}_2}\varphi^{\left(\nu_3\right)}_{\vec{k}_3} \rangle^\prime\Big|_{k_2 \rightarrow 0,\, \vec{k_1}\sim-\vec{k}_3} = \frac{\mathcal{N}_2}{4\,\pi}\,\sin \left(\tfrac{\pi}{4}  (d+2 \ell+2 i (\nu_1+\nu_2+\nu_3))\right)\\\nonumber
    \times\,\frac{ \left(\tfrac{d-2 \ell-2 i (\nu_1-\nu_2-\nu_3)}{4}\right)_\ell \left(\tfrac{d-2 \ell+2 i (\nu_1+\nu_2+\nu_3)}{4}\right)_\ell}{\left(\frac{d}{2}+i \nu_3-1\right)_\ell}\,\left(\frac{k_3}{2}\right)^{-\tfrac{d}{2}-\ell+i (\nu_1-\nu_2+\nu_3)}\,(i\vec{\xi} \cdot \vec{k}_3)^{\ell}\\\nonumber
    \times\, \int^{i\infty}_{-i\infty}\frac{ds_1}{2\pi i}\,\Gamma \left(s_1-\tfrac{i \nu_1}{2}\right) \Gamma \left(s_1+\tfrac{i \nu_1}{2}\right) \Gamma \left(\tfrac{d+2\ell-4 s_1+2 i (\nu_2-\nu_3)}{4}\right) \Gamma \left(\tfrac{d-2\ell-4 s_1+2 i (\nu_2+\nu_3)}{4}\right)\,.
\end{multline}
This Mellin integral can be easily lifted using Barnes' first lemma, which gives
\begin{multline}
   \langle \phi^{\left(\nu_1\right)}_{\vec{k}_1}\phi^{\left(\nu_2\right)}_{\vec{k}_2}\varphi^{\left(\nu_3\right)}_{\vec{k}_3} \rangle^\prime\Big|_{k_2 \rightarrow 0,\, \vec{k_1}\sim-\vec{k}_3} =  \,\frac{\mathcal{N}_2}{4\,\pi\,\Gamma\left(\tfrac{d}{2}+i\nu_2\right)}\,\sin \left(\tfrac{\pi}{4}  (d+2 \ell+2 i (\nu_1+\nu_2+\nu_3))\right)\\
   \times\,\Gamma \left(\tfrac{d-2 \ell+2 i \nu_1+2 i (\nu_2+\nu_3)}{4}\right)\Gamma \left(\tfrac{d-2 \ell-2 i \nu_1+2 i (\nu_2+\nu_3)}{4}\right)\Gamma \left(\tfrac{d+2 \ell+2 i \nu_1+2 i (\nu_2-\nu_3)}{4}\right)\Gamma \left(\tfrac{d+2 \ell-2 i \nu_1+2 i (\nu_2-\nu_3)}{4}\right)\\
    \times \frac{\left(\tfrac{d-2 \ell-2 i (\nu_1-\nu_2-\nu_3)}{4}\right)_\ell \left(\tfrac{d-2 \ell+2 i (\nu_1+\nu_2+\nu_3)}{4}\right)_\ell}{\left(\frac{d}{2}+i \nu_3-1\right)_\ell} \left(\frac{k_3}{2}\right)^{-\tfrac{d}{2}-\ell+i (\nu_1-\nu_2+\nu_3)}\,(i \vec{\xi} \cdot \vec{k}_3)^{\ell}\,,
\end{multline}
where we also divided by the two-point function of the leg with respect to which we are taking the soft limit.

\paragraph{Inflationary two-point function.} The inflationary two-point function can be obtained from the above by giving the soft leg a small mass: $\nu_2=i(\tfrac{d}{2}-\epsilon)$, and collecting the terms linear in $\epsilon$:
\begin{shaded}
\noindent \emph{Inflationary two-point function}
\begin{multline}\label{twoPointInf}
    \langle \phi^{\left(\nu_1\right)}_{\vec{k}_1}\varphi^{\left(\nu_3\right)}_{\vec{k}_3} \rangle^\prime_{(\text{Infl.})}= \epsilon\,\frac{\mathcal{N}_2}{4\pi}\,\sin \left(\tfrac{\pi}{2} (\ell+i \nu_1+i\nu_3)\right)\, \prod_{\pm {\hat \pm}}\Gamma \left(\tfrac{\ell\pm i \nu_1{\hat\pm} i \nu_3}{2}\right)\\
    \times \frac{1}{\left(\frac{d}{2}+i \nu_3-1\right)_\ell}\left(\frac{k_3}2\right)^{-\ell+i \nu_1+i \nu_3}(i\,\vec{\xi} \cdot \vec{k}_3)^{\ell}\,,
\end{multline}
\end{shaded}
\noindent This matches and generalises equation (C.211) in \cite{Arkani-Hamed:2015bza} where it was given for $\nu_3=\frac{di}2$ (massless scalar) in $d=3$. 

\paragraph{Soft limit of spinning leg.} In the previous part we took the soft limit of a scalar leg. It is also straightforward to take the soft limit of the spinning leg in \eqref{00lltmb}, i.e. $k_3 \rightarrow 0$. In this case it is useful to note that
\begin{subequations}\label{rhok3soft}
\begin{align}
    {\sf p}^{\left(\ell\right)}_{\nu_1,\nu_2,\nu_3}(\vec{\xi} \cdot \vec{k}_1,\vec{\xi} \cdot \vec{k}_2,\vec{\xi} \cdot \vec{k}_3|s_1,s_2,s_3)\Big|_{k_3\to0}&=\mathcal{Y}^{(\ell)}_{\nu_1,\nu_2,\nu_3|0,0}(\vec{\xi} \cdot \vec{k}_1,-\vec{\xi} \cdot \vec{k}_1)\,\\
    &=  (-i\vec{\xi}\cdot \vec{k}_1)^{\ell}
\end{align}
\end{subequations}
which is independent of the Mellin variables $s_i$, and we used that $k_1 \sim k_2$ as $k_3 \rightarrow 0$. At the level of the full correlator \eqref{00lltmb}, like for the soft limit of the scalar leg considered earlier, the leading term in the limit $k_3\to0$ is given by the residue of the leading pole encoding the non-analytic dependence on $k_3$ in the Mellin-Barnes representation \eqref{MBbubosc} of the corresponding propagator, which is at $s_3 = -\tfrac{i\nu_3}{2}$. Together with the behaviour \eqref{rhok3soft}, this gives:
\begin{multline}
     \langle \phi^{\left(\nu_1\right)}_{\vec{k}_1}\phi^{\left(\nu_2\right)}_{\vec{k}_2}\varphi^{\left(\nu_3\right)}_{\vec{k}_3} \rangle^\prime\Big|_{k_3 \rightarrow 0}  = \frac{{\cal N}_3}{4\pi} \frac{\sin\left(\pi \left(\tfrac{d}{4}+\tfrac{i(\nu_1+\nu_2+\nu_3+\ell)}{2}\right)\right)}{\Gamma\left(\frac{d}{2}+i\nu_3+\ell\right)}
   \prod_{\pm {\hat \pm}} \Gamma\left(\tfrac{d+2\ell}{4}+\tfrac{i\left(\nu_3\pm\nu_1{\hat \pm}\nu_2\right)}{2}\right)
    \\
    \times (-i\vec{\xi} \cdot \vec{k}_1)^{\ell} \left(\frac{k_1}{2}\right)^{i\left(\nu_1+\nu_2-\nu_3\right)-\tfrac{d}{2}-\ell}\frac{\Gamma\left(-i\nu_3\right)}{2\sqrt{\pi}}\left(\frac{k_3}{2}\right)^{2i\nu_3}.
\end{multline}

\section{Four-point Exchange Diagrams}\label{sec:FourPoint}

In this section we consider late-time exchange four-point functions in dS$_{d+1}$ at tree-level. Throughout we shall consider four-point functions with general external scalars, though the approach is applicable to general external spinning fields which shall be detailed elsewhere \cite{ToAppear2}. We start in section \ref{subsec:Generalscalar} with the derivation of the Mellin-Barnes representation for the exchange of a general scalar field in Fourier-space, and then the exchange of a general field with integer spin $\ell$ in section \ref{subsecc::spinlexch}. 

The remaining sections are dedicated to the discussion of various properties of the Mellin-Barnes representation for exchange diagrams in Fourier-Space. In sections \ref{subsec::OPElimit} and \ref{subsec::EFTexp} we detail how the representation encodes the Operator-Product- and Effective-Field-Theory-expansions of the exchange four-point function. In section \ref{subsec::recursion} we show how the representation makes manifest recursion relations between correlation functions with fields of different scaling dimensions and spins, which can be reformulated as the action of weight-shifting operators. In section \ref{subsec::someparticcases} we discuss the simplifications that occur for certain scaling dimensions, and the extra care that needs to be taken when analytically continuing the Mellin-Barnes representation away from the Principal Series. We conclude in section \ref{dSvsAdS} by comparing the Mellin-Barnes representations of exchange four-point functions in anti-de Sitter and de Sitter space.

\subsection{Exchange of a General Scalar}\label{subsec:Generalscalar}

In this section we derive first the Mellin-Barnes representation for the late-time limit of a general tree-level four-point scalar exchange in de Sitter space. We shall detail how this result can be obtained simply from the knowledge of the associated tree-level three-point Witten diagrams in EAdS$_{d+1}$ -- i.e. those generated by the cubic vertices participating in the exchange under consideration -- and enforcing causality as we go from Euclidean anti-de Sitter space to de Sitter. This approach lends itself to the extension to spinning fields, which we consider in \S \ref{subsecc::spinlexch}.

The defining feature of the four-point exchange diagram is the bulk-to-bulk propagator for the exchanged particle. As we saw in \S \ref{subsec::SKdic}, this is expressed in terms of EAdS Harmonic functions \eqref{scalarharm} which are appropriately analytically continued on the various branches of the in-in contour. The boundary dual of a bulk Harmonic function is a Conformal Partial Wave (CPW), which gives the contribution of the Harmonic function to the boundary correlation function. These comprise a complete basis of single-valued orthogonal Eigenfunctions of the Casimir invariants for the Conformal group \cite{Ferrara:1972xe,Ferrara:1972uq,Dobrev:1977qv,Caron-Huot:2017vep}. In momentum space they are completely factorised \cite{Polyakov:1974gs}:
\begin{equation}\label{cpwfact}
     {\cal F}^\prime_{\nu,0}(\vec{k}_i;\vec{k}) =\langle \mathcal{O}_1(\vec{k}_1)\mathcal{O}_2(\vec{k}_2)\mathcal{O}_{\Delta}(\vec{k})\rangle \langle \tilde{\mathcal{O}}_{d-\Delta}(-\vec{k}\,)\mathcal{O}_3(\vec{k}_3)\mathcal{O}_4(\vec{k}_4)\rangle,
\end{equation}
where $\tilde{\mathcal{O}}_{d-\Delta}$ is the scaling dimension $d-\Delta$ of the shadow of the boundary operator $\mathcal{O}_{\Delta}$. Momentum conservation implies $\vec{k}=\vec{k}_1+\vec{k}_2=-\vec{k}_3-\vec{k}_4$. The scaling dimension $\Delta$ of the exchanged massive representations (in the ${\sf s}$-channel in this case) is taken to lie on the Principal Series: $\Delta=\tfrac{d}{2}+i\nu$, $\nu \in \mathbb{R}$ for dS exchange. Other representations including complementary series, can be obtained with due care on the analytic continuation of the Principal Series result, which we discuss in more detail in section \ref{subsec::someparticcases}. The three-point function factors are given explicitly as a Mellin-Barnes integral \eqref{mb3ptwitten}, which implies the following Mellin-Barnes representation for a CPW in momentum space
\begin{subequations}\label{Momentum CPW}
\begin{align}
   \mathcal{F}^\prime_{\nu,0}(\vec{k}_i;\vec{k})&=\int_{-i\infty}^{+i\infty}[ds]_4\frac{du \,d\bar{u}}{(2\pi i)^2}\, {\cal F}^\prime_{\nu,0}(s_i;u,{\bar u}|z,{\bar z}|\vec{k}_i;\vec{k})\\
   {\cal F}^\prime_{\nu,0}(s_i;u,{\bar u}|\vec{k}_i;\vec{k})&=
   \rho_{\nu_1,\nu_2,\nu}(s_1,s_2,u)\rho_{\nu_3,\nu_4,-\nu}(s_3,s_4,\bar{u})\left(\tfrac{k}{2}\right)^{-2\left(u+{\bar u}\right)}\prod^4_{j=1}\left(\tfrac{k_j}{2}\right)^{-2s_j+i\nu_j}\\ \nonumber & \hspace*{3cm}\times\,(i\pi) \delta(\tfrac{d}4-s_1-s_2-u)\, (i\pi) \delta(\tfrac{d}4-s_3-s_4-\bar{u})\,.
\end{align}
\end{subequations}
The Mellin-Barnes representation makes manifest the duality between CPWs and bulk Harmonic function $\Omega_{\nu,\vec{k}}$. In particular, we have
\begin{align}\label{dualharmcpw}
    {\cal F}^\prime_{\nu,0}(\vec{k}_i;\vec{k}) &= \int^{\infty}_0\frac{dz}{z^{d+1}}\,\frac{d{\bar z}}{{\bar z}^{d+1}}\, {\cal F}^\prime_{\nu,0}(z,{\bar z}|\vec{k}_i;\vec{k}),
\end{align}
with
\begin{subequations}
\begin{align}\label{harmfunc}
    {\cal F}^\prime_{\nu,0}(z,{\bar z}|\vec{k}_i;\vec{k})&=  {\cal K}_{\frac{d}{2}+i\nu_1}(z;\vec{k}_1){\cal K}_{\frac{d}{2}+i\nu_2}(z;\vec{k}_2)\Omega_{\nu,\vec{k}}\left(z;{\bar z}\right){\cal K}_{\frac{d}{2}+i\nu_3}({\bar z};\vec{k}_3){\cal K}_{\frac{d}{2}+i\nu_4}({\bar z};\vec{k}_4)\nonumber \\
    &=\int_{-i\infty}^{+i\infty}\left[ds\right]_4\frac{du \,d\bar{u}}{(2\pi i)^2}\, {\cal F}^\prime_{\nu,0}(s_i;u,{\bar u}|z,{\bar z}|\vec{k}_i;\vec{k}),\\
    {\cal F}^\prime_{\nu,0}(s_i;u,{\bar u}|z,{\bar z}|\vec{k}_i;\vec{k}) &=    \rho_{\nu_1,\nu_2,\nu}(s_1,s_2,u)\rho_{\nu_3,\nu_4,-\nu}(s_3,s_4,\bar{u})\left(\tfrac{k}{2}\right)^{-2\left(u+{\bar u}\right)}\prod^4_{j=1}\left(\tfrac{k_j}{2}\right)^{-2s_j+i\nu_j}\nonumber\\
    & \hspace*{3cm} \times z^{\tfrac{3d}{2}-2(s_1+s_2+u)}\bar{z}^{\tfrac{3d}{2}-2(s_3+s_4+\bar{u})},  \label{mellinharm}
\end{align}
\end{subequations}
recalling the Mellin-Barnes representation for the bulk propagators (given in \S \ref{MSMBrepprops}), where the Mellin variables $u$ and ${\bar u}$ are associated to the internal legs of the Harmonic function. The integrals over the radial co-ordinates $z$ and ${\bar z}$ generate the delta functions in \eqref{Momentum CPW}, just as for the three-point functions \eqref{Mellin Delta}. The identification \eqref{dualharmcpw}, which is unique owing to the on-shell uniqueness of cubic interactions with two scalars,\footnote{This statement is strictly true only at the level of Conformal Partial Waves/bulk Harmonic functions. The full exchange amplitude includes contact terms, which are sensitive to the choice of improvements of each cubic coupling. We shall comment on this point shortly when discussing the full exchange amplitude and the associated contact interactions.} will be instrumental for the extension of the results in this section to the exchange of spinning particles -- where the CPW \eqref{cpwfact} is also known \cite{Sleight:2017fpc}.  

To go from Euclidean AdS to Lorentzian de Sitter we send $\left(z, {\bar z}\right) \rightarrow \left(-\eta, -{\bar \eta}\right)$ and dress the Mellin representation of each propagator with the appropriate phase as prescribed by the dictionary in \S \ref{subsec::SKdic}. Recall that for the external legs the phase depends on the branch of the in-in contour while for the internal legs it depends on the path ordering of $\eta$ and ${\bar \eta}$. In particular:
\begin{subequations}\label{analcont}
\begin{align}
\mathcal{F}^{\left(+-\right)}(s_i;u,{\bar u}|\eta,\bar{\eta}|\vec{k}_i)&=e^{\delta^{+}_{\nu_1}\left(s_1\right)+\delta^{+}_{\nu_2}\left(s_2\right)+\delta^{-}_{\nu_3}\left(s_3\right)+\delta^{-}_{\nu_4}\left(s_4\right)+\delta_{\prec}\left(u_1,u_2\right)}  \\ & \hspace*{5.7cm} \times \mathcal{F}^\prime(s_i;u,{\bar u}|-\eta,-\bar{\eta}|\vec{k}_i)\,, \nonumber \\
   \mathcal{F}^{\left(-+ \right)}(s_i;u,{\bar u}|\eta,\bar{\eta}|\vec{k}_i)&=e^{\delta^{-}_{\nu_1}\left(s_1\right)+\delta^{-}_{\nu_2}\left(s_2\right)+\delta^{+}_{\nu_3}\left(s_3\right)+\delta^{+}_{\nu_4}\left(s_4\right)+\delta_{\succ}\left(u_1,u_2\right)}  \\ & \hspace*{5.7cm} \times \mathcal{F}^\prime(s_i;u,{\bar u}|-\eta,-\bar{\eta}|\vec{k}_i)\,, \nonumber \\
     \mathcal{F}^{\left(\pm \pm\right)}_{\succ}(s_i;u,{\bar u}|\eta,\bar{\eta}|\vec{k}_i)&=e^{\delta^{\pm}_{\nu_1}\left(s_1\right)+\delta^{\pm}_{\nu_2}\left(s_2\right)+\delta^{\pm}_{\nu_3}\left(s_3\right)+\delta^{\pm}_{\nu_4}\left(s_4\right)+\delta_{\succ}\left(u_1,u_2\right)} \\ & \hspace*{5.7cm} \times  \mathcal{F}^\prime(s_i;u,{\bar u}|-\eta,-\bar{\eta}|\vec{k}_i)\,,\nonumber \\
     \mathcal{F}^{\left(\pm \pm\right)}_{\prec}(s_i;u,{\bar u}|\eta,\bar{\eta}|\vec{k}_i)&=e^{\delta^{\pm}_{\nu_1}\left(s_1\right)+\delta^{\pm}_{\nu_2}\left(s_2\right)+\delta^{\pm}_{\nu_3}\left(s_3\right)+\delta^{\pm}_{\nu_4}\left(s_4\right)+\delta_{\prec}\left(u_1,u_2\right)} \\ & \hspace*{5.7cm} \times  \mathcal{F}^\prime(s_i;u,{\bar u}|-\eta,-\bar{\eta}|\vec{k}_i)\,, \nonumber
\end{align}
\end{subequations}
where the subscripts $\prec$ and $\succ$ denote the path orderings $\eta \prec {\bar \eta}$ and $\eta \succ {\bar \eta}$, respectively on the in-in contour and the phases $\delta^{\pm}_\nu(s)$ and $\delta_{\prec}(u,\bar{u})$ and $\delta_{\succ}(u,\bar{u})$ are defined in \eqref{IntLegPh} and \eqref{ExtLegPh} respectively.  The superscripts denote the branch of the in-in contour, i.e. $++$, $+-$, $-+$ or $--$. Note that these analytic continuations also hold for spinning fields, owing to the independence of the phase factors from the spin as discussed in \S \ref{subsec::spin2pt}. With \eqref{analcont} we can construct the full late-time exchange amplitude:\footnote{In the following we focus on the contribution from the ${\sf s}$-channel exchange. Expressions for the ${\sf t}$- and ${\sf u}$-channel contributions can be obtained in the same way (or just by permuting the external legs). From this point onwards, for ease of presentation we shall leave the contributions from the ${\sf t}$- and ${\sf u}$-channel exchanges implicit.
}
\begin{multline}\label{exchamp0}
    \langle \phi^{\left(\nu_1\right)}_{\vec{k}_1}\phi^{\left(\nu_2\right)}_{\vec{k}_2}\phi^{\left(\nu_3\right)}_{\vec{k}_3}\phi^{\left(\nu_4\right)}_{\vec{k}_4} \rangle^\prime \\= {\cal N}_4 \lim_{\eta_0\to0} \sum_{\pm {\hat \pm}}\left(\pm i\right)\left({\hat \pm} i\right){\cal A}_{\pm {\hat \pm}|\nu_1,\nu_2,\nu_3,\nu_4}(\vec{k}_i;\vec{k})+{\sf t}- \text{and\:} {\sf u-}\text{channels} ,
\end{multline}
where we sum over all pieces of the in-in contour, with:
\begin{subequations}\label{exchcontributions}
\begin{align} \label{A++}
   \hspace*{-0.65cm} {\cal A}_{++|\nu_1,\nu_2,\nu_3,\nu_4}(\vec{k}_i;\vec{k}) &=  \int^{\eta_0}_{-\infty}d\eta\,d{\bar \eta} \left[\theta\left(\eta-{\bar \eta}\right) \mathcal{F}^{\left(++\right)}_{\succ}(\eta,\bar{\eta}|\vec{k}_i)+\theta\left({\bar \eta}-\eta\right) \mathcal{F}^{\left(++\right)}_{\prec}(\eta,\bar{\eta}|\vec{k}_i)\right],\\ \label{A+-}
  \hspace*{-0.65cm} {\cal A}_{+-|\nu_1,\nu_2,\nu_3,\nu_4}(\vec{k}_i;\vec{k}) &=  \int^{\eta_0}_{-\infty}d\eta\,d{\bar \eta}\,   \mathcal{F}^{\left(+-\right)}(\eta,\bar{\eta}|\vec{k}_i),\\ \label{A-+}
   \hspace*{-0.65cm} {\cal A}_{-+|\nu_1,\nu_2,\nu_3,\nu_4}(\vec{k}_i;\vec{k}) &=  \int^{\eta_0}_{-\infty}d\eta\,d{\bar \eta}\,   \mathcal{F}^{\left(-+\right)}(\eta,\bar{\eta}|\vec{k}_i),\\
   \hspace*{-0.65cm} {\cal A}_{--|\nu_1,\nu_2,\nu_3,\nu_4}(\vec{k}_i;\vec{k}) &=  \int^{\eta_0}_{-\infty}d\eta\,d{\bar \eta} \left[\theta\left(\eta-{\bar \eta}\right) \mathcal{F}^{\left(--\right)}_{\prec}(\eta,\bar{\eta}|k_i)+\theta\left({\bar \eta}-\eta\right) \mathcal{F}^{\left(--\right)}_{\succ}(\eta,\bar{\eta}|\vec{k}_i)\right]. \label{A--}
\end{align}
\end{subequations}
c.f. equation \eqref{SKprop} for the de Sitter bulk-to-bulk propagator in terms of analytically continued EAdS Harmonic functions. When writing this expression for the exchange we want to stress that we are reconstructing the dS exchange from first principles via our dictionary. In particular, on the bulk side, this implicitly entails making a choice of improvement terms in the bulk cubic couplings. This naturally corresponds to the physical freedom of adding contact interactions to the exchange amplitude by modifying cubic couplings with terms proportional to the equations of motion. In the following we shall stick to the above minimal choice to define a basis of exchange amplitudes due to its strikingly simple relation to conformal partial waves.

The integrals over conformal time in \eqref{exchcontributions} are simple combinations of the following basic integrals:
\begin{subequations}\label{Qkernints}
\begin{align}\label{Qkernintso}
    \mathcal{Q}_\odot^{(x,\bar{x})}\left(s_i;u,{\bar u}\right)&=\int_{-\infty}^{\eta_0}\int_{-\infty}^{\eta_0}d\eta d{\bar \eta}(-\eta)^{\frac{x}2-2(s_1+s_2+u)-1}(-\bar{\eta})^{\frac{\bar{x}}2-2(s_1+s_2+u)-1}\,\\ \nonumber
    &=\frac{4 (-\eta_0)^{\tfrac12[x+\bar{x}-4 (s_1+s_2+s_3+s_4+u+\bar{u})]}}{(x-4 (s_1+s_2+u)) (\bar{x}-4 (s_3+s_4+\bar{u}))},
    \\  \label{Qkernints>}
    \mathcal{Q}_>^{(x,\bar{x})}\left(s_i;u,{\bar u}\right)&=\int_{-\infty}^{\eta_0}\int_{-\infty}^{\eta_0}d\eta d{\bar \eta}(-\eta)^{\frac{x}2-2(s_1+s_2+u)-1}(-\bar{\eta})^{\frac{\bar{x}}2-2(s_1+s_2+u)-1}\theta(\eta-\bar{\eta})\,\\
    &=\frac{4 (-\eta_0)^{\tfrac{1}{2}[x+\bar{x}-4(s_1+s_2+ s_3+ s_4+ u+\bar{u})]}}{(\bar{x}-4 (s_3+s_4+\bar{u})) (x+\bar{x}-4(s_1+ s_2+ s_3+s_4 +u+\bar{u}))},\nonumber \\ \label{Qkernints<}
    \mathcal{Q}_<^{(x,\bar{x})}\left(s_i;u,{\bar u}\right)&=\int_{-\infty}^{\eta_0}\int_{-\infty}^{\eta_0}(-\eta)^{\frac{x}2-2(s_1+s_2+u)-1}(-\bar{\eta})^{\frac{\bar{x}}2-2(s_1+s_2+u)-1}\theta(\bar{\eta}-{\eta})\,\\ \nonumber
    &=\frac{4 (-\eta_0)^{\tfrac{1}{2}[x+\bar{x}-4(s_1+s_2+ s_3+ s_4+ u+\bar{u})]}}{(x-4 (s_1+s_2+u)) (x+\bar{x}-4(s_1+ s_2+ s_3+s_4 +u+\bar{u}))}.
    \end{align}
\end{subequations}
where the contour prescription we used to evaluate the $\eta$ and $\bar{\eta}$ integrals is\footnote{We assume that the integrand is well behaved at infinity. This puts constraints on the real part of the exponents of $\eta$ and $\bar{\eta}$ in \eqref{Qkernints} which constrains the Mellin integration contour. After performing the $\eta$ and $\bar{\eta}$ integrals we can move the contour around, but we will have to pick residues according to the initial choice of the integration contour.} 
\begin{align}\label{contprescript}
    \Re(s_1+s_2+u)&>\tfrac{x}4\,,& \Re(s_3+s_4+\bar{u})&>\tfrac{\bar{x}}4\,, 
\end{align}
so that the integration contour passes to the right of poles.
For scalar exchange diagrams we have: $x={\bar x}=d$, as can be read off from \eqref{mellinharm}. For spinning exchange diagrams, $x$ and ${\bar x}$ will also depend on the spin, which can already be anticipated from the expression \eqref{00lltmb} for a three-point function involving one spinning field. Since, as we shall see, many of the results presented in this section can be recycled when we consider spinning exchanges, we shall often keep $x$ and ${\bar x}$ as arbitrary real numbers in the following -- the dependence on which we display in the superscripts.

It is useful to organise the exchange four-point function in terms of contributions that are given by the same basic integral in conformal time as in the kernels \eqref{Qkernints}. These are:
{\allowdisplaybreaks
\begin{subequations}\label{Ao<>}
\begin{align}\label{Ao}
    \mathcal{A}^{\left(x,{\bar x}\right)}_{\odot|\nu_1,\nu_2,\nu_3,\nu_4}(\vec{k}_i;\vec{k})    &=2\int_{-i\infty}^{+i\infty}\left[ds\right]_4\frac{du \,d\bar{u}}{(2\pi i)^2}\,\mathcal{Q}^{(x,{\bar x})}_\odot\left(s_i;u,{\bar u}\right)\,\\& \hspace*{0.5cm} \times \cos\left(\pi(\tfrac{i\nu_1+i\nu_2-i\nu_3-i\nu_4}2+ s_1+s_2+u-s_3-s_4- \bar{u})\right)\, \nonumber \\  \nonumber & \hspace*{0.5cm} \times \rho_{\nu_1,\nu_2,\nu}(s_1,s_2,u)\rho_{\nu_3,\nu_4,-\nu}(s_3,s_4,\bar{u}) \left(\frac{k}{2}\right)^{-2\left(u+{\bar u}\right)}\prod^4_{j=1}\left(\frac{k_j}{2}\right)^{-2s_j+i\nu_j},\\
    \label{A>}
    \mathcal{A}^{\left(x,{\bar x}\right)}_{>|\nu_1,\nu_2,\nu_3,\nu_4}(\vec{k}_i;\vec{k}) 
    &=2\int_{-i\infty}^{+i\infty}\left[ds\right]_4\frac{du \,d\bar{u}}{(2\pi i)^2}\,\mathcal{Q}^{(x,{\bar x})}_>\left(s_i;u,{\bar u}\right)\,\\
    & \hspace*{0.5cm} \times \cos\left(\pi(\tfrac{i\nu_1+i\nu_2+i\nu_3+i\nu_4}2+s_1+ s_2+ s_3+ s_4- u+ \bar{u})\right)\,\nonumber \\ & \hspace*{0.5cm}  \times \rho_{\nu_1,\nu_2,\nu}(s_1,s_2,u)\rho_{\nu_3,\nu_4,-\nu}(s_3,s_4,\bar{u}) \left(\frac{k}{2}\right)^{-2\left(u+{\bar u}\right)}\prod^4_{j=1}\left(\frac{k_j}{2}\right)^{-2s_j+i\nu_j}\,,\nonumber \\
    \label{A<}
    \mathcal{A}^{\left(x,{\bar x}\right)}_{<|\nu_1,\nu_2,\nu_3,\nu_4}(\vec{k}_i;\vec{k}) &=2\int_{-i\infty}^{+i\infty}\left[ds\right]_4\frac{du \,d\bar{u}}{(2\pi i)^2}\,\mathcal{Q}^{(x,{\bar x})}_<\left(s_i;u,{\bar u}\right)\,\\
    & \hspace*{0.5cm}  \times \cos\left(\pi(\tfrac{i\nu_1+i\nu_2+i\nu_3+i\nu_4}2+s_1+s_2+s_3+s_4+ u- \bar{u})\right)\,\nonumber \\ \nonumber & \hspace*{0.5cm} \times \rho_{\nu_1,\nu_2,\nu}(s_1,s_2,u)\rho_{\nu_3,\nu_4,-\nu}(s_3,s_4,\bar{u}) \left(\frac{k}{2}\right)^{-2\left(u+{\bar u}\right)}\prod^4_{j=1}\left(\frac{k_j}{2}\right)^{-2s_j+i\nu_j}\,,
\end{align}
\end{subequations}}
\! 
where the subscript ``$\odot$" denotes the sum of the $+-$ and $-+$ contributions \eqref{A+-} and \eqref{A-+}, while the subscripts $<$ and $>$ denote the sum of $++$ and $--$ contributions \eqref{A++} and \eqref{A--} for $\eta < {\bar \eta}$ and $\eta > {\bar \eta}$ respectively. The appearance of the sinusoidal factors in the Mellin integrands originate from the combination of contributions from different branches of the in-in contour, which have relative phases given by \eqref{analcont}. The Mellin-Barnes representation for exchange diagrams makes it simple to take the late-time limit $\eta_0 \rightarrow 0$, whose leading contribution is controlled by the Mellin-poles in the integrals \eqref{Qkernints} over conformal time. For the total contribution \eqref{Ao} from the $+-$ and $-+$ contours, the leading term in the late-time limit is given by the residues of both poles in \eqref{Qkernintso} and the resulting expression is completely factorised:
\begin{subequations}\label{Io}
\begin{align}
    \lim_{\eta_0\to0}\mathcal{A}^{\left(x,{\bar x}\right)}_{\odot|\nu_1,\nu_2,\nu_3,\nu_4}(\vec{k}_i;\vec{k}) &=\frac{1}2\int_{-i\infty}^{+i\infty}[ds]_4\,\cos\left(\tfrac{\pi}2  (\nu_1+ \nu_2-\nu_3-\nu_4)\right)\\& \hspace*{-2.25cm} \times \rho_{\nu_1,\nu_2,\nu}(s_1,s_2,w)\rho_{\nu_3,\nu_4,-\nu}(s_3,s_4,\bar{w})\left(\frac{k}{2}\right)^{-2\left(w+{\bar w}\right)}\prod^4_{j=1}\left(\frac{k_j}{2}\right)^{-2s_j+i\nu_j}\Big|_{{}^{w=\frac{x}4-s_1-s_2}_{{\bar w}=\frac{{\bar x}}4-s_3-s_4}}\nonumber
    \\&\overset{x={\bar x}=d}{=}\frac{1}2\cos\left(\tfrac{\pi}2  (\nu_1+ \nu_2-\nu_3-\nu_4)\right)\\ \nonumber & \hspace*{2cm}\times \langle \mathcal{O}_{\nu_1}(\vec{k}_1)\mathcal{O}_{\nu_2}(\vec{k}_2)\mathcal{O}_{\nu}(\vec{k})\rangle^\prime \langle \tilde{\mathcal{O}}_{-\nu}(-\vec{k})\mathcal{O}_{\nu_3}(\vec{k}_3)\mathcal{O}_{\nu_4}(\vec{k}_4)\rangle^\prime, \nonumber
\end{align}
\end{subequations}
which, setting $x={\bar x}=d$, is proportional to a single conformal partial wave \eqref{cpwfact}, as shown in the second equality where we used equation \eqref{mb3ptwitten}. 

For the remaining contributions \eqref{A>} and \eqref{A<}, the leading contribution in the late time limit is of the same order and is given by the residue of the pole at $x+\bar{x}-4(s_1+ s_2+ s_3+s_4 +u+\bar{u}) \sim 0$ in \eqref{Qkernints>} and \eqref{Qkernints<}, respectively. This gives,
{\allowdisplaybreaks
\begin{subequations}\label{Ilatetime<>}
\begin{align}\nonumber
    \lim_{\eta_0\to0}\mathcal{A}^{\left(x,{\bar x}\right)}_{>|\nu_1,\nu_2,\nu_3,\nu_4}(\vec{k}_i;\vec{k})&=\frac{1}2\int_{-i\infty}^{+i\infty}\frac{du}{2\pi i}\int_{-i\infty}^{+i\infty}[ds]_4\,\frac{\cos \left(\frac{\pi}{2}  (4(s_1+s_2+u)+i \nu_1+i\nu_2+i\nu_3+i\nu_4)\right)}{u+\epsilon}\\ &  \hspace*{-3cm}\times \rho_{\nu_1,\nu_2,\nu}(s_1,s_2,w\tcr{-u})\rho_{\nu_3,\nu_4,-\nu}(s_3,s_4,{\bar w}\tcr{+u})\left(\tfrac{k}{2}\right)^{-2\left(w+{\bar w}\right)}\prod^4_{j=1}\left(\tfrac{k_j}{2}\right)^{-2s_j+i\nu_j}\Big|_{{}^{w=\frac{x}4-s_1-s_2}_{{\bar w}=\frac{{\bar x}}4-s_3-s_4}}, \label{I>} \\\nonumber
    \lim_{\eta_0\to0}\mathcal{A}^{\left(x,{\bar x}\right)}_{<|\nu_1,\nu_2,\nu_3,\nu_4}(\vec{k}_i;\vec{k})&=\frac{1}2\int_{-i\infty}^{+i\infty}\frac{du}{2\pi i}\int_{-i\infty}^{+i\infty}[ds]_4\,\frac{\cos \left(\frac{\pi}{2}  (4(s_3+s_4+u)+i \nu_1+i\nu_2+i\nu_3+i\nu_4)\right)}{u+\epsilon}\\ \label{I<} & \hspace*{-3cm} \times \rho_{\nu_1,\nu_2,\nu}(s_1,s_2,w\tcr{+u})\rho_{\nu_3,\nu_4,-\nu}(s_3,s_4,{\bar w}\tcr{-u})\left(\tfrac{k}{2}\right)^{-2\left(w+{\bar w}\right)}\prod^4_{j=1}\left(\tfrac{k_j}{2}\right)^{-2s_j+i\nu_j}\Big|_{{}^{w=\frac{x}4-s_1-s_2}_{{\bar w}=\frac{{\bar x}}4-s_3-s_4}}. 
\end{align}
\end{subequations}}
where, due to the restrictions \eqref{contprescript} on the Mellin-variables, the $u$-integrals run over the imaginary axis to the right of the pole at $u \sim 0$, as indicated by the $\epsilon$-prescription. These contributions, which originate from the $++$ and $--$ branches of the in-in contour, are not factorised due to the presence of bulk contact terms.\footnote{I.e. contributions generated by the collision of the points on the same branch of the in-in contour between which the particle is exchanged.} Factorised terms within these contributions are however generated by the residues of the poles at $u=0$.

The leftover $u$-integral in the $++$ and $--$ contributions can in fact be lifted to give a Mellin-Barnes representation for the exchange which employs the same number of Mellin-variables as the factorised contribution \eqref{Io}.\footnote{In general this is the minimal number of Mellin variables to represent a function of four variables.} We give the details for the evaluation of this integral in appendix \ref{app:u-integral}. The resulting expression for the exchange \eqref{exchamp0} after combining all terms of the in-in contour acquires the following general form:
\begin{align} \nonumber
   & \hspace*{-0.5cm} \langle \phi^{\left(\nu_1\right)}_{\vec{k}_1}\phi^{\left(\nu_2\right)}_{\vec{k}_2}\phi^{\left(\nu_3\right)}_{\vec{k}_3}\phi^{\left(\nu_4\right)}_{\vec{k}_4} \rangle^\prime ={\cal N}_4 \lim_{\eta_0\to0}\left(\mathcal{A}^{\left(d,d\right)}_{\odot|\nu_1,\nu_2,\nu_3,\nu_4}(\vec{k}_i;\vec{k})-\mathcal{A}^{\left(d,d\right)}_{<|\nu_1,\nu_2,\nu_3,\nu_4}(\vec{k}_i;\vec{k})-\mathcal{A}^{\left(d,d\right)}_{>|\nu_1,\nu_2,\nu_3,\nu_4}(\vec{k}_i;\vec{k})\right)\\ \label{exchuint}
    &\hspace*{3cm}=\int [ds]_4\,\csc(\pi(u+\bar{u}))\delta^{(d,d)}(u,\bar{u})\\
    &  \hspace*{2cm} \times \rho_{\nu_1,\nu_2,\nu}(s_1,s_2,u)\rho_{\nu_3,\nu_4,-\nu}(s_3,s_4,\bar{u})\left(\frac{k}{2}\right)^{-2\left(u+{\bar u}\right)}\prod^4_{j=1}\left(\frac{k_j}{2}\right)^{-2s_j+i\nu_j}\Big|_{{}^{u=\frac{d}4-s_1-s_2}_{{\bar u}=\frac{d}4-s_3-s_4}}. \nonumber
\end{align}
The poles of the Mellin integrand are manifest and the zeros are given by the function\footnote{There are two further equivalent representations of this function depending on how we evaluate the $u$-integral, which we give in appendix \ref{app:u-integral}. The representation \eqref{symmmellinzeros} is the most symmetric under exchange of $u$ and $\bar{u}$.}

{\footnotesize \begin{align}\nonumber
    \delta^{(x,\bar{x})}(u,\bar{u})&=\frac12\sin (\pi  (u+\bar{u})) \Big[\sin \left(\tfrac{\pi}{4}  (x+2 i (\nu_1+\nu_2)-4 u)\right) \sin \left(\tfrac{\pi}{4}  (\bar{x}+2 i (\nu_3+\nu_4)-4 u)\right)+u \rightarrow {\bar u}\Big]\\ \label{symmmellinzeros}
    &-\frac14\sin \left(\tfrac{\pi}{4} (x+\bar{x}+2 i (\nu_1+\nu_2+\nu_3+\nu_4)-4 (u+\bar{u}))\right) (\cos (2 \pi  u)-\cosh (\pi  \nu )+u \rightarrow {\bar u})\,,
\end{align}}

\noindent which encodes the interference between the different physical processes as dictated by the early-time boundary conditions. The final line of the expression \eqref{exchuint} for the exchange should be recognised as the Mellin-Barnes representation \eqref{Momentum CPW} for the dual Conformal Partial Wave, which implies the following more compact expression for the exchange:
\begin{shaded}
\noindent \emph{Mellin-Barnes representation for a general tree-level four-point exchange diagram}
\begin{multline}\label{exchamp}
    \langle \phi^{\left(\nu_1\right)}_{\vec{k}_1}\phi^{\left(\nu_2\right)}_{\vec{k}_2}\phi^{\left(\nu_3\right)}_{\vec{k}_3}\phi^{\left(\nu_4\right)}_{\vec{k}_4} \rangle^\prime = {\cal N}_4\int_{-i\infty}^{+i\infty}\frac{du\,d\bar{u}}{(2\pi i)^2}\int [ds]_4\,4 \csc(\pi(u+\bar{u}))\delta^{\left(d,d\right)}(u,\bar{u}) \\ \times {\cal F}^\prime_{\nu,0}(s_i;u,{\bar u}|\vec{k}_i;\vec{k})\,,
\end{multline}
\end{shaded}
\noindent which is manifestly in terms of the dual Conformal Partial Wave \eqref{cpwfact}. This expression was also derived in \cite{Charlotte} using a direct bulk approach which employs the Mellin-Barnes representation of the propagators. As we shall see, this form of the exchange four-point function is universal, extending to spin-$\ell$ exchanges (section \ref{subsecc::spinlexch}), exchanges in anti-de Sitter space (section \ref{dSvsAdS}), and external spinning fields \cite{ToAppear2}. Each case is characterised by the interference factor $\delta^{\left(x,{\bar x}\right)}(u,\bar{u})$. The expression \eqref{exchamp} neatly encodes various properties of the exchange-four-point function, as we discuss in the comments below.
\\

\begin{figure}
    \centering
    \captionsetup{width=0.95\textwidth}
    \includegraphics[width=\textwidth]{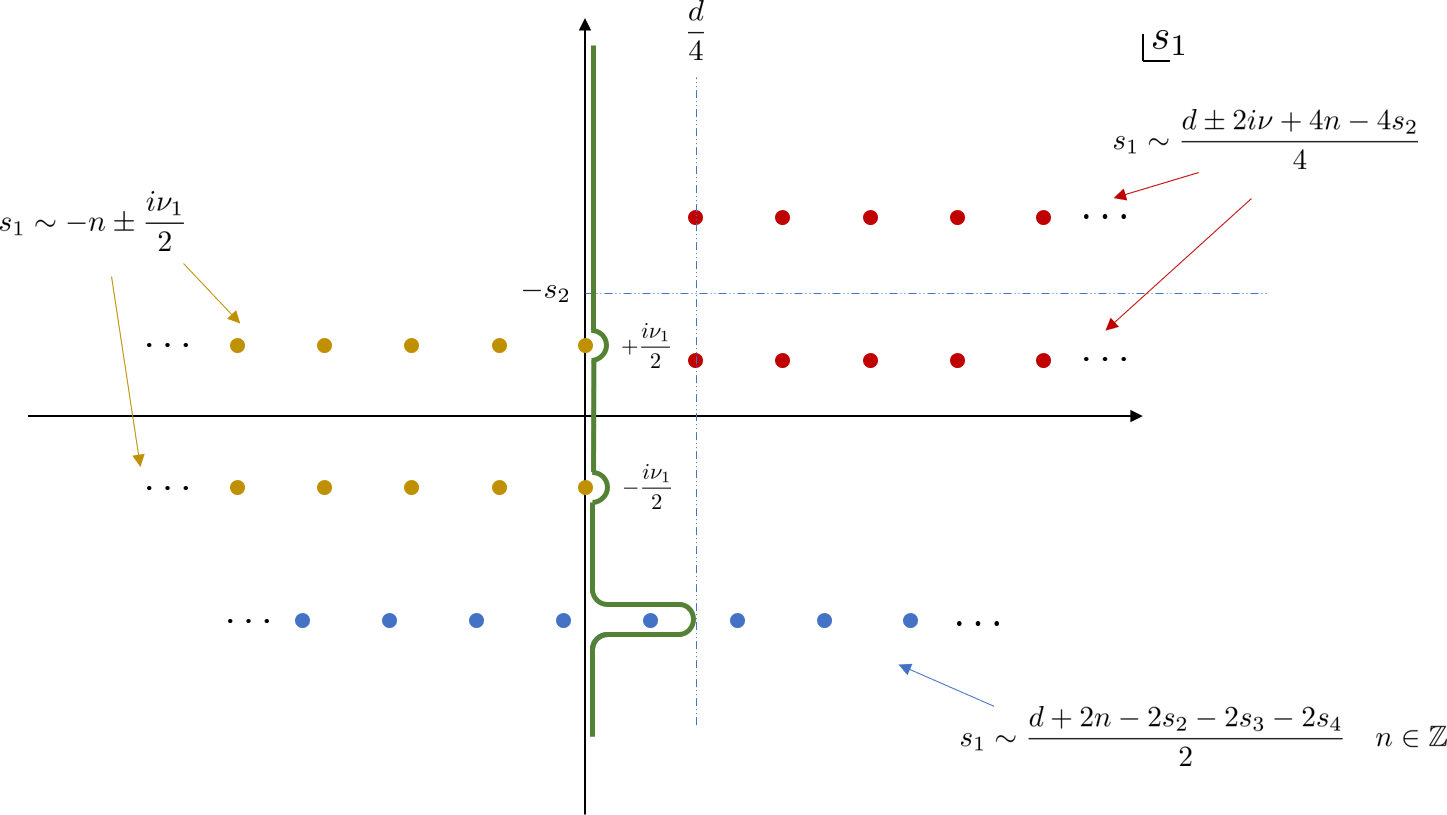}
    \caption{Pole structure of the Mellin-Barnes representation \eqref{exchamp} for the scalar exchange four-point function, focusing w.l.o.g. on that of the Mellin variable $s_1$. The different coloured ``$\bullet$" (red, yellow and blue) denote the different sets of Gamma function poles in $s_1$, while the green line represents the integration contour which w.l.o.g. we take to be indented along the imaginary axis with $\mathfrak{Re}(s_j)=0$. Notice that the poles of the $\csc$ factor, which are in blue, run from $-\infty$ to $+\infty$ and are split by the integration contour according to \eqref{cscident} with $q=0$.
    For Principal Series representations, where $\nu_1, \nu \in \mathbb{R}$, the different sets of Gamma function poles do not collide as the red and yellow poles can only move vertically along the imaginary axis as one varies the scaling dimensions. Away from the Principal Series these poles can move horizontally, which for certain scaling dimensions pinches the integration contour, generating singularities. Such cases can be treated by regulating the contour pinching to obtain the analytic continuation of the exchange four-point function for these values of the scaling dimensions, as we shall see in section \ref{subsec::someparticcases}.}
    \label{fig:Poles_s1}
\end{figure}

\noindent $\bullet$ The Mellin-Barnes integral \eqref{exchamp} is a general expression for a late-time scalar exchange in dS$_{d+1}$, where all scaling dimensions are generic and on the Principal Series. Other representations e.g. the complementary and discrete series can be reached with due care about the analytic continuation away from the Principal Series. For generic scaling dimensions on the Principal Series, the exchange four-point function is a function of four variables $k_j/k$ and is accordingly described by a quadruple Mellin-Barnes integral of the Mejer $G$ type. For certain scaling dimensions there are simplifications. For example, in analytically continuing some or all of the external legs to be conformally coupled, some of the Mellin-Barnes integrals can be lifted and the exchange is accordingly a function of fewer variables (see equation \eqref{confCMellin}). Further simplifications arise when the exchanged field lies on the Discrete Series, as we discuss in section \ref{subsec::someparticcases}, which requires extra care in the analytic continuation. \\

\noindent $\bullet$ The cosecant factor $\csc(\pi(u+\bar{u}))$ gives contact contributions to the exchange four-point function. In particular, the residues of the poles at
\begin{equation}\label{eftpoles}
    u+{\bar u}=-n, \qquad n=0,1,2,... 
\end{equation}
generate only analytic contributions in the exchanged momentum $k$ in \eqref{exchuint}:
\begin{equation}
    u+{\bar u}=-n\qquad\rightarrow\qquad \left(\frac{k}2\right)^{2n}\,.
\end{equation}
These are not factorised and thus give the EFT expansion of the four-point function. The non-perturbative corrections to the EFT expansion are encoded in the remaining poles, which are those of the Mellin representation \eqref{Momentum CPW} for the Conformal Partial Wave. On these poles, by construction, the interference factor \eqref{symmmellinzeros} factorises so that these terms just generate factorised contributions to the exchange four-point function, associated to the genuine exchange of a single-particle state. This in particular includes non-analytic terms in the exchanged momentum, which are characteristic of particle production \cite{Assassi:2012zq,Arkani-Hamed:2015bza}. We shall discuss these contributions in more detail towards the end of section \ref{subsecc::spinlexch}, where we consider the OPE expansion of exchange four-point functions, and section \ref{subsec::EFTexp} where we derive the EFT expansion from the Mellin-Barnes representation of the four-point function.\\

\noindent $\bullet$ It is interesting to note that the EFT expansion is entirely specified by the CPW \eqref{Momentum CPW} multiplied by the interference factor \eqref{symmmellinzeros} in a \emph{minimal} way through the overall $\csc(\pi(u+\bar{u}))$ function. This is not a priori required due to the contact term ambiguity. It is however interesting to point out that the most general EFT expansion would only differ by our minimal choice by a finite number of pure contact terms. Furthermore, the discontinuity in $\mathsf{s}=k^2$ precisely compensates the $\csc(\pi(u+\bar{u}))$ factor in \eqref{exchamp}, setting to zero all EFT terms:
\begin{equation}
    \text{Disc}_{s}\left(k^{2(u+\bar{u})}\right)=\sin(\pi(u+\bar{u}))\,k^{2(u+\bar{u})}\,,
\end{equation}
with
\begin{equation}
    2i\,\text{Disc}_{\mathsf{s}}[f(\mathsf{s})]=f\left(e^{i\pi}\mathsf{s} \right)-f\left(e^{-i\pi}\mathsf{s} \right)\,.
\end{equation}
One is then left with the factorised contribution to the exchange:
\begin{multline}
    \text{Disc}_{\mathsf{s}}\left[\langle \phi^{\left(\nu_1\right)}_{\vec{k}_1}\phi^{\left(\nu_2\right)}_{\vec{k}_2}\phi^{\left(\nu_3\right)}_{\vec{k}_3}\phi^{\left(\nu_4\right)}_{\vec{k}_4} \rangle^\prime\right]\\ = {\cal N}_4\int_{-i\infty}^{+i\infty}\frac{du\,d\bar{u}}{(2\pi i)^2}\int [ds]_4\,4\, \delta^{\left(d,d\right)}(u,\bar{u})  {\cal F}^\prime_{\nu,0}(s_i;u,{\bar u}|\vec{k}_i;\vec{k})\,.
\end{multline}

\noindent $\bullet$ It is important to stress that the integral \eqref{exchamp} does not, a priori, specify an integration contour.\footnote{In contrast, Mellin-Barnes integrals which are given explicitly in terms of $\Gamma$-functions automatically specify an integration contour by requiring that all Gamma function poles accumulating at $+\infty$ are separated by the poles accumulating at $-\infty$ \cite{MellinBook}.} In particular, the $\csc$-function in the Mellin integrand has an infinite series of poles spanning from $-\infty$ to $+\infty$ and it is necessary to provide the location where the contour cuts across them. The various possible choices for the contour correspond to the identities:
\begin{subequations}\label{cscident}
\begin{align}
    \pi\csc \left(\pi (u+\bar{u})\right)&=\Gamma (1-u-\bar{u}) \Gamma (u+\bar{u})\\&=(-1)^q\Gamma (1-u-\bar{u}-q) \Gamma (u+\bar{u}+q)\,,\quad \forall\; q\in \mathbb{N}\,,
\end{align}
\end{subequations}
which arise from the periodicity of the $\csc$-function, where for each $q$ the contour is chosen to separate the poles accumulating at $+\infty$ from the poles accumulating at $-\infty$. The different ways of splitting the $\csc$-function given above differ by contact terms in the exchange four-point function and correspond to the freedom of including improvement (on-shell trivial) terms in the bulk cubic vertices. This is discussed in further detail at the end of appendix \ref{app:u-integral}. In this work we fix the latter contact term ambiguity by making the minimal choice of improvement terms corresponding to $q=0$. 
\\

\noindent $\bullet$ Another reflection of the freedom to add improvement terms to the bulk cubic vertex is the possibility of including terms which are proportional to the argument of the Dirac delta function in the Mellin representation \eqref{mb3ptwitten} for the corresponding three-point conformal structures, which are $s_1+s_2+u-\tfrac{x}4$ or $s_3+s_4+\bar{u}-\tfrac{x}4$. At the three-point function level, such terms would vanish identically. However, the same terms will give a non-vanishing contributions to the exchange four-point function along the $++$ and $--$ branches of the in-in contour, where the internal leg is off-shell. For such terms, it is possible to show that when taking the residue $x+\bar{x}-4(s_1+s_2+s_3+s_4+u+\bar{u})$ in \eqref{Qkernints>} and \eqref{Qkernints<}, one recovers the following integration rule:
\begin{equation}
    \left(s_1+s_2+u-\tfrac{x}4\right)^n\left(s_3+s_4+\bar{u}-\tfrac{x}4\right)^m\rightarrow u^{n+m}\,,
\end{equation}
so that the corresponding $\mathcal{A}_>$ and $\mathcal{A}_<$ take the same form as in \eqref{I>} and \eqref{I<} but with the integrand multiplied by a power $u^{n+m}$, which cancels the single pole at $u=0$. This turns out to be the physical counterpart of the standard fact that adding cubic couplings which vanish on-shell generates contact terms in the exchange amplitude, which here can be neatly associated with polynomial contributions $p(u)$ in \eqref{I>} and \eqref{I<} in addition to the single pole at $u=0$:
\begin{equation}
    \frac{1}{u+\epsilon}\rightarrow \frac{1}{u+\epsilon}+p(u)\,.
\end{equation}
In the following we shall set $p(u)=0$ without loss of generality. It is perhaps useful to keep in mind the existence of such contact term ambiguities, especially when considering exchange four-point functions involving fields of non-zero spin, where it might be used to simplify the expression by removing potentially complicated contact terms -- allowing to focus on the singular part of the exchange.\\

\noindent $\bullet$ The representation \eqref{exchamp} for the exchange makes manifest the relation between the original bulk Harmonic function \eqref{harmfunc} and the exchange amplitude. In particular, via \eqref{dualharmcpw}, we can write:
\begin{multline}\label{ExchangeFromCPW}
    \langle \phi^{\left(\nu_1\right)}_{\vec{k}_1}\phi^{\left(\nu_2\right)}_{\vec{k}_2}\phi^{\left(\nu_3\right)}_{\vec{k}_3}\phi^{\left(\nu_4\right)}_{\vec{k}_4} \rangle^\prime={\cal N}_4 \lim_{\eta_0\to0}\int_{-i\infty}^{+i\infty}\frac{du\,d\bar{u}}{(2\pi i)^2}\int[ds]_4 \,4\csc(\pi(u+\bar{u}))\delta^{(d,d)}(u,\bar{u})\,\\ \times \int_{-\infty}^{\eta_0}\int_{-\infty}^{\eta_0}\frac{d\eta}{(-\eta)^{d+1}}\frac{d\bar{\eta}}{(-\bar{\eta})^{d+1}}\,\mathcal{F}_{\nu,0}(s_i;u,{\bar u}|-\eta,-\bar{\eta}|\vec{k}_i;\vec{k}).
\end{multline}
In the above all $\theta$-function insertions along the various in-in contours in the expression \eqref{exchcontributions} have been replaced/mapped into an integral kernel in the Mellin variables $u$ and ${\bar u}$. These are encoded into the zeros of $\delta^{(x,\bar{x})}(u,\bar{u})$.\\

All of the above points carry over to exchange four-point functions involving spinning fields, which we consider in the following section.

\subsection{Exchange of a Spin-$\ell$ Field between General Scalars}
\label{subsecc::spinlexch}

The approach presented in the previous section naturally extends to exchange four-point functions involving fields with spin. In the following we shall demonstrate this for the exchange of a field with integer spin-$\ell$ between two pairs of general scalar operators. The result, given in \eqref{exchampl}, can be expressed in the same form as the expression \eqref{exchamp} for the scalar exchange diagram but with $x={\bar x}=d+2\ell$. 

When considering fields with spin, the only difference with respect to the scalar exchange is a technical one due to the tensorial structure of the each three-point function factor in the Conformal Partial Wave. The extension of \eqref{cpwfact} to a spin-$\ell$ exchange is:
\begin{equation}\label{spinlCPW}
     {\cal F}^\prime_{\nu,\ell}(\vec{k}_i;\vec{k}) = \frac{1}{\ell!\left(\frac{d}{2}-1\right)_{\ell}} \langle \mathcal{O}_{\nu_1}(\vec{k}_1)\mathcal{O}_{\nu_2}(\vec{k}_2)\mathcal{O}_{\nu,\ell}(\vec{k};{\hat \partial}_{\xi})\rangle \langle \tilde{\mathcal{O}}_{-\nu,\ell}(-\vec{k};\xi)\mathcal{O}_{\nu_3}(\vec{k}_3)\mathcal{O}_{\nu_3}(\vec{k}_4)\rangle,
\end{equation}
where the explicit form of the three-point functions was derived in \S \ref{subsec::spinningads3pt}, which gives the following Mellin-Barnes representation for the CPW \eqref{spinlCPW}:
{\allowdisplaybreaks
\begin{subequations}\label{CPWspinlexpMB}
\begin{align}
     {\cal F}^\prime_{\nu,\ell}(\vec{k}_i;\vec{k}) =& \int_{-i\infty}^{+i\infty}\left[ds\right]_4\frac{du \,d\bar{u}}{(2\pi i)^2}\, {\cal F}^\prime_{\nu,\ell}(s_i;u,{\bar u}|\vec{k}_i;\vec{k}),\\
     {\cal F}^\prime_{\nu,\ell}(s_i;u,{\bar u}|\vec{k}_i;\vec{k})&=(i\pi) \delta\left(\tfrac{d+2\ell}4-s_1-s_2-u\right) (i\pi) \delta\left(\tfrac{d+2\ell}4-s_3-s_4-\bar{u}\right)\\ \nonumber & \times 
     \rho_{\nu_1,\nu_2,\nu}(s_1,s_2,u)\rho_{\nu_3,\nu_4,-\nu}(s_3,s_4,\bar{u})\left(\frac{k}{2}\right)^{-2\left(u+{\bar u}\right)} \prod^4_{j=1}\left(\frac{k_j}{2}\right)^{-2s_j+i\nu_j} \nonumber\\ \nonumber
     &  \hspace*{-1.75cm} \times \underbrace{\frac1{\ell!\left(\tfrac{d}{2}-1\right)_\ell}{\sf p}^{\left(\ell\right)}_{\nu_1,\nu_2,\nu}({\hat \partial}_{\xi}\cdot \vec{k}_1,{\hat \partial}_{\xi}\cdot \vec{k}_2,{\hat \partial}_{\xi}\cdot \vec{k}|s_1,s_2,u){\sf p}^{\left(\ell\right)}_{\nu_3,\nu_4,-\nu}(\vec{\xi}\cdot \vec{k}_3,\vec{\xi}\cdot \vec{k}_4,-\vec{\xi}\cdot \vec{k}|s_3,s_4,{\bar u})}_{\Theta^{\left(\ell\right)}_{\nu_1,\nu_2,\nu_3,\nu_4;\nu}(\vec{k}_i;\vec{k}|s_i,u,\bar{u})},
\end{align}
\end{subequations}}
and where we introduced the function $\Theta^{\left(\ell\right)}_{\nu_1,\nu_2,\nu_3,\nu_4;\nu}(k_i;k|s_i,u,\bar{u})$ which encodes the trace-less contraction of the three-point tensor structures given in \eqref{00lmbfseads}.

To obtain the spin-$\ell$ exchange four-point function, one can proceed much in the same way as for the scalar exchange in the previous section. In this case the identification with the dual bulk Harmonic function,
\begin{equation}
    {\cal F}^\prime_{\nu,\ell}(\vec{k}_i;\vec{k}) = \int^{\infty}_0 \frac{dz}{z^{d+1}}\frac{d{\bar z}}{{\bar z}^{d+1}}  {\cal F}^\prime_{\nu,\ell}(z,{\bar z}|\vec{k}_i;\vec{k}),
\end{equation}
is given by
\begin{subequations}\label{harmfuncspinl}
\begin{align}
     {\cal F}^\prime_{\nu,\ell}(z,{\bar z}|\vec{k}_i;\vec{k}) =& \int_{-i\infty}^{+i\infty}\left[ds\right]_4\frac{du \,d\bar{u}}{(2\pi i)^2}\, {\cal F}^\prime_{\nu,\ell}(s_i;u,{\bar u}|z,{\bar z}|\vec{k}_i;\vec{k}),\\
     {\cal F}^\prime_{\nu,\ell}(s_i;u,{\bar u}|z,{\bar z}|\vec{k}_i;\vec{k}) &= z^{\tfrac{3d}{2}+\ell-2(s_1+s_2+u)-1}\bar{z}^{\tfrac{3d}{2}+\ell-2(s_3+s_4+\bar{u})-1}\Theta^{\left(\ell\right)}_{\nu_1,\nu_2,\nu_3,\nu_4;\nu}(\vec{k}_i;\vec{k}|s_i,u,\bar{u}) 
     \nonumber \\  
     & \hspace{-1cm}\times \rho_{\nu_1,\nu_2,\nu}(s_1,s_2,u)\rho_{\nu_3,\nu_4,-\nu}(s_3,s_4,\bar{u})\left(\frac{k}{2}\right)^{-2\left(u+{\bar u}\right)} \prod^4_{j=1}\left(\frac{k_j}{2}\right)^{-2s_j+i\nu_j}, 
\end{align}
\end{subequations}
recalling the integrand \eqref{00lmbfseads} of the $0$-$0$-$\ell$ Witten diagrams in the bulk radial co-ordinate. The analytic continuations from EAdS to the various branches of the in-in contour in de Sitter were given in equation \eqref{analcont}.\footnote{Recall that the phases in \eqref{analcont} do not depend on the spin, as discussed in \S \ref{subsec::spin2pt}.} It is convenient to express the contraction $\Theta^{\left(\ell\right)}_{\nu_1,\nu_2,\nu_3,\nu_4;\nu}(k_i;k|s_i,u,\bar{u})$ in the form
\begin{align}\label{fullthetacontra}
  \Theta^{\left(\ell\right)}_{\nu_1,\nu_2,\nu_3,\nu_4;\nu}(\vec{k}_i;\vec{k}|s_i,u,\bar{u})&=\sum_{\alpha_i=0}^\ell\sum_{\beta_i=0}^{\alpha_i}\binom{\ell}{\alpha_1}\binom{\ell}{\alpha_2}\binom{\alpha_1}{\beta_1}\binom{\alpha_2}{\beta_2}\Theta_{\alpha_1,\beta_1;\alpha_2,\beta_2}^{(\ell)}(\vec{k}_i;\vec{k})
    \\ & \hspace*{2.25cm} \times  H_{\nu_1,\nu_2,\nu|\alpha_1,\beta_1}(s_1,s_2,u)\,H_{\nu_3,\nu_4,-\nu|\alpha_2,\beta_2}(s_3,s_4,\bar{u}),  \nonumber
\end{align}
where we used the definition \eqref{00lmbfseads} of the three-point structures and we introduced the contraction 
\begin{multline}\label{ThetaPol}
    \Theta^{\left(\ell\right)}_{\nu_1,\nu_2,\nu_3,\nu_4;\nu|\alpha_1, \beta_1;\alpha_2,\beta_2}(\vec{k}_i;\vec{k}) = \frac{1}{\ell!\left(\frac{d}{2}-1\right)_{\ell}} (-{\hat \partial}_{\xi}\cdot \vec{k}\,)^{\alpha_1}\, \mathcal{Y}^{(\ell)}_{\nu_1,\nu_2,\nu|\alpha_1,\beta_1}({\hat \partial}_{\xi} \cdot \vec{k}_1,{\hat \partial}_{\xi} \cdot \vec{k}_2) \\ \times (\vec{\xi}\cdot \vec{k}\,)^{\alpha_2}\, \mathcal{Y}^{(\ell)}_{\nu_3,\nu_4,-\nu|\alpha_2,\beta_2}(\vec{\xi}\cdot \vec{k}_3,\vec{\xi}\cdot \vec{k}_4)\Big|_{\xi=0},
\end{multline}
which is independent of the Mellin variables. In this way the contributions to the spin-$\ell$ exchange four-point function 
\begin{multline}
   \langle \phi^{\left(\nu_1\right)}_{\vec{k}_1}\phi^{\left(\nu_2\right)}_{\vec{k}_2}\phi^{\left(\nu_3\right)}_{\vec{k}_3}\phi^{\left(\nu_4\right)}_{\vec{k}_4} \rangle^\prime \\ ={\cal N}_4 \lim_{\eta_0\to0}\left[\mathcal{A}^{\left(d+2\ell,d+2\ell\right)}_{\odot|\nu_1,\nu_2,\nu_3,\nu_4}(\vec{k}_i;\vec{k})-\mathcal{A}^{\left(d+2\ell,d+2\ell\right)}_{<|\nu_1,\nu_2,\nu_3,\nu_4}(\vec{k}_i;\vec{k})-\mathcal{A}^{\left(d+2\ell,d+2\ell\right)}_{>|\nu_1,\nu_2,\nu_3,\nu_4}(\vec{k}_i;\vec{k})\right],
\end{multline}
can be decomposed as
\begin{multline}\label{expspinl}
  {\cal A}^{\left(d+2\ell,d+2\ell\right)}_{\bullet|\nu_1,\nu_2,\nu_3,\nu_4}(\vec{k}_i;\vec{k}) = \sum^\ell_{\alpha_1, \alpha_2=0}\binom{\ell}{\alpha_1}\binom{\ell}{\alpha_2}\sum^{\alpha_1, \alpha_2}_{\beta_1, \beta_2=0} \binom{\alpha_1}{\beta_1}\binom{\alpha_2}{\beta_2}\\ \times\, \Theta^{\left(\ell\right)}_{\nu_1,\nu_2,\nu_3,\nu_4;\nu|\alpha_1, \beta_1;\alpha_2,\beta_2}(\vec{k}_i;\vec{k})  {\cal A}^{\left(d+2\ell,d+2\ell\right)}_{\bullet|\nu_1,\nu_2,\nu_3,\nu_4|\alpha_1, \beta_1;\alpha_2,\beta_2}(\vec{k}_i;\vec{k}),
\end{multline}
where 
{\allowdisplaybreaks 
\begin{subequations}\label{Ao<>spinl}
\begin{align}\label{Aospinl}
    \mathcal{A}^{\left(x,{\bar x}\right)}_{\odot|\nu_1,\nu_2,\nu_3,\nu_4|\alpha_1, \beta_1;\alpha_2,\beta_2}(\vec{k}_i;\vec{k})    &=2\int_{-i\infty}^{+i\infty}\left[ds\right]_4\frac{du \,d\bar{u}}{(2\pi i)^2}\,\mathcal{Q}^{\left(x,{\bar x}\right)}_\odot\left(s_i;u,{\bar u}\right)\,\\& \times \cos\left(\pi(\tfrac{i\nu_1+i\nu_2-i\nu_3-i\nu_4}2+ s_1+s_2+u-s_3-s_4- \bar{u})\right)\, \nonumber \\  \nonumber & \hspace*{-1cm} \times H_{\nu_1,\nu_2,\nu|\alpha_1,\beta_1}(s_1,s_2,u)\rho_{\nu_1,\nu_2,\nu}(s_1,s_2,u)\left(\frac{k}{2}\right)^{-2u}\prod^2_{j=1}\left(\frac{k_j}{2}\right)^{-2s_j+i\nu_j}
     \\  \nonumber & \hspace*{-1.5cm} \times H_{\nu_3,\nu_4,-\nu|\alpha_2,\beta_2}(s_3,s_4,\bar{u}) \rho_{\nu_3,\nu_4,-\nu}(s_3,s_4,\bar{u}) \left(\frac{k}{2}\right)^{-2{\bar u}}\prod^4_{j=3}\left(\frac{k_j}{2}\right)^{-2s_j+i\nu_j},\\
    \label{A>spinl}
    \mathcal{A}^{\left(x,{\bar x}\right)}_{>|\nu_1,\nu_2,\nu_3,\nu_4|\alpha_1, \beta_1;\alpha_2,\beta_2}(\vec{k}_i;\vec{k})
    &=2\int_{-i\infty}^{+i\infty}\left[ds\right]_4\frac{du \,d\bar{u}}{(2\pi i)^2}\,\mathcal{Q}^{\left(x,{\bar x}\right)}_>\left(s_i;u,{\bar u}\right)\,\\
    &  \times \cos\left(\pi(\tfrac{i\nu_1+i\nu_2+i\nu_3+i\nu_4}2+s_1+ s_2+ s_3+ s_4- u+ \bar{u})\right)\,\nonumber\\  \nonumber & \hspace*{-1cm} \times H_{\nu_1,\nu_2,\nu|\alpha_1,\beta_1}(s_1,s_2,u)\rho_{\nu_1,\nu_2,\nu}(s_1,s_2,u)\left(\frac{k}{2}\right)^{-2u}\prod^2_{j=1}\left(\frac{k_j}{2}\right)^{-2s_j+i\nu_j}
     \\  \nonumber & \hspace*{-1.5cm} \times H_{\nu_3,\nu_4,-\nu|\alpha_2,\beta_2}(s_3,s_4,\bar{u}) \rho_{\nu_3,\nu_4,-\nu}(s_3,s_4,\bar{u}) \left(\frac{k}{2}\right)^{-2{\bar u}}\prod^4_{j=3}\left(\frac{k_j}{2}\right)^{-2s_j+i\nu_j},\nonumber \\
    \label{A<spinl}
   \mathcal{A}^{\left(x,{\bar x}\right)}_{<|\nu_1,\nu_2,\nu_3,\nu_4|\alpha_1, \beta_1;\alpha_2,\beta_2}(\vec{k}_i;\vec{k})&=2\int_{-i\infty}^{+i\infty}\left[ds\right]_4\frac{du \,d\bar{u}}{(2\pi i)^2}\,\mathcal{Q}^{\left(x,{\bar x}\right)}_<\left(s_i;u,{\bar u}\right) \\
    &  \times \cos\left(\pi(\tfrac{i\nu_1+i\nu_2+i\nu_3+i\nu_4}2+s_1+s_2+s_3+s_4+ u- \bar{u})\right)\,\nonumber \\  \nonumber & \hspace*{-1cm} \times H_{\nu_1,\nu_2,\nu|\alpha_1,\beta_1}(s_1,s_2,u)\rho_{\nu_1,\nu_2,\nu}(s_1,s_2,u)\left(\frac{k}{2}\right)^{-2u}\prod^2_{j=1}\left(\frac{k_j}{2}\right)^{-2s_j+i\nu_j}
     \\  \nonumber & \hspace*{-1.5cm} \times H_{\nu_3,\nu_4,-\nu|\alpha_2,\beta_2}(s_3,s_4,\bar{u}) \rho_{\nu_3,\nu_4,-\nu}(s_3,s_4,\bar{u}) \left(\frac{k}{2}\right)^{-2{\bar u}}\prod^4_{j=3}\left(\frac{k_j}{2}\right)^{-2s_j+i\nu_j}.
\end{align}
\end{subequations}}
This way of decomposing the exchange is advantageous as the functions $H_{\nu_1,\nu_2,\nu_3|\alpha,\beta}\left(s_1,s_2,s_3\right)$ telescopically combine with the function $\rho_{\nu_1,\nu_2,\nu_3}\left(s_1,s_2,s_3\right)$ to shift the arguments of the Mellin integral by integers $\alpha, \beta$ (see section \ref{subsec::spinningads3pt}):
\begin{align}\label{telescopicH}
H_{\nu_1,\nu_2,\nu_3|\alpha,\beta}(s_1,s_2,s_3)\rho_{\nu_1,\nu_2,\nu_3}\left(s_1,s_2,s_3\right)=\rho_{\nu_1-i\left(\alpha-\beta\right),\nu_2-i\beta,\nu_3+i\alpha}(s^\prime_1,s^\prime_2,s^\prime_3),
\end{align}
where  $s^\prime_1= s_1+\tfrac{\alpha-\beta}{2}$, $s^\prime_2=s_2+\tfrac{\beta}{2}$ and $s^\prime_3=s_3-\tfrac{\alpha}{2}$. We thus see that the leading term in the decomposition \eqref{expspinl} (i.e. that with $\alpha_i=\beta_i=0$) is equal to the corresponding contribution \eqref{Ao<>} to the scalar exchange but where now $x={\bar x}=d+2\ell$, while the sub-leading terms differ only by integer shifts in the arguments.

From the decomposition \eqref{expspinl}, the steps to lift the $u$ and ${\bar u}$ integrals in the late-time limit are therefore the same as for the scalar exchange four-point function -- the details of which we give in appendix \ref{app:u-integral}. The resulting expression for the spin-$\ell$ exchange four-point function is
\begin{multline}\label{exchdecompspinl}
    \langle \phi^{\left(\nu_1\right)}_{\vec{k}_1}\phi^{\left(\nu_2\right)}_{\vec{k}_2}\phi^{\left(\nu_3\right)}_{\vec{k}_3}\phi^{\left(\nu_4\right)}_{\vec{k}_4} \rangle^\prime = {\cal N}_4 \sum^\ell_{\alpha_1, \alpha_2=0}\binom{\ell}{\alpha_1}\binom{\ell}{\alpha_2}\sum^{\alpha_1, \alpha_2}_{\beta_1, \beta_2=0} \binom{\alpha_1}{\beta_1}\binom{\alpha_2}{\beta_2} \\  
    \times \Theta^{\left(\ell\right)}_{\nu_1,\nu_2,\nu_3,\nu_4;\nu|\alpha_1, \beta_1;\alpha_2,\beta_2}(\vec{k}_i;\vec{k}) {\cal A}^{(d+2\ell,d+2\ell)}_{\nu_1,\nu_2,\nu_3,\nu_4|\alpha_1,\beta_1;\alpha_2,\beta_2}(\vec{k}_i;\vec{k})\,,
\end{multline}
which is a finite sum of the Mellin-Barnes integrals: 
\begin{align}\label{exch_function2}
  & {\cal A}_{\nu_j|\alpha_1,\beta_1;\alpha_2,\beta_2}^{(x,\bar{x})}(\vec{k}_i;\vec{k}) \\ \nonumber
  & \hspace*{1.5cm}\equiv 
    \lim_{\eta_0\to0}\left[\mathcal{A}^{(x,\bar{x})}_{\odot|\nu_j|\alpha_1,\beta_1;\alpha_2,\beta_2}(\vec{k}_i;\vec{k})-\mathcal{A}^{(x,\bar{x})}_{<|\nu_j|\alpha_1,\beta_1;\alpha_2,\beta_2}(\vec{k}_i;\vec{k})-\mathcal{A}^{(x,\bar{x})}_{>|\nu_j|\alpha_1,\beta_1;\alpha_2,\beta_2}(\vec{k}_i;\vec{k})\right]
    \\ \nonumber & \hspace*{1.5cm} =\int [ds]_4\, \csc(\pi(u+\bar{u}))\delta^{(x,\bar{x})}(u,\bar{u})\,\\
    \nonumber & \hspace*{2cm}    \times H_{\nu_1,\nu_2,\nu|\alpha_1,\beta_1}(s_1,s_2,u)\rho_{\nu_1,\nu_2,\nu}(s_1,s_2,u)\left(\frac{k}{2}\right)^{-2u}\prod^2_{j=1}\left(\frac{k_j}{2}\right)^{-2s_j+i\nu_j}\\
    & \hspace*{1.5cm} \times H_{\nu_3,\nu_4,-\nu|\alpha_2,\beta_2}(s_3,s_4,\bar{u}) \rho_{\nu_3,\nu_4,-\nu}(s_3,s_4,\bar{u})\left(\frac{k}{2}\right)^{-2{\bar u}}\prod^4_{j=3}\left(\frac{k_j}{2}\right)^{-2s_j+i\nu_j}\Big|_{{}^{u=\frac{x}4-s_1-s_2}_{{\bar u}=\frac{{\bar x}}4-s_3-s_4}}. \nonumber
\end{align}
Due to the telescopic nature \eqref{telescopicH} of  $H_{\nu_1,\nu_2,\nu_3|\alpha_1,\beta_1}(s_1,s_2,s_3)$, these take the same form as the Mellin-Barnes representation \eqref{exchuint} for the scalar exchange four-point function but with integer-shifted arguments.\footnote{This key property of the Mellin-Barnes representation can be used to establish recursion relations between exchange four-point functions with scaling dimensions differing by integers, which we consider in section \ref{subsec::recursion}.} Fascinatingly, since the dependence on the spin and boundary dimension $d$ always enters in the combination $d+2\ell$ through $x$ and ${\bar x}$, the knowledge of the scalar exchange four-point function for general $d$ is equivalent to knowing the exchange four-point function for general spin $\ell$ when combined with the Mellin-independent $\Theta$-polynomials \eqref{ThetaPol} in the momenta.

By comparing with the expression \eqref{CPWspinlexpMB} for the Mellin-Barnes representation of the spin-$\ell$ Conformal Partial Wave, it follows that the spin-$\ell$ exchange four-point function can moreover be equivalently expressed in the more compact form:\footnote{The factor of $4$ arises from the compensation of the factor $(2\pi i)^2$ in the measure for the $u$ and $\bar{u}$-integrals, together with the factor of $(i\pi)^2$ included in the definition \eqref{Momentum CPW} of the CPW.}
\begin{shaded}
\noindent \emph{Mellin-Barnes representation for a spin-$\ell$ exchange diagram in dS$_{d+1}$}
\begin{multline}\label{exchampl}
    \langle \phi^{\left(\nu_1\right)}_{\vec{k}_1}\phi^{\left(\nu_2\right)}_{\vec{k}_2}\phi^{\left(\nu_3\right)}_{\vec{k}_3}\phi^{\left(\nu_4\right)}_{\vec{k}_4} \rangle^\prime = {\cal N}_4\int [ds]_4\int_{-i\infty}^{+i\infty}\frac{du\,d\bar{u}}{(2\pi i)^2}\,4 \csc(\pi(u+\bar{u}))\delta^{(d+2\ell,d+2\ell)}(u,\bar{u})\\ \times {\cal F}^\prime_{\nu,\ell}(s_i;u,{\bar u}|\vec{k}_i;\vec{k})\,,
\end{multline}
\end{shaded}
\noindent which displays explicitly the relation to the corresponding spin-$\ell$ Conformal Partial Wave mellin representation. This confirms that the expression \eqref{exchamp} for the scalar exchange extends straightforwardly to the exchange of spinning fields. This universal form also carries over to external spinning fields, the details of which will be presented in \cite{ToAppear2}. 

Before exploring various properties of this expression in the subsequent sections, below we discuss the helicity decomposition, which can be performed entirely at the level of the Conformal Partial Wave. 

\paragraph{Helicity Decomposition.} 

The helicity decomposition of the four-point exchange \eqref{exchampl} can be obtained directly at the level of the $\Theta$-polynomials \eqref{ThetaPol}. Using momentum conservation these can be expanded in the form:
\begin{align}
    \Theta^{(\ell)}_{\nu_1,\nu_2,\nu_3,\nu_4;\nu|\alpha_1,\beta_1;\alpha_2,\beta_2}(\vec{k}_i;\vec{k})=\sum_{n=0}^{\ell}c^{(n)}_{\nu_1,\nu_2,\nu_3,\nu_4;\nu|\alpha_1,\beta_1;\alpha_2,\beta_2}(\vec{k}_i;\vec{k})\,(\vec{q}_{12}\cdot \vec{q}_{34})^n\,,
\end{align}
where \footnote{Note that $|\vec{q}_{12}|\ne q_{12}$, where instead $q_{12}$ is defined in this work according to \eqref{pqdef} as $q_{12}=\frac{k_1-k_2}{k}$.}
\begin{equation}
    \vec{q}_{12}=\tfrac12(\vec{k}_1- \vec{k}_2), \qquad \vec{q}_{34}=\tfrac12(\vec{k}_3- \vec{k}_4).
\end{equation}
The helicity decomposition then simply amounts to a projection of each monomial $(\vec{q}_{12}\cdot \vec{q}_{34})^n$ onto spherical harmonics $\Xi_m$ for the rotation group of the plane orthogonal to the exchanged momentum $\vec{k}$, which in our case is $SO(d-1)$. We first decompose $\vec{q}_{12}$ and $\vec{q}_{34}$ in components transverse and longitudinal to $\vec{k}$ 
\begin{align}
    \vec{q}_{12}&=\vec{q}^\perp_{12}+\left(\vec{q}_{12} \cdot {\hat k}\right){\hat k}, \qquad \vec{q}^\perp_{12} \cdot \hat{k}=0,\\
    \vec{q}_{34}&=\vec{q}^\perp_{34}+\left(\vec{q}_{34} \cdot {\hat k}\right){\hat k}, \qquad \vec{q}^\perp_{34} \cdot \hat{k}=0,
\end{align}
so that 
\begin{align}
    \vec{q}_{12}\cdot \vec{q}_{34}&=\sigma\, \hat{q}^\perp_{12}\cdot \hat{q}^\perp_{34}+\rho\,,
\end{align}
where
\begin{align}
    \sigma=|\vec{q}^\perp_{12}||\vec{q}^\perp_{34}|, \qquad
    \rho=(\vec{q}_{12}\cdot\hat{k})(\vec{q}_{34}\cdot\hat{k}).
\end{align}
We then expand
\begin{align}\label{qmonomialexp}
\left(\vec{q}_{12}\cdot\vec{q}_{34}\right)^n=(\sigma\, \hat{q}^\perp_{12}\cdot \hat{q}^\perp_{34}+\rho)^n&=\sum_{m=0}^n c^{\left(n\right)}_m(\sigma,\rho)\,\Xi_m\left(\hat{q}^\perp_{12}\cdot \hat{q}^\perp_{34}\right)\,,
\end{align}
where the coefficients are determined using the orthogonality properties of the Gegenbauer polynomial:
\begin{align}
    c^{\left(n\right)}_m(\sigma,\rho)&=\frac{2^{m-1}}{m!} \rho^{n-1} \sigma^{-m} \Gamma \left(\tfrac{d-3}{2}+m+1\right)\\\nonumber&\hspace{50pt}\times\left\{\begin{matrix}-\frac{n\, \sigma}{\Gamma\left(\frac{3-m}{2}\right)\Gamma\left(\frac{m+d}{2}\right)} {}_3F_2\left(\begin{matrix}1,\frac{1-n}{2},\frac{2-n}{2}\\\frac{3-m}{2},\frac{m+d}{2} \end{matrix};\frac{\sigma^2}{\rho^2}\right),\qquad m\: \text{odd},\\
\frac{2\,\rho}{\Gamma\left(\frac{2-m}{2}\right)\Gamma\left(\frac{m+d-1}{2}\right)} {}_3F_2\left(\begin{matrix}1,\frac{1-n}{2},-\frac{n}{2}\\\frac{2-m}{2},\frac{m+d-1}{2} \end{matrix};\frac{\sigma^2}{\rho^2}\right), \qquad m\: \text{even}.
\end{matrix}\right.
\end{align}
Using the above it is straightforward to obtain the helicity decomposition of each exchange amplitude. 

It may be also useful to note that the helicity decomposition above can also be obtained from the helicity decomposition of the corresponding three-point conformal structures \eqref{HD00l}. We have \begin{subequations}
\begin{align}
   \mathcal{Y}_{\nu_1,\nu_2,\nu|\alpha_1,\beta_1}^{(\ell)}\left(\vec{\xi}\cdot \vec{k}_1,\vec{\xi}\cdot \vec{k}_2\right)&=\sum_{m=0}^{\ell}\mathfrak{p}_{\nu_1,\nu_2,\nu|\alpha_1,\beta_1|m}^{(\ell)}\left(p_{12},q_{12}\right)\,\Xi_m\left({\xi}_\perp\cdot \vec{q}_{12}\right),\\
    \mathcal{Y}_{\nu_3,\nu_4,-\nu|\alpha_2,\beta_2}^{(\ell)}\left(\vec{\xi}\cdot \vec{k}_3,\vec{\xi}\cdot \vec{k}_4\right)&=\sum_{m=0}^{\ell}\mathfrak{p}_{\nu_3,\nu_4,-\nu|\alpha_2,\beta_2|m}^{(\ell)}\left(p_{34},q_{34}\right)\,\Xi_m\left({\xi}_\perp\cdot \vec{q}_{34}\right).
\end{align}
\end{subequations}
In terms of the above helicity components, the contribution of helicity-$m$ to \eqref{ThetaPol} is given by: 
\begin{multline}
    \Theta^{(\ell,m)}_{\nu_1,\nu_2,\nu_3,\nu_4;\nu|\alpha_1,\beta_1;\alpha_2,\beta_2}=
    \left(-ik\right)^{\alpha_1} \left(ik\right)^{\alpha_2}
    \mathfrak{p}_{\nu_1,\nu_2,\nu|\alpha_1,\beta_1|m}^{(\ell)}\left(p_{12},q_{12}\right)\\ \times \mathfrak{p}_{\nu_3,\nu_4,-\nu|\alpha_2,\beta_2|m}^{(\ell)}\left(p_{34},q_{34}\right)\Xi_m\left(\hat{q}^\perp_{12}\cdot \hat{q}^\perp_{34}\right)\,,
\end{multline}
where
\begin{equation}\label{HDthetapol}
    \Theta_{\nu_1,\nu_2,\nu_3,\nu_4;\nu|\alpha_1,\beta_1;\alpha_2,\beta_2}^{(\ell)}=\sum_{m=0}^\ell \Theta^{(\ell,m)}_{\nu_1,\nu_2,\nu_3,\nu_4;\nu|\alpha_1,\beta_1;\alpha_2,\beta_2}\,,
\end{equation}
and we recall the definition \eqref{pqdef} of $p_{12},\,p_{34}$ and $q_{12},\,q_{34}$:
\begin{subequations}\label{1234pqdef}
\begin{align}
    p_{12}&=\frac{k_1+k_2}{k}, \qquad q_{12}=\frac{k_1-k_2}{k}, \\
    p_{34}&=\frac{k_3+k_4}{k}, \qquad q_{34}=\frac{k_3-k_4}{k}.
\end{align}
\end{subequations}

The highest-helicity component receives contributions only from the leading theta polynomial (\eqref{ThetaPol} with $\alpha_i=\beta_i=0$) and is universal:
\begin{align}\label{Theta}
    {\Theta}^{\left(\ell,\ell\right)}_{\nu_1,\nu_2,\nu_3,\nu_4;\nu|0,0;0,0}=\left(-\frac{1}{4}\right)^\ell\Xi_\ell\,.
\end{align}
Lower helicity components instead depend on the external operator dimension.

\subsection{OPE limit} 
\label{subsec::OPElimit}

From the Mellin-Barnes representation \eqref{exchampl} it is straightforward to obtain Operator Product Expansion (OPE) limit of the exchange four-point function, which is the limit $k \to 0$ whilst keeping all external momenta hard.\footnote{Note that the OPE limit is often referred to in the literature as the ``collapsed limit".} In this limit, non-analytic terms in $k$ are characteristic signatures for the exchange of the single-particle state \cite{Arkani-Hamed:2015bza}. These can be extracted from the Mellin-Barnes representation in a systematic fashion, as we detail in the following.

\begin{figure}
    \centering
    \captionsetup{width=0.95\textwidth}
    \includegraphics[width=0.8\textwidth]{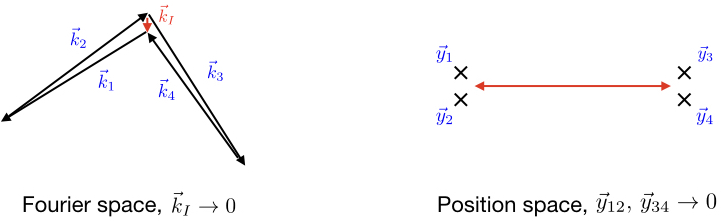}
    \caption{OPE limit in momentum space vs OPE limit in position space.}
    \label{fig:OPElimit}
\end{figure}

As discussed at the end of section \ref{subsec:Generalscalar}, only the Conformal Partial Wave \eqref{CPWspinlexpMB} encodes non-analytic contributions in the exchanged momentum, which in this limit reads:
\begin{multline}
 \lim_{k\to 0}{\cal F}^\prime_{\nu,\ell}(s_i;u,{\bar u}|\vec{k}_i;\vec{k}) = {\bar \Theta}^{(\ell)}(\vec{k}_i;\vec{k}) \left(\frac{4}{k^2}\right)^{\frac{d}{2}+\ell}\left(\frac{k_{12}}{4}\right)^{i(\nu_1+\nu_2)}\left(\frac{k_{34}}{4}\right)^{i(\nu_3+\nu_4)}\\ 
  \times (i\pi) \delta\left(\tfrac{d+2\ell}4-s_1-u\right) (i\pi) \delta\left(\tfrac{d+2\ell}4-s_3-\bar{u}\right) \\\times 
     \rho_{\nu_1,\nu_2,\nu}(s_1-s_2,s_2,u) \rho_{\nu_3,\nu_4,-\nu}(s_3-s_4,s_4,{\bar u}) \left(\frac{2k}{k_{12}}\right)^{2s_1}\left(\frac{2k}{k_{34}}\right)^{2s_3},
\end{multline}
where we used that $\vec{k}_1 \sim -\vec{k}_2$ and $\vec{k}_3 \sim -\vec{k}_4$ as $k \to 0$ due to momentum conservation, and that only the leading term (with $\alpha_i=\beta_i=0$) in the contraction \eqref{fullthetacontra} survives in the limit $k \to 0$, where it reduces to
\begin{subequations}
\begin{align}
 {\bar \Theta}^{(\ell)}\left(\vec{q}_{12},\vec{q}_{34}\right) &=  \lim_{k\to 0} \Theta^{(\ell)}_{\nu_1,\nu_2,\nu_3,\nu_4;\nu|0,0;0,0}(\vec{k}_i;\vec{k})\\
 &=\frac{\ell!(-1)^\ell}{2^{\ell} \left(\frac{d-2}{2} \right)_\ell}\,\left(\frac{k_{12} k_{34}}{4}\right)^\ell\,C_\ell^{\left(\frac{d-2}{2} \right)}\left(\cos\theta\right),
\end{align}
\end{subequations}
where $\cos\theta=\hat{k}_{1}\cdot \hat{k}_{3}$. For convenience we also made the change of variables $s_1 \to s_1 - s_2$ and $s_3 \to s_3 - s_4$, so that the expansion in $k$ is obtained from the integrals in $s_1$ and $s_3$ by closing the contours to the right. This encircles the following poles of the CPW:
\begin{align}\label{opepoles}
    s_1=\frac{d}{4}+\frac{\ell}{2}\pm\frac{i\nu}{2}+n, \qquad  s_3=\frac{d}{4}+\frac{\ell}{2}{\hat \pm}\frac{i\nu}{2}+m, \qquad n,m \in \mathbb{N}_0.
\end{align}
Since the exponent of $k$ depends on $s_1$ and $s_3$ through their sum, only the residues of the poles \eqref{opepoles} with the same sign for $\nu$ generate non-analytic terms:
\begin{equation}
    s_1=\frac{d}{4}+\frac{\ell}{2}\pm\frac{i\nu}{2}+n, \quad  s_3=\frac{d}{4}+\frac{\ell}{2} \pm\frac{i\nu}{2}+m, \quad \to \quad \left(k^2\right)^{\pm i\nu+n+m},
\end{equation}
where only the terms generated by the leading poles (those with $n=m=0$) survive in the limit $k \to 0$. The remaining Mellin integrals in $s_2$ and $s_4$ can be lifted using Barnes' first lemma, which gives 
\begin{shaded}\noindent \emph{OPE limit of a general spin-$\ell$ exchange in dS$_{d+1}$}
\begin{multline}\label{OPElimspinl}
  \lim_{k \to 0}\,  \langle \phi^{\left(\nu_1\right)}_{\vec{k}_1}\phi^{\left(\nu_2\right)}_{\vec{k}_2}\phi^{\left(\nu_3\right)}_{\vec{k}_3}\phi^{\left(\nu_4\right)}_{\vec{k}_4} \rangle^\prime = (-1)^\ell\frac{{\cal N}_4}{(4\pi)^3}\, \left(\frac{16}{k_{12}k_{34}}\right)^{\tfrac{d}{2}}  \prod^4_{j=1} \left(\frac{k_j}{2}\right)^{i\nu_j} 
  \\ \times   \frac{2^{\ell}\ell!}{\left(\frac{d-2}{2} \right)_\ell} C_\ell^{\left(\frac{d-2}{2} \right)}\left(\cos\theta\right)\left[\left(\frac{4k^2}{k_{12}k_{34}}\right)^{i\nu}\frac{\Gamma(-i\nu)^2}{\Gamma\left(\tfrac{d}{2}+i\nu+\ell\right)^2}\text{csch}(\pi\nu)\delta^{(d+2\ell,d+2\ell)}\left(-\tfrac{i\nu}{2},-\tfrac{i\nu}{2}\right)  \right. \\ \left. \times \prod_{\pm {\hat \pm}} \Gamma\left(\tfrac{d+2i\nu+2\ell}{4} \pm \tfrac{i\nu_1}2{\hat \pm}\tfrac{i\nu_2}2\right) \Gamma\left(\tfrac{d+2i\nu+2\ell}{4} \pm \tfrac{i\nu_3}2{\hat \pm}\tfrac{i\nu_4}2\right)+ \nu \rightarrow -\nu\right].
\end{multline}
\end{shaded}
\noindent Recall the definition \eqref{1234pqdef} of $p_{12}$ and $p_{34}$ in terms of $k_i$ and $k$. As far as we are aware, this is the first expression available for general external scalars both for $d=3$ and general $d>3$. 

Note that the expression \eqref{OPElimspinl} contains oscillatory terms in $\log\left(\frac{k^2}{k_{12}k_{34}}\right)$ when the exchanged particle is a massive particle on the Principal Series, $\nu \in \mathbb{R}$, which was previously worked out explicitly for external conformally coupled and massless scalars when $d=3$ in \cite{Arkani-Hamed:2015bza}. The phase of the oscillations in particular depends on the interference factor \eqref{symmmellinzeros}. The angular dependence is encoded in the Gegenbauer polynomial, which signals that we are exchanging a spin-$\ell$ particle. When $d=3$ the Gegenbauer polynomial reduces to a Legendre polynomial, as consistent with the $d=3$ analysis in \cite{Arkani-Hamed:2015bza}.

As a consistency check, for $d=3$ and for external conformally coupled and massless scalars this expression coincides respectively with equations (5.120) and (5.123) in \cite{Arkani-Hamed:2015bza}. For example, for external massless scalars in $d=3$ (where $\nu_j=\frac{3i}{2}$) we have\footnote{Note that there is a typo in (5.123) of \cite{Arkani-Hamed:2015bza}, where the factor $\left(\tfrac{3}2-\ell-i\nu\right)^2$ has a $``+"$ in front of $\ell$ instead of $``-"$.} 
\begin{multline}
\lim_{k \to 0}\langle \phi^{\left(3i/2\right)}_{\vec{k}_1}\phi^{\left(3i/2\right)}_{\vec{k}_2}\phi^{\left(3i/2\right)}_{\vec{k}_3}\phi^{\left(3i/2\right)}_{\vec{k}_4} \rangle^\prime =  \frac{\mathcal{N}_4}{2\pi}\,\frac{\left(k_{12} k_{34}\right)^{\frac{3}{2}}}{k_1^3k_2^3k_3^3k_4^3}\,\frac{\left(-1\right)^{\ell}\ell!}{8^{\ell} \left(\frac{d-2}{2} \right)_\ell}\,C_\ell^{\left(\frac{d-2}{2} \right)}\left(\cos\theta\right)\\\times  \left(\frac{k^2}{4k_{12}k_{34}}\right)^{i\nu}\frac{\left(\tfrac{5}2+\ell+i \nu \right)^2}{\left(\tfrac{3}2-\ell-i\nu\right)^2}\,\Big[1+i(-1)^\ell\sinh (\pi \nu )\Big] \Gamma (-i \nu )^2\Gamma \left(\ell+i \nu +\tfrac{1}{2}\right)^2\\+(\nu\to-\nu)\,.
\end{multline}
The factor $\left(-1\right)^\ell$ is not observed in the analysis of \cite{Arkani-Hamed:2015bza} since equal external scalars are assumed from the beginning, which precludes the exchange of odd spin $\ell$.

\subsection{Recursion relations}
\label{subsec::recursion}

A nice feature of the Mellin-Barnes representation is that it makes manifest certain recursion relations between late-time correlators with different scaling dimensions -- as we briefly saw in \S\ref{sec::3pt} at the level of three-point functions. Such recursion relations are valid also at the level of four-point functions, and can be useful for the scaling dimensions where the initial (or ``seed") correlator has a simpler form with respect to the Mellin-Barnes representation for generic scaling dimensions.\footnote{We shall consider some examples of this type in \S \ref{subsec::someparticcases}.} For this reason similar recursion relations are often used in the study of scattering processes on (anti-)de Sitter space, in particular for the technically more complicated case of correlators involving spinning fields, see e.g. \cite{Sleight:2016dba}\cite{Sleight:2016hyl}\cite{Sleight:2017krf}\cite{Castro:2017hpx}\cite{Sleight:2017fpc}\cite{Chen:2017yia}\cite{Costa:2018mcg}\cite{Isono:2018rrb}\cite{Arkani-Hamed:2018kmz}\cite{Isono:2019ihz}.

As we saw in \S \ref{subsecc::spinlexch}, a spin-$\ell$ exchange four-point function is defined by the collection \eqref{exch_function2} of Mellin-Barnes integrals. The external scaling dimensions in these Mellin-Barnes integrals can then be raised and lowered by acting with simple operators in the external momenta. In particular, the raising operators
\begin{subequations}\label{raisingop}
\begin{align} \nonumber
   & {\cal A}_{\nu_1-i,\nu_2,\nu_3,\nu_4|\alpha_1,\beta_1;\alpha_2,\beta_2}^{(x,\bar{x})}(\vec{k}_i;\vec{k})\\
   & \hspace*{2cm}=\frac{1}{2} k_1^{1+2 i \nu_1+2(\alpha_1-\beta_1)}\partial_{k_1}\left[k_1^{-2i\nu_1-2(\alpha_1-\beta_1)}{\cal A}_{\nu_1,\nu_2,\nu_3,\nu_4|\alpha_1,\beta_1;\alpha_2,\beta_2}^{(x-2,\bar{x})}(\vec{k}_i;\vec{k})\right]\,,\\ \nonumber
   & {\cal A}_{\nu_1,\nu_2-i,\nu_3,\nu_4|\alpha_1,\beta_1;\alpha_2,\beta_2}^{(x,\bar{x})}(\vec{k}_i;\vec{k})\\
   & \hspace*{2cm}=\frac{1}{2} k_2^{1+2 i \nu_2+2\beta_1}\partial_{k_2}\left[k_2^{-2i\nu_2-2\beta_1}{\cal A}_{\nu_1,\nu_2,\nu_3,\nu_4|\alpha_1,\beta_1;\alpha_2,\beta_2}^{(x-2,\bar{x})}(\vec{k}_i;\vec{k})\right]\,,\\ \nonumber
   & {\cal A}_{\nu_1,\nu_2,\nu_3-i,\nu_4|\alpha_1,\beta_1;\alpha_2,\beta_2}^{(x,\bar{x})}(\vec{k}_i;\vec{k})\\
   & \hspace*{2cm}=\frac{1}{2} k_3^{1+2 i \nu_3+2(\alpha_2-\beta_2)}\partial_{k_3}\left[k_3^{-2i\nu_3-2(\alpha_2-\beta_2)}{\cal A}_{\nu_1,\nu_2,\nu_3,\nu_4|\alpha_1,\beta_1;\alpha_2,\beta_2}^{(x,\bar{x}-2)}(\vec{k}_i;\vec{k})\right]\,,\\ \nonumber
   & {\cal A}_{\nu_1,\nu_2,\nu_3,\nu_4-i|\alpha_1,\beta_1;\alpha_2,\beta_2}^{(x,\bar{x})}(\vec{k}_i;\vec{k})\\ 
   & \hspace*{2cm}=\frac{1}{2} k_4^{1+2 i \nu_4+2\beta_2}\partial_{k_4}\left[k_4^{-2i\nu_4-2\beta_2}{\cal A}_{\nu_1,\nu_2,\nu_3,\nu_4|\alpha_1,\beta_1;\alpha_2,\beta_2}^{(x,\bar{x}-2)}(\vec{k}_i;\vec{k})\right]\,,
\end{align}
\end{subequations}
increase the external scaling dimensions by an integer, while the lowering operators are given by:
\begin{subequations}\label{recursion1}
\begin{align}
    {\cal A}_{\nu_1+i,\nu_2,\nu_3,\nu_4|\alpha_1,\beta_1;\alpha_2,\beta_2}^{(x,\bar{x})}(\vec{k}_i;\vec{k})&=-\frac{2}{k_1}\partial_{k_1}{\cal A}_{\nu_1,\nu_2,\nu_3,\nu_4|\alpha_1,\beta_1;\alpha_2,\beta_2}^{(x-2,\bar{x})}(\vec{k}_i;\vec{k})\,,\\
    {\cal A}_{\nu_1,\nu_2+i,\nu_3,\nu_4|\alpha_1,\beta_1;\alpha_2,\beta_2}^{(x,\bar{x})}(\vec{k}_i;\vec{k})&=-\frac{2}{k_2}\partial_{k_2}{\cal A}_{\nu_1,\nu_2,\nu_3,\nu_4|\alpha_1,\beta_1;\alpha_2,\beta_2}^{(x-2,\bar{x})}(\vec{k}_i;\vec{k})\,,\\
    {\cal A}_{\nu_1,\nu_2,\nu_3+i,\nu_4|\alpha_1,\beta_1;\alpha_2,\beta_2}^{(x,\bar{x})}(\vec{k}_i;\vec{k})&=-\frac{2}{k_3}\partial_{k_3}{\cal A}_{\nu_1,\nu_2,\nu_3,\nu_4|\alpha_1,\beta_1;\alpha_2,\beta_2}^{(x,\bar{x}-2)}(\vec{k}_i;\vec{k})\,,\\
    {\cal A}_{\nu_1,\nu_2,\nu_3,\nu_4+i|\alpha_1,\beta_1;\alpha_2,\beta_2}^{(x,\bar{x})}(\vec{k}_i;\vec{k})&=-\frac{2}{k_4}\partial_{k_4}{\cal A}_{\nu_1,\nu_2,\nu_3,\nu_4|\alpha_1,\beta_1;\alpha_2,\beta_2}^{(x,\bar{x}-2)}(\vec{k}_i;\vec{k})\,.
\end{align}
\end{subequations}
These relations are straightforward to establish due to the simple power-law dependence of the momenta on the Mellin variables. Since the dependence of exchange four-point functions on the internal spin $\ell$ and boundary dimension $d$ enters through $x$ and ${\bar x}$ in the combination $d+2\ell$, the above operators can also be used relate exchange four-point functions with different internal spins $\ell$ and/or $d$. We shall see some examples of this type in sections \ref{subsec::recursion} and \ref{subsec::someparticcases}.

In a similar way we can lift the Pochhammer factors carried by the functions $H_{\nu_1,\nu_2,\nu_3|\alpha_1,\beta_1}$ by replacing them with the action of a differential operator in the external momentum. For example, the shifts (Pochhammer factors) associated to the external scaling dimensions can be lifted from \eqref{exch_function2} via
\begin{multline}\label{alphabeta_fromgamma}
   {\cal A}_{\nu_1,\nu_2,\nu_3,\nu_4|\alpha_1,\beta_1;\alpha_2,\beta_2}^{(x,\bar{x})}(\vec{k}_i;\vec{k})=(-1)^{\alpha_1+\alpha_2} k_1^{2 (\alpha_1-\beta_1+i \nu_1)}k_2^{2 (\beta_1+i \nu_2)}k_3^{2 (\alpha_2-\beta_2+i \nu_3)}k_4^{2 (\beta_2+i \nu_4)}\\\times\pl_{k_1^2}^{\alpha_1-\beta_1}\pl_{k_2^2}^{\beta_1}\pl_{k_3^2}^{\alpha_2-\beta_2}\pl_{k_4^2}^{\beta_2}\left[k_1^{-2i\nu_1}k_2^{-2i\nu_2}k_3^{-2i\nu_3}k_4^{-2i\nu_4}{\cal A}_{\nu_1,\nu_2,\nu_3,\nu_4|\alpha_1;\alpha_2}^{(x,\bar{x})}(\vec{k}_i;\vec{k})\right]\,,
\end{multline}
where 
\begin{multline}\label{seedalphabeta}
    {\cal A}_{\nu_1,\nu_2,\nu_3,\nu_4|\alpha_1;\alpha_2}^{(x,\bar{x})}(\vec{k}_i;\vec{k})\equiv \int [ds]_4\, \csc(\pi(u+\bar{u}))\delta^{(x,\bar{x})}(u,\bar{u})\\\times\frac{\rho_{\nu_1,\nu_2,\nu}(s_1,s_2,u)\rho_{\nu_3,\nu_4,-\nu}(s_3,s_4,\bar{u})}{\left(u+\tfrac{i\nu}2-\alpha_1\right)_{\alpha_1}\left(\bar{u}-\tfrac{i\nu}2-\alpha_2\right)_{\alpha_2}}\left(\frac{k}{2}\right)^{-2\left(u+{\bar u}\right)}\prod^4_{j=1}\left(\frac{k_j}{2}\right)^{-2s_j+i\nu_j}\Big|_{{}^{u=\frac{x}4-s_1-s_2}_{{\bar u}=\frac{{\bar x}}4-s_3-s_4}},
\end{multline}
so that only the shifts associated to the internal scaling dimensions remain. This recursion relation will turn out to be useful for the scaling dimensions where some of the Mellin-Barnes integrals can be straightforwardly lifted from the seed integral \eqref{seedalphabeta}, as we shall see in section \ref{subsec::someparticcases}.

The seed integral \eqref{seedalphabeta} can furthermore be expressed recursively in terms of that with $\alpha_1=\alpha_2=\ell$ via:
\begin{multline}\label{seed_recursion_full}
    {\cal A}_{\nu_1,\nu_2,\nu_3,\nu_4|\alpha_1;\alpha_2}^{(x,\bar{x})}(\vec{k}_i;\vec{k})=\pl_{\lambda_1}^{\ell-\alpha_1}\pl_{\lambda_2}^{\ell-\alpha_2}\Big[\lambda_1 ^{-\alpha_1+\frac{i \nu }{2}-\frac{i \nu_1}{2}-\frac{i \nu_2}{2}+\frac{x}{4}-1} \lambda_2^{-\alpha_2-\frac{i \nu }{2}-\frac{i \nu_3}{2}-\frac{i \nu_4}{2}+\frac{\bar{x}}{4}-1}\\\times{\cal A}_{\nu_1,\nu_2,\nu_3,\nu_4|\ell;\ell}^{(x,\bar{x})}(\lambda_1^{1/2} \vec{k}_1,\lambda_1^{1/2} \vec{k}_2,\lambda_2^{1/2} \vec{k}_3,\lambda_2^{1/2} \vec{k}_4;\vec{k}\,)\Big]_{\lambda_1,\lambda_2=1}\,.
\end{multline}
In this way the entire spin-$\ell$ exchange four-point function can be generated from the single seed integral \eqref{seedalphabeta} with $\alpha_1=\alpha_2=\ell$, combined with the $\Theta$-polynomials \eqref{ThetaPol} in the momenta.

\subsection{EFT expansion}
\label{subsec::EFTexp}

As discussed at the end of section \ref{subsec:Generalscalar}, the effective field theory expansion of the exchange four-point function is encoded in the poles \eqref{eftpoles} of the $\csc$-factor in the Mellin-Barnes representation \eqref{exchampl}. We expand upon this in the following section, for simplicity focusing mostly on correlators with external conformally coupled scalars and external massless scalars, though all the steps carry over to the general case with a few minor technical complications.

\paragraph{External Conformally Coupled Scalars.} A useful example to consider is when all external scalars are conformally coupled. In this case, two of the four Mellin-Barnes integrals can be lifted by virtue of the Legendre duplication formula, which simplifies the extraction of the expansion coefficients. Furthermore, when $d$ is odd, the result for external massless scalars (or scalars anywhere on the discrete series) can be obtained by acting on the result for external conformally coupled scalars a finite number of times with the raising operators \eqref{raisingop}, which we consider towards the end of this section. 

Let us first review the case where the exchanged field is a scalar, which was presented in \cite{Charlotte}. In this case, the Mellin-Barnes representation of the exchange four-point function reads
\begin{multline}\label{confCMellin}
    \langle \phi^{\left(i/2\right)}_{\vec{k}_1}\phi^{\left(i/2\right)}_{\vec{k}_2}\phi^{\left(i/2\right)}_{\vec{k}_3}\phi^{\left(i/2\right)}_{\vec{k}_4} \rangle^\prime=\frac1{\pi}\frac{1}{k_1 k_2 k_3 k_4}\left(\frac{k}2\right)^{2-d}\int_{-i\infty}^{+i\infty}\frac{du d{\bar u}}{(2\pi i)^2}\,(2p_{12})^{1-\frac{d}{4}+u}(2p_{34})^{1-\frac{d}{4}+{\bar u}}\\\times \csc \left(\pi (u+{\bar u})\right)\,\delta^{(d,d)}(u,{\bar u})\,\Gamma \left(\tfrac{d-2}{2}-2u\right)\Gamma \left(\tfrac{d-2}{2}-2{\bar u}\right) \\\times\, \Gamma \left(u+\tfrac{i \nu}{2}\right) \Gamma \left(u-\tfrac{i \nu}{2}\right) \Gamma \left({\bar u}+\tfrac{i \nu}{2}\right) \Gamma \left({\bar u}-\tfrac{i \nu}{2}\right)\,,
\end{multline}
where the interference factor simplifies to:
\begin{multline}
    \delta^{(d,d)}(u,{\bar u})=\cos ^2\left(\pi \left(\tfrac{d}{4}-{\bar u}\right) \right) \sin \left(\pi \left(u+{\bar u}\right)\right)\\-\sin \left(\pi  \left(\tfrac{d}{2}-u-{\bar u}\right)\right) \sin \left(\pi\left({\bar u}-\tfrac{i \nu}{2}\right)\right) \sin \left(\pi\left({\bar u}+\tfrac{i \nu}{2}\right)\right).
\end{multline}

Re-defining $u \to u-{\bar u}$ so that the $\csc$ factor is just a function of $u$, the above Mellin-Barnes integral can be expressed as a series expansion in $\frac{1}{p_{12}}$ and $\frac{p_{34}}{p_{12}}$ by evaluating the residues of the $\csc$ poles \eqref{eftpoles} in $u$ and the poles ${\bar u} \sim \frac{d-2}{4}+m+\frac{n}{2}$, with $m,n \in \mathbb{N}_0$, which gives\footnote{Similarly one can obtain an expansion in $\frac{1}{p_{34}}$ and $\frac{p_{12}}{p_{34}}$ by instead re-defining ${\bar u} \to {\bar u}-u$.}
\begin{equation}
     \langle \phi^{\left(i/2\right)}_{\vec{k}_1}\phi^{\left(i/2\right)}_{\vec{k}_2}\phi^{\left(i/2\right)}_{\vec{k}_3}\phi^{\left(i/2\right)}_{\vec{k}_4} \rangle^\prime\Big|_{\text{EFT}} = -\mathcal{N}_4\frac{\sin \left(\frac{\pi  d}{2}\right)}{k_1 k_2 k_3 k_4}\left(\frac{k}2\right)^{2-d}\sum_{n,m=0}^{\infty}c^{\left(d\right)}_{mn}\,p_{12}^{2-d-2 m} \left(\frac{p_{34}}{p_{12}}\right)^n,
\end{equation}
where 
\begin{equation}\label{cmndS}
    c^{\left(x\right)}_{mn}=\frac{\left(-1\right)^n}{2^{x-1+2m}n!}\frac{(x+2 m+n-3)!}{ \left(\tfrac{x+2 n+2 i \nu -2}{4}\right)_{m+1}\left(\tfrac{x+2 n-2 i \nu -2}{4}\right)_{m+1}},
\end{equation}
and recall the definition \eqref{1234pqdef} of $p_{12}$ and $p_{34}$ in terms of $k_i$ and $k$. The non-perturbative corrections to the EFT expansion are generated by the remaining poles \eqref{opepoles} of the Conformal Partial Wave, which gives the factorised contributions: 
\begin{align}
     \langle \phi^{\left(i/2\right)}_{\vec{k}_1}\phi^{\left(i/2\right)}_{\vec{k}_2}\phi^{\left(i/2\right)}_{\vec{k}_3}\phi^{\left(i/2\right)}_{\vec{k}_4} \rangle^\prime&\Big|_{\text{non-pert.}}   \\ \nonumber
     =  \frac{\pi{\cal N}_4 }{k_1k_2k_3k_4}\left(\frac{k}{2}\right)^{2-d}\Big[&\text{csch}^2(\pi  \nu ) \cos ^2\left(\tfrac{\pi}{4} (d-2 i \nu )\right)\left(\mathfrak{F}^{\left(d\right)}_+(p_{12})-\mathfrak{F}^{\left(d\right)}_-(p_{12})\right)\mathfrak{F}^{\left(d\right)}_-(p_{34})\\ \nonumber
     +&\text{csch}^2(\pi  \nu ) \cos ^2\left(\tfrac{\pi}{4} (d+2 i \nu )\right)\left(\mathfrak{F}^{\left(d\right)}_-(p_{12})-\mathfrak{F}^{\left(d\right)}_+(p_{12})\right)\mathfrak{F}^{\left(d\right)}_+(p_{34})\\ \nonumber
     + &i\cosh(\pi\nu)\sin\left(\tfrac{\pi d}2\right)\left(\mathfrak{F}^{\left(d\right)}_-(p_{12})\mathfrak{F}^{\left(d\right)}_+(p_{34})-\mathfrak{F}^{\left(d\right)}_+(p_{12})\mathfrak{F}^{\left(d\right)}_-(p_{34})\right)\Big]
\end{align}
where\footnote{In the language of \cite{Arkani-Hamed:2015bza,Arkani-Hamed:2018kmz}, the functions \eqref{homo3pt} are homogeneous solutions to the conformal invariance condition on exchange four-point functions.}
\begin{align}\label{homo3pt}
    \mathfrak{F}^{\left(x\right)}_{\pm}(z)&= (2 z)^{-\frac{x}{2}\mp i \nu +1} \frac{\Gamma \left(\tfrac{x}{2}\pm i \nu -1\right)}{\Gamma\left(1\pm i \nu\right)}  \, {}_2F_1\left(\begin{matrix}\frac{x\pm2 i \nu -2}{4},\frac{x\pm2 i \nu}{4}\\1\pm i \nu \end{matrix};\frac{1}{z^2}\right)\,,
\end{align}
which are three-point conformal structures contributed by a single series of poles \eqref{opepoles} in the Conformal Partial Wave. Note that for certain (imaginary) values of $\nu$ the above expressions exhibit singularities that require regularisation, as we discuss in section \ref{subsec::someparticcases}.

In the same way one obtains the EFT expansion for a general spin-$\ell$ exchange \eqref{exchampl}. In fact, for the leading helicity component, the above result for the scalar exchange can be recycled using that the leading term in the decomposition \eqref{exchdecompspinl} (with $\alpha_i=\beta_i=0$) is proportional to the scalar exchange four-point function but with $d \to d+2\ell$. In particular,
\begin{shaded}
\begin{multline}\label{spinlEFT}
     \langle \phi^{\left(i/2\right)}_{\vec{k}_1}\phi^{\left(i/2\right)}_{\vec{k}_2}\phi^{\left(i/2\right)}_{\vec{k}_3}\phi^{\left(i/2\right)}_{\vec{k}_4} \rangle^\prime\Big|_{\text{EFT, helicity-}\ell}  = -\mathcal{N}_4\left(-\frac{1}4\right)^\ell\,\Xi_\ell\,\frac{\sin \left(\frac{\pi \left(d+2\ell\right)}{2}\right)}{k_1 k_2 k_3 k_4}\left(\frac{k}2\right)^{2-d-2\ell}\\ \times \sum_{n,m=0}^{\infty}c^{\left(d+2\ell\right)}_{mn}\,p_{12}^{2-d-2\left(\ell+m\right)} \left(\frac{p_{34}}{p_{12}}\right)^n
\end{multline}
\end{shaded}
\noindent and 
\begin{shaded}
\begin{align}\label{nonpertspinl}
     \langle \phi^{\left(i/2\right)}_{\vec{k}_1}&\phi^{\left(i/2\right)}_{\vec{k}_2}\phi^{\left(i/2\right)}_{\vec{k}_3}\phi^{\left(i/2\right)}_{\vec{k}_4} \rangle^\prime\Big|_{\text{non-pert., helicity-}\ell} = {\cal N}_4\left(-\frac{1}4\right)^\ell\,\Xi_\ell\,\frac{\pi}{k_1k_2k_3k_4}\left(\frac{k}{2}\right)^{2-d-2\ell}\\\nonumber
     \times \Big[&\text{csch}^2(\pi  \nu ) \cos ^2\left(\tfrac{\pi}{4} (d+2\ell-2 i \nu )\right)\left(\mathfrak{F}^{\left(d+2\ell\right)}_+(p_{12})-\mathfrak{F}^{\left(d+2\ell\right)}_-(p_{12})\right)\mathfrak{F}^{\left(d+2\ell\right)}_-(p_{34})\\\nonumber
     +&\text{csch}^2(\pi  \nu ) \cos ^2\left(\tfrac{\pi}{4} (d+2\ell+2 i \nu )\right)\left(\mathfrak{F}^{\left(d+2\ell\right)}_-(p_{12})-\mathfrak{F}^{\left(d+2\ell\right)}_+(p_{12})\right)\mathfrak{F}^{\left(d+2\ell\right)}_+(p_{34})\\\nonumber
     +&i\cosh(\pi\nu)\sin\left(\tfrac{\pi \left(d+2\ell\right)}2\right)\left(\mathfrak{F}^{\left(d+2\ell\right)}_-(p_{12})\mathfrak{F}^{\left(d+2\ell\right)}_+(p_{34})-\mathfrak{F}^{\left(d+2\ell\right)}_+(p_{12})\mathfrak{F}^{\left(d+2\ell\right)}_-(p_{34})\right)\Big].
\end{align}
\end{shaded}
\noindent Similar expressions for the sub-leading terms in \eqref{exchdecompspinl} can be easily obtained using the recursion relations in section \ref{subsec::recursion}. When the exchanged particle is massless, these sub-leading terms are just local contact-terms since they contribute only to the lower helicity components of the exchange four-point function -- which, for the exchange of massless particles, do not encode the propagating degrees of freedom.

\paragraph{External Massless Scalars.} The EFT expansion for external scaling dimensions differing from that of the conformally coupled scalar by an integer can be obtained from the expressions \eqref{spinlEFT} and \eqref{nonpertspinl} by acting with the raising operators \eqref{raisingop}. When $d$ is odd this includes external massless scalars.\footnote{This is because conformally coupled scalars have scaling dimension $\nu=\frac{i}{2}$ while massless scalars have $\nu=\frac{di}{2}$, so one can only be reached from the other in integer steps if $d$ is odd.} We shall focus on $d=3$, for which massless scalars have $\nu=\frac{3i}{2}$, so we only need to act once with each raising operator \eqref{raisingop}. In particular,
\begin{multline}\label{masslessextfromcc}
    \langle \phi^{\left(3i/2\right)}_{\vec{k}_1}\phi^{\left(3i/2\right)}_{\vec{k}_2}\phi^{\left(3i/2\right)}_{\vec{k}_3}\phi^{\left(3i/2\right)}_{\vec{k}_4} \rangle^\prime\Big|_{\text{helicity-}\ell}\\=\frac{16}{k_1k_2k_3k_4} \partial_{k_1}\partial_{k_2}\partial_{k_3}\partial_{k_4} \left[\langle \phi^{\left(i/2\right)}_{\vec{k}_1}\phi^{\left(i/2\right)}_{\vec{k}_2}\phi^{\left(i/2\right)}_{\vec{k}_3}\phi^{\left(i/2\right)}_{\vec{k}_4} \rangle^\prime\Big|_{\text{helicity-}\ell,\,x=d-4+2\ell}\right].
\end{multline}
Evaluating the action of the differential operators gives\footnote{Note that in $d=3$ the sum over $m$ should start from $m=2$. This is possible up to contact terms making use of the ambiguity in the splitting of the poles in the $\csc$ factor, as discussed towards the end of \S\ref{subsecc::spinlexch}.} 
\begin{multline}
    \langle \phi^{\left(3i/2\right)}_{\vec{k}_1}\phi^{\left(3i/2\right)}_{\vec{k}_2}\phi^{\left(3i/2\right)}_{\vec{k}_3}\phi^{\left(3i/2\right)}_{\vec{k}_4} \rangle^\prime\Big|_{\text{EFT, helicity-}\ell}=-\mathcal{N}_4\left(-\frac{1}4\right)^\ell\,\Xi_\ell\,\left(\frac{k}{2}\right)^{6-d-2\ell}\frac{\sin \left(\frac{\pi \left(d+2\ell\right)}{2}\right)}{(k_1 k_2 k_3 k_4)^3}\\ \times \sum^\infty_{n,m}\frac{(n-1)  \Gamma (d+2\ell+2 m+n-4)}{p_{34}^2 2^{d+2\ell+2m-5} n! \left(\tfrac{d+2\ell+2 n-2 i \nu -6}{4}\right)_{m+1} \left(\tfrac{d+2\ell+2 n+2 i \nu -6}{4}\right)_{m+1}}\nonumber\\\times 
    \left((n-4) p_{34}^2-n q_{34}^2\right)\left(p_{12}^2 \frac{d+2\ell+2 m+n-2}{d+2\ell+2 m+n-6}
     -q_{12}^2\right)p_{12}^{4-d-2\left(\ell+m\right)} \left(-\frac{p_{34}}{p_{12}}\right)^n.
\end{multline}
Setting $d=3$, this gives the EFT expansion of the helicity-$\ell$ component of the spin-$\ell$ exchange four-point function with massless external scalars.

Let us stress that in obtaining the above EFT expansion using the raising operators \eqref{raisingop} we are not requiring that the corresponding cubic coupling for the massless scalars is shift-symmetric. To impose shift-symmetry, these raising operators need to be modified. This only affects the EFT expansion by a finite number of additional contact terms since the field redefinition relating the couplings are local.\footnote{The fact that our couplings are related to shift-symmetric couplings by a local field re-definition follows from the fact that couplings of two scalar fields to a spin-$\ell$ field are unique on-shell.} The non-perturbative corrections are only changed by an overall factor which is a polynomial in the scaling dimensions.\footnote{These polynomials are straightforward to work out in each case using the formalism developed in \cite{Joung:2012fv,Sleight:2017fpc}.} For $d=3$, the operator corresponding to the shift-symmetric coupling $\sigma(\nabla\phi)^2$ of two massless scalars $\phi$ to a massive scalar $\sigma$ was given in \cite{Arkani-Hamed:2015bza} and reads
\begin{align}\label{operO}
    O_{ij}=\frac{k^2}8[(q_{ij}^2-p_{ij}^2)(1-p_{ij}^2)\pl_{p_{ij}}^2+2(p_{ij}^2+q_{ij}^2-2)(1-p_{ij}\pl_{p_{ij}})]\,,
\end{align}
so that the corresponding exchange four-point function is generated from the result \eqref{confCMellin} for external conformaly coupled scalars via:
\begin{align}\label{masslessextfromcc2}
    \langle \phi^{\left(3i/2\right)}_{\vec{k}_1}\phi^{\left(3i/2\right)}_{\vec{k}_2}\phi^{\left(3i/2\right)}_{\vec{k}_3}\phi^{\left(3i/2\right)}_{\vec{k}_4} \rangle^\prime\Big|_{\ell=0,d=3}=O_{12}O_{34} \left[\langle \phi^{\left(i/2\right)}_{\vec{k}_1}\phi^{\left(i/2\right)}_{\vec{k}_2}\phi^{\left(i/2\right)}_{\vec{k}_3}\phi^{\left(i/2\right)}_{\vec{k}_4} \rangle^\prime\Big|_{\ell=0,d=3}\right].
\end{align}
For shift-symmetric couplings of massless fields to spin-$\ell$ fields, differential operators for spin-1 and spin-2 were obtained in \cite{Arkani-Hamed:2018kmz}. An equivalent systematic approach to obtain the result for shift-symmetric couplings is to add a finite number of contact terms either via a field redefinition or by fixing the contact term ambiguity through the requirement of reproducing the correct Adler zero.

\begin{figure}
    \centering
    \captionsetup{width=0.95\textwidth}
    \includegraphics[width=0.7\textwidth]{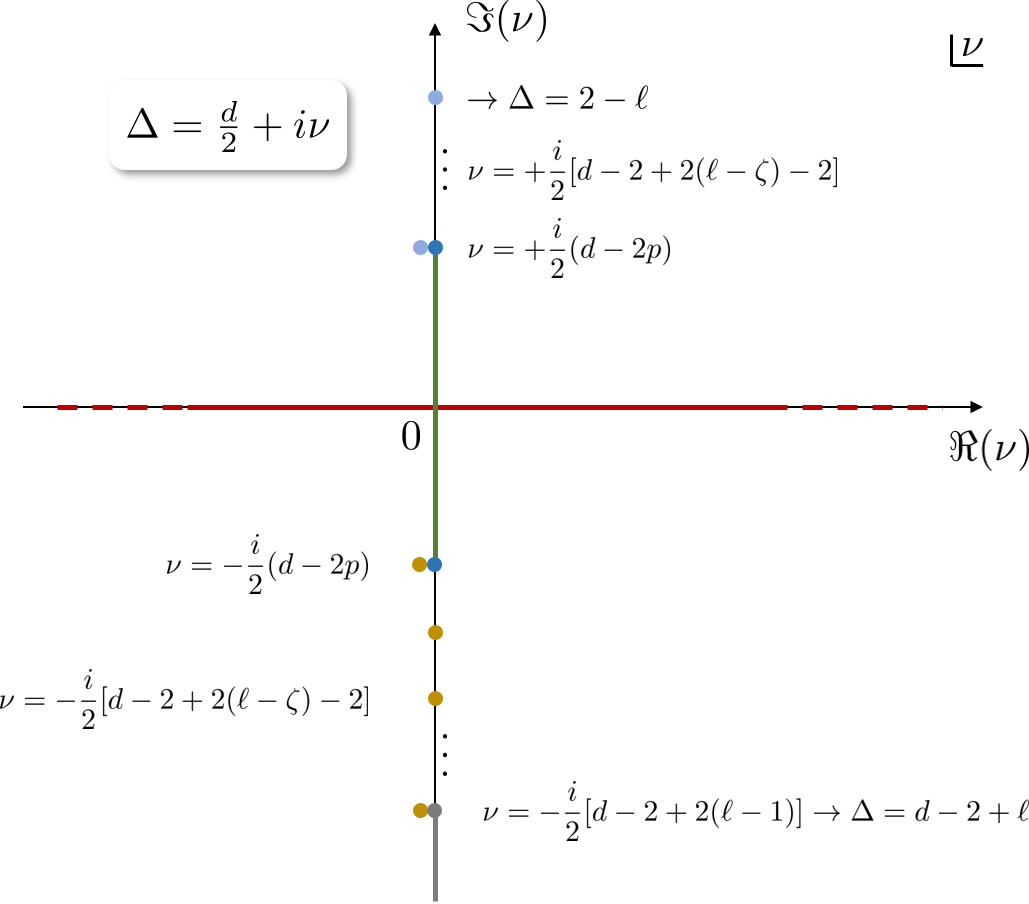}
    \caption{Depiction of the Unitary representations in dS$_{d+1}$ for symmetric tensors ($p=1$) and scalar ($p=0$) representations. Representations differing by $\nu\to -\nu$ are shadow of each other and are partially-equivalent. In red we display the \emph{principal series}, in green the \emph{complementary series} and the dark-blue dots are the boundary of the complementary series known as the \emph{exceptional series}. All boundary operators dual to curvature tensors of bulk (partially-)massless gauge fields sit at this blue point \cite{Basile:2016aen}. The yellow dots denote the (partially-)massless points $\nu=-\frac{i}{2} [d-2+2 (\ell-\zeta)-2]$ for totally symmetric representations where $\zeta=0,1,\ldots,\ell-1$, with $\zeta=0$ corresponding to the massless representation. The maximal depth partially-massless point coincides with the exceptional series point. In dS$_4$ the exceptional series is supplemented by the discrete series, which gives a unitary representation for each dual operator to (partially-)massless gauge fields. In grey we instead highlight the usual unitary representations in AdS$_{d+1}$, on the boundary of which (the grey dot) lie the massless representations. In light-blue we highlight instead the shadow-dual of the yellow dots. It is worth stressing that for generic $d$ the discrete series does not involve totally symmetric fields (and is thus absent from our diagram), and remarkably in $d=3$ it coincides with the yellow dots furnishing a unitary representation for each gauge field otherwise absent in generic dimensions. It is remarkable that the existence of short unitary representations with the quantum numbers dual to Fronsdal fields is an accident of dS$_3$ as opposed to what happens in AdS$_{d+1}$, where massless representations, instead of maximal-depth partially-massless fields, can be always found at the boundary of the unitary region (the grey dot at the end of the grey line).}
    \label{fig:dS_unitary}
\end{figure}

\subsection{Simplifications and Subtleties away from the Principal Series}
\label{subsec::someparticcases}

\begin{figure}[h]
    \centering
    \captionsetup{width=0.95\textwidth}
    \includegraphics[width=0.8\textwidth]{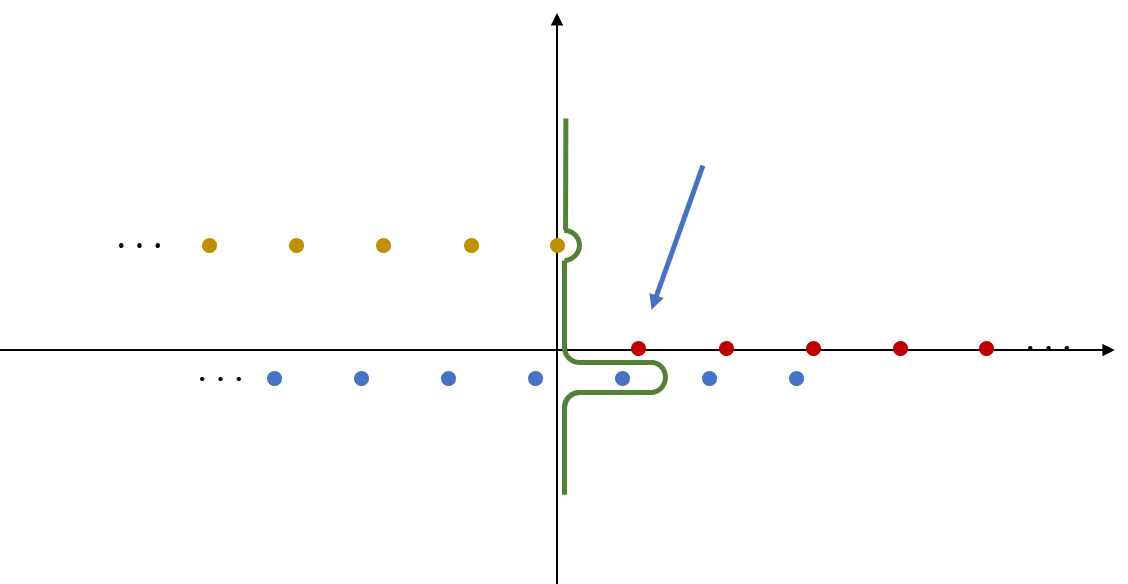}
    \caption{Depiction of contour pinching. When the poles in the red sequence collide with poles in the blue or yellow sequence, the integration contour (which is prescribed to separate such poles) is no longer defined.}
    \label{fig:contour_pinching}
\end{figure}

A crucial aspect of the Mellin-Barnes representation is the contour prescription \cite{MellinBook}. In writing a Mellin-Barnes representation for a late-time correlator, one is implicitly assuming values of the parameters for which the poles of $\Gamma$-functions of the type $\Gamma\left(a+s\right)$ do not collide with those of the type $\Gamma\left(b-s\right)$, where $a$ and $b$ depend on the scaling dimensions, $\ell$ and $d$. Otherwise, the integration contour gets ``pinched", leading to divergences for the values of the parameters where such poles collide.\footnote{The canonical example of such a singularity is:
\begin{align}
    \int_{-i\infty}^{+i\infty} \frac{ds}{2\pi i}\frac{1}{(s+a)(s-b)}=-\frac1{a+b}
\end{align}
which has a single pole at $a+b\sim0$.} This is depicted in figure \ref{fig:contour_pinching}. In these cases, one introduces a regulator to separate the poles, which can then be set to zero after evaluating the Mellin integral. This gives the analytic continuation of the late-time correlator to such values of the parameters.

For particles on the Principal Series (fig.\ref{fig:dS_unitary}), where the scaling dimensions are of the form $\Delta=\frac{d}{2}+i\nu$ with $\nu \in \mathbb{R}$, there is no possible pinching of the integration contour (see figure \ref{fig:Poles_s1}) in the Mellin-Barnes representation \eqref{exchampl} for the exchange four-point function. Away from the Principal Series however, where $\nu$ is now imaginary and the poles move horizontally, extra care should be taken since contour pinching can occur -- in which case regularisation is required.\footnote{Note that in section \ref{subsec::EFTexp} we took the external fields to be either conformally coupled scalars or massless scalars, which are on the Complementary and Discrete Series respectively. The value of $\nu$ associated to the exchanged field was taken to be generic, though divergences associated to the pinching of the integration contour arise when $\nu$ is on the discrete series. For these values of $\nu$ the results presented should be re-visited using the regularisation described above.} To illustrate how the Mellin-Barnes technology works in these cases we shall consider some examples of this type below. This includes physically interesting cases like the graviton and (partially-)massless exchanges \cite{Deser:2001us,Deser:2003gw,Joung:2012rv,deRham:2012kf,Deser:2013uy}, which in $d=3$ lie on the Discrete Series.

Correlators for representations away from the Principal Series furthermore exhibit simplifications for certain special values of the scaling dimensions, so it is furthermore interesting to understand how these arise from the Mellin-Barnes representation \eqref{exchampl}. An example we have already touched upon in section \ref{subsec::EFTexp} and \ref{subsec::3ptexamples} is when the external scalars are conformally coupled or massless (which we shall also review below), for which some of the Mellin-Barnes integrals can be lifted. Further simplifications can occur when the internal field is also conformally coupled or lies on the discrete series,\footnote{In particular, for $d=3$, tree-exchanges of conformally coupled and massless scalars are given by (Di)-Logarithms in the momenta \cite{Arkani-Hamed:2015bza,Arkani-Hamed:2018kmz}, while exchanges of (partially)-massless fields (which are further constrained by the corresponding gauge-symmetries) have been observed to be given by rational functions of the momenta \cite{Seery:2008ax,Bzowski:2013sza,Ghosh:2014kba,Baumann:2017jvh,Arkani-Hamed:2018kmz,Goon:2018fyu}.} which appear to arise naturally from the Mellin-Barnes representation. To see this, it is most convenient to employ the expression \eqref{Ao<>spinl}\footnote{I.e. before we evaluated the $u$-integral.} for the contributions \eqref{expspinl} to the exchange four-point function and combine them with the recursion relation \eqref{alphabeta_fromgamma}. This gives:  \begin{multline}\label{variecontrib}
    \mathcal{A}^{\left(x,{\bar x}\right)}_{\bullet|\nu_1,\nu_2,\nu_3,\nu_4|\alpha_1,\beta_1;\alpha_2,\beta_2}(\vec{k}_i;\vec{k})=(-1)^{\alpha_1+\alpha_2} k_1^{2 (\alpha_1-\beta_1+i \nu_1)}k_2^{2 (\beta_1+i \nu_2)}k_3^{2 (\alpha_2-\beta_2+i \nu_3)}k_4^{2 (\beta_2+i \nu_4)}\\\times\pl_{k_1^2}^{\alpha_1-\beta_1}\pl_{k_2^2}^{\beta_1}\pl_{k_3^2}^{\alpha_2-\beta_2}\pl_{k_4^2}^{\beta_2}\left[k_1^{-2i\nu_1}k_2^{-2i\nu_2}k_3^{-2i\nu_3}k_4^{-2i\nu_4} \mathcal{A}^{\left(x,{\bar x}\right)}_{\bullet|\nu_1,\nu_2,\nu_3,\nu_4|\alpha_1;\alpha_2}(\vec{k}_i;\vec{k})\right]
\end{multline}
with
\begin{subequations}\label{altcontrib<>}
\begin{align} 
   & \lim_{\eta_0\to0}\mathcal{A}^{\left(x,{\bar x}\right)}_{>|\nu_1,\nu_2,\nu_3,\nu_4|\alpha_1;\alpha_2}(\vec{k}_i;\vec{k})=\frac{1}2\int_{-i\infty}^{+i\infty}\frac{du}{2\pi i}\frac{1}{u+\epsilon}\, \widetilde{\mathfrak{A}}^{\left(u,x,\alpha_1\right)}_{\nu_1,\nu_2,\nu}(\vec{k}_1,\vec{k}_2,\vec{k})\mathfrak{A}^{\left(u,{\bar x},\alpha_2\right)}_{\nu_3,\nu_4,-\nu}(\vec{k}_3,\vec{k}_4,\vec{k})\,,  
    \\
  &  \lim_{\eta_0\to0}\mathcal{A}^{\left(x,{\bar x}\right)}_{<|\nu_1,\nu_2,\nu_3,\nu_4|\alpha_1;\alpha_2}(\vec{k}_i;\vec{k})=\frac{1}2\int_{-i\infty}^{+i\infty}\frac{du}{2\pi i}\frac{1}{u+\epsilon}\mathfrak{A}^{\left(u,x,\alpha_1\right)}_{\nu_1,\nu_2,\nu}(\vec{k}_1,\vec{k}_2,\vec{k})\widetilde{\mathfrak{A}}^{\left(u,{\bar x},\alpha_2\right)}_{\nu_3,\nu_4,-\nu}(\vec{k}_3,\vec{k}_4,\vec{k}), \label{altcontrib<>2}
\end{align}
\end{subequations}
where we defined the three-point structures:
\begin{subequations}\label{Amathfraksub}
\begin{align}\label{Amathfraksub1}
   &  \mathfrak{A}^{\left(u,x,\alpha_1\right)}_{\nu_1,\nu_2,\nu}(\vec{k}_1,\vec{k}_2,\vec{k})=\int[ds]_2\int_{-i\infty}^{+i\infty}dw\,\delta\left(\tfrac{x}4-s_1-s_2-w\right)\\& \hspace*{5cm} \times \frac{\rho_{\nu_1,\nu_2,\nu}(s_1,s_2,w+u)}{\left(w+u+\tfrac{i\nu}{2}-\alpha_1\right)_{\alpha_1}}\left(\frac{k}{2}\right)^{-2\left(w+u\right)+i\nu}\prod^2_{j=1}\left(\frac{k_j}{2}\right)^{-2s_j+i\nu_j}, \nonumber \\ 
    \label{Amathfraksub2}
   & \widetilde{\mathfrak{A}}^{\left(u,x,\alpha_1\right)}_{\nu_1,\nu_2,\nu}(\vec{k}_1,\vec{k}_2,\vec{k})=\int[ds]_2\int_{-i\infty}^{+i\infty}dw\,\delta\left(\tfrac{x}4-s_1-s_2-w\right)\\ \nonumber
   & \hspace*{3cm} \times \cos\left(\tfrac{\pi}2(4(s_1+s_2+u)+i\nu_1+i\nu_2+i\nu_3+i\nu_4)\right)\\
   & \hspace*{5cm} \times 
   \frac{\rho_{\nu_1,\nu_2,\nu}(s_1,s_2,w-u)}{\left(w-u+\tfrac{i\nu}{2}-\alpha_1\right)_{\alpha_1}}\left(\frac{k}{2}\right)^{-2\left(w-u\right)+i \nu}\prod^2_{j=1}\left(\frac{k_j}{2}\right)^{-2s_j+i\nu_j}. \nonumber
\end{align}
\end{subequations}
In terms of these, the remaining contribution to exchange four-point function is given by:
\begin{equation}\label{AAfact}
    \lim_{\eta_0\to0}\mathcal{A}^{\left(x,{\bar x}\right)}_{\odot|\nu_1,\nu_2,\nu_3,\nu_4|\alpha_1;\alpha_2}(\vec{k}_i;\vec{k}) = \frac{1}{2} \mathfrak{A}^{\left(\tcr{0},x,\alpha_1\right)}_{\nu_1,\nu_2,\nu|\alpha_1,\beta_1}(\vec{k}_1,\vec{k}_2,\vec{k}\,)\mathfrak{A}^{\left(\tcr{0},{\bar x},\alpha_2\right)}_{\nu_3,\nu_3,-\nu}(\vec{k}_3,\vec{k}_4,\vec{k}).
\end{equation}
This representation is convenient because when the external scalars are conformally coupled the three-point structures \eqref{Amathfraksub} reduce to Gauss Hypergeometric functions (see \S\ref{subsec::3ptexamples}):
{\allowdisplaybreaks\begin{subequations}\label{Afunctions}
\begin{align}
    \mathfrak{A}^{\left(u,x,\alpha_1\right)}_{\nu_1,\nu_2,\nu}(\vec{k}_1,\vec{k}_2,\vec{k})&=\frac{2^{2\alpha_1-i \nu -2 u-\frac{x}{2}+2} k^{i \nu -2 u-\frac{x}{2}+1}}{k_1 k_2}\\ & \hspace*{1.5cm} \times \frac{ \Gamma \left(2 u-2\alpha_1+\tfrac{x}{2}+i \nu -1\right)\Gamma \left(2 u+\tfrac{x}{2}-i \nu -1\right)}{\Gamma\left(\frac{x+4 u-2 \alpha_1-1}{2} \right)}\nonumber\\\nonumber
    &\hspace{2.5cm}\times\, {}_2F_1\left(\begin{matrix}\tfrac{x+4 u-4\alpha_1+2 i \nu -2}{2}, \tfrac{x+4 u-2 i \nu -2}{2}\\\tfrac{x+4 u-2 \alpha_1-1}{2} \end{matrix};\frac{1-p_{12}}{2}\right),\\
   \widetilde{\mathfrak{A}}^{\left(u,x,\alpha_1\right)}_{\nu_1,\nu_2,\nu}(\vec{k}_1,\vec{k}_2,\vec{k})&=\frac{2^{2\alpha_1-i \nu +2 u-\frac{x}{2}+2} k^{i \nu +2 u-\frac{x}{2}+1}}{k_1 k_2} \\ & \hspace*{1.5cm} \times \frac{\Gamma \left(-2\alpha_1-2 u+\tfrac{x}{2}+i \nu -1\right)\Gamma \left(-2 u+\tfrac{x}{2}-i \nu -1\right)}{\Gamma\left(\tfrac{x-2\alpha_1-4 u-1}{2}\right)}\nonumber\\\nonumber
    &\hspace{2.5cm}\times \cos (2 \pi  u)\, {}_2F_1\left(\begin{matrix}\tfrac{x-4\alpha_1-4 u+2 i \nu -2}{2},\tfrac{x-4 u-2 i \nu -2}{2}\\\tfrac{x-2\alpha_1-4 u-1}{2} \end{matrix};\frac{1+p_{12}}{2}\right)\,,
\end{align}
\end{subequations}}
so that the contributions \eqref{altcontrib<>} take the following simple form: 
\begin{subequations}\label{2f1prodrep}
\begin{multline}
   \hspace*{-0.75cm} \lim_{\eta_0\to0}\mathcal{A}^{(x,{\bar x})}_{<|\nu_1,\nu_2,\nu_3,\nu_4|\alpha_1;\alpha_2}(\vec{k}_i;\vec{k})\sim
    \int^{i\infty}_{-i\infty} \frac{du}{2\pi i}\,\frac{\cos(\pi u)}{u+\epsilon}\,\frac{\Gamma(a-2\alpha_1+u)\Gamma(b+u)\Gamma(a-2\alpha_2-u)\Gamma(b-u)}{\Gamma(c-\alpha_1+u)\Gamma(c-\alpha_2-u)}\\\times\,_2F_1\left(\begin{matrix}a-2\alpha_1+u,b+u\\c-\alpha_1+u\end{matrix};\frac{1-p_{12}}2\right)\,_2F_1\left(\begin{matrix}a-2\alpha_2-u,b-u\\c-\alpha_2-u\end{matrix};\frac{1+p_{34}}2\right),
\end{multline}
\begin{multline}
   \hspace*{-0.75cm}  \lim_{\eta_0\to0}\mathcal{A}^{(x,{\bar x})}_{>|\nu_1,\nu_2,\nu_3,\nu_4|\alpha_1;\alpha_2}(\vec{k}_i;\vec{k})\sim
    \int^{i\infty}_{-i\infty} \frac{du}{2\pi i}\,\frac{\cos(\pi u)}{u+\epsilon}\,\frac{\Gamma(a-2\alpha_1-u)\Gamma(b-u)\Gamma(a-2\alpha_2+u)\Gamma(b+u)}{\Gamma(c-\alpha_1-u)\Gamma(c-\alpha_2+u)}\\\times\,_2F_1\left(\begin{matrix}a-2\alpha_1-u,b-u\\c-\alpha_1-u\end{matrix};\frac{1+p_{12}}2\right)\,_2F_1\left(\begin{matrix}a-2\alpha_2+u,b+u\\c-\alpha_2+u\end{matrix};\frac{1-p_{34}}2\right),
\end{multline}
\end{subequations}
where 
\begin{align}
    a&=\frac{x+2i\nu-2}2\,,& b&=\frac{x-2i\nu-2}2\,,& c&=\frac{a+b+1}2=\frac{x-1}2\,.
\end{align}
The arguments of the Gamma functions on the first line coincide with those of the Hypergeometric functions on the second line. There are further simplifications for certain values of $\nu$. For example, when 
\begin{equation}\label{nutrunc}
    \nu=i\left(n+\tfrac{1}{2}\right), \qquad n \in \mathbb{Z},
\end{equation}
the Gauss Hypergeometric functions truncate to a rational function. These values of $\nu$ include the conformally coupled scalar ($n=0$) and (partially-)massless fields for $d$ odd. In the following, we shall study the exchange four-point function for such representations, focusing on regularisation of divergences which arise from the pinching of the integration contour. To this end we shall restrict to $d=3$, which provides the simplest setting to demonstrate the approach. We start with the exchange of conformally coupled and massless scalars, before considering (partially)-massless fields of arbitrary integer spin. At the end of this section we also consider the graviton exchange between massless scalars.

\paragraph{Exchange of a Conformally Coupled Scalar.} When the exchanged particle is a scalar there are only terms with $\alpha_i=\beta_i=0$ in the contributions \eqref{expspinl}. For a conformally coupled scalar we have $\nu=\frac{i}{2}$, in which case the $u$-integral \eqref{altcontrib<>} with $d=3$ reduces to:  
\begin{subequations}\label{ccscafrak}
\begin{align} \nonumber
   & \lim_{\eta_0\to0}\mathcal{A}^{(3+{\bar \epsilon},3+{\bar \epsilon})}_{>|\frac{i}{2},\frac{i}{2},\frac{i}{2},\frac{i}{2}}(\vec{k}_i;\vec{k})=\frac{1}{2k_1k_2k_3k_4}\frac{1}{k^{1+\epsilon}}\int_{-i\infty}^{+i\infty}\frac{du}{2\pi i}\frac{\cos (2 \pi  u)}{u+\epsilon}\, \Gamma \left(\tfrac{{\bar \epsilon}}{2}-2 u\right) \Gamma \left(\tfrac{{\bar \epsilon}}{2}+2u\right)\\
   & \hspace*{8cm} \times (1-p_{12})^{2 u-\frac{{\bar \epsilon}}{2}} (1+p_{34})^{-2 u-\frac{{\bar \epsilon}}{2}},  
    \\ \nonumber
  &  \lim_{\eta_0\to0}\mathcal{A}^{(3+{\bar \epsilon},3+{\bar \epsilon})}_{<|\frac{i}{2},\frac{i}{2},\frac{i}{2},\frac{i}{2}}(\vec{k}_i;\vec{k})=\frac{1}{2k_1k_2k_3k_4}\frac{1}{k^{1+{\bar \epsilon}}}\int_{-i\infty}^{+i\infty}\frac{du}{2\pi i}\frac{ \cos (2 \pi  u)}{u+\epsilon}\Gamma \left(\tfrac{{\bar \epsilon}}{2}-2 u\right) \Gamma \left(\tfrac{{\bar \epsilon}}{2}+2u\right)\\
   & \hspace*{8cm} \times (p_{12}+1)^{-2 u-\frac{{\bar \epsilon}}{2}} (1-p_{34})^{2 u-\frac{{\bar \epsilon}}{2}}. 
\end{align}
\end{subequations}
We see that the poles of the two $\Gamma$-functions overlap at $u \sim 0$, for which we set $d=3+{\bar \epsilon}$ to regulate the pinching of the integration contour. Recall that the latter runs to the right of the single pole $\frac{1}{u+\epsilon}$, so to evaluate the integrals \eqref{ccscafrak} it is simplest to close the integration contour to the right of the imaginary axis. Evaluating the residues and expanding then gives:\footnote{Although here we expanded in $\epsilon$, the Mellin integral can be evaluated for arbitrary $\epsilon$ very easily. In this case one can recognise the Mellin representation of a standard Gauss Hypergeometric function giving the corresponding general $\epsilon=d-3$ result valid in arbitrary dimension.}
\begin{subequations}\label{ccscexexp}
\begin{align} 
  \hspace*{-0.25cm} \lim_{\eta_0\to0}\mathcal{A}^{(3+{\bar \epsilon},3+{\bar \epsilon})}_{>|\frac{i}{2},\frac{i}{2},\frac{i}{2},\frac{i}{2}}(\vec{k}_i;\vec{k})&=\frac{1}{k_1k_2k_3k_4}\frac{1}{k}\left[\frac{2}{{\bar \epsilon}^2}-\frac{2 (\log (k)+\log (p_{34}+1)+\gamma )}{{\bar \epsilon}}-\frac{\pi ^2}{12}+\gamma ^2\right.\\ & \left.+(\log (k)+\log (p_{34}+1)+2 \gamma ) \log (k (p_{34}+1))+ \text{Li}_2\left(\frac{1-p_{12}}{p_{34}+1}\right)+O\left({\bar \epsilon}\right) \right], \nonumber
    \\
   \hspace*{-0.25cm} \lim_{\eta_0\to0}\mathcal{A}^{(3+{\bar \epsilon},3+{\bar \epsilon})}_{<|\frac{i}{2},\frac{i}{2},\frac{i}{2},\frac{i}{2}}(\vec{k}_i;\vec{k})&=\frac{1}{k_1k_2k_3k_4}\frac{1}{k}\left[\frac{2}{{\bar \epsilon}^2}-\frac{2(\log (k)+\log (p_{12}+1)+\gamma )}{{\bar \epsilon}}-\frac{\pi ^2}{12}+\gamma ^2\right.\\&\left.+(\log (k)+\log (p_{12}+1)+2 \gamma ) \log (k (p_{12}+1))+\text{Li}_2\left(\frac{1-p_{34}}{p_{12}+1}\right)+O\left({\bar \epsilon}\right)\right], \nonumber
\end{align}
\end{subequations}
which exhibits poles at ${\bar \epsilon}=0$, as anticipated. All of these poles are however cancelled when we include the contribution \eqref{AAfact}, which also has poles at ${\bar \epsilon}=0$ and this case reads:
\begin{multline}
  \hspace*{-0.75cm}  \lim_{\eta_0\to0}\mathcal{A}^{(3+{\bar \epsilon},3+{\bar \epsilon})}_{\odot|\frac{i}{2},\frac{i}{2},\frac{i}{2},\frac{i}{2}}(\vec{k}_i;\vec{k}) = \frac{1}{k_1k_2k_3k_4}\frac{1}{k} \left[ \frac{4}{{\bar \epsilon}^2}-\frac{2 (2 \log (k)+\log (p_{12}+1)+\log (p_{34}+1)+2 \gamma )}{{\bar \epsilon}}+ 2\gamma^2\right.\\\left.+\frac{\pi^2}{6}+\frac{(2 \log (k)+\log (p_{12}+1)(p_{34}+1)+4 \gamma ) (2 \log (k)+\log ((p_{12}+1) (p_{34}+1)))}{2}+O\left({\bar \epsilon}\right)\right]. \nonumber
\end{multline}
This gives the following finite result for the exchange four-point function:
\begin{align}\label{combinccsc}
   &  \langle \phi^{\left(i/2\right)}_{\vec{k}_1}\phi^{\left(i/2\right)}_{\vec{k}_2}\phi^{\left(i/2\right)}_{\vec{k}_3}\phi^{\left(i/2\right)}_{\vec{k}_4} \rangle^\prime \\ \nonumber  & \hspace*{1.5cm} ={\cal N}_4 \lim_{\eta_0\to0}\left(\mathcal{A}^{(3+{\bar \epsilon},3+{\bar \epsilon})}_{\odot|\frac{i}{2},\frac{i}{2},\frac{i}{2},\frac{i}{2}}(\vec{k}_i;\vec{k})-\mathcal{A}^{(3+{\bar \epsilon},3+{\bar \epsilon})}_{<|\frac{i}{2},\frac{i}{2},\frac{i}{2},\frac{i}{2}}(\vec{k}_i;\vec{k})-\mathcal{A}^{(3+{\bar \epsilon},3+{\bar \epsilon})}_{>|\frac{i}{2},\frac{i}{2},\frac{i}{2},\frac{i}{2}}(\vec{k}_i;\vec{k})\right)\Big|_{{\bar \epsilon}=0}\\
     & \hspace*{1.5cm} =\frac{\mathcal{N}_4}{k k_1k_2k_3k_4}\,\left[\frac{ \pi ^2}3- \text{Li}_2\left(\frac{1-p_{34}}{1+p_{12}}\right)- \text{Li}_2\left(\frac{1-p_{12}}{1+p_{34}}\right)-\frac12 \log ^2\left(\frac{1+p_{34}}{1+p_{12}}\right)\right], \nonumber
\end{align}
which matches equation (5.74) of \cite{Arkani-Hamed:2015bza}.\footnote{To compare with (5.74) of \cite{Arkani-Hamed:2015bza} one needs to massage the above expression using some properties of the Dilogarithm function, including the identity:
\begin{equation}
    \text{Li}_2(\text{z})= -\text{Li}_2\left(\frac{z}{z-1}\right)-\frac{1}{2} \log ^2(1-z).
\end{equation}} 

\paragraph{Exchange of massless scalar.} The exchange of a massless scalar in $d=3$ presents an interesting example. This is the simplest case in which the divergences of the individual contributions \eqref{altcontrib<>} and \eqref{AAfact} do not cancel among each other, so it is necessary to add a counter-term. For the massless scalar with $d=3$ we have $\nu=\tfrac{3i}{2}$ and in this case the $u$-integrals reduce to,
\begin{subequations}\label{A<>divmassles}
\begin{align} 
\hspace*{-0.25cm}  \lim_{\eta_0\to0}\mathcal{A}^{(3+{\bar \epsilon},3+{\bar \epsilon})}_{>|\frac{i}{2},\frac{i}{2},\frac{i}{2},\frac{i}{2}}(\vec{k}_i;\vec{k})=&\frac{1}{4k_1 k_2 k_3 k_4}\frac{1}{k^{1+{\bar \epsilon}}}\int_{-i\infty}^{+i\infty}\frac{du}{2\pi i}\,\frac{\cos(2\pi u)}{u+\epsilon}\, (1-p_{12})^{2 u-\frac{{\bar \epsilon}}{2}}(1+p_{34})^{-2 u-\frac{{\bar \epsilon}}{2}}\nonumber\\\nonumber & \hspace*{-0.5cm}\times (-4 u-2 p_{12}+{\bar \epsilon}) (4 u+2 p_{34}+{\bar \epsilon}) \Gamma \left(\tfrac{{\bar \epsilon}}{2}-2 u-1\right) \Gamma \left(\tfrac{{\bar 
\epsilon}}{2}+2 u-1\right)\\
=&-\frac{2}{{\bar \epsilon}^2}\frac{p_{12} p_{34}}{k k_1 k_2 k_3 k_4}\\ \nonumber
&+\frac{1}{2{\bar \epsilon}}\frac{4 p_{12} p_{34} \log (k (p_{34}+1))+4 (\gamma -1) p_{12} p_{34}-p_{12} (p_{12}+4)+p_{34}^2}{k k_1 k_2 k_3 k_4}\\
& +O\left(1\right), \nonumber
 \end{align}
\begin{align} \nonumber
 \hspace*{-0.25cm}  \lim_{\eta_0\to0}\mathcal{A}^{(3+{\bar \epsilon},3+{\bar \epsilon})}_{>|\frac{i}{2},\frac{i}{2},\frac{i}{2},\frac{i}{2}}(\vec{k}_i;\vec{k})=&\frac{1}{4k_1 k_2 k_3 k_4}\frac{1}{k^{1+{\bar \epsilon}}}\int_{-i\infty}^{+i\infty}\frac{du}{2\pi i}\,\frac{\cos(2\pi u)}{u+\epsilon}\, (1-p_{12})^{2 u-\frac{{\bar \epsilon}}{2}}(1+p_{34})^{-2 u-\frac{{\bar \epsilon}}{2}}\\&\hspace*{-0.5cm} \times (-4 u-2 p_{12}+{\bar \epsilon}) (4 u+2 p_{34}+{\bar \epsilon}) \Gamma \left(\tfrac{{\bar \epsilon}}{2}-2 u-1\right) \Gamma \left(\tfrac{{\bar \epsilon}}{2}+2 u-1\right) \nonumber
    \\ 
    =&-\frac{2}{{\bar \epsilon}^2}\frac{p_{12} p_{34}}{k k_1 k_2 k_3 k_4}\\ \nonumber
&+\frac{1}{2{\bar \epsilon}}\frac{4 p_{12} p_{34} \log (k (p_{12}+1))+4 (\gamma -1) p_{12} p_{34}-p_{34} (p_{34}+4)+p_{12}^2}{k k_1 k_2 k_3 k_4}\\
& +O\left(1\right), \nonumber
\end{align}
\end{subequations}
\noindent where, as before, we set $d=3+{\bar \epsilon}$ to regulate the pinching of poles and we closed the integration contour to the right of the imaginary axis. As in the conformally coupled case the integral is convergent at infinity and no further singularity is generated. The constant and higher order terms in the ${\bar \epsilon}$-expansion are quite cumbersome so we do not display them here. For the remaining contribution, we have
\begin{align}\nonumber
   \lim_{\eta_0\to0}\mathcal{A}^{(3+{\bar \epsilon},3+{\bar \epsilon})}_{\odot|\frac{i}{2},\frac{i}{2},\frac{i}{2},\frac{i}{2}}(\vec{k}_i;\vec{k}) =& \frac{\Gamma \left(\tfrac{{\bar \epsilon}}{2}-1\right)^2}{4k_1k_2k_3k_4} \frac{1}{k^{1+{\bar \epsilon}}} \left(2p_{12}+{\bar \epsilon}\right)\left(2p_{34}+{\bar \epsilon}\right)\left(p_{12}+1\right)^{-\frac{{\bar \epsilon}}{2}} \left(p_{34}+1\right)^{-\frac{{\bar \epsilon}}{2}}\\ \label{Aodivmassles}
    =&\frac{4 p_{12}p_{34}}{k k_1 k_2 k_3 k_4 {\bar \epsilon}^2}\\ \nonumber
    &+\frac{2}{{\bar \epsilon}}\frac{-p_{12}p_{34} \log (k^2(p_{12}+1) (p_{34}+1))-2 (\gamma -1) p_{12}p_{34}+p_{12}+p_{34}}{k k_1 k_2 k_3 k_4}\\
& +O\left(1\right). \nonumber
\end{align}
Upon combining these contributions as in \eqref{combinccsc}, one finds that the leftover poles are proportional to those of the factorised contribution \eqref{Aodivmassles}:
\begin{align}
   & \lim_{\eta_0\to0}\left[\mathcal{A}^{(3+{\bar \epsilon},3+{\bar \epsilon})}_{\odot|\frac{i}{2},\frac{i}{2},\frac{i}{2},\frac{i}{2}}(\vec{k}_i;\vec{k})-\mathcal{A}^{(3+{\bar \epsilon},3+{\bar \epsilon})}_{>|\frac{i}{2},\frac{i}{2},\frac{i}{2},\frac{i}{2}}(\vec{k}_i;\vec{k})-\mathcal{A}^{(3+{\bar \epsilon},3+{\bar \epsilon})}_{<|\frac{i}{2},\frac{i}{2},\frac{i}{2},\frac{i}{2}}(\vec{k}_i;\vec{k})\right]\\ \nonumber
   &\hspace*{1cm}=\frac{8 p_{12}p_{34}}{k k_1 k_2 k_3 k_4 {\bar \epsilon}^2}+\frac{4}{{\bar \epsilon}}\frac{-p_{12}p_{34} \log (k^2(p_{12}+1) (p_{34}+1))-2(\gamma -1) p_{12}p_{34}+p_{12}+p_{34}}{k k_1 k_2 k_3 k_4}\\
&\hspace*{1cm}+O\left(1\right). \nonumber
\end{align}
A minimal conformally invariant counter-term with the right singularities is therefore given by
\begin{equation}
  \lambda    \lim_{\eta_0\to0}\mathcal{A}^{(3+{\bar \epsilon},3+{\bar \epsilon})}_{\odot|\frac{i}{2},\frac{i}{2},\frac{i}{2},\frac{i}{2}}(\vec{k}_i;\vec{k}), \label{counterterm}
\end{equation}
where complete cancellation of the poles fixes $\lambda=-2$. This gives
\begin{align}\label{mlesssccc}
    & \langle \phi^{\left(i/2\right)}_{\vec{k}_1}\phi^{\left(i/2\right)}_{\vec{k}_2}\phi^{\left(i/2\right)}_{\vec{k}_3}\phi^{\left(i/2\right)}_{\vec{k}_4} \rangle^\prime\\
    &\hspace*{1.5cm}=-{\cal N}_4 \lim_{\eta_0\to0}\left(\mathcal{A}^{(3+{\bar \epsilon},3+{\bar \epsilon})}_{\odot|\frac{i}{2},\frac{i}{2},\frac{i}{2},\frac{i}{2}}(\vec{k}_i;\vec{k})+\mathcal{A}^{(3+{\bar \epsilon},3+{\bar \epsilon})}_{<|\frac{i}{2},\frac{i}{2},\frac{i}{2},\frac{i}{2}}(\vec{k}_i;\vec{k})+\mathcal{A}^{(3+{\bar \epsilon},3+{\bar \epsilon})}_{>|\frac{i}{2},\frac{i}{2},\frac{i}{2},\frac{i}{2}}(\vec{k}_i;\vec{k})\right)\Big|_{{\bar \epsilon}=0}\nonumber\\ \nonumber
    & \hspace*{1.5cm}= \frac{\mathcal{N}_4}{k\,k_1 k_2 k_3 k_4}\Bigg[p_{12} p_{34} \left(\text{Li}_2\left(\frac{1-p_{34}}{1+p_{12}}\right)+\text{Li}_2\left(\frac{1-p_{12}}{1+p_{34}}\right)\right)\\\nonumber
    &+p_{12} p_{34}\left(\frac{1}{2}  \log^2 \left(\frac{1+p_{12}}{1+p_{34}}\right)-\frac{\pi ^2}{3}  \right)+2 p_{12} \log \left(\frac{p_{12}+1}{p_{12}+p_{34}}\right)+2 p_{34} \log \left(\frac{p_{34}+1}{p_{12}+p_{34}}\right)-2\Bigg]\,,
\end{align}
which matches equation (4.59) obtained in \cite{Arkani-Hamed:2018kmz}. 

\paragraph{Massless spin-$1$ exchange.} Similarly we can consider the exchange of spinning fields on the discrete series. In this case it is necessary to combine the contributions \eqref{variecontrib} with the $\Theta$-polynomials \eqref{ThetaPol}. The simplest example is the mass-less spin-1 exchange, which in $d=3$ corresponds to $\nu=\frac{i}{2}$. The $\Theta$-polynomials in this case have the following helicity decomposition (see the end of section \ref{subsecc::spinlexch}): 
\begin{subequations}
\begin{align}\label{leadinghel}
  \hspace*{-0.7cm}  \Theta^{(1,1)}_{\frac{i}{2},..,\frac{i}{2};\frac{i}{2}|0,0;0,0}&=-\frac1{4}\,\Xi_1\,, \qquad
    \Theta^{(1,0)}_{\frac{i}{2},..,\frac{i}{2};\frac{i}{2}|0,0;0,0}=\frac{1}{4} k^2 p_{12} p_{34} q_{12} q_{34}\,\Xi_0\,,\\
   \hspace*{-0.75cm} \Theta^{(1,0)}_{\frac{i}{2},..,\frac{i}{2};\frac{i}{2}|1,0;0,0}&=\Theta^{(1,0)}_{\frac{i}{2},..,\frac{i}{2};\frac{i}{2}|0,0;1,0}=-\Theta^{(1,0)}_{\frac{i}{2},..,\frac{i}{2};\frac{i}{2}|0,0;1,1}=-\Theta^{(1,0)}_{\frac{i}{2},..,\frac{i}{2};\frac{i}{2}|1,1;0,0}=\frac{1}{4} k^2 p_{12} q_{12}\,\Xi_0\,,\\
  \hspace*{-0.7cm}  \Theta^{(1,0)}_{\frac{i}{2},..,\frac{i}{2};\frac{i}{2}|1,0;1,0}&=\Theta^{(1,0)}_{\frac{i}{2},..,\frac{i}{2};\frac{i}{2}|1,1;1,1}=-\Theta^{(1,0)}_{\frac{i}{2},..,\frac{i}{2};\frac{i}{2}|1,0;1,1}=-\Theta^{(1,0)}_{\frac{i}{2},..,\frac{i}{2};\frac{i}{2}|1,1;1,0}=\frac{k^2}{4}\,\Xi_0\,,
\end{align}
\end{subequations}
where we recall that only the leading terms \eqref{leadinghel} contribute to the component with helicity one.

In contrast to the scalar exchanges considered above, when the exchanged field has non-zero spin the leading terms in the exchange four-point function are finite. In particular, for the massless spin-1 exchange we have
\begin{subequations}
\begin{align} \nonumber
  \lim_{\eta_0\to0}\mathcal{A}^{(5,5)}_{>|\frac{i}{2},\frac{i}{2},\frac{i}{2},\frac{i}{2}|0,0,0,0}(\vec{k}_i;\vec{k})&=\frac{1}{k_1k_2k_3k_4}\frac{1}{k^3}\int_{-i\infty}^{+i\infty}\frac{du}{2\pi i}\frac{\cos (2 \pi  u)}{u+\epsilon}\,\Gamma \left(1-2u\right) \Gamma \left(1+2u\right)\\
   & \hspace*{4cm} \times (1-p_{12})^{2 u-1} (p_{34}+1)^{-2 u-1}   \nonumber \\
   &=-\frac{1}{k_1k_2k_3k_4}\frac{1}{k^3} \frac{1}{(p_{34}+1) (p_{12}+p_{34})},   \\
   \lim_{\eta_0\to0}\mathcal{A}^{(5,5)}_{<|\frac{i}{2},\frac{i}{2},\frac{i}{2},\frac{i}{2}|0,0,0,0}(\vec{k}_i;\vec{k})&=\frac{1}{k_1k_2k_3k_4}\frac{1}{k^3}\int_{-i\infty}^{+i\infty}\frac{du}{2\pi i}\frac{\cos (2 \pi  u)}{u+\epsilon}\, \Gamma \left(1-2 u\right) \Gamma \left(2 u+1\right) \nonumber \\
   & \hspace*{4cm}
 \times  (p_{12}+1)^{-2 u-1} (1-p_{34})^{2 u-1}   \nonumber \\
   &=-\frac{1}{k_1k_2k_3k_4}\frac{1}{k^3} \frac{1}{(p_{12}+1) (p_{12}+p_{34})}, \\
  \lim_{\eta_0\to0}\mathcal{A}^{(5,5)}_{\odot|\frac{i}{2},\frac{i}{2},\frac{i}{2},\frac{i}{2}|0,0,0,0}(\vec{k}_i;\vec{k})&=-\frac{1}{k_1k_2k_3k_4}\frac{1}{k^3} \frac{1}{(p_{12}+1) (p_{34}+1)},\label{odotspin1mass}
\end{align}
\end{subequations}
where we note that the $\Gamma$-function poles in the $u$-integrals do not overlap. 

For the sub-leading terms, the $u$-integrals \eqref{altcontrib<>} take a similar form to those we evaluated for the scalar exchange diagrams and they all have divergences due to contour pinching. For example, to obtain the contributions \eqref{A<spinl} with $\alpha_1=0$ and $\alpha_2=1$, the basic integral \eqref{altcontrib<>} is:
\begin{align}\label{subleadingegspin1}
   &  \lim_{\eta_0\to0}\mathcal{A}^{(5+{\bar \epsilon},5+{\bar \epsilon})}_{<|\frac{i}{2},\frac{i}{2},\frac{i}{2},\frac{i}{2}|0;1}(\vec{k}_i;\vec{k})\\ \nonumber
   & \hspace*{2cm} = \frac{1}2\int_{-i\infty}^{+i\infty}\frac{du}{2\pi i}\frac{1}{u+\epsilon}\mathfrak{A}^{\left(u,5+{\bar \epsilon},\alpha_1\right)}_{\frac{i}{2},\frac{i}{2},\frac{i}{2}}\left(k_1,k_2,k\right)\widetilde{\mathfrak{A}}^{\left(u,5+{\bar \epsilon},\alpha_2\right)}_{\frac{i}{2},\frac{i}{2},-\frac{i}{2}}(k_3,k_4,k)\Big|_{\alpha_1=0,\alpha_2=1}\\& \hspace*{2cm}=\frac{1}{k_1k_2k_3k_4}\frac{1}{k^{3+{\bar \epsilon}}}\int_{-i\infty}^{+i\infty}\frac{du}{2\pi i}\frac{\cos{2\pi u}}{u+\epsilon} \Gamma \left(\tfrac{{\bar \epsilon}}{2}-2 u\right) \Gamma \left(2 u+\tfrac{{\bar \epsilon}}{2}+1\right) \nonumber \\ \nonumber & \hspace*{6cm}\times \left(p_{12}+1\right)^{-2 u-\frac{{\bar \epsilon}}{2}-1} \left(1-p_{34}\right)^{2 u-\frac{{\bar \epsilon}}{2}} \\
    &\hspace*{2cm}=\frac{1}{p_{12}+1}\frac{1}{k^3k_1k_2k_3k_4 }\left[\frac{4}{{\bar \epsilon}}-4 \log (k)-2 \log ((p_{12}+1) (p_{12}+p_{34}))-4 \gamma +O\left({\bar \epsilon}\right) \right], \nonumber
\end{align}
which has a simple pole at ${\bar \epsilon}=0$. One then evaluates the derivatives as prescribed in \eqref{variecontrib}. Upon combining all contributions to the exchange four-point function, these divergences in the helicity-0 components do not cancel and it is necessary to add a counter-term. The conformally invariant counter-term with the correct singularity structure is the spin-1 analogue \eqref{expspinl} of the counter-term \eqref{counterterm} we used for the massless scalar,\footnote{In particular it is given by equation \eqref{expspinl} with $\bullet = \odot$, $\ell=1$ and $\nu_i=\nu=\frac{i}{2}$.} with the same coefficient $\lambda=-2$. This contains the term \eqref{odotspin1mass} and therefore also induces a finite correction to the helicity-1 component. The result for the exchange four-point function is:
\begin{multline}\label{spin1mless}
   \langle \phi^{\left(i/2\right)}_{\vec{k}_1}\phi^{\left(i/2\right)}_{\vec{k}_2}\phi^{\left(i/2\right)}_{\vec{k}_3}\phi^{\left(i/2\right)}_{\vec{k}_4} \rangle^\prime\\=\frac12\frac{\mathcal{N}_4}{k^3\,k_1 k_2 k_3 k_4}\left[\frac{\Xi_1}{(1+p_{12}) (1+p_{34}) (p_{12}+p_{34})}+\frac{k^2 q_{12} q_{34}}{(p_{12}+p_{34})}\,\Xi_0\right]\,. 
\end{multline}
Notice that the helicity-1 component is singular not only at $p_{12}+p_{34} \to 0$, but also at $1+p_{12} \to 0$ and $1+p_{34} \to 0$ which correspond to setting either of the cubic interactions on-shell (for energy conservation at the vertex) and thus carries information about the propagating degrees of freedom. The helicity-0 component does not have these singularities and is only singular for $p_{12}+p_{34} \to 0$. The helicity-0 component is therefore a contact contribution to the exchange four-point function, consistent with the fact that a massless spin-one particle has only helicity $|\lambda|=1$. It is therefore tempting to interpret the divergences we encountered in the subleading terms, which contribute only helicity-0 components, as a consequence of the decoupling of the lower-helicity modes in the massless limit.

The helicity-1 component matches with that in equation (4.55) of \cite{Arkani-Hamed:2018kmz}. The helicity-0 component differs by a local contact term which can be accounted for by a choice of improvement (on-shell vanishing) terms in the cubic vertex.\footnote{More generally, the contact terms in any helicity component can be changed ad libitum by adding improvements to the cubic vertex.}

\paragraph{Massless spin-2 exchange.} In this case, we have $\nu=\frac{3i}{2}$ for $d=3$. Similar to the massless spin-1 example above, the leading terms in the exchange four-point function are finite:
{\allowdisplaybreaks\begin{subequations}
\begin{align}
   & \lim_{\eta_0\to0}\mathcal{A}^{(7,7)}_{>|\frac{i}{2},\frac{i}{2},\frac{i}{2},\frac{i}{2}|0,0,0,0}(\vec{k}_i;\vec{k})\\ \nonumber
   & \hspace*{2cm}=\frac{1}{k_1k_2k_3k_4}\frac{1}{k^5}\int_{-i\infty}^{+i\infty}\frac{du}{2\pi i}\frac{\cos (2 \pi  u)}{u+\epsilon}\,\Gamma \left(1-2u\right) \Gamma \left(1+2u\right)\\
   & \hspace*{5cm} \times
   \left(2-p_{12}-2 u\right) \left(2+p_{34}+2u\right)
   \left(1-p_{12}\right)^{2 u-2} \left(p_{34}+1\right)^{-2 u-2}   \nonumber \\
   &\hspace*{2cm} =\frac{1}{k_1k_2k_3k_4}\frac{1}{k^5} \frac{-(p_{34}+2)p_{12}^2-2 p_{12} p_{34} (p_{34}+2)-(p_{34}-1) p_{34} (p_{34}+3)+p_{12}+2}{\left(p_{34}+1\right)^2 \left(p_{12}+p_{34}\right)^3},  \nonumber \\
   & \lim_{\eta_0\to0}\mathcal{A}^{(7,7)}_{<|\frac{i}{2},\frac{i}{2},\frac{i}{2},\frac{i}{2}|0,0,0,0}(\vec{k}_i;\vec{k})\\
   &\hspace*{2cm}=\frac{1}{k_1k_2k_3k_4}\frac{1}{k^5}\int_{-i\infty}^{+i\infty}\frac{du}{2\pi i}\frac{\cos (2 \pi  u)}{u+\epsilon}\, \Gamma \left(1-2 u\right) \Gamma \left(1+2 u\right) \nonumber \\
   & \hspace*{5cm}
   \times  \left(2-p_{34}-2u\right) \left(2+p_{12}+2u\right) (p_{12}+1)^{-2 u-2}(1-p_{34})^{2 u-2} \nonumber \\
   &\hspace*{2cm}=\frac{1}{k_1k_2k_3k_4}\frac{1}{k^5} \frac{-(p_{12}+2) p_{34}^2-2 p_{12} p_{34}(p_{12}+2)-(p_{12}-1) p_{12} (p_{12}+3)+p_{34}+2}{\left(p_{12}+1\right)^2 \left(p_{12}+p_{34}\right)^3},  \nonumber
\end{align}
\end{subequations}}
where the integration contour is not pinched by the Gamma function poles, and
\begin{equation}
    \lim_{\eta_0\to0}\mathcal{A}^{(7,7)}_{\odot|\frac{i}{2},\frac{i}{2},\frac{i}{2},\frac{i}{2}|0,0,0,0}(\vec{k}_i;\vec{k})=\frac{1}{k_1k_2k_3k_4}\frac{1}{k^5} \frac{(p_{12}+2) (p_{34}+2)}{(p_{12}+1)^2 (p_{34}+1)^2}.
\end{equation}

The sub-leading terms, which contribute only to the helicity-1 and -0 components, instead diverge. As before, a conformal invariant counter-term with the right singularity structure is provided by \eqref{expspinl} with $\bullet=\odot$ and coefficient $\lambda=-2$. Combined with the helicity decomposition \eqref{HDthetapol} of the $\Theta_{\alpha_1,\beta_1;\alpha_2,\beta_2}$-polynomials \eqref{ThetaPol} this gives
\begin{multline}
    \langle \phi^{\left(i/2\right)}_{\vec{k}_1}\phi^{\left(i/2\right)}_{\vec{k}_2}\phi^{\left(i/2\right)}_{\vec{k}_3}\phi^{\left(i/2\right)}_{\vec{k}_4} \rangle^\prime=\frac14\frac{\mathcal{N}_4}{k^5\,k_1 k_2 k_3 k_4}\\ \times \left[\frac{p_{12} p_{34}+2 (p_{12}+p_{34})+1}{(1+p_{12})^2 (1+p_{34})^2 (p_{12}+p_{34})^3}\,\Xi_2+\frac{2k^2q_{12} q_{34}}{(p_{12}+p_{34})^3}\,\Xi_1+\ldots\right]\,, 
\end{multline}
where the $\ldots$ are the helicity-0 components which are a bit cumbersome so we do not display them here. As before, the counter-term induces a finite correction to the helicity-2 component. The helicity-1 and -0 components are only singular for $p_{12}+p_{34} \to 0$ and are therefore contact terms which are sensitive to the choice of improvements in the cubic vertex. The highest helicity component on the other hand also has singularities at $1+p_{12} \to 0$ and $1+p_{34} \to 0$, as consistent with the propagation of a helicity-2 degree of freedom. Accordingly, the helicity-2 component matches that of equation (4.61) in \cite{Arkani-Hamed:2018kmz} and differs by a local contact term in the lower helicity components.

\paragraph{Massless spin-$\ell$ exchange.} The four-point exchange of any given massless spin-$\ell$ field follows in the same way as for the low spin $\ell \leq 2$ examples detailed above. The leading terms in the contributions \eqref{variecontrib} (with $\alpha_i=\beta_i=0$), which encode the highest helicity component, are always finite. The subleading terms instead exhibit divergences, which can be interpreted as a consequence of the decoupling of the lower helicity modes in the massless limit. Interestingly, for any spin $\ell$, these divergences can all be cancelled with the conformally invariant counter-term \eqref{expspinl} where $\bullet = \odot$ and coefficient $\lambda=-2$.\footnote{It is tempting to explain this as a consequence of higher-spin symmetry, which would relate all counterterms together into a single higher-spin symmetric counterterm.} 

The helicity-$\ell$ component has the following general form 
\begin{align}
    \langle \phi^{\left(i/2\right)}_{\vec{k}_1}\phi^{\left(i/2\right)}_{\vec{k}_2}\phi^{\left(i/2\right)}_{\vec{k}_3}\phi^{\left(i/2\right)}_{\vec{k}_4} \rangle^\prime=\frac{\mathcal{N}_4}{k^{2\ell+1}\,k_1 k_2 k_3 k_4}\left[\frac{f_\ell(p_{12},p_{34})}{(1+p_{12})^ \ell(1+p_{34})^ \ell(p_{12}+p_{34})^{2\ell-1}}\,\Xi_{\ell}+\ldots\right]\,,
\end{align}
where the $\ldots$ denote non-singular contact interactions and $f_\ell(p_{12},p_{34})$ is a polynomial in $p_{12}$ and $p_{34}$ of degree $\ell-1$. Below we list a few examples for higher-spins $\ell > 2$:
\begin{subequations}
\begin{align}
    f_3(p_{12},p_{34})&=8 (p_{12}+p_{34})^2+9 (p_{12} p_{34}+1) (p_{12}+p_{34})+3 (p_{12} p_{34}+1)^2\,,\\
    f_4(p_{12},p_{34})&=9 \Big(16 (p_{12}+p_{34})^3+29 (p_{12} p_{34}+1) (p_{12}+p_{34})^2\\\nonumber
    &\hspace{100pt}+20 (p_{12} p_{34}+1)^2 (p_{12}+p_{34})+5 (p_{12} p_{34}+1)^3\Big)\,,\\
    f_5(p_{12},p_{34})&=9 \Big(128 (p_{12}+p_{34})^4+325 (p_{12} p_{34}+1) (p_{12}+p_{34})^3\\\nonumber
    &+345 (p_{12} p_{34}+1)^2 (p_{12}+p_{34})^2+175 (p_{12} p_{34}+1)^3 (p_{12}+p_{34})+35 (p_{12} p_{34}+1)^4\Big)\\\nonumber
    \ldots,
\end{align}
\end{subequations}
and so on for higher and higher spins.

\paragraph{Partially-massless spin-2} Similarly we can consider the exchange of the partially mass-less spin-2 field, which has $\nu=\frac{i}{2}$. In this case the $\Gamma$-function poles in the $u$-integrals \eqref{altcontrib<>} do not overlap and there are no divergences. We obtain 
\begin{multline}\label{pmspin2exch}
   \langle \phi^{\left(i/2\right)}_{\vec{k}_1}\phi^{\left(i/2\right)}_{\vec{k}_2}\phi^{\left(i/2\right)}_{\vec{k}_3}\phi^{\left(i/2\right)}_{\vec{k}_4} \rangle^\prime=\frac{\mathcal{N}_4}{k^5\,k_1 k_2 k_3 k_4}\Big[\frac{(p_{12}+p_{34})^2+2(p_{12}+p_{34})+p_{12} p_{34}+1}{4 \left((p_{12}+1)^2 (p_{34}+1)^2 (p_{12}+p_{34})^3\right)}\,\Xi_2\\+\frac{k^2 q_{12} q_{34} \left((p_{12}+p_{34})^2-(p_{12}+p_{34})-p_{12} p_{34}-1\right)}{2 (p_{12}+1) (p_{34}+1) (p_{12}+p_{34})^3}\,\Xi_1+\ldots\Big]\,,
\end{multline}
where the $\ldots$ is the helicity-$0$ contribution, which is singular only for $p_{12}+p_{34} \to 0$ and so does not encode propagating degrees of freedom. Note that partially massless spin-2 fields also propagate a helicity-1 degree of freedom, which is manifest in the above expression from the singularities as $1+p_{12} \to 0$ and $1+p_{34} \to 0$ for the helicity-1 component, in addition to those in the helicity-2 component.

The expression \eqref{pmspin2exch} agrees with equation (4.56) in \cite{Arkani-Hamed:2018kmz} up to local contact terms (i.e. those singular only for $p_{12}+p_{34} \to 0$) in the helicity-1 and -0 components, corresponding to the freedom of adding improvements (on-shell vanishing terms) to the cubic vertices.

\paragraph{Maximal-depth partially-massless fields} More generally one can consider partially-massless fields of spin $\ell > 2$. The simplest are those of maximal depth which lie on the boundary with the complementary series, where the scaling dimension is spin-independent and given by $\nu=\frac{i}{2}$.

For the contributions \eqref{altcontrib<>}, the $u$-integrals for the leading terms are given for $\ell>0$ by:
{\allowdisplaybreaks\begin{subequations}
\begin{align} \nonumber
   & \lim_{\eta_0\to0}\mathcal{A}^{\left(3+2\ell,3+2\ell\right)}_{>|\frac{i}{2},\frac{i}{2},\frac{i}{2},\frac{i}{2}|0,0,0,0}(\vec{k}_i;\vec{k})=\frac{1}{k^{2\ell +1}k_1k_2k_3k_4}\int_{-i\infty}^{+i\infty}\frac{du}{2\pi i} \frac{\cos (2 \pi  u)}{u+\epsilon} \Gamma \left(\ell-2 u\right) \Gamma \left(\ell+2 u\right)\\ \nonumber
   & \hspace*{6cm} \times (1-p_{12})^{-\ell+2 u} (1+p_{34})^{-\ell-2 u} 
    \\
   &\hspace*{4.5cm}=\frac{\cos (\pi  \ell) \Gamma (2 \ell)}{\ell k^{2\ell +1}k_1k_2k_3k_4}\frac{1}{(1+p_{34})^{2 \ell}}
   \,{}_2F_1\left(\begin{matrix}\ell,2\ell\\\ell+1 \end{matrix};\frac{1-p_{12}}{1+p_{34}}\right),
    \\  \nonumber
  &  \lim_{\eta_0\to0}\mathcal{A}^{\left(3+2\ell,3+2\ell\right)}_{<|\frac{i}{2},\frac{i}{2},\frac{i}{2},\frac{i}{2}|0,0,0,0}(\vec{k}_i;\vec{k})=\frac{1}{k^{2\ell +1}k_1k_2k_3k_4}\int_{-i\infty}^{+i\infty}\frac{du}{2\pi i} \frac{\cos (2 \pi  u)}{u+\epsilon} \Gamma \left(\ell-2 u\right) \Gamma \left(\ell+2 u\right)\\ \nonumber
   & \hspace*{6cm} \times (1+p_{12})^{-\ell-2 u} (1-p_{34})^{-\ell+2 u} 
    \\
   &\hspace*{4.5cm}=\frac{\cos (\pi  \ell) \Gamma (2 \ell)}{\ell k^{2\ell +1}k_1k_2k_3k_4}\frac{1}{(1+p_{12})^{2 \ell}}
   \,{}_2F_1\left(\begin{matrix}\ell,2\ell\\\ell+1 \end{matrix};\frac{1-p_{34}}{1+p_{12}}\right),
\end{align}
\end{subequations}}
which, like for the $\ell=2$ case given above, do not exhibit any divergence for $\ell>0$. The above expressions however do not hold for $\ell=0$, where the integration contour becomes pinched, which manifests itself in the pole at $\ell=0$. The analytic continuation to $\ell=0$, which corresponds to the exchange of a conformally coupled scalar, was treated earlier in equation \eqref{ccscafrak}. The remaining contribution \eqref{AAfact} reads
\begin{equation}
    \lim_{\eta_0\to0}\mathcal{A}^{\left(3+2\ell,3+2\ell\right)}_{\odot|\frac{i}{2},\frac{i}{2},\frac{i}{2},\frac{i}{2}|0,0,0,0}(\vec{k}_i;\vec{k})=\frac{\Gamma (\ell)^2}{k^{2\ell +1}k_1k_2k_3k_4}\frac{1}{(1+p_{12})^{\ell}(1+p_{34})^{\ell}},
\end{equation}
so that the leading-helicity contribution to the exchange of a maximal depth partially massless spin-$\ell$ field is:
\begin{multline}
     \langle \phi^{\left(i/2\right)}_{\vec{k}_1}\phi^{\left(i/2\right)}_{\vec{k}_2}\phi^{\left(i/2\right)}_{\vec{k}_3}\phi^{\left(i/2\right)}_{\vec{k}_4} \rangle^\prime=\left(-\frac{1}{4}\right)^\ell \Xi_\ell\, \frac{{\cal N}_4}{k^{2\ell +1}k_1k_2k_3k_4}\left[\frac{\Gamma (\ell)^2}{(1+p_{12})^{\ell}(1+p_{34})^{\ell}}
     \right. \\ \left.
    + \frac{\cos (\pi  \ell) \Gamma (2 \ell)}{\ell}
   \left(\frac{1}{(1+p_{34})^{2 \ell}}\,{}_2F_1\left(\begin{matrix}\ell,2\ell\\\ell+1 \end{matrix};\frac{1-p_{12}}{1+p_{34}}\right)+\frac{1}{(1+p_{12})^{2 \ell}}
   \,{}_2F_1\left(\begin{matrix}\ell,2\ell\\\ell+1 \end{matrix};\frac{1-p_{34}}{1+p_{12}}\right)\right)\right]+...\,. \nonumber
\end{multline}
Setting $\ell=1,2$ this recovers the leading-helicity contributions in the massless spin-1 exchange \eqref{spin1mless} and the partially massless spin-2 exchange \eqref{pmspin2exch} respectively. In writing this expression we are assuming that there are no divergences in the lower-helicity components -- as was the case for the partially massless spin-2 exchange \eqref{pmspin2exch} -- which can be extracted in a similar way.

\paragraph{Graviton Exchange between massless scalars.} Using the raising operators \eqref{raisingop}, from the above results for external conformally coupled scalars one can obtain expressions for external fields with integer shifted scaling dimensions. In the following we shall consider the example of the graviton exchange between massless scalars. The leading helicity component, which encodes the propagating degrees of freedom, can be simply obtained from the contributions \eqref{A<>divmassles} and \eqref{Aodivmassles} to the massless scalar exchange by acting once with each raising operator \eqref{raisingop}. In particular,
\begin{align}
    \lim_{\eta_0\to0}\mathcal{A}^{(7+{\bar \epsilon},7+{\bar \epsilon})}_{\bullet|\frac{3i}{2},\frac{3i}{2},\frac{3i}{2},\frac{3i}{2}|0,0,0,0}(\vec{k}_i;\vec{k})=\lim_{\eta_0\to0}\left[\frac{16}{k_1k_2k_3k_4} \partial_{k_1}\partial_{k_2}\partial_{k_3}\partial_{k_4}\mathcal{A}^{(3+{\bar \epsilon},3+{\bar \epsilon})}_{\bullet|\frac{i}{2},\frac{i}{2},\frac{i}{2},\frac{i}{2}}(\vec{k}_i;\vec{k})\right].
\end{align}
Interestingly, upon acting with the above operators, each of the contributions above are non-singular in the limit ${\bar \epsilon} \to 0$. In particular,
\begin{multline}
   \lim_{\eta_0\to0}\mathcal{A}^{(7,7)}_{\bullet|\frac{3i}{2},\frac{3i}{2},\frac{3i}{2},\frac{3i}{2}|0,0,0,0}(\vec{k}_i;\vec{k})\\=\frac{1}{(k_1 k_2 k_3 k_4)^3\,k^5}\left[\frac{f_\bullet\,\Xi_2}{(1+p_{12})^2(1+p_{34})^2(p_{12}+p_{34})^3}+\ldots\right]\,,
\end{multline}
where the $\ldots$ represent lower helicity components which are proportional to contact terms, and the functions $f_\bullet$ are given by:
\begin{multline}
    f_\odot=(p_{12}+p_{34})^3\left(3 p_{12}^3+6 p_{12}^2+p_{12} \left(q_{12}^2+8\right)+2 \left(q_{12}^2+2\right)\right)\\\times \left(3 p_{34}^3+6 p_{34}^2+p_{34} \left(q_{34}^2+8\right)+2 \left(q_{34}^2+2\right)\right)\,,
\end{multline}
and 
\begin{align}
    f_\gtrless&=f_{<}+f_{>}=9 p_{34}^3 p_{12}^6+18 p_{34}^2 p_{12}^6+3 p_{34} q_{34}^2 p_{12}^6+6 q_{34}^2 p_{12}^6+24 p_{34} p_{12}^6+12 p_{12}^6+27 p_{34}^4 p_{12}^5\nonumber\\\nonumber
    &+72 p_{34}^3 p_{12}^5+108 p_{34}^2 p_{12}^5+9 p_{34}^2 q_{34}^2 p_{12}^5+24 p_{34} q_{34}^2 p_{12}^5+12 q_{34}^2 p_{12}^5+84 p_{34} p_{12}^5+24 p_{12}^5\\\nonumber
    &+27 p_{34}^5 p_{12}^4+108 p_{34}^4 p_{12}^4+204 p_{34}^3 p_{12}^4+188 p_{34}^2 p_{12}^4+3 p_{34}^3 q_{12}^2 p_{12}^4+6 p_{34}^2 q_{12}^2 p_{12}^4+8 p_{34} q_{12}^2 p_{12}^4\\\nonumber
    &+4 q_{12}^2 p_{12}^4+9 p_{34}^3 q_{34}^2 p_{12}^4+36 p_{34}^2 q_{34}^2 p_{12}^4+p_{34} q_{12}^2 q_{34}^2 p_{12}^4+2 q_{12}^2 q_{34}^2 p_{12}^4+44 p_{34} q_{34}^2 p_{12}^4+16 q_{34}^2 p_{12}^4\\\nonumber
    &+56 p_{34} p_{12}^4-8 p_{12}^4+9 p_{34}^6 p_{12}^3+72 p_{34}^5 p_{12}^3+204 p_{34}^4 p_{12}^3+228 p_{34}^3 p_{12}^3+40 p_{34}^2 p_{12}^3+9 p_{34}^4 q_{12}^2 p_{12}^3\\\nonumber
    &+24 p_{34}^3 q_{12}^2 p_{12}^3+36 p_{34}^2 q_{12}^2 p_{12}^3+28 p_{34} q_{12}^2 p_{12}^3+8 q_{12}^2 p_{12}^3+3 p_{34}^4 q_{34}^2 p_{12}^3+24 p_{34}^3 q_{34}^2 p_{12}^3\\\nonumber
    &+60 p_{34}^2 q_{34}^2 p_{12}^3+3 p_{34}^2 q_{12}^2 q_{34}^2 p_{12}^3+8 p_{34} q_{12}^2 q_{34}^2 p_{12}^3+4 q_{12}^2 q_{34}^2 p_{12}^3+64 p_{34} q_{34}^2 p_{12}^3\\\nonumber
    &+32 q_{34}^2 p_{12}^3-56 p_{34} p_{12}^3+16 p_{12}^3+18 p_{34}^6 p_{12}^2+108 p_{34}^5 p_{12}^2+188 p_{34}^4 p_{12}^2+40 p_{34}^3 p_{12}^2-100 p_{34}^2 p_{12}^2\\\nonumber
    &+9 p_{34}^5 q_{12}^2 p_{12}^2+36 p_{34}^4 q_{12}^2 p_{12}^2+60 p_{34}^3 q_{12}^2 p_{12}^2+68 p_{34}^2 q_{12}^2 p_{12}^2+40 p_{34} q_{12}^2 p_{12}^2+8 q_{12}^2 p_{12}^2+6 p_{34}^4 q_{34}^2 p_{12}^2\\\nonumber
    &+36 p_{34}^3 q_{34}^2 p_{12}^2+68 p_{34}^2 q_{34}^2 p_{12}^2+3 p_{34}^3 q_{12}^2 q_{34}^2 p_{12}^2+12 p_{34}^2 q_{12}^2 q_{34}^2 p_{12}^2+12 p_{34} q_{12}^2 q_{34}^2 p_{12}^2+64 p_{34} q_{34}^2 p_{12}^2\\\nonumber
    &+20 q_{34}^2 p_{12}^2-48 p_{34} p_{12}^2+24 p_{34}^6 p_{12}+84 p_{34}^5 p_{12}+56 p_{34}^4 p_{12}-56 p_{34}^3 p_{12}-48 p_{34}^2 p_{12}+3 p_{34}^6 q_{12}^2 p_{12}\\\nonumber
    &+24 p_{34}^5 q_{12}^2 p_{12}+44 p_{34}^4 q_{12}^2 p_{12}+64 p_{34}^3 q_{12}^2 p_{12}+64 p_{34}^2 q_{12}^2 p_{12}+24 p_{34} q_{12}^2 p_{12}+8 p_{34}^4 q_{34}^2 p_{12}\\\nonumber
    &+28 p_{34}^3 q_{34}^2 p_{12}+40 p_{34}^2 q_{34}^2 p_{12}+p_{34}^4 q_{12}^2 q_{34}^2 p_{12}+8 p_{34}^3 q_{12}^2 q_{34}^2 p_{12}+12 p_{34}^2 q_{12}^2 q_{34}^2 p_{12}\\\nonumber
    &-4 p_{34} q_{12}^2 q_{34}^2 p_{12}-8 q_{12}^2 q_{34}^2 p_{12}+24 p_{34} q_{34}^2 p_{12}+12 p_{34}^6+24 p_{34}^5-8 p_{34}^4-16 p_{34}^3+6 p_{34}^6 q_{12}^2\\\nonumber
    &+16 p_{34}^4 q_{12}^2+32 p_{34}^3 q_{12}^2+20 p_{34}^2 q_{12}^2+4 p_{34}^4 q_{34}^2+8 p_{34}^3 q_{34}^2+8 p_{34}^2 q_{34}^2+2 p_{34}^4 q_{12}^2 q_{34}^2+4 p_{34}^3 q_{12}^2 q_{34}^2\\
    &-8 p_{34} q_{12}^2 q_{34}^2-4 q_{12}^2 q_{34}^2+12 p_{34}^5 q_{12}^2\,.
\end{align}
In this case it is therefore not necessary to add a counter-term, and the corresponding leading helicity component to the graviton exchange is given by
\begin{multline}
     \langle \phi^{\left(3i/2\right)}_{\vec{k}_1}\phi^{\left(3i/2\right)}_{\vec{k}_2}\phi^{\left(3i/2\right)}_{\vec{k}_3}\phi^{\left(3i/2\right)}_{\vec{k}_4} \rangle^\prime \\ = \frac{{\cal N}_4}{(k_1 k_2 k_3 k_4)^3\,k^5}\left[\frac{\left(f_{\odot}-f_\gtrless\right)\,\Xi_2}{(1+p_{12})^2(1+p_{34})^2(p_{12}+p_{34})^3}+\ldots\right],
\end{multline}
which we confirmed matches with equation (2.26) of \cite{Seery:2008ax}. 

\subsection{Exchanges in dS vs. exchanges in AdS}
\label{dSvsAdS}

It is instructive to compare momentum space exchange diagrams in de Sitter and anti-de Sitter at the level of the Mellin-Barnes representation considered in this work, to better appreciate differences and similarities between these two cases. In the following without loss of generality we shall consider the tree-level exchange of scalar fields. AdS spinning exchanges can be obtained following the same steps we have taken in the dS case. The Witten diagram for the exchange of a general scalar field in AdS$_{d+1}$ in terms of the corresponding Conformal Partial Wave reads (see e.g. \cite{Polyakov:1974gs,Moschella:2007zza,Penedones:2010ue}):\footnote{Here we used $\mu$ as the spectral parameter in the spectral integral to distinguish it from the spectral dimension of the exchanged particle, which in this work is given by $\nu$. Note that in the AdS literature the spectral parameter is instead usually taken to be $\nu$.} 
\begin{align}\label{spectralAdS}
    \mathcal{A}_{\text{AdS}}(\vec{k}_i;\vec{k})=\int_{-\infty}^{\infty} d\mu\,\frac{1}{\mu^2-\nu^2+i\epsilon}\,\frac{\mu^2}{\pi}\,{\cal F}_{\text{AdS}}(\vec{k}_i;\vec{k})\,,
\end{align}
where the exchanged scalar has scaling dimension $\Delta=\tfrac{d}{2}+i(\nu-i\epsilon)$, $\epsilon > 0$, and where we introduced an epsilon prescription to avoid crossing the contour of the spectral integral (i.e the real line) as we analytically continue $\nu$ from the imaginary axis to the real line. The Conformal Partial Wave for the AdS exchange is:
\begin{align}
{\cal F}_{\text{AdS}}(\vec{k}_i;\vec{k})&=[\mathcal{O}_{\nu_1}(\vec{k}_1)\mathcal{O}_{\nu_2}(\vec{k}_2)\mathcal{O}_{\mu}(\vec{k})]\, [\mathcal{O}_{-\mu}(-\vec{k})\mathcal{O}_{\nu_3}(\vec{k}_3)\mathcal{O}_{\nu_4}(\vec{k}_4)],
\end{align}
where the three-point factors are given by scalar three-point Witten diagrams (see equation \eqref{BAdS}):
\begin{subequations}
\begin{align}
[\mathcal{O}_{\nu_1}(\vec{k}_1)\mathcal{O}_{\nu_2}(\vec{k}_2)\mathcal{O}_{\nu_3}(\vec{k})]&=\mathsf{B}_{\nu_1,\nu_2,\nu_3} \langle \langle\mathcal{O}_{\nu_1}(\vec{k}_1)\mathcal{O}_{\nu_2}(\vec{k}_2)\mathcal{O}_{\nu_3}(\vec{k}) \rangle \rangle. \\ 
&=\left(\prod^3_{j=1} \frac{\sqrt{\pi}}{\Gamma(1+i\nu_j)}\right)
    \int \left[ds\right]_3\,i\pi \delta\left(\tfrac{d}{4}-s_1-s_2-s_3\right)\,\\ & \hspace*{4cm}\times \rho_{\nu_1,\nu_2,\nu_3}(s_1,s_2,s_3)\prod^3_{j=1}\left(\frac{k_j}{2}\right)^{-2s_j+i\nu_j}, \nonumber
\end{align}
\end{subequations}
which gives the following Mellin-Barnes representation for the exchange Witten diagram:
\begin{align}\nonumber
    \mathcal{A}_{\text{AdS}}(\vec{k}_i;\vec{k})&=\mathcal{N}_4^{\text{AdS}} \int^{i\infty}_{-i\infty} \left[ds\right]_4\,\frac{du d{\bar u}}{\left(2\pi i\right)^2}
    \,i\pi \delta\left(\tfrac{d}{4}-s_1-s_2-u\right)i\pi \delta\left(\tfrac{d}{4}-s_3-s_4-{\bar u}\right)\\  & \times
   \int_{-\infty}^{\infty}\frac{d\mu}{\mu^2-\nu^2+i\epsilon}\,
       \left(\frac{k}{2}\right)^{-2\left(u+{\bar u}\right)}\prod^4_{j=1}\left(\frac{k_j}{2}\right)^{-2s_j+i\nu_j} \label{mellinadsmom} \\\nonumber
       &\hspace{100pt}\times\frac{\rho_{\nu_1,\nu_2,\mu}(s_1,s_2,u)\rho_{\nu_3,\nu_4,-\mu}(s_3,s_4,\bar{u})}{\Gamma (i \mu ) \Gamma (-i \mu )},
\end{align}
where we have factored out the normalisation:
\begin{equation}
    \mathcal{N}_4^{\text{AdS}}=\left(\prod^4_{j=1} \frac{\sqrt{\pi}}{\Gamma(1+i\nu_j)}\right)\,.
\end{equation}
In the above form the spectral integral in $\mu$ is factorised and can be easily evaluated using the following identity:
\begin{multline}\label{specident}
    \int_{-\infty}^\infty d\mu \frac{\Gamma \left(a_1+\frac{i \mu }{2}\right)\Gamma \left(a_1-\frac{i \mu }{2}\right)\Gamma \left(a_2+\frac{i \mu }{2}\right)\Gamma \left(a_2-\frac{i \mu }{2}\right)\Gamma \left(a_3+\frac{i \mu }{2}\right)\Gamma \left(a_3-\frac{i \mu }{2}\right) }{\Gamma (-i \mu ) \Gamma (i \mu ) \Gamma \left(a_4+\frac{i \mu }{2}+1\right) \Gamma \left(a_4-\frac{i \mu }{2}+1\right)}\\=\frac{8 \pi  \Gamma (a_1+a_2) \Gamma (a_1+a_3) \Gamma (a_2+a_3) \Gamma (-a_1-a_2-a_3+a_4+1)}{\Gamma (1-a_1+a_4) \Gamma (1-a_2+a_4) \Gamma (1-a_3+a_4)},
\end{multline}
the proof of which can be found in \cite{Sleight:2018ryu}. In particular, we have that\footnote{To apply the identity \eqref{specident} to the spectral integral in \eqref{mellinadsmom} one notes that 
\begin{equation}
    \frac{1}{\mu^2-\nu^2+i\epsilon} = \frac{\Gamma\left(i\mu-i\nu+\epsilon\right)\Gamma\left(-i\mu-i\nu+\epsilon\right)}{\Gamma\left(i\mu-i\nu+\epsilon+1\right)\Gamma\left(-i\mu-i\nu+\epsilon+1\right)}
\end{equation}
}${}^{,}$\footnote{We have explicitly checked this to be equivalent to using the more standard representation of the momentum-space AdS bulk-to-bulk propagator:
\begin{align}
    \Pi_{\vec{k}}(z,\bar{z})&=z^{d/2} \bar{z}^{d/2} K_{i \nu }(k z) I_{i \nu }(k \bar{z})\theta(z-\bar{z})+ z^{d/2} \bar{z}^{d/2} I_{i \nu }(k z) K_{i \nu }(k \bar{z})\theta(\bar{z}-z)\,.
\end{align}} 
\begin{multline}
    \int_{-\infty}^{+\infty} d\mu\,\frac{1}{\mu^2-\nu^2+i\epsilon}\,\frac{\Gamma(u+\tfrac{i\mu}2)\Gamma(u-\tfrac{i\mu}2)\Gamma({\bar u}+\tfrac{i\mu}2)\Gamma({\bar u}-\tfrac{i\mu}2)}{\Gamma (i \mu ) \Gamma (-i \mu )}\\
    =\frac{2 \pi ^2 \csc (\pi  (u+\bar{u}))  \Gamma \left(u+\frac{i \nu }{2}\right)\Gamma \left(\bar{u}+\frac{i \nu }{2}\right)}{\Gamma \left(1-u+\frac{i \nu }{2}\right) \Gamma \left(1-\bar{u}+\frac{i \nu }{2}\right)}\,.
\end{multline}
Using the Mellin-Barnes representation \eqref{Momentum CPW} for Conformal Partial Waves, the resulting expression for the AdS exchange is then
\begin{shaded}
\noindent \emph{Four-point exchange Witten diagram for generic scalar fields in AdS$_{d+1}$}
\begin{equation}\label{adsexchMB}
    \mathcal{A}_{\text{AdS}}(\vec{k}_i;\vec{k})=4\mathcal{N}^{\text{AdS}}_4\int_{-i\infty}^{+i\infty}\frac{du\,d\bar{u}}{(2\pi i)^2}\int [ds]_4\,\csc{(\pi(u+\bar{u}))}\, \delta_{\text{AdS}}(u,\bar{u})\,\mathcal{F}^\prime(s_i;u,\bar{u}|\vec{k}_i,\vec{k})\,,
\end{equation}
\end{shaded}
\noindent where the function $\delta_{\text{AdS}}(u,\bar{u})$ in the Mellin-integrand in this case reads simply:
\begin{align}\label{AdSzeros}
    \delta_{\text{AdS}}(u,\bar{u})=\frac{1}{2} \sin \left(\pi  \left(u-\tfrac{i \nu }{2}\right)\right)\sin \left(\pi  \left(\bar{u}-\tfrac{i \nu }{2}\right)\right)\,.
\end{align}
We thus see that the representation \eqref{adsexchMB} for the exchange diagram is universal in both de Sitter and anti-de Sitter, which differ only in the interference factors $\delta(u,{\bar u})$ which encode the zeros of the Mellin integrand. In AdS, the factor \eqref{AdSzeros} precisely cancel the poles
\begin{equation}
    \Gamma\left(u-\frac{i\nu}{2}\right)\Gamma\left({\bar u}-\frac{i\nu}{2}\right),
\end{equation}
in the CPW \eqref{Momentum CPW} associated to contributions from the shadow conformal multiplet (labelled by scaling dimension $d-\Delta$), which go as: 
\begin{equation}
    \left(k^2\right)^{-i\nu}\left[\#+O\left(k\right)\right].
\end{equation}
The role of the spectral integral in the expression \eqref{spectralAdS} for the EAdS exchange is to precisely project away the shadow contributions, which is translated into the zeros of the function \eqref{AdSzeros} in the Mellin-Barnes representation for the exchange. In the dS case both the shadow and non-shadow multiplets contribute, as can be seen from the OPE limit \eqref{OPElimspinl}.

As for the de Sitter case, the poles \eqref{eftpoles} of the $\csc$-function in the Mellin-Barnes representation \eqref{adsexchMB} encode the EFT expansion for the exchange Witten diagram. Considering for example external conformally coupled scalars, for which the Mellin-Barnes integral \eqref{adsexchMB} takes the same form as \eqref{confCMellin} but with interference factor \eqref{AdSzeros}, proceeding as in \S \ref{subsec::EFTexp} one obtains the EFT expansion:
\begin{align}
     \mathcal{A}_{\text{AdS}}(\vec{k}_i;\vec{k})\Big|_{\text{EFT}}=\frac{\mathcal{N}^{\text{AdS}}_4}{2 k_1 k_2 k_3 k_4}\left(\frac{k}{2}\right)^{2-d}
     \sum_{n,m=0}^\infty c^{\left(d\right)}_{mn}\,(p_{12})^{2-d-2 m} \left(\frac{p_{34}}{p_{12}}\right)^n\,,
\end{align}
where the coefficients $c^{\left(d\right)}_{mn}$ are the same as in the de Sitter case and are given by equation \eqref{cmndS}, while for the non-perturbative corrections we have 
\begin{align}\label{adsfactor}
   \mathcal{A}_{\text{AdS}}(\vec{k}_i;\vec{k})\Big|_{\text{non-pert.}}&=i\,\text{csch}(\pi  \nu )\frac{\pi \,\mathcal{N}^{\text{AdS}}_4}{2 k_1 k_2 k_3 k_4}\left(\frac{k}{2}\right)^{2-d}\Big[\mathfrak{F}_+(p_{12})\mathfrak{F}_+(p_{34})\\\nonumber
    &\hspace{120pt}-\mathfrak{F}_+(p_{12})\mathfrak{F}_-(p_{34})-\mathfrak{F}_-(p_{12})\mathfrak{F}_+(p_{34})\Big]\,,
\end{align}
where one notes the absence of terms $\mathfrak{F}_-(p_{12})\mathfrak{F}_-(p_{34})$, which would give non-analytic contributions corresponding to the shadow conformal multiplet. The zeros of the interference factor \eqref{AdSzeros} ensure that such terms do not arise when evaluating the residues of the poles \eqref{opepoles}, in contrast to the de Sitter case.

We note that, just as for the de Sitter exchange, the EFT contribution is entirely specified by the factorised contribution \eqref{adsfactor}. The latter can be recovered from the discontinuity of the exchange four-point function in $\mathsf{s}=k^2$. 

In would be interesting to explore Witten diagrams in AdS further using this formalism. To date there are only a handful of works on momentum space approaches to Witten diagrams see e.g. \cite{Chalmers:1998wu,Chalmers:1999gc,Raju:2010by,Raju:2012zr,Ghosh:2014kba,Anninos:2014lwa,Albayrak:2018tam,Albayrak:2019asr}.

\section{Inflationary Correlators}\label{sec:InflationaryCorr}

\begin{figure}
    \centering
    \captionsetup{width=0.95\textwidth}
    \includegraphics[width=0.8\textwidth]{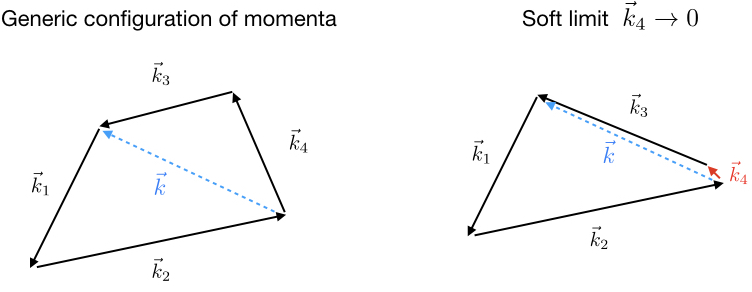}
    \caption{The soft limit is closer in spirit to the collapsed triangle limit in which $p_{34}\to1$. In the Mellin-Barnes representation this is manifestly non-singular.}
    \label{fig:collapsed_triangle}
\end{figure}

The Mellin framework can also be used to study inflationary three-point functions. These can be obtained from de Sitter four-point functions under the assumption of slightly broken conformal symmetry by taking the soft momentum limit of a scalar external leg with a small mass $\nu=i\left(\tfrac{d}{2}-\epsilon \right)$, where $\epsilon$ is related to the slow-roll parameter \cite{Kundu:2014gxa,Arkani-Hamed:2015bza,Kundu:2015xta}. A simpler example of this idea was followed in \S \ref{subsec::softlimit2pt}, where we obtained the corresponding inflationary two-point function from the late-time de Sitter tree-level three-point functions \eqref{00lltmb}. In contrast, inflationary correlators obtained from de Sitter exchange diagrams contain signals for particle exchanges, which manifest themselves in non-analytic contributions in the squeezed limit $k_3 \sim k \rightarrow 0$. This limit has some similarities to the OPE limit considered towards the end of section \ref{subsecc::spinlexch}, but a key difference is that there one takes $k \ll k_3$ while for the squeezed limit we have $k_3 \sim k$ (see e.g. figure \ref{fig:collapsed_triangle}).

Before proceeding let us note that in this setting the inflaton $\phi$, which we are approximating by a massless scalar, is usually assumed to couple in a shift-symmetric fashion -- i.e. in a way which is invariant under $\phi \to \phi+c$ where $c$ is a constant. The shift-symmetry implies that there are no $O\left(1\right)$ terms in $\epsilon$ in the soft limit of the exchange four-point function. I.e. there is an Adler zero. It should be kept in mind that, while we do not explicitly consider shift symmetric couplings in this work, since any non-trivial coupling of two scalar fields and a spin-$\ell$ field is unique on-shell, the exchange four-point function differs from that generated by a shift symmetric coupling only by a finite number of contact terms and an overall polynomial in the scaling dimensions which multiplies the factorised contributions. This in particular means that, even for couplings which are not shift symmetric, there are no $O\left(1\right)$ terms in $\epsilon$ when focusing on the non-analytic terms in the exchanged momentum in the soft limit. 

\paragraph{Soft limit.} First we take the soft momentum limit of a scalar external leg, say $\phi^{(\nu_4)}_{\vec{k_4}}$, in the spin-$\ell$ exchange four-point function \eqref{exchampl}. Like in section \ref{subsec::softlimit2pt}, this is given by the residue of the leading pole $s_4=-\tfrac{i\nu_4}{2}$ in the corresponding bulk-to-boundary propagator \eqref{buboFsc}. At the level of the Conformal Partial Wave, which carries the pole in question, this reads: 
\begin{multline}\label{softcpw}
   \lim_{k_4 \to 0}  {\cal F}^\prime_{\nu,\ell}(s_i;u,{\bar u}|\vec{k}_i;\vec{k}) =   \left(\frac{k_4}{2}\right)^{2i\nu_4}
     (i\pi) \delta\left(\tfrac{d+2\ell}4-s_1-s_2-u\right) (i\pi) \delta\left(\tfrac{d+2\ell}4-s_3+\tfrac{i\nu_4}2-\bar{u}\right)
  \\\times  \rho_{\nu_3,\nu}(s_3,{\bar u})\rho_{\nu_1,\nu_2,\nu}(s_1,s_2,u)
\left(\frac{k_3}{2}\right)^{-2(u+{\bar u})}\prod^3_{j=1}\left(\frac{k_j}{2}\right)^{-2s_j+i\nu_j}
\\\times  \Theta^{\left(\ell\right)}_{\nu_1,\nu_2,\nu_3,\nu_4;\nu}(\vec{k}_i;\vec{k}|s_i,u,\bar{u})\Big|_{s_4=-\tfrac{i\nu_4}{2},\,k_4=0,\,\vec{k}\sim-\vec{k}_3},
\end{multline}
where we used that $\vec{k}\sim-\vec{k}_3$ as $\vec{k}_4 \to 0$ and defined
\begin{subequations}\label{rho2}
\begin{align}
    \rho_{\nu_3,\nu}(s_3,{\bar u})&\equiv \frac{\text{Res}_{s_4=-\tfrac{i\nu_4}2}\Big[\rho_{\nu_3,\nu_4,\nu}(s_3,s_4,\bar{u})\Big]}{\text{Res}_{s_4=-\tfrac{i\nu_4}2}\Big[\frac{1}{2\sqrt{\pi}}\Gamma(s_4-\tfrac{i\nu_4}2)\Gamma(s_4+ \tfrac{i\nu_4}2)\Big]}\,\\
    &=\frac{1}{4\pi}\Gamma(s_3+ \tfrac{i\nu_3}2)\Gamma(s_3-\tfrac{i\nu_3}2)\Gamma({\bar u}+ \tfrac{i\nu}2)\Gamma({\bar u}-\tfrac{i\nu}2).
\end{align}
\end{subequations}
This is equivalent to taking the soft limit of one of the three-point functions in the factorised expression \eqref{spinlCPW} for the Conformal Partial Wave. The soft limit of three-point functions was considered in \S \ref{subsec::softlimit2pt}, where it was in particular shown that 
\begin{multline}\label{softsoft}
    {\sf p}^{\left(\ell\right)}_{\nu_3,\nu_4,-\nu}(\vec{\xi}\cdot \vec{k}_3,0,\vec{\xi}\cdot \vec{k}_3|s_3,s_4,{\bar u})\Big|_{s_4=-\tfrac{i\nu_4}{2}}\\=\underbrace{\frac{\left(\tfrac{d-2 \ell-2 i (\nu_3- \nu_4+\nu)}{4}\right)_\ell \left(\tfrac{2s_3+2{\bar u}-2\ell+i (\nu_3-\nu)}{2}\right)_\ell}{\left(\tfrac{d}{2}-i \nu-1\right)_\ell \left({\bar u}-\tfrac{i \nu}{2}-\ell\right)_\ell}}_{{\widetilde H}_{\nu_3,-\nu;\nu_4|\ell}(s_3,\bar{u})}\,(-i\,\vec{\xi}\cdot \vec{k}_3)^\ell,
\end{multline}
where for convenience we introduced the function $\gamma_\ell(s_3,\bar{u})$, so that in the limit $k_4 \to 0$ the contraction \eqref{fullthetacontra} reads
\begin{multline}
   \Theta^{\left(\ell\right)}_{\nu_1,\nu_2,\nu_3,\nu_4;\nu}(\vec{k}_i;\vec{k}|s_i,u,\bar{u})\Big|_{s_4=-\tfrac{i\nu_4}{2},\,k_4=0,\,\vec{k}\sim-\vec{k}_3}
    \\ = \sum_{\alpha=0}^\ell\sum_{\beta=0}^\alpha\binom{\ell}{\alpha}\binom{\alpha}{\beta}H_{\nu_1,\nu_2,\nu|\alpha,\beta}(s_1,s_2,u){\widetilde H}_{\nu_3,-\nu;\nu_4|\ell}(s_3,\bar{u})\,\widetilde{\Theta}_{\nu_1,\nu_2,\nu|\alpha,\beta}^{(\ell)}(\vec{k}_1,\vec{k}_2,\vec{k}_3),
\end{multline}
where we defined
\begin{align}
    \widetilde{\Theta}_{\nu_1,\nu_2,\nu|\alpha,\beta}^{(\ell)}(\vec{k}_1,\vec{k}_2,\vec{k}_3)=\frac{1}{\ell!\left(\tfrac{d}{2}-1\right)_\ell}(-i\vec{k}_3\cdot\hat{\pl}_\xi)^\ell\left[(-\vec{\xi}\cdot \vec{k}_3)^\alpha\mathcal{Y}_{\nu_1,\nu_2,\nu|\alpha,\beta}^{(\ell)}(\vec{\xi}\cdot \vec{k}_1,\vec{\xi}\cdot \vec{k}_2)\right].
\end{align}
This gives the soft limit $k_4 \to 0$ of the Conformal Partial Wave \eqref{CPWspinlexpMB} which, via \eqref{exchampl}, gives the soft limit of the spin-$\ell$ exchange four-point function:
\begin{multline}\label{spinning_infla}
  \lim_{k_4 \to 0}  \langle \phi^{\left(\nu_1\right)}_{\vec{k}_1}\phi^{\left(\nu_2\right)}_{\vec{k}_2}\phi^{\left(\nu_3\right)}_{\vec{k}_3}\phi^{\left(\nu_4\right)}_{\vec{k}_4} \rangle^\prime=\left(\frac{k_4}{2}\right)^{2 i \nu_4}{\cal N}_4\frac{\Gamma\left(-i\nu_4\right)}{2\sqrt{\pi}} \sum_{\alpha=0}^\ell\sum_{\beta=0}^\alpha\binom{\ell}{\alpha}\binom{\alpha}{\beta}\widetilde{\Theta}^{(\ell)}_{\nu_1,\nu_2,\nu|\alpha,\beta}(\vec{k}_1,\vec{k}_2,\vec{k}_3)\\ \times \tilde{\mathcal{A}}^{(d+2\ell)}_{\nu_1,\nu_2,\nu_3;\nu_4;\nu,\ell|\alpha,\beta}(\vec{k}_1,\vec{k}_2,\vec{k}_3)\,,
\end{multline}
where 
\begin{multline}\label{spinning_inflaalphabeta}
    \tilde{\mathcal{A}}^{(x)}_{\nu_1,\nu_2,\nu_3;\nu_4;\nu,\ell|\alpha,\beta}(\vec{k}_1,\vec{k}_2,\vec{k}_3)= \int [ds]_3\, \csc(\pi(u+\bar{u}))\delta^{(x,d+2\ell)}(u,\bar{u})\\\times\,\,H_{\nu_1,\nu_2,\nu|\alpha,\beta}(s_1,s_2,u)\rho_{\nu_1,\nu_2,\nu}(s_1,s_2,u){\widetilde H}_{\nu_3,-\nu;\nu_4|\ell}(s_3,\bar{u})\rho_{\nu_3,\nu}(s_3,\bar{u})\\ \times \left(\frac{k_3}{2}\right)^{-2\left(u+{\bar u}\right)}\prod^3_{j=1}\left(\frac{k_j}{2}\right)^{-2s_j+i\nu_j}\Big|_{{}^{u=\frac{x}4-s_1-s_2}_{{\bar u}=\frac{{d+2\ell}}4-s_3+\tfrac{i\nu_4}2}}\,.
\end{multline}
Setting $\nu_4=i\left(\frac{d}{2}-\epsilon\right)$ and collecting the terms linear in $\epsilon$ gives the leading slow roll correction to the inflationary three-point function with three general external scalars due to a general spin-$\ell$ field, which we shall consider in more detail in the following sections.

Note that, strictly speaking, one should also include contributions from exchange diagrams in the other channels (i.e. the ${\sf t}$- and ${\sf u}$-channels). These can be computed in the same way as we did for the ${\sf s}$-channel exchange (or simply by permuting the external legs), and they contribute only analytic terms in $k\sim k_3$.\footnote{Recall that $k$ is the exchanged momentum in the ${\sf s}$-channel.} Non-analytic terms in $k\sim k_3$, on the other hand, signal the presence of new particles in the inflationary three-point functions \cite{Chen:2009we,Chen:2009zp,Byrnes:2010xd,Baumann:2011nk,Assassi:2012zq,Noumi:2012vr,Arkani-Hamed:2015bza,Mirbabayi:2015hva,Chen:2015lza,Lee:2016vti}. In the following sections we focus on the non-analytic terms generated by the exchange of a general spin-$\ell$ particle, starting in section \ref{subsec::squeezlimit} with the leading terms in the squeezed limit $k_3 \sim k \to 0$. In section \ref{subsec::beyondthesq} we consider the full non-analytic contribution.

\subsection{Squeezed limit} 
\label{subsec::squeezlimit}
If we take the squeezed limit $k_3 \sim k \rightarrow 0$, at the level of the CPW \eqref{softcpw} this corresponds to taking the soft limit of the spin-$\ell$ leg in the other three-point function factor, which we also carried out in \S \ref{subsec::softlimit2pt}. In particular, using \eqref{rhok3soft} and \eqref{softlimitpl}, in the squeezed limit we have
\begin{subequations}
\begin{multline}
 \hspace*{-0.5cm} \Theta^{\left(\ell\right)}_{\nu_1,\nu_2,\nu_3,\nu_4;\nu}(\vec{k}_i;\vec{k}|s_i,u,\bar{u})\sim {\widetilde H}_{\nu_3,-\nu;\nu_4|\ell}(s_3,\bar{u})\frac{1}{\ell!\left(\tfrac{d}{2}-1\right)_\ell} (-i\vec{k}_3\cdot\hat{\pl}_\xi)^\ell (-i\,\vec{\xi}\cdot \vec{k}_1)^\ell
    \\=  {\widetilde H}_{\nu_3,-\nu;\nu_4|\ell}(s_3,\bar{u})\frac{(-1)^\ell\ell!}{2^\ell \left(\frac{d}{2}-1\right)_\ell}\,C_\ell^{\left(\frac{d}{2}-1\right)}\left(\cos\theta\right)\,k_1^\ell k_3^\ell,
\end{multline}
\end{subequations}
where we used that $\vec{k_2}\sim-\vec{k}_1$ as $\vec{k} \to 0$ and defined $\cos\theta=\hat{k}_1\cdot \hat{k}_3$.

In the squeezed limit the integrals in $s_2$ and $s_3$ can be lifted in the usual way with the residue theorem. The easiest is the integral in $s_2$ which, after using that $k_2\sim k_1$ as $k \to 0$ and re-defining $s_1 \to s_1-s_2$, can be lifted using Barnes' first lemma, which gives
\begin{multline}
    \langle \phi^{\left(\nu_1\right)}_{\vec{k}_1}\phi^{\left(\nu_2\right)}_{\vec{k}_2}\phi^{\left(\nu_3\right)}_{\vec{k}_3}\phi^{\left(\nu_4\right)}_{\vec{k}_4} \rangle^\prime \\ \sim  {\cal N}_4 \frac{\Gamma\left(-i\nu_4\right)}{2\sqrt{\pi}}\left(\frac{k_4}{2}\right)^{2i\nu_4}\frac{(-1)^\ell\ell!}{2^\ell \left(\frac{d}{2}-1\right)_\ell}\,C_\ell^{\left(\frac{d}{2}-1\right)}\left(\cos\theta\right)k_1^\ell k_3^\ell 
    \,{\cal I}_{\nu_1,\nu_2,\nu_3;\nu_4;\nu,\ell}(\vec{k}_1,\vec{k}_3),
\end{multline}
where 
\begin{multline}
  \hspace*{-0.75cm} {\cal I}_{\nu_i;\nu_4;\nu,\ell}(\vec{k}_1,\vec{k}_3)= \frac{1}{4\pi} \left(\frac{k_3}{2}\right)^{-(d+i(\nu_4-\nu_3)+2\ell)}\left(\frac{k_1}{2}\right)^{i(\nu_1+\nu_2)}\frac{\left(\tfrac{d-2 \ell+2 i (-\nu_3+\nu_4-\nu)}{4}\right)_\ell \left(\tfrac{d-2\ell+2i\left(\nu_3+\nu_4-\nu\right)}{4}\right)_\ell}{\left(\tfrac{d}{2}-i \nu-1\right)_\ell}\\ 
   \times \int^{i\infty}_{-i\infty} \frac{ds_1}{2\pi i}\frac{ds_3}{2\pi i}\, \csc\left(\pi\left(\tfrac{d}{2}+\tfrac{i\nu_4}{2}+\ell-s_1-s_3\right)\right)\delta^{(d+2\ell,d+2\ell)}(\tfrac{d+2\ell}4-s_1,\tfrac{d+2\ell}4-s_3+\tfrac{i\nu_4}2)
  \\\times \frac{1}{\Gamma\left(2s_1\right)}\prod_{\pm {\hat \pm}} \Gamma\left(s_1\pm\tfrac{i\nu_1}{2}{\hat \pm}\tfrac{i\nu_2}{2}\right) \rho_{\nu_3,\nu-i\ell,\nu}(s_3,\tfrac{d+2\ell}4-s_3+\tfrac{i\nu_4}2,\tfrac{d+2\ell}4-s_1)
\left(\frac{k_1}{k_3}\right)^{-2s_1}.
\end{multline}
The integral in $s_3$ can be evaluated by closing the contour on the positive real axis, which encloses the following sequences of poles:\footnote{It is also important to keep in mind that the behaviour of the integrand for $s_3=R e^{i\theta}$ with $R\to\infty$ is (see \S\ref{App:convergence}):
\begin{equation}
    e^{\left(\tfrac{d}2-2- \Im(\nu_4)\right)\,\log (R) }\,.
\end{equation}
This means that there are no potential divergences when $\nu_4=i\left(\tfrac{d}{2}-\epsilon\right)$ for $\epsilon\to0$.}
\begin{subequations}\label{nonanalsqu}
\begin{align}\label{nonanalsqu1}
   s_3&\sim\frac{d-2 \ell\pm2 i \nu +2i\nu_4}{4}+n, \hspace*{0.7cm} n \in \mathbb{Z}_{\geq0},\\
    s_3&\sim\frac{d+2 \ell+i \nu_4+2 n-2 s_1}{2}+m\,, \hspace*{0.5cm} m \in \mathbb{Z}_{\geq0},\label{nonanalsqu2}
\end{align}
\end{subequations}
where the second sequence originates from the $\csc$-factor. Closing the contour on the left, the re-summation of the residues from the first two sequences \eqref{nonanalsqu1} is given by a Gauss Hypergeometric function $_2F_1$ at argument $z=1$, which is a simple ratio of $\Gamma$-functions by virtue of Gauss' theorem.\footnote{Gauss' theorem is the identity:
\begin{equation}
    {}_2F_1\left(a,b;c;z=1\right)=\frac{\Gamma\left(c\right)\Gamma\left(c-a-b\right)}{\Gamma\left(c-a\right)\Gamma\left(c-b\right)}.
\end{equation}
} Instead the re-summation of the residues from the $\csc$-poles \eqref{nonanalsqu2} gives a generalised Hypergeometric function $_3F_2$ at argument $z=1$ with parameters depending on the remaining Mellin variable $s_1$.

The remaining integral in $s_1$ can be evaluated as a series expansion in $k_3/k_1$. In the squeezed limit we are required to close the integration contour to the right of the imaginary axis, where the non-analytic terms in $k$ are encoded in the residues of the sequence of poles:
\begin{equation}\label{coses3squ}
    s_1=\frac{d+2\ell \pm 2 i \nu}{4}-n^\prime, \hspace*{0.5cm} n^\prime \in \mathbb{Z}_{\geq0},
\end{equation}
where the leading term is given by the residue of the poles with $n^\prime=0$.\footnote{It is interesting to note that while in the OPE limit, where $k \to 0$ and $k << k_3$, we observed that the cosecant function generates only analytic terms in $k$, for the squeezed limit we instead have $k \sim k_3 \to 0$ so that non-analytic terms may also be generated from the residues in \eqref{nonanalsqu2} -- as we have just confirmed.} Nicely, the $_3F_2$ Hypergeometric function generated by the $s_3$ integral has simple residues on the above poles, which are given by a basic ratio of $\Gamma$-functions.

Setting $\nu_4=i\left(\tfrac{d}{2}-\epsilon\right)$, one can first verify that the terms of order in $\epsilon$ vanish identically. One can then collect the terms linear in $\epsilon$, which give the following leading slow-roll correction in the squeezed limit:\footnote{It may be useful to note that, in order to arrive to the simplified expression \eqref{squeezed General}, in the last step we used the identity
\begin{equation}
    \csc \left(\tfrac{\pi}{2}  (\ell+i (\nu -\nu_3))\right) \sin \left(\tfrac{\pi}{2}  (-\ell+i(\nu -\nu_3))\right)=(-1)^\ell\,,
\end{equation}
which is valid only for $\ell\in\mathbb{N}$.}
\begin{samepage}
\begin{shaded}
\noindent \emph{Squeezed limit of the correction to the inflationary 3pt function from a spin-$\ell$ exchange}
\begin{multline}\label{squeezed General}
    \langle \phi^{\left(\nu_1\right)}_{\vec{k}_1}\phi^{\left(\nu_2\right)}_{\vec{k}_2}\phi^{\left(\nu_3\right)}_{\vec{k}_3} \rangle^\prime_{(\text{infl.})}\sim -\frac{\epsilon\, \mathcal{N}_3}{32 \sqrt{\pi }}\frac{(-2)^\ell\ell! }{\left(\tfrac{d}{2}-1\right)_\ell}\,\left(\frac{k_1}{2}\right)^{i (\nu_1+\nu_2)}\left(\frac{k_3}{2}\right)^{-\frac{d}{2}+i \nu_3}  \\\times\,C_{\ell}^{\left(\frac{d-2}{2}\right)}(\cos\theta)\Bigg[\left(\frac{k_3}{k_1}\right)^{\frac{d}{2}-i \nu }\frac{\Gamma (i \nu )\left(-\tfrac{\ell}{2}-\tfrac{i \nu }{2}\pm\tfrac{i \nu_3}{2}+1\right)_{\ell-1}}{ \left(\tfrac{d}{2}-i \nu -1\right)_\ell \Gamma \left(\frac{d}{2}+\ell-i \nu \right)} \\
    \times\, \sin \left(\tfrac{\pi}{4}  (d+2 \ell-2 i (\nu -\nu_1-\nu_2))\right)\text{csch}\left(\tfrac{\pi}{2}  (i \ell+\nu +\nu_3)\right)\prod_{\pm\,{\hat \pm}} \Gamma \left(\tfrac{d+2 \ell-2 i (\nu \pm\nu_1{\hat \pm}\nu_2)}{4}\right)\\
    + \nu \to -\nu \Bigg],
\end{multline}
\end{shaded}
\end{samepage}
\noindent where we divided by the two-point function of the leg with respect to which we are taking the soft limit. To the best of our knowledge the above result for general external scalars is new, even when $d=3$, and its generality demonstrates the effectiveness of the Mellin space tools developed in this work. In the next section we shall moreover consider the subleading corrections to the squeezed limit.

The expression \eqref{squeezed General} is consistent with the signatures described in \cite{Noumi:2012vr,Arkani-Hamed:2015bza} when $d=3$ for new particles in inflationary three-point functions, whose form is constrained by the slightly broken conformal symmetry \cite{Arkani-Hamed:2015bza}. In particular, for exchanged massive particles on the Principal Series, $\nu \in \mathbb{R}$, we have oscillatory terms in $\log \left(\frac{k_3}{k_1}\right)$, while the fact that the exchanged particle has spin $\ell$ is indicated by the Gegenbauer polynomial (which reduces to a Legendre polynomial when $d=3$). 

Considering external massless scalars (i.e. $\nu_j=\frac{di}{2}$) and setting $d=3$, one recovers equation (6.144) of \cite{Arkani-Hamed:2015bza}:
\begin{multline}
 \hspace*{-0.5cm}   \langle \phi^{\left(\nu_1\right)}_{\vec{k}_1}\phi^{\left(\nu_2\right)}_{\vec{k}_2}\phi^{\left(\nu_3\right)}_{\vec{k}_3} \rangle^\prime_{(\text{infl.})}\sim -\epsilon\,\frac{(-8)^{-\ell} \,\ell!}{4\,k_1^3 k_3^3\, \Gamma \left(\ell+\frac{1}{2}\right)}\,P_\ell(\cos\theta)\,\left(\frac{k_3}{4 k_1}\right)^{\frac{3}{2}-i \nu }\frac{\Gamma \left(\ell+i \nu +\frac{1}{2}\right) \Gamma \left(\ell-i \nu +\frac{1}{2}\right)}{\left(\ell-\frac{3}{2}\right)^2+\nu ^2}\\\times\frac{\pi}{\cosh (\pi  \nu ) }\frac{\left(\tfrac{5}{2}+\ell-i \nu\right) (1-i(-1)^\ell \sinh (\pi  \nu )) \Gamma (i \nu ) }{\left(\tfrac{3}{2}-\ell+i \nu \right) \Gamma \left(i \nu +\frac{1}{2}\right)}+\nu \to -\nu\,,
\end{multline}
Note that the $\left(-1\right)^\ell$ was not predicted in \cite{Arkani-Hamed:2015bza} since their analysis assumed equal external scalars from the beginning which precludes the exchange of odd spins.

We emphasise that the above result was obtained by considering the canonical $\ell$-derivative coupling of a spin $\ell$ field with two scalars. Higher derivative couplings change the above result (and the subleading corrections) just by an overall polynomial factor in the scaling dimensions.

\subsection{Beyond the squeezed limit}
\label{subsec::beyondthesq}

In this section we shall consider the subleading corrections to the squeezed limit. For simplicity we focus on the case where two of the external scalars, $\phi^{\left(\nu_1\right)}$ and $\phi^{\left(\nu_2\right)}$, are conformally coupled while the remaining external scalar $\phi^{\left(\nu_3\right)}$ is kept general. By acting with the raising operators \eqref{raisingop} we then obtain the result for $\nu_{1,2}=\frac{3i}{2}$, which for $d=3$ correspond to massless scalars. We compute all subleading corrections to the squeezed limit, focusing on the non-analytic terms in $k=k_3$ which, quite remarkably, re-sum to a Gauss Hypergeometric function which moreover provides the analytic continuation away from the regime $k \sim k_3 \ll k_{1,2}$.

To this end we consider the basic seed integral: 
\begin{multline}
  \tilde{\mathcal{A}}^{(x)}_{\nu_1,\nu_2,\nu_3;\nu_4;\nu,\ell|\alpha}(\vec{k}_1,\vec{k}_2,\vec{k}_3)=\frac{\left(\tfrac{d-2 \ell+2 i (-\nu_3+\nu_4-\nu)}{4}\right)_\ell \left(\tfrac{d-2\ell+2i\left(\nu_3+\nu_4-\nu\right)}{4}\right)_\ell}{\left(\tfrac{d}{2}-i \nu-1\right)_\ell}\\
 \times \int [ds]_3\, \csc(\pi(u+\bar{u}))\delta^{(x,d+2\ell)}(u,\bar{u})\\
 \hspace*{-1cm} \times\frac{\rho_{\nu_1,\nu_2,\nu}(s_1,s_2,u)\rho_{\nu_3,\nu}(s_3,\bar{u})}{\left(u+\tfrac{i\nu}2-\alpha\right)_{\alpha}\left({\bar u}-\tfrac{i \nu}{2}-\ell\right)_\ell}\left(\frac{k_3}{2}\right)^{-2\left(u+{\bar u}\right)}\prod^3_{j=1}\left(\frac{k_j}{2}\right)^{-2s_j+i\nu_j}\Big|_{{}^{u=\frac{x}4-s_1-s_2}_{{\bar u}=\frac{{d+2\ell}}4-s_3+\tfrac{i\nu_4}2}}\,,
\end{multline}
from which all contributions \eqref{spinning_inflaalphabeta} to \eqref{spinning_infla} are given through the recursion relations \eqref{alphabeta_fromgamma}:
\begin{multline}\label{alphabeta_fromgamma_infl}
  \tilde{\cal A}_{\nu_1,\nu_2,\nu_3;\nu_4;\nu,\ell|\alpha,\beta}^{(x)}(\vec{k}_1,\vec{k}_2,\vec{k}_3) \\= (-1)^{\alpha} k_1^{2 (\alpha-\beta+i \nu_1)}k_2^{2 (\beta+i \nu_2)}\pl_{k_1^2}^{\alpha-\beta}\pl_{k_2^2}^{\beta}\left[k_1^{-2i\nu_1}k_2^{-2i\nu_2}\tilde{\cal A}_{\nu_1,\nu_2,\nu_3;\nu_4;\nu,\ell|\alpha}^{(x)}(\vec{k}_1,\vec{k}_2,\vec{k}_3)\right]\,.
\end{multline}

When $\phi^{\left(\nu_1\right)}$ and $\phi^{\left(\nu_2\right)}$ are conformally coupled, $\nu_{1,2}=\frac{i}{2}$, both the integrals in $s_2$ and $s_3$ can be lifted. As before, the $s_2$ integral is the easiest after re-defining $s_1 \to s_1-s_2$, and is a simple application of the residue theorem:
\begin{align}\nonumber
   & \hspace*{-0.5cm} \tilde{\mathcal{A}}^{(x)}_{\nu_i;\nu_4;\nu,\ell|\alpha}(\vec{k}_1,\vec{k}_2,\vec{k}_3)=\frac{1}{\sqrt{\pi} k_1 k_2}\,\left(\frac{k_3}{2}\right)^{i \nu_3-i \nu_4-\frac{x}{2}-\frac{\bar{x}}{2}+1}\tfrac{ \left(\tfrac{\bar{x}-4 \ell-2 i (\nu -\nu_3-\nu_4)}{4}\right)_\ell \left(\tfrac{\bar{x}-4 \ell-2 i (\nu +\nu_3-\nu_4)}{4}\right)_\ell}{ \left(\tfrac{\bar{x}}{2}-\ell-i \nu -1\right)_\ell}\\\nonumber
    & \hspace*{1.5cm}\times\int^{+i\infty}_{-i\infty} \frac{ds_1}{2\pi i}\frac{ds_3}{2\pi i}\,\csc \left(\pi\left(\tfrac{x+\bar{x}}{4} - s_1-s_3+\tfrac{i\nu_4}2\right)\right)\delta^{(x,\bar{x})}(\tfrac{x}4-s_1,\tfrac{\bar{x}}4-s_3+\tfrac{i\nu_4}2)\\
    &\hspace*{3.5cm} \times\Gamma(2s_1-1)\Gamma \left(\tfrac{x-2 i \nu }{4}-s_1\right) \Gamma \left(\tfrac{x-4 \alpha+2 i \nu}{4}-s_1\right)
    (2 p_{12})^{1-2 s_1}\\ \nonumber
    &\hspace*{1.5cm} \times \Gamma \left(s_3+\tfrac{i \nu_3}{2}\right) \Gamma \left(s_3-\tfrac{i\nu_3}{2}\right) \Gamma \left(\tfrac{{\bar x}+2 i (\nu_4+\nu)}{4}-s_3\right) \Gamma \left(\tfrac{{\bar x}+2 i (\nu_4-\nu)}{4}-\ell-s_3\right)\Bigg|_{\bar{x}=d+2\ell}.
\end{align}
The integral over $s_3$ can be evaluated by closing the integration contour on the positive real axis, which encloses the following sequences of poles:
\begin{subequations}
\begin{align}
    s_3&\sim \frac{\bar{x}-4 \ell-2 i (\nu -\nu_4)}{4}+n,\qquad n\in\mathbb{N}\,,\\
    s_3&\sim \frac{2 i (\nu +\nu_4)+\bar{x}}{4}+n,\qquad n\in\mathbb{N}\,,\\
    s_3&\sim \frac{2 i \nu_4-4 s_1+x+\bar{x}}{4}+n,\qquad n\in\mathbb{N}\,,
\end{align}
\end{subequations}
where the final sequence comes from the $\csc$-function. The re-summation of the residues for the first two sequences of poles above is given by Gauss $_2F_1$ hypergeometric functions at argument $z=1$. The resummation of the residues of the last series of poles gives instead a generalised Hypergeometric function ${}_3F_2$ at argument $z=1$.

The remaining integral in $s_1$ can be evaluated as a series in $p_{12}^{-1}=\tfrac{k_3}{k_1+k_2}$. In the following we shall focus on the non-analytic terms, which are encoded in the residues of the poles
\begin{equation}\label{non-anal-squeezed}
    s_1\sim\frac{x}4\pm\frac{i\nu}2+n\qquad n\in\mathbb{N}\,.
\end{equation}
It is further possible to resum the corresponding series using the fact that the residues of the $_3F_2$ hypergeometric function collapse to a $_2F_1$ at argument $z=1$, so that all terms can be simplified using Gauss' theorem to a ratio of $\Gamma$-functions.

Setting $\nu_4=i\left(\tfrac{d}{2}-\epsilon\right)$ and expanding in $\epsilon$, it is lengthy but straightforward to check that the order zero result in $\epsilon$ cancels identically. For the terms linear in $\epsilon$, the residues of each sequence of the poles \eqref{non-anal-squeezed} re-sum to a Gauss Hypergeometric function:\footnote{In obtaining \eqref{AtildeResFin}, only for the last step we used the identity $\sinh \left(\tfrac{\pi}{2}  (i \ell+\nu -\nu_3)\right)\text{csch}\left(\tfrac{\pi}{2} (-i \ell+\nu -\nu_3)\right)=(-1)^\ell$, which requires the assumption $\ell\in\mathbb{N}$.}
\begin{multline}
 \hspace*{-0.35cm} \tilde{\mathcal{A}}_{\frac{i}{2},\frac{i}{2},\nu_3;i\left(\frac{d}{2}-\epsilon\right);\nu,\ell|\alpha}^{(x)}(\vec{k}_1,\vec{k}_2,\vec{k}_3)\\=\epsilon(-1)^{\ell+\alpha}\frac{\pi ^{3/2}}{4 k_1 k_2}\, \left(\frac{k_3}{2}\right)^{-\ell+i \nu_3-\frac{x}{2}+1}(2 p_{12})^{i \nu -\frac{x}{2}+1} \, _2F_1\left(\begin{matrix}\tfrac{x-2 i \nu -2}{4},\tfrac{x-2 i \nu }{4}\\\alpha-i \nu +1\end{matrix};\frac{1}{p_{12}^2}\right)\\
    \hspace*{-0.4cm} \times \frac{\left(\tfrac{-\ell-i \nu +i \nu_3+2}{2}\right)_{\ell-1} \left(\tfrac{-\ell+i \nu +i \nu_3+2}{2}\right)_{\ell-1}}{\left(\tfrac{d}{2}-i \nu -1\right)_\ell \left(\frac{x}{2}-i \nu -1\right)_{\alpha-\frac{x}{2}+2}}\csc (i \pi \nu  ) \csc \left(\tfrac{\pi}{2}  (\ell-i (\nu +\nu_3))\right) \cos \left(\tfrac{\pi}{4}  (x-2 i \nu )\right)\\ \label{AtildeResFin}
    +(\nu\to-\nu)+\text{local}\,.
\end{multline}
\noindent Through the recursion relation \eqref{alphabeta_fromgamma_infl}, this gives all the contributions \eqref{spinning_inflaalphabeta} to the leading slow-roll correction from a general spin-$\ell$ field to the inflationary three-point function with two conformally coupled scalars and a general scalar $\phi^{\left(\nu_3\right)}$:
\begin{shaded}
\noindent \emph{Leading slow roll correction induced by a general spin-$\ell$ field to the inflationary 3pt function}
\begin{multline}\label{infl3ptspinl}
 \hspace*{-0.5cm} \langle \phi^{\left(i/2\right)}_{\vec{k}_1}\phi^{\left(i/2\right)}_{\vec{k}_2}\phi^{\left(\nu_3\right)}_{\vec{k}_3} \rangle^\prime_{\left(\text{infl.}\right)}={\cal N}_3 \sum_{\alpha=0}^\ell\sum_{\beta=0}^\alpha\binom{\ell}{\alpha}\binom{\alpha}{\beta}\widetilde{\Theta}^{(\ell)}_{\nu_1,\nu_2,\nu|\alpha,\beta}(\vec{k}_1,\vec{k}_2,\vec{k}_3)(-1)^{\alpha} k_1^{2 (\alpha-\beta+i \nu_1)}k_2^{2 (\beta+i \nu_2)}\\ \times \pl_{k_1^2}^{\alpha-\beta}\pl_{k_2^2}^{\beta}\left[k_1^{-2i\nu_1}k_2^{-2i\nu_2}\tilde{\cal A}_{i/2,i/2,\nu_3;i\left(\frac{d}{2}-\epsilon\right);\nu,\ell|\alpha}^{(d+2\ell)}(\vec{k}_1,\vec{k}_2,\vec{k}_3)\right]\,.
\end{multline}
\end{shaded}
\noindent Furthermore, by acting with the raising operators \eqref{raisingop},
\begin{align}\label{raisingfinfl}
    \tilde{\mathcal{A}}_{\nu_1+i,\nu_2+i,\nu_3;\nu,\ell|\alpha}^{(x)}=\frac{4}{k_1 k_2}\pl_{k_1}\pl_{k_2}\,\tilde{\mathcal{A}}_{\nu_1,\nu_2,\nu_3;\nu,\ell|\alpha}^{(x-4)}\,,
\end{align}
one obtains the corresponding result for when the two scalars have $\nu_{1,2}=\frac{3i}{2}$, which in $d=3$ is the massless scalar.

When the correction is induced by a scalar field, we see that the contribution \eqref{spinning_infla} with $\alpha,\beta=0$ gives the full inflationary correlator \eqref{infl3ptspinl} at leading order in slow roll, so in that case there is no need to apply the recursion relation \eqref{alphabeta_fromgamma_infl} and the result takes a simpler form: 
\begin{samepage}
\begin{shaded}
\noindent \emph{Leading slow roll correction induced by a general scalar field to the inflationary 3pt function}
\begin{multline}\label{2Conf1arbInfla}
    \langle \phi^{\left(i/2\right)}_{\vec{k}_1}\phi^{\left(i/2\right)}_{\vec{k}_2}\phi^{\left(\nu_3\right)}_{\vec{k}_3} \rangle^\prime_{\text{infl.}}={\cal N}_3\tilde{\mathcal{A}}_{\frac{i}{2},\frac{i}{2},\nu_3;\nu,0|0}^{(d)}(\vec{k}_1,\vec{k}_2,\vec{k}_3)\\=i\epsilon\,{\cal N}_3\frac{\sqrt{\pi}}{2 k_1 k_2} \left(\frac{k_3}{2}\right)^{-\frac{d}{2}+i \nu_3+1} (2 p_{12})^{-\frac{d}{2}+i \nu +1}\\\times\frac{\Gamma(i\nu)\Gamma \left(\frac{d}{2}-i \nu -1\right) \cos \left(\tfrac{\pi}{4} (d-2 i \nu )\right) \text{csch}\left(\tfrac{\pi}{2}  (\nu +\nu_3)\right)}{(\nu -\nu_3) (\nu +\nu_3)}\\\times\, _2F_1\left(\begin{matrix}\tfrac{d-2 i \nu -2}{4},\tfrac{d-2 i \nu}{4}\\1-i \nu\end{matrix} ;\frac{1}{p_{12}^2}\right)+(\nu\to-\nu)+\text{local}\,.
\end{multline}
\end{shaded}
\end{samepage}
\noindent By acting with the raising operators \eqref{raisingop} as in \eqref{raisingfinfl}, from the above expression we can obtain the corresponding result for when the two scalars have $\nu_{1,2}=\frac{3i}{2}$: 
\begin{samepage}
{\allowdisplaybreaks\begin{shaded}
\noindent \emph{Inflationary 3pt function of a general scalar and two scalars with $\nu_{1,2}=\frac{3i}{2}$}
\begin{multline}
    \langle \phi^{\left(3i/2\right)}_{\vec{k}_1}\phi^{\left(3i/2\right)}_{\vec{k}_2}\phi^{\left(\nu_3\right)}_{\vec{k}_3} \rangle^\prime_{\text{infl.}}={\cal N}_3\frac{4}{k_1 k_2}\pl_{k_1}\pl_{k_2}\tilde{\mathcal{A}}_{\frac{i}{2},\frac{i}{2},\nu_3;\nu,0|0}^{(d-4)}(\vec{k}_1,\vec{k}_2,\vec{k}_3)\\
    \hspace*{0.5cm}=i\epsilon\,{\cal N}_3\frac{\sqrt{\pi}}{2}\left(\frac{k_3}{2}\right)^{-\frac{d}{2}+i \nu_3+3}\frac{\Gamma(i\nu)\Gamma \left(\frac{d}{2}-i \nu -3\right) \cos \left(\tfrac{\pi}{4} (d-4-2 i \nu )\right) \text{csch}\left(\tfrac{\pi}{2}  (\nu +\nu_3)\right)}{(\nu -\nu_3) (\nu +\nu_3)}
    \\
   \times  \left[\frac{k_1 k_2 (d-2 i \nu -6) (d-2 i \nu -4)+2 k_3 p_{12}^2 (d-2 i \nu -6)+4 k_3^2 p_{12}^2}{k_1^3 k_2^3 k_3^2}\right.\\\hspace{50pt}\times(2p_{12})^{i \nu -\frac{d}{2}+1}\, _2F_1\left(\begin{matrix}\tfrac{d-6-2 i \nu +2}{4},\tfrac{d-4-2 i \nu}{4}\\1-i \nu\end{matrix} ;\frac{1}{p_{12}^2}\right)\\\nonumber
    +\frac{(d-2 i \nu -6) (d-2 i \nu -4) (k_1 k_2 (d-2 i \nu -3)+ k_3^2 p_{12}^2)}{k_1^3 k_2^3 k_3^2(1-i\nu)}\\\nonumber
    \hspace{100pt} \times(2p_{12})^{i \nu -\frac{d}{2}-1} \, _2F_1\left(\begin{matrix}\tfrac{d-2-2 i \nu}{4},\tfrac{d-2 i \nu}{4}\\2-i \nu \end{matrix} ;\frac{1}{p_{12}^2}\right)\\\nonumber
    +\frac{(d-2 i \nu -6) (d-2 i \nu -4) (d-2 i \nu -2) (d-2 i \nu )}{k_1^2 k_2^2 k_3^2(1-i\nu)(2-i\nu)} \\\nonumber
    \hspace{100pt}\left.\times(2p_{12})^{i \nu -\frac{d}{2}-3}\, {}_2F_1\left(\begin{matrix}\tfrac{d+2-2 i \nu}{4},\tfrac{d+4-2 i \nu}{4}\\3-i \nu \end{matrix};\frac{1}{p_{12}^2}\right)\right]\\
    + \left(\nu \to -\nu\right) + \text{local}.
\end{multline}
\end{shaded}}
\end{samepage}
\noindent When $d=3$, the two scalars with $\nu_{1,2}=\frac{3i}{2}$ are massless.

\section{Conclusions and Outlook}

In this work we have presented a new systematic approach to de Sitter and Inflationary correlators based on the Mellin-Barnes representation in momentum space. The machinery of Mellin space gives a firm grasp on the analytic structure of momentum space correlators, through which we have identified key general features of tree-level exchange four-point functions, that are common to both de Sitter and anti-de Sitter space, which appear to be universal consequences of conformal symmetry and the boundary conditions. 

This framework has allowed us to obtain, in a straightforward manner, analytic expressions for late-time exchange four-point functions with generic external scalars and a generic exchanged spin-$\ell$ field, moreover for general boundary dimension $d$. As an intermediate step we also obtained analytic expressions for late-time three-point functions of two general scalars and a general spin-$\ell$ field. From these results, assuming the weak breaking of the de Sitter isometries, we extracted the corresponding correction to the inflationary three-point function of general external scalars induced by a general spin-$\ell$ field at leading order in slow roll. In the squeezed limit this exhibits the power-law/oscillatory behaviour expected from the presence of new particles \cite{Chen:2009we,Byrnes:2010xd,Baumann:2011nk,Assassi:2012zq,Noumi:2012vr,Arkani-Hamed:2015bza}, with the angular dependence associated to the exchange of a spin-$\ell$ field encoded in a Gegenbauer polynomial for general $d$.

We conclude by highlighting a few interesting future directions which naturally follow:

\begin{itemize}

\item The universal form \eqref{exchampl} of the exchange four-point function admits a natural extension to the case of spinning external fields, directly in terms of the corresponding Conformal Partial Waves for spinning external legs \cite{Sleight:2017fpc}. These details will be presented in \cite{ToAppear2}.

\item The central role of the Mellin formalism in uncovering analytic properties of late-time correlators at tree-level motivates the exploration of quantum corrections within this framework. It would be desirable to obtain a better understanding of quantum corrections at both the perturbative and non-perturbative level, which a bootstrap approach to de Sitter correlators may facilitate.

\item Having considered four-point correlators both in AdS and dS, it would be interesting to explore the more challenging case of flat space holography in this framework. In this context various new ideas have been set forth recently \cite{Pasterski:2016qvg,Pasterski:2017kqt,Cardona:2017keg,Stieberger:2018onx,Nandan:2019jas} and it would be interesting to further understand the possible relation between the results obtained in this paper and the corresponding flat space analysis which lies in between. The differences between dS and AdS cases with a time-like and a space-like boundary respectively, compared to the light-like nature of null-infinity in flat space, make clear that key differences are expected and that these dualities cannot be smoothly connected to each other. On the other hand our analysis clarifies quantitatively how some specific signatures of conformal symmetry are in fact universal. It would be interesting to clarify if these universal features do play a role for flat-space holography.

\item Another advantage of the Mellin formalism presented in this work is that higher-point correlators may be easily constructed by multiplying together three-point structures in a way that generalises the construction of exchange four-point functions. What remains is to determine the corresponding interference factor $\delta$, which may be fixed by demanding the correct singularity structure.

\item Utility of the Mellin-Barnes representation extends beyond the context of boundary correlation functions in (anti-)de Sitter space, and may more generally be used to study conformal structures in momentum space. It would be especially interesting to explore possible applications of this framework to the Conformal Bootstrap,\footnote{For a recent comprehensive review see \cite{Poland:2018epd}.} particularly its incarnation due to Polyakov \cite{Polyakov:1974gs}, who in the 70s proposed expanding CFT four-point functions in a crossing-symmetric basis of building blocks in momentum space and using consistency with the Operator Product Expansion as a constraint on the CFT data. At the time this approach was little explored due to the technical complications in implementing conformal symmetry in momentum space, but recently it has experienced a revival due to the observation that these building blocks are essentially boundary exchange diagrams (i.e. exchange Witten diagrams) in anti-de Sitter space \cite{Gopakumar:2016wkt}. The success of this revival came from combining this observation with the Mellin representation of Witten diagrams in position space \cite{Costa:2012cb}, which has led to important advances in analytic approaches to the Conformal Bootstrap (see e.g. \cite{Costa:2014kfa,Gopakumar:2016cpb,Aharony:2016dwx,Dey:2017fab,Sleight:2018epi,Sleight:2018ryu,Gopakumar:2018xqi}). Given the pivotal role of the Mellin space machinery in this progress, it would be very interesting to understand whether the tools presented in this work could help overcome the difficulties of implementing the Polyakov Bootstrap in momentum space, potentially offering new perspectives on the analytic bootstrap.
\end{itemize}

We leave these questions for the near future.

\section*{Acknowledgments}
We would like to thank Dionysios Anninos, Thomas Basile, Ruben Monten
for useful discussions. We are also grateful to Dionysios Anninos and Ruben Monten for comments on the draft.  C.S. gratefully acknowledges Universit\'e Libre de Bruxelles and Princeton University for support and hospitality during various stages of this work, and the  University  of  Naples  Federico  II and the Erwin Schr\"odinger International Institute for Mathematics and Physics, where this work was finalised. The research of C.S. is supported by the European Union's Horizon 2020 research and innovation programme under the Marie Sk\l odowska-Curie grant agreement No 793661 and, until October 2018, by a Marina Solvay Fellowship. M.T. thanks Caltech and the Simons Bootstrap Collaboration for hospitality and support during the Bootstrap workshop in July 2018, where part of this work was carried out. The research of M.T. was partially supported by the European Union's Horizon 2020 research and innovation programme under the Marie Sklodowska-Curie grant agreement No 747228.

\begin{appendix}

\section{Wick rotation}\label{Wick Rotation}

In this appendix we briefly review the Wick rotation from Euclidean to Lorentzian signature. For definiteness, we will focus on the leading short distance singularity of the 2pt function, which reproduces the flat space singularity and in Euclidean signature reads:
\begin{equation}
    \left\langle\phi(\tau_1,\vec{x}_1)\phi(\tau_2,\vec{x}_2)\right\rangle=\frac{c_d}{\left[(\tau_1-\tau_2)^2+|\vec{x}_1-\vec{x}_2|^2\right]^{\tfrac{d-1}2}}+\ldots\,,
\end{equation}
where the $\ldots$ give less singular contributions. The Wick rotation to Lorentzian signature is defined by specifying the corresponding domain of analyticity. In the case of Wightman functions, positivity of the energy implies the following $i\epsilon$ prescription:\footnote{More generally a Wightman correlator defines an analytic functions:
\begin{equation}\label{analytic 2pt domain}
    \left\langle\phi(t_1,\vec{x}_1)\cdots\phi(t_n,\vec{x}_n)\right\rangle=\lim_{\epsilon_i\to0}\left\langle\phi(t_1-i\epsilon_1,\vec{x}_1)\cdots\phi(t_2-i\epsilon_n,\vec{x}_2)\right\rangle\,,
\end{equation}
in the domain $\epsilon_1>\epsilon_2>\ldots>\epsilon_n$ which corresponds to the Euclidean ordering of the operators. This analyticity region in terms of the complex coordinates $z_j=x_j-i\eta_j$ can be extended to the ``tube" $\eta_j-\eta_{j+1}\in V_+$ where $V_+$ is the forward cone. Analyticity can be proven starting from the Laplace transform of the correlator and using positivity of the energy to show that the integral representation for the position space correlator is exponentially suppressed in the tube above (see e.g. Theorem 2-8 and 3-5 of \cite{Streater:1989vi}).
}
\begin{subequations}
\begin{align}
        \left\langle\phi(t_1-\tfrac{i\epsilon}2,\vec{x}_1)\phi(t_2+\tfrac{i\epsilon}2,\vec{x}_2)\right\rangle &=\frac{c_d}{\left[-(t_1-t_2-i\epsilon)^2+|\vec{x}_1-\vec{x}_2|^2\right]^{\tfrac{d-1}2}}+\ldots\,,\\
        \left\langle\phi(t_2-\tfrac{i\epsilon}2,\vec{x}_2)\phi(t_1+\tfrac{i\epsilon}2,\vec{x}_1)\right\rangle &=\frac{c_d}{\left[-(t_1-t_2+i\epsilon)^2+|\vec{x}_1-\vec{x}_2|^2\right]^{\tfrac{d-1}2}}+\ldots\,,
\end{align}
\end{subequations}
One can therefore see that the Wightman correlator has an $\epsilon$-prescription which depends on time-ordering:
\begin{equation}
        \left\langle\,\phi(t_1,\vec{x}_1)\phi(t_2,\vec{x}_2)\right\rangle=\frac{c_d}{\left[-(t_1-t_2)^2+|\vec{x}_1-\vec{x}_2|^2 +i\,\text{sig}(t_1-t_2)\epsilon\right]^{\tfrac{d-1}2}}+\ldots\,.
\end{equation}
The above analytic continuations for time-like separation $x_{12}^2<0$ can be summarised as
\begin{subequations}
\begin{align}
   \phi(t_1)\phi(t_2)&:\qquad x_{12}^2\to |x_{12}|^2\,e^{+i\pi\,\text{sig}(t_1-t_2)},\\
   \phi(t_2)\phi(t_1)&:\qquad x_{12}^2\to |x_{12}|^2\,e^{-i\pi\,\text{sig}(t_1-t_2)},
\end{align}
\end{subequations}
one for each operator ordering.

From the above it is manifest that the non-singularity of the Harmonic function in EAdS is equivalent to the Hadamard condition for the corresponding analytic continuation. Indeed, following the above prescription, the dS short distance singularity reads:
\begin{align}
    \Omega_\nu(X_1,X_2)&\sim -\frac12\frac1{\Gamma(i\nu)\Gamma(- i\nu)}\,\frac{\Gamma(\tfrac{d-1}2)}{(2\pi)^{(d+1)/2}}\left(\frac{1-X_1\cdot X_2}2\right)^{\tfrac{1-d}2}+\ldots\\
    &\hspace{-50pt}=-\frac12\frac1{\Gamma(i\nu)\Gamma(- i\nu)}\,\frac{\Gamma(\tfrac{d-1}2)}{(2\pi)^{(d+1)/2}}\left(\frac{-(\eta_1-\eta_2)^2+|\vec{x}_1-\vec{x}_2|^2\pm i\,\text{sgn}(\eta_1-\eta_2)\,\epsilon}2\right)^{\tfrac{1-d}2}+\ldots\,,\nonumber
\end{align}
thus recovering the dS Wightman functions directly from the EAdS Harmonic function up to an overall coefficient. 

\section{Convergence of Mellin-Barnes Integrals}\label{App:convergence}

In this Appendix we review the convergence of the Mellin-Barnes integrals of the type:
\begin{equation}
    \mathcal{I}=\int^{i\infty}_{-i\infty}\frac{ds}{2\pi i}\,g(s) z^{-s},
\end{equation}
where the function $g(s)$ is assumed to have the form
\begin{equation}\label{genmellinbarnsint}
    g(s)= \frac{\Gamma\left(a_1+A_1 s\right)...\Gamma\left(a_{n}+A_{n} s\right)\Gamma\left(b_1-B_1 s\right)...\Gamma\left(b_{m}-B_{m} s\right)}{\Gamma\left(c_1+C_1 s\right)...\Gamma\left(c_{p}+C_{p} s\right)\Gamma\left(d_1-D_1 s\right)...\Gamma\left(d_{q}-D_{q} s\right)}\,z^s,
\end{equation}
where $A_j$, $B_j$, $C_j$, $D_j >0 $.\footnote{In the case of multiple Mellin-Barnes integrals, the parameters $a_j$, $b_j$, $c_j$, $d_j$ may depend on the other Mellin variables.} The integration contour runs parallel to the imaginary axis with suitable indentations to separate the poles from Gamma functions of the type $\Gamma\left(a_j+A_j s\right)$ from those of the type $\Gamma\left(b_j-B_j s\right)$. In this work all Mellin-Barnes integrals considered are of this type, usually with $A_j=B_j=C_j=D_j=1$. In the following we mostly follow the treatment of section 2.4 in \cite{MellinBook}. 

In order to study the convergence of integrals of the above type, one has to determine the behaviour of the integrand as $|s|\to \infty$. There are two useful parametrisations for this limit. One is
\begin{equation}\label{param2}
    s=\sigma+i t\,,\qquad t\to\pm\infty\,,
\end{equation}
and the other
\begin{equation}\label{param1}
    s=R\,e^{i\theta}\qquad R\to\infty\,.
\end{equation}
The first parameterisation \eqref{param2} can probe the convergence of the integral with contour as prescribed along the imaginary axis, while the second parameterisation \eqref{param1} can also probe the convergence for completions of the contour by an arc of infinite radius.

Using Stirling's approximation
\begin{equation}
    \log\Gamma(z)\sim(z-\tfrac12)\log z-z+\tfrac12 \log(2\pi)\,,\qquad |\arg{z}|<\pi,
\end{equation}
and the standard identity:
\begin{equation}
    \Gamma(1-z)=\frac{\pi}{\sin(\pi z)}\frac1{\Gamma(z)}\,,
\end{equation}
for each Gamma function factor in the integrand \eqref{genmellinbarnsint}, for \eqref{param1} one has
\begin{subequations}
\begin{align}
    \log|\Gamma(\alpha+\beta R e^{i\theta})|&\sim \beta  R \cos (\theta ) \log (\beta  R)\\\nonumber
    &\hspace{40pt}-R\beta  \Big[\theta  \sin (\theta )+\cos (\theta )\Big]+\log R\left(\mathfrak{Re}(\alpha)-\tfrac12\right)+O(1)\,,\\
    \log|\Gamma(\alpha-\beta R e^{i\theta})|&\sim -\beta  R \cos (\theta ) \log (\beta  R)\\\nonumber
    &\hspace{40pt}+\beta  R \Big[\theta  \sin (\theta )+\cos (\theta )-\pi  \left| \sin (\theta )\right|\Big]+\log R\left(\mathfrak{Re}(\alpha)-\tfrac12\right)+O(1)\,,
\end{align}
\end{subequations}
where we have assumed $\beta>0$ and $|\theta|<\pi$, while for \eqref{param2} we have 
\begin{subequations}
\begin{align}
    \log|\Gamma(\alpha+\beta \,\sigma+i\,\beta t )|&\sim \log (\beta  \left| t\right|)  \Big[\mathfrak{Re}(\alpha )+\beta\,  \sigma -\tfrac12\Big]-\frac{\pi  \beta  \left| t\right| }{2}+O(1)\,,\\
    \log|\Gamma(\alpha-\beta \sigma-i\,\beta t)|&\sim \log (\beta  \left| t\right|)  \Big[\mathfrak{Re}(\alpha )-\beta\,  \sigma-\tfrac12\Big]-\frac{\pi  \beta  \left| t\right| }{2}+O(1)\,.
\end{align}
\end{subequations}
Combined with
\begin{equation}
    \log|z^{\alpha+\beta R e^{i\theta}}|\sim \mathfrak{Re}\Big[\beta R (\cos \theta +i \sin \theta )(\log (\left| z\right| )+i \arg (z))\Big]+O(1)\,,
\end{equation}
using the above formulas we can estimate 
\begin{equation}
    \log|g(s)z^{-s}|=\mathfrak{Re}[\log g(s)z^{-s}]
\end{equation}
by combining the behaviour of each Gamma function in the integrand \eqref{genmellinbarnsint}. 
 
For the parametrisation \eqref{param2}, or equivalently \eqref{param1} for $\theta=\pm \frac{\pi i}{2}$, exponential decay of the integrand ensures that it converges. Unless stated explicitly otherwise, this is the case for all Mellin-Barnes integrals evaluated in the main text.

When the integrand decays exponentially for the parametrisation \eqref{param1}, the integration contour can be closed and Cauchy's residue theorem can then be applied to evaluate the integral. For all Mellin-Barnes integrals considered in this work, it is possible to close the Mellin integration contour.

\section{Various integrals}\label{App:Various Integrals}

In this appendix we give further details on some of the more involved integrals appearing in the main text.

\subsection{Fourier transform of spinning two-point conformal structures}\label{Appendix: Fourier Transform0}

Here we detail how to derive the expression in \eqref{GenD_2pt} for the Fourier transform of the spinning two-point conformal structure \eqref{spin2ptpos}. This entails evaluating the integral
\begin{equation}
    \mathcal{I}(\,\vec{k}_1,\vec{k}_2;\vec{\xi}_1,\vec{\xi}_2\,)=\int d^d\vec{x}_1d^d\vec{x}_2\,\frac{e^{-i\vec{k}_1\cdot \vec{x}_1-i\vec{k}_2\cdot \vec{x}_2}}{(\vec{x}_{12}^2)^\Delta}\left(\vec{\xi}_1\cdot\vec{\xi}_2+\frac{2\vec{\xi}_1\cdot \vec{x}_{12}\,\vec{\xi}_2\cdot \vec{x}_{12}}{\vec{x}_{12}^2}\right)^\ell\,,
\end{equation}
where $\vec{y}_{12}=\vec{y}_1-\vec{y}_2$. Translation invariance allows to perform one of the two integrals for free, replacing it with a momentum-conserving delta-function:
\begin{subequations}\label{ftspinning2pt}
\begin{align}
    \mathcal{I}(\vec{k}_1,\vec{k}_2;\vec{\xi}_1,\vec{\xi}_2)&=(2\pi)^d\delta^{(d)}(\vec{k}_1+\vec{k}_2)\, \mathcal{I}^\prime(\vec{k}_1;\vec{\xi}_1,\vec{\xi}_2)\,,\\
    \mathcal{I}^\prime(\vec{k};\vec{\xi}_1,\vec{\xi}_2)&=\int d^d\vec{x}_{12}\, \frac{e^{-i\vec{k}\cdot \vec{x}_{12}}}{(\vec{x}_{12}^2)^\Delta}\left(\vec{\xi}_1\cdot\vec{\xi}_2+\frac{2\vec{\xi}_1\cdot \vec{x}_{12}\,\vec{\xi}_2\cdot \vec{x}_{12}}{\vec{x}_{12}^2}\right)^\ell\,.
\end{align}
\end{subequations}
Using that
\begin{equation}
    (2\vec{\xi}_1\cdot\vec{x}\,\vec{\xi}_2\cdot\vec{x}\,)^n e^{-i\vec{k}\cdot \vec{x}} = (-2\vec{\xi}_1\cdot\vec{\pl}_k\,\vec{\xi}_2\cdot\vec{\pl}_k)^n e^{-i\vec{k}\cdot \vec{x}},
\end{equation}
this can be evaluated by expressing it in terms of the basic integral
\begin{equation}
    \int d^d \vec{x}\,\frac{e^{-i \vec{k}\cdot \vec{x}}}{(\vec{x}^2)^{\Delta+n}}=\pi^{d/2}\left(\frac{k^2}4\right)^{\Delta+n-\frac{d}{2}}\frac{\Gamma\left(\frac{d}{2}-\Delta-n\right)}{\Gamma(\Delta+n)}.
\end{equation}
To wit,
\begin{align}
  \hspace*{-0.25cm}  \mathcal{I}^\prime(\vec{k};\vec{\xi}_1,\vec{\xi}_2)&=\sum_{n=0}^\ell\binom{\ell}{n}(\vec{\xi}_1\cdot\vec{\xi}_2)^{\ell-n}(-2\vec{\xi}_1\cdot\vec{\pl}_k\,\vec{\xi}_2\cdot\vec{\pl}_k)^n\,\int d^d \vec{x}\,\frac{e^{-i \vec{k}\cdot \vec{x}}}{(\vec{x}^2)^{\Delta+n}}\\ \nonumber
  \hspace*{-0.5cm}   &=\pi ^{\frac{d}{2}}\sum_{n=0}^\ell  \binom{\ell}{n} \frac{(\Delta -1)_n\Gamma \left(\frac{d}{2}+\ell-n-\Delta \right)}{\Gamma (\ell+\Delta )}\left(\frac{k}{2}\right)^{2 \Delta -d}(\vec{\xi}_1\cdot\vec{\xi}_2)^n\,\left(-\frac{2 \vec{\xi}_1\cdot \vec{k} \vec{\xi}_2\cdot\vec{k}}{k^2}\right)^{\ell-n}\,.
\end{align}
Carrying out the sum over $n$ gives the Gauss Hypergeometric function in \eqref{GenD_2pt}.

\subsection{Fourier transform of three-point conformal structures}\label{Appendix: Fourier Transform}

Here we give further details on the Fourier transform of three-point conformal structures, expressing the result as a Mellin-Barnes integral. Conformal structures in Fourier space have been widely studied in the literature, see e.g. \cite{Polyakov:1974gs,Maldacena:2011nz,Mata:2012bx,Coriano:2013jba,Bzowski:2013sza,Arkani-Hamed:2015bza,Bzowski:2015pba,Bzowski:2015yxv,Coriano:2018bbe,Bzowski:2018fql,Coriano:2018bsy,Isono:2018rrb,Gillioz:2018mto,Arkani-Hamed:2018kmz,Isono:2019ihz,Maglio:2019grh}. In the following we will emphasise some simple properties of the Mellin-Barnes representation, which appear to have been little explored (to the best of our knowledge). Basic integral one considers is of the form:
\begin{align}\label{basicFT}
    {\mathcal{I}}_{\alpha_1,\alpha_2,\alpha_3}(\vec{k}_1,\vec{k}_2,\vec{k}_3)=\int d^dx_1d^dx_2d^dx_3\frac{e^{-i\vec{k}_1\cdot \vec{x}_1-i\vec{k}_2\cdot \vec{x}_2-i\vec{k}_3\cdot \vec{x}_3}}{(x_{12}^2)^{\alpha_3}(x_{23}^2)^{\alpha_1}(x_{31}^2)^{\alpha_2}}\,,
\end{align}
which for
\begin{equation}\label{scalarcorparaalph}
    \alpha_1=\tfrac{\Delta_2+\Delta_3-\Delta_1}2, \qquad \alpha_2=\tfrac{\Delta_1+\Delta_3-\Delta_2}2, \qquad \alpha_3=\tfrac{\Delta_1+\Delta_2-\Delta_3}2,
\end{equation}
is precisely the Fourier transform of the three-point conformal structure \eqref{scwd3ptb} for scalar fields. Spinning three-point conformal structures such as \eqref{00lposdb} can be decomposed in terms of a finite sum of the integrals of the type \eqref{basicFT}, as we shall see below.

Any one of the three integrals in \eqref{basicFT} can be replaced by a momentum conserving delta function. For instance, introducing $\vec{v}_1=\vec{x}_{13}$ and $\vec{v}_2=\vec{x}_{23}$ where $\vec{x}_{12}=\vec{v}_{12}$, we have\footnote{With this change of variables the delta function arises from
\begin{equation}
    \left(2\pi\right)^d \delta^{(d)}\left(\vec{k}_1+\vec{k}_2+\vec{k}_3\right)=\int d^dx_3\,e^{-i\vec{x}_3 \cdot \left(\vec{k}_1+\vec{k}_2+\vec{k}_3\right)}.
\end{equation}

} 
\begin{subequations}
\begin{align}
{\mathcal{I}}_{\alpha_1,\alpha_2,\alpha_3}(\vec{k}_1,\vec{k}_2,\vec{k}_3)&=(2\pi)^d\delta^{(d)}\left(\vec{k}_1+\vec{k}_2+\vec{k}_3\right) {\mathcal{I}}_{\alpha_1,\alpha_2,\alpha_3}^\prime(\vec{k}_1,\vec{k}_2,\vec{k}_3),\\ 
    {\mathcal{I}}^\prime_{\alpha_1,\alpha_2,\alpha_3}(\vec{k}_1,\vec{k}_2,\vec{k}_3)&=\int d^dq_1d^dq_2\,\frac{e^{-i\vec{k}_1\cdot \vec{v}_1-i\vec{k}_2\cdot \vec{v}_2}}{\left(v^2_1\right)^{\alpha_2}\left(v^2_2\right)^{\alpha_1}\left(v^2_{12}\right)^{\alpha_3}}
\end{align}
\end{subequations}
The remaining integrals can be reduced to Gaussian integrals by employing Schwinger parameterisation. In particular,
\begin{align}\nonumber
    {\mathcal{I}}^\prime_{\alpha_1,\alpha_2,\alpha_3}(\vec{k}_1,\vec{k}_2,\vec{k}_3)&=\frac{1}{\Gamma\left(\alpha_1\right)\Gamma\left(\alpha_2\right)\Gamma\left(\alpha_3\right)}\int^\infty_0 \frac{dt_1dt_2dt_3}{t_1t_2t_3}t^{\alpha_1}_1t^{\alpha_2}_2t^{\alpha_3}_3\\& \hspace*{3.5cm} \times \int d^dv_1d^dv_2 e^{-t_2v_1^2-t_1v_2^2-t_3v_{12}^2-i\vec{k}_1\cdot \vec{v}_1-i\vec{k}_2\cdot \vec{v}_2}\,,
\end{align}
which upon evaluating the Gaussian integrals becomes
\begin{multline}
    {\mathcal{I}}^\prime_{\alpha_1,\alpha_2,\alpha_3}(\vec{k}_1,\vec{k}_2,\vec{k}_3)=\frac{\pi^d}{\Gamma\left(\alpha_1\right)\Gamma\left(\alpha_2\right)\Gamma\left(\alpha_3\right)}\int_0^\infty \frac{dt_1dt_2dt_3}{t_1t_2t_3}t^{\alpha_1}_1t^{\alpha_2}_2t^{\alpha_3}_3\, T^{d/2-\alpha_1-\alpha_2-\alpha_3}\\
    \times e^{-\tfrac{t_1k_1^2+t_2k_2^2+t_3k_3^2}{4}}
\end{multline}
where $T=t_1t_2+t_2t_3+t_3t_1$. To obtain the Mellin-Barnes representation of the integral \eqref{basicFT} we use the following representation of the exponential function
\begin{equation}
    e^{-x}=\int_{-i\infty}^{+i\infty}\frac{ds}{2\pi i}\,\Gamma(s)\,x^{-s}\,,
\end{equation}
which gives
\begin{multline}
   {\mathcal{I}}^\prime_{\alpha_1,\alpha_2,\alpha_3}(\vec{k}_1,\vec{k}_2,\vec{k}_3)=\frac{\pi^d}{\Gamma\left(\alpha_1\right)\Gamma\left(\alpha_2\right)\Gamma\left(\alpha_3\right)}\int_0^\infty \frac{dt_1dt_2dt_3}{t_1t_2t_3}t^{\alpha_1}_1t^{\alpha_2}_2t^{\alpha_3}_3\, \\
   \times \int \left[ds\right]_3 \Gamma\left(s_1\right)\Gamma\left(s_2\right)\Gamma\left(s_3\right)\left(\frac{t_1k^2_1}{4}\right)^{-s_1}\left(\frac{t_2k^2_2}{4}\right)^{-s_2}\left(\frac{t_3k^2_3}{4}\right)^{-s_3}\\
   \times \frac{1}{\Gamma\left(\alpha_1+\alpha_2+\alpha_3-\tfrac{d}{2}\right)} \int^\infty_0\frac{d\lambda}{\lambda}\lambda^{\alpha_1+\alpha_2+\alpha_3-\tfrac{d}{2}}e^{-\lambda T},
\end{multline}
where we also used Schwinger parameterisation to exponentiate the dependence on $T$. The integral in the $t_i$ can be performed after re-scaling $t_i \to t_i/\sqrt{\lambda}$ and making the change of variables
\begin{equation}
    t_1=\sqrt{\frac{m_2m_3}{m_1}}, \qquad t_2=\sqrt{\frac{m_1m_3}{m_2}}, \qquad t_3=\sqrt{\frac{m_1m_2}{m_3}},
\end{equation}
where the resulting integrals in $m_i$ are of the form:
\begin{equation}
    \Gamma\left(z\right)=\int^\infty_0 \frac{dm_i}{m_i}\,m_i^z e^{-m_i}.
\end{equation}
The resulting integral in $\lambda$ is 
\begin{equation}
    \int^\infty_0\frac{d\lambda}{\lambda}\, \lambda^{\frac{\alpha_1+\alpha_2+\alpha_3+s_1+s_2+s_3-d}{2}} 
    = 2\pi i\, \delta\left(\frac{\alpha_1+\alpha_2+\alpha_3+s_1+s_2+s_3-d}{2}\,\right)
\end{equation}
which is the origin of the Dirac delta function in \eqref{mb3ptwitten}. This gives the following Mellin-Barnes representation for the integral \eqref{basicFT}:
\begin{multline}\label{Corr000}
    {\mathcal{I}}^\prime_{\alpha_1,\alpha_2,\alpha_3}(\vec{k}_1,\vec{k}_2,\vec{k}_3)=\frac{\pi^d}{\Gamma(\alpha_1)\Gamma(\alpha_2)\Gamma(\alpha_3)\Gamma\left(\alpha_1+\alpha_2+\alpha_3-\tfrac{d}{2}\right)}\int \left[ds\right]_3\,i\pi \delta\left(\tfrac{d}{4}-s_1-s_2-s_3\right)\\\times
    \,\prod_{\pm}\Gamma \left(s_1\pm\tfrac{d-2\alpha_2-2\alpha_3}{4}\right) \Gamma \left(s_2\pm\tfrac{d-2\alpha_3-2\alpha_1}{4}\right) \Gamma \left(s_3\pm\tfrac{d-2\alpha_1-2\alpha_2}{4}\right)\\ \times \left(\frac{k_1}{2}\right)^{-2s_1+\alpha_2+\alpha_3-\frac{d}{2}}\,\left(\frac{k_2}{2}\right)^{-2s_2+\alpha_1+\alpha_3-\frac{d}{2}}\left(\frac{k_3}{2}\right)^{-2s_3+\alpha_1+\alpha_2-\frac{d}{2}}\,.
\end{multline}
For the parameters \eqref{scalarcorparaalph} this gives the Mellin-Barnes representation \eqref{mb3ptwitten} for scalar three-point conformal correlators in momentum space. The expression \eqref{Corr000} can also be used to derive the Mellin-Barnes representation \eqref{00lmbfseads} for the Fourier transform of the spinning conformal structure \eqref{00lposdb}. To this end we find it useful to expand the tensor structure as follows
\begin{align}
   & \langle  \langle \mathcal{O}_{\Delta_1}(\vec{x}_1)\mathcal{O}_{\Delta_2}(\vec{x}_2)\mathcal{O}_{\Delta_3}(\vec{x}_3;\vec{\xi}\,)\rangle\rangle\\&\hspace*{0.5cm}=I_{\Delta_1,\Delta_2,\Delta_3-\ell}\left(\vec{x}_1,\vec{x}_2,\vec{x}_3\right)\left[\vec{\xi}\cdot \vec{x}_3\,\left(\frac{1}{x_{31}^2}-\frac1{x_{23}^2}\right)+\left(\frac{\vec{\xi}\cdot \vec{x}_2}{x_{23}^2}-\frac{\vec{\xi}\cdot \vec{x}_1}{x_{31}^2}\right)\right]^\ell \nonumber \\
   &\hspace*{0.5cm}=\sum_{\alpha=0}^\ell\sum_{\beta=0}^\alpha(-1)^\beta\,\binom{\ell}{\alpha}\binom{\alpha}{\beta}(\vec{\xi} \cdot \vec{x}_3)^\alpha I_{\Delta_1+\alpha-\beta,\Delta_2+\beta,\Delta_3-\ell+\alpha}\left(\vec{x}_1,\vec{x}_2,\vec{x}_3\right)\left(\frac{\vec{\xi} \cdot \vec{x}_2}{x_{23}^2}-\frac{\vec{\xi} \cdot \vec{x}_1}{x_{31}^2}\right)^{\ell-\alpha}, \nonumber
\end{align}
We can then Fourier transform of each term in the sum, where the tensor structure translates into a differential operator via
\begin{align}
    \vec{\xi} \cdot x_j\to i \vec{\xi} \cdot \pl_{k_j}\,,
\end{align}
which acts on the result \eqref{Corr000} for the Fourier transform of the three-point structures \eqref{basicFT}. The above expansion and the Mellin-Barnes representation simplify the action of the differential operator. Avoiding trivial details one finds (where the symbol ${\sf F}$ denotes the Fourier transform):
\begin{multline}
    (-1)^\beta{\cal B}\left(0,0,\ell;{\bf 0};\Delta_1,\Delta_2,\Delta_3-\ell\right)\\\times\mathsf{F}\left[\left(\vec{\xi}\cdot \vec{x}_3\right)^\alpha I_{\Delta_1+\alpha-\beta,\Delta_2+\beta,\Delta_3-\ell+\alpha}\left(\vec{x}_1,\vec{x}_2,\vec{x}_3\right)\left(\frac{\vec{\xi}\cdot \vec{x}_2}{\vec{x}_{23}^2}-\frac{\vec{\xi}\cdot \vec{x}_1}{\vec{x}_{31}^2}\right)^{\ell-\alpha}\right]\\=
    \int \left[ds\right]_3\, i\pi \delta\left(\tfrac{d+2\ell}{4}-s_1-s_2-s_3\right)\,(-\vec{\xi}\cdot k_3)^\alpha\,H_{\alpha,\beta}(s_1,s_2,s_3)\\\times\, \mathcal{Y}^{(\ell)}_{\alpha,\beta}(\vec{\xi}\cdot \vec{k}_1,\vec{\xi} \cdot \vec{k}_2)\,\rho_{\nu_1,\nu_2,\nu_3}(s_1,s_2,s_3)\prod^3_{j=1}\left(\frac{k_j}{2}\right)^{-2s_j+i\nu_j}\,,
\end{multline}
which yields the Mellin-Barnes representation \eqref{00lmbfseads} for the Fourier transform of the spinning conformal structure \eqref{00leadswitten}. 

\subsection{The $u$-integral}
\label{app:u-integral}

To obtain the expression \eqref{exchuint} for the exchange diagram we evaluated $u$-integrals appearing in the contributions \eqref{I>} and \eqref{I<}. We give the details on how to do this in the following, encompassing also the contributions \eqref{Ao<>spinl} to the spin-$\ell$ exchange four-point functions.

It is useful to combine the contributions of the type \eqref{I>} and \eqref{I<} as it simplifies the evaluation of the $u$-integral, as we shall see below. One way to do this is to re-define $u \to -u$ in either contribution, so that they share the same $\rho$ factor. This manipulation changes the $\epsilon$-prescription of the contribution concerned. Maintaining the same $\epsilon$-prescription for both contributions requires to subtract the residue of the pole at $u=-\epsilon$ in the contribution concerned, which generates a factorised contribution which can then be combined with \eqref{Aospinl}. There are two ways to carry out this manipulation, denoted by $\sharp$ and $\flat$, which correspond to the choice of re-defining $u \to -u$ in \eqref{I>} and \eqref{I<} respectively. We have\footnote{For the scalar exchange four-point function, $\alpha_i=\beta_i=0$.}
{\allowdisplaybreaks\begin{subequations}\label{Isharp}
\begin{align} \label{odotsharp}
  & \lim_{\eta_0\to0} {}^{\sharp}\mathcal{A}^{\left(x,{\bar x}\right)}_{\odot|\nu_1,\nu_2,\nu_3,\nu_4|\alpha_1, \beta_1;\alpha_2,\beta_2}\left(k_i;k\right)\\ \nonumber
  & \hspace*{2cm}=\int[ds]_4\,\sin \left(\pi\left(s_3+s_4+\tfrac{i(\nu_1+\nu_2)}2\right)\right)\sin \left(\pi\left(s_3+s_4+\tfrac{i(\nu_3+\nu_4)}2\right)\right)\\ \nonumber
   & \hspace*{1cm}\times H_{\nu_1,\nu_2,\nu|\alpha_1,\beta_1}(s_1,s_2,u)H_{\nu_3,\nu_4,-\nu|\alpha_2,\beta_2}(s_3,s_4,\bar{u})\rho_{\nu_1,\nu_2,\nu}(s_1,s_2,u)\rho_{\nu_3,\nu_4,-\nu}(s_3,s_4,\bar{u})
    \\& \hspace*{7cm} \times
   \left(\frac{k}{2}\right)^{-2\left(u+{\bar u}\right)}\prod^4_{j=1}\left(\frac{k_j}{2}\right)^{-2s_j+i\nu_j}\Big|_{{}^{u=\frac{x}4-s_1-s_2}_{{\bar u}=\frac{{\bar x}}4-s_3-s_4}} \nonumber
   \\ \label{sharpremin}
   & \lim_{\eta_0\to0} {}^{\sharp}\mathcal{A}^{\left(x,{\bar x}\right)}_{\gtrless|\nu_1,\nu_2,\nu_3,\nu_4|\alpha_1, \beta_1;\alpha_2,\beta_2}\left(k_i;k\right)\\ \nonumber
   & \hspace*{2cm}=\int_{-i\infty}^{+i\infty} \frac{du}{2\pi i}\frac1{u+\epsilon}\int[ds]_4\,\sin (\pi  (s_1+s_2-s_3-s_4+2 u))\\
   & \hspace*{7cm} \times \sin \left(\pi  \left(s_1+s_2+ s_3+s_4+\tfrac{i(\nu_1+\nu_2+\nu_3+\nu_4)}{2}\right)\right) \nonumber \\
    &  \hspace*{2cm} \times\,H_{\nu_1,\nu_2,\nu|\alpha_1,\beta_1}(s_1,s_2,w-u)H_{\nu_3,\nu_4,-\nu|\alpha_2,\beta_2}(s_3,s_4,{\bar w}+u)\nonumber \\
    & \times
  \rho_{\nu_1,\nu_2,\nu}(s_1,s_2,w-u)\rho_{\nu_3,\nu_4,-\nu}(s_3,s_4,{\bar w}+u)  \left(\frac{k}{2}\right)^{-2\left(w+{\bar w}\right)}\prod^4_{j=1}\left(\frac{k_j}{2}\right)^{-2s_j+i\nu_j}\Big|_{{}^{w=\frac{x}4-s_1-s_2}_{{\bar w}=\frac{{\bar x}}4-s_3-s_4}}, \nonumber
   \end{align}
\end{subequations}}
where in \eqref{odotsharp} we combined the residue of the pole at $u=-\epsilon$ in \eqref{I>} with the contribution \eqref{Io}, with the remaining contribution to the exchange given by \eqref{sharpremin}. If we instead send $u \to -u$ in \eqref{I<} we have
\begin{subequations}\label{Iflat}
\begin{align}\label{appIflatodot}
   & \lim_{\eta_0\to0} {}^{\flat}\mathcal{A}^{\left(x,{\bar x}\right)}_{\odot|\nu_1,\nu_2,\nu_3,\nu_4|\alpha_1, \beta_1;\alpha_2,\beta_2}\left(k_i;k\right)\\ \nonumber
  & \hspace*{2cm}=\int[ds]_4\,\sin \left(\pi\left(s_1+s_2+\tfrac{i(\nu_1+\nu_2)}2\right)\right)\sin \left(\pi\left(s_1+s_2+\tfrac{i(\nu_3+\nu_4)}2\right)\right)\\ \nonumber
   & \hspace*{1cm}\times H_{\nu_1,\nu_2,\nu|\alpha_1,\beta_1}(s_1,s_2,u)H_{\nu_3,\nu_4,-\nu|\alpha_2,\beta_2}(s_3,s_4,\bar{u})\rho_{\nu_1,\nu_2,\nu}(s_1,s_2,u)\rho_{\nu_3,\nu_4,-\nu}(s_3,s_4,{\bar u})
    \\& \hspace*{7cm} \times
   \left(\frac{k}{2}\right)^{-2\left(u+{\bar u}\right)}\prod^4_{j=1}\left(\frac{k_j}{2}\right)^{-2s_j+i\nu_j}\Big|_{{}^{u=\frac{x}4-s_1-s_2}_{{\bar u}=\frac{{\bar x}}4-s_3-s_4}} \nonumber,\\ \label{Iflat<>}
  & \lim_{\eta_0\to0} {}^{\flat}\mathcal{A}^{\left(x,{\bar x}\right)}_{\gtrless|\nu_1,\nu_2,\nu_3,\nu_4|\alpha_1, \beta_1;\alpha_2,\beta_2}\left(k_i;k\right)\\ \nonumber
   & \hspace*{2cm}=\int_{-i\infty}^{+i\infty} \frac{du}{2\pi i}\frac1{u+\epsilon}\int[ds]_4\,\sin (\pi  (s_3+s_4-s_1-s_2+2 u))\\
   & \hspace*{7cm} \times \sin \left(\pi  \left(s_1+s_2+ s_3+s_4+\tfrac{i(\nu_1+\nu_2+\nu_3+\nu_4)}{2}\right)\right) \nonumber\\
    &  \hspace*{2cm} \times\,H_{\nu_1,\nu_2,\nu|\alpha_1,\beta_1}(s_1,s_2,w+u)H_{\nu_3,\nu_4,-\nu|\alpha_2,\beta_2}(s_3,s_4,{\bar w}-u)\nonumber \\
    & \times
  \rho_{\nu_1,\nu_2,\nu}(s_1,s_2,w+u)\rho_{\nu_3,\nu_4,-\nu}(s_3,s_4,{\bar w}-u) \left(\frac{k}{2}\right)^{-2\left(w+{\bar w}\right)}\prod^4_{j=1}\left(\frac{k_j}{2}\right)^{-2s_j+i\nu_j}\Big|_{{}^{w=\frac{x}4-s_1-s_2}_{{\bar w}=\frac{{\bar x}}4-s_3-s_4}}. \nonumber
   \end{align}
\end{subequations}
Both \eqref{Isharp} and \eqref{Iflat} give equivalent expressions for the exchange. The $\epsilon$-prescription is such that closing the $u$-contour on the positive real axis is more natural -- as it avoids the residue at $u \sim 0$. 

The $u$-integrals appearing in \eqref{sharpremin} and \eqref{Iflat<>} are of examples of the general type
\begin{multline}\label{genintu}
    \mathfrak{R}^{(\alpha,\beta)}_{A,B;t}=\int^{i\infty}_{-i\infty} \frac{du}{2\pi i}\,\frac{\sin(\pi(t+2u))}{u+\epsilon}\,\Gamma \left(A-u-\tfrac{i \nu }{2}\right)\Gamma \left(B+u+\tfrac{i \nu }{2}\right) \\\times \Gamma \left(A-\alpha-u+\tfrac{i \nu }{2}\right)\Gamma \left(B-\beta+u-\tfrac{i \nu }{2}\right) \,,
\end{multline}
where $\alpha=\alpha_1$ or $\alpha_2$ and $\beta=\alpha_2$ or $\alpha_1$. To evaluate this integral we close the contour on the positive real axis, which encloses the following two series of poles
\begin{subequations}\label{uintpoles}
\begin{align}
    u&=A-\tfrac{i\nu}{2}+p, \\
    u&=A+\tfrac{i\nu}{2}-\alpha+p, 
\end{align}
\end{subequations}
where $p\in \mathbb{Z}_{\geq 0}$. If we instead closed on the negative real axis we would in addition capture the pole at $u=-\epsilon$. Each series in \eqref{uintpoles} contributes a ${}_3F_2$ Hypergeometric function,\footnote{I.e. the re-summation of the residues from each series is a ${}_3F_2$ of argument $z=1$.} giving:
\begin{align}\nonumber
    \mathfrak{R}^{(\alpha,\beta)}_{A,B;t}&=\frac{\Gamma (\alpha-i \nu ) \Gamma (A+B-\beta-\alpha) \Gamma (A+B-\alpha+i \nu )}{A+\tfrac{i \nu}2 - \alpha}\,\sin (2 \pi  A+i \pi  \nu -2 \pi  \alpha+\pi  t)\\\label{Rres1}&\hspace{3cm}\times\,_3F_2\left(\begin{matrix}A+B-\beta-\alpha,A-\alpha+\frac{i \nu }{2},A+B-\alpha+i \nu \\A-\alpha+\frac{i \nu }{2}+1,-\alpha+i \nu +1\end{matrix};1\right)\\\nonumber&+\frac{\Gamma (A+B) \Gamma (i \nu -\alpha)  \Gamma (A+B-\beta-i \nu )}{A-\tfrac{i \nu}2 }\,\sin (2 \pi  A-i \pi  \nu +\pi  t)\\\nonumber&\hspace{3cm}\times\, _3F_2\left(\begin{matrix}A+B,A-\frac{i \nu }{2},A+B-\beta-i \nu \\A-\frac{i \nu }{2}+1,\alpha-i \nu +1\end{matrix};1\right)\,.
\end{align}
A priori with this expression we do not get much further than the original Mellin-Barnes integral in the variable $u$, since Hypergeometric functions themselves are generically defined by a Mellin-Barnes integral. Fortunately, there exist three-term relations for the function ${}_3F_2$ (see e.g. \S 3.5-\S 3.8 of \cite{Baileybook}) which in particular allow us to relate the above combination of two ${}_3F_2$\,s to a single ${}_3F_2$ which can be re-summed as a simple ratio of Gamma functions. The relevant identity is:
\begin{multline}
    \frac{\sin(\pi\beta_{45})}{\Gamma(\alpha_{012})\Gamma(\alpha_{013})\Gamma(\alpha_{023})}\,F_p(0;4,5)+\frac{\sin(\pi\beta_{50})}{\Gamma(\alpha_{124})\Gamma(\alpha_{134})\Gamma(\alpha_{234})}\,F_p(4;0,5)\\+\frac{\sin(\pi\beta_{04})}{\Gamma(\alpha_{125})\Gamma(\alpha_{135})\Gamma(\alpha_{235})}\,F_p(5)=0,
\end{multline}
where, adopting the notation of \S 3.5 in \cite{Baileybook},
\begin{align}\label{Id3F2}
    F_p(u;v,w)&=\frac{1}{\Gamma(\alpha_{xyz})\Gamma(\beta_{vu})\Gamma(\beta_{wu})}\,_3F_2\left(\begin{matrix}\alpha_{vwx},\alpha_{vwy},\alpha_{vwz}\\\beta_{vu},\beta_{wu}\end{matrix};1\right)\,,
\end{align}
and $u$, $v$, $w$, $x$, $y$ and $z$ are unequal integers in the range $0,\ldots, 5$ and with
\begin{align}
    \alpha_{lmn}&=\frac12+r_l+r_m+r_n\,,& \beta_{mn}&=1+r_m-r_n\,,& \sum_{i=0}^5r_i=0\,,
\end{align}
and the $r_i$ are six free parameters. Each of these identities is a linear combination of the fundamental three term relation which can be obtained from the following Mellin-Barnes integral:
\begin{align}
    \int_{-i\infty}^{+i\infty}\frac{ds}{2\pi i}\,\sin(\pi(s-a_1))\,\frac{\Gamma(s)\Gamma(a_1-s)\Gamma(a_2-s)\Gamma(a_3-s)}{\Gamma(b_1-s)\Gamma(b_2-s)}\,,
\end{align}
where closing the contour on the left gives a single $_3F_2$ while closing on the right gives a sum of two $_3F_2$. When the parameters $r_i$ are tuned so that:
\begin{equation}
    F_p(0;4,5)=\, _3F_2\left(\begin{matrix}A+B,A-\frac{i \nu }{2},A+B-\beta-i \nu \\A-\frac{i \nu }{2}+1,1+\alpha-i \nu\end{matrix};1\right)\,,
\end{equation}
we have
\begin{equation}
    F_p(4;0,5)=\,_3F_2\left(\begin{matrix}A+B,A-\alpha+\frac{i \nu }{2},A+B-\alpha+i \nu \\A-\alpha+\frac{i \nu }{2}+1,1-\alpha+i \nu\end{matrix};1\right)\,,
\end{equation}
and
\begin{equation}
    F_p(5;0,4)=\frac{1}{\Gamma \left(1-A+\frac{i \nu }{2}\right) \Gamma \left(1-A+\alpha-\frac{i \nu }{2}\right) \Gamma (2-2 A-2 B+\beta+\alpha)}\,.
\end{equation}
The above three term relation becomes useful to evaluate \eqref{Rres1} when $t=B-A$. In this case the three term relation allows to sum the two Hypergeometric functions in \eqref{Rres1} into a simple ratio of Gamma-functions obtaining:
\begin{shaded}
\begin{align}\label{Rfinal}
   \mathfrak{R}^{(\alpha,\beta)}_{A,B}\equiv \mathfrak{R}^{(\alpha,\beta)}_{A,B;B-A}=(-1)^\alpha\,\frac{\pi ^2 \csc (\pi  (A+B))\Gamma \left(A-\frac{i \nu }{2}\right) \Gamma \left(A-\alpha+\frac{i \nu }{2}\right)}{\Gamma \left(1-B-\frac{i \nu }{2}\right) \Gamma \left(1-B+\beta+\frac{i \nu }{2}\right)}\,.
\end{align}
\end{shaded}
\noindent This result allows us to lift the $u$-integral in any exchange diagram appearing in this work, where
\begin{align}
    A&=\frac{d+2\ell}{4}-s_1-s_2\,,& B&=\frac{d+2\ell}{4}-s_3-s_4\,,
\end{align}
with $t=B-A=s_3+s_4-s_1-s_2$.

Since the expression for the exchange diagram is still given as a Mellin-Barnes integral upon lifting the $u$-integral, extra care has to be taken with the contour prescription for the poles encoded in the cosecant function. The standard prescription for the integration contour of a Mellin-Barnes integral is that it should separate the Gamma function poles which extend along the positive real axis from those which extend along the negative real axis (see e.g. \cite{MellinBook}). Naively, there thus appears to be an ambiguity in the contour prescription when there is a cosecant function in the Mellin-integrand, since there are infinitely many ways one can split it into a product of Gamma-functions:
\begin{subequations}\label{cosecdiffsplit}
\begin{align}
    \pi\csc \left(\pi (A+B)\right)&=\Gamma (1-A-B) \Gamma (A+B)\\&=(-1)^q\Gamma (1-A-B-q) \Gamma (A+B+q)\,,\quad \forall q\in \mathbb{N}\,,
\end{align}
\end{subequations}
where each possible splitting (labelled by $q$) requires a different contour. This freedom has an interesting interpretation within exchange diagrams, where it corresponds to the possibility of adding contact interactions/improvement contributions.\footnote{I.e. interactions which are trivial on-shell.} 

For scalar exchange diagrams, for which $\alpha=0$ and $\beta=0$, the correct splitting is 
\begin{align}\label{prescr1}
    \pi\csc \left(\pi (A+B)\right)=\Gamma (1-A-B) \Gamma (A+B)\,,
\end{align}
 which can be understood from the expression \eqref{Rres1}, where the first $_3F_2$ is proportional to $\Gamma(A+B-\beta-\alpha)$ and the second $_3F_2$ to $\Gamma(A+B)$, so to have a consistent contour prescription for both terms we require the splitting \eqref{prescr1}.

For the exchange of spinning particles 
we need to consider $\alpha,\beta\neq0$ and in this case the contour prescription for the two terms in \eqref{Rres1} has to be chosen compatibly with both terms in the sum \eqref{Rres1}. With this proviso, and up to simple contact terms, we can still use \eqref{Rfinal} with the minimal prescription \eqref{prescr1} for the integration contour regardless of $\alpha$ and $\beta$, fixing a choice of contact terms/improvements in the exchange amplitude. In this work we use the different prescriptions for the splitting of the cosecant function as an organising principle for the improvement terms that can be included in an exchange amplitude.

Before concluding this Appendix we take the opportunity to give two equivalent representations for the interference factor \eqref{symmmellinzeros}. The following representations arise directly from the $u$ integrals \eqref{sharpremin} and \eqref{Iflat<>} for ${}^\sharp\mathcal{A}$ and ${}^\flat\mathcal{A}$ respectively:
\begin{subequations}\label{PdS}
\begin{align}\label{PdS1}
    \delta^{\left(x,\bar{x}\right)}_\sharp(u,\bar{u})&=\sin (\pi  (u+\bar{u})) \sin \left(\tfrac{\pi}{4}  (x+2 i (\nu_1+\nu_2)-4 \bar{u})\right) \sin \left(\tfrac{\pi}{4}  (\bar{x}+2 i (\nu_3+\nu_4)-4 \bar{u})\right)\nonumber\\
    &\hspace{-30pt}+\sin \left(\pi  \left(\bar{u}+\tfrac{i \nu }{2}\right)\right) \sin \left(\pi  \left(\bar{u}-\tfrac{i \nu }{2}\right)\right) \sin \left(\tfrac{\pi}{4}  (x+\bar{x}+2 i \nu_1+2 i \nu_2+2 i \nu_3+2 i \nu_4-4 u-4 \bar{u})\right)\,,
\end{align}
and
\begin{align}\label{PdS2}
   \delta^{\left(x,\bar{x}\right)}_\flat(u,\bar{u})&=\sin (\pi  (u+\bar{u})) \sin \left(\tfrac{\pi}{4}  (x+2 i (\nu_1+\nu_2)-4 u)\right) \sin \left(\tfrac{\pi}{4}  (\bar{x}+2 i (\nu_3+\nu_4)-4 u)\right)\nonumber\\
    &\hspace{-30pt}+\sin \left(\pi  \left(u+\tfrac{i \nu }{2}\right)\right) \sin \left(\pi  \left(u-\tfrac{i \nu }{2}\right)\right) \sin \left(\tfrac{\pi}{4}  (x+\bar{x}+2 i \nu_1+2 i \nu_2+2 i \nu_3+2 i \nu_4-4 u-4 \bar{u})\right)\,.
\end{align}
\end{subequations}
The first line of both expressions in \eqref{PdS} comes from the factorised contributions \eqref{odotsharp} and \eqref{appIflatodot}, while the second line from the corresponding $u$-integral \eqref{Rfinal}.
The interference factor we gave in \eqref{symmmellinzeros} is then obtained by making the above manifestly symmetric under the exchange of $u$ and ${\bar u}$.

\end{appendix}

\bibliographystyle{JHEP}

\providecommand{\href}[2]{#2}\begingroup\raggedright\endgroup

\end{document}